\documentclass[iop,twocolumn]{emulateapj} 
\usepackage{color}
\usepackage{natbib}
\usepackage{subfigure}
\usepackage{verbatim}
\usepackage{graphicx}
\usepackage{bm}              
\usepackage[amssymb]{SIunits}
\usepackage{multirow}
\usepackage{tabularx}
\usepackage{booktabs}
\usepackage{float}
\usepackage[outercaption]{sidecap} 
\usepackage{threeparttable}
\usepackage[colorlinks=true, allcolors=blue]{hyperref}
\usepackage[version=4]{mhchem}
\hypersetup{linktocpage}
\newcommand{\ack}[3]{#3}
\newcommand{\code}[1]{\texttt{#1}}

\DeclareFixedFont{\ttb}{T1}{txtt}{bx}{n}{9} 
\DeclareFixedFont{\ttm}{T1}{txtt}{m}{n}{9}  
\newcommand{\map}{{\sl WMAP}}
\newcommand{\planck}{{\it Planck}}

\newcommand{\ba}{\begin{eqnarray}}
\newcommand{\ea}{\end{eqnarray}}

\newcommand  \gtsim  {\lower.5ex\hbox{$\; \buildrel > \over \sim \;$}}
\newcommand  \ltsim  {\lower.5ex\hbox{$\; \buildrel < \over \sim \;$}}

\newcommand{\LCDM}   {$\Lambda$CDM}

\newcommand{\be}{\begin{equation}}
\newcommand{\ee}{\end{equation}}

\newcommand{\neff}  {$N_{\rm eff}$}
\newcommand{\mnu} {\Sigma m_\nu}

\newcommand{\so}{SO}

\newcommand{\pb}{\textsc{Polarbear}}

\newcommand{\polarbear}{\textsc{Polarbear}}
\newcommand{\act}{ACT}

\bibpunct{(}{)}{;}{a}{}{,}

\slugcomment{}
\shorttitle{SO Science Goals}
\begin{document}

\title{The Simons Observatory: Science goals and forecasts}
\author{%
The~Simons~Observatory~Collaboration:$^{1}$
Peter~Ade$^{2}$,
James~Aguirre$^{3}$,
Zeeshan~Ahmed$^{4,5}$,
Simone~Aiola$^{6,7}$,
Aamir~Ali$^{8}$,
David~Alonso$^{2,9}$,
Marcelo~A.~Alvarez$^{8,10}$,
Kam~Arnold$^{11}$,
Peter~Ashton$^{8,12,13}$
% ...
}

\address{$^{1}$~\href{mailto:so_tac@simonsobservatory.org}{Correspondence address: so\_tac@simonsobservatory.org}}
\address{$^{2}$~School of Physics and Astronomy, Cardiff University, The Parade, Cardiff, CF24 3AA, UK}
\address{$^{3}$~Department of Physics and Astronomy, University of Pennsylvania, 209 South 33rd Street, Philadelphia, PA, USA 19104}
\address{$^{4}$~Kavli Institute for Particle Astrophysics and Cosmology, Menlo Park, CA 94025}
\address{$^{5}$~SLAC National Accelerator Laboratory, Menlo Park, CA 94025}
\address{$^{6}$~Joseph Henry Laboratories of Physics, Jadwin Hall, Princeton University, Princeton, NJ, USA 08544}
\address{$^{7}$~Center for Computational Astrophysics, Flatiron Institute, 162 5th Avenue, New York, NY 10010, USA}
\address{$^{8}$~Department of Physics, University of California, Berkeley, CA, USA 94720}
\address{$^{9}$~University of Oxford, Denys Wilkinson Building, Keble Road, Oxford OX1 3RH, UK}
\address{$^{10}$~Computational Cosmology Center, Lawrence Berkeley National Laboratory, Berkeley, CA 94720, USA}
\address{$^{11}$~Department of Physics, University of California San Diego, CA, 92093 USA}
\address{$^{12}$~Physics Division, Lawrence Berkeley National Laboratory, Berkeley, CA 94720, USA}
\address{$^{13}$~Kavli Institute for The Physics and Mathematics of The Universe (WPI), The University of Tokyo, Kashiwa, 277- 8583, Japan}
% ...

\begin{abstract}
The Simons Observatory (SO) is a new cosmic microwave background experiment being built on Cerro Toco in Chile, due to begin observations in the early 2020s. We describe the scientific goals of the experiment, motivate the design, and forecast its performance. SO will measure the temperature and polarization anisotropy of the cosmic microwave background in six frequency bands centered at: 27, 39, 93, 145, 225 and 280 GHz. The initial configuration of SO will have three small-aperture 0.5-m telescopes and one large-aperture 6-m telescope, with a total of 60,000 cryogenic bolometers. Our key science goals are to characterize the primordial perturbations, measure the number of relativistic species and the mass of neutrinos, test for deviations from a cosmological constant, improve our understanding of galaxy evolution, and constrain the duration of reionization. The small aperture telescopes will target the largest angular scales observable from Chile, mapping $\approx 10\%$ of the sky to a white noise level of 2~$\mu$K-arcmin in combined 93 and 145 GHz bands, to measure the primordial tensor-to-scalar ratio, $r$, at a target level of $\sigma(r)=0.003$. The large aperture telescope will map $\approx 40\%$ of the sky at arcminute angular resolution to an expected white noise level of 6 $\mu$K-arcmin in combined 93 and 145 GHz bands, overlapping with the majority of the Large Synoptic Survey Telescope sky region and partially with the Dark Energy Spectroscopic Instrument.  With up to an order of magnitude lower polarization noise than maps from the \planck{} satellite, the high-resolution sky maps will constrain cosmological parameters derived from the damping tail, gravitational lensing of the microwave background, the primordial bispectrum, and the thermal and kinematic Sunyaev--Zel'dovich effects, and will aid in delensing the large-angle polarization signal to measure the tensor-to-scalar ratio. The survey will also provide a legacy catalog of 16,000 galaxy clusters and more than 20,000 extragalactic sources\footnote{A supplement describing author contributions to this paper can be found at \url{https://simonsobservatory.org/publications.php}}.
\end{abstract}

\maketitle
\tableofcontents

\section{Introduction}
\label{sec:intro}
Fifteen years have passed between the first release of full-sky microwave maps from the {\it WMAP} satellite \citep{2003ApJS..148....1B}, and the final legacy maps from the {\it Planck} satellite \citep{planck2018:overview}. During that time, ground-based observations of the cosmic microwave background (CMB) anisotropies have made enormous strides. Maps from these experiments over hundreds of square degrees of the sky currently surpass all balloon and satellite experiments in sensitivity \citep{2016PhRvL.116c1302B,2017JCAP...06..031L,2018ApJ...852...97H,2014ApJ...794..171P,2017ApJ...848..121P}, and have set new standards for the mitigation of systematic errors. A rich legacy of scientific discovery has followed.

The rapid progress of ground-based measurements in the last decade has been driven by the development of arrays of superconducting transition-edge sensor (TES) bolometers coupled to multiplexed readout electronics \citep{henderson/etal:2016,posada/etal:2016,2016JLTP..184..805S,2016SPIE.9914E..0TH}. The current generation of detector arrays simultaneously measure linear polarization at multiple frequency bands in each focal-plane pixel. 
Two other innovations have also been essential: optical designs with minimal optical distortions and sizeable focal planes; and sophisticated computational techniques for extracting small sky signals in the presence of far larger atmosphere signals and noise sources. 

Ground-based experiments have targeted a range of angular scales. 
High-resolution experiments such as the Atacama Cosmology Telescope (ACT) and South Pole Telescope (SPT) have arcminute resolution, sufficient to measure not only the power spectrum of primary perturbations, but also secondary perturbations including gravitational lensing and the thermal and kinematic Sunyaev--Zel'dovich (SZ) effects. The low-resolution Background Imaging of Cosmic Extragalactic Polarization experiment and successors (BICEP, BICEP2, Keck Array) have pursued the polarization signal from primordial gravitational waves, which should be most prominent at scales of several degrees. The mid-resolution experiment Simons Array (SA), the successor to \pb{}, aims to probe both primordial gravitational waves and gravitational lensing.

Over the past decade, ground-based experimental efforts have made the first-ever detections of the power spectrum of gravitational lensing of the microwave background in both temperature ~\citep{2011PhRvL.107b1301D,2012ApJ...756..142V} and polarization~\citep{2013PhRvL.111n1301H,pbear-herschel/2013,pbear-eeeb/2013,2015ApJ...810...50S,2017PhRvD..95l3529S}, lensing by galaxy clusters~\citep{2015PhRvL.114o1302M,2015ApJ...806..247B}, the kinematic Sunyaev--Zel'dovich effect~\citep{2012PhRvL.109d1101H},  and the thermal Sunyaev--Zel'dovich effect associated with radio galaxies \citep{Gralla2014}. They have compiled catalogs of SZ-selected galaxy clusters~\citep{2015ApJS..216...27B,2018ApJS..235...20H} comparable in size to that extracted from the full-sky maps made by {\it Planck}~\citep{2016A&A...594A..27P}, including many of the most extreme-mass and highest redshift clusters known \citep[e.g.,][]{2012ApJ...748....7M}, demonstrated quasar feedback from the thermal Sunyaev--Zel'dovich effect~\citep{2016MNRAS.458.1478C}, and found numerous lensed high-redshift dusty galaxies~\citep{vieira/etal:2010}. Cosmological parameter constraints from ground-based primary temperature and polarization power spectra are close, both in central values and uncertainties, to the definitive ones produced by {\it Planck}~\citep{planck2018:parameters}. The {\it Planck} full-sky maps provide important information at the largest angular scales and in frequency bands which are inaccessible from the ground~\citep{planck2018:overview}. 

This scientific legacy is remarkable, but more is yet to come (see, e.g., \citet{CMBS42016} for a detailed overview). 
Measurements of gravitational lensing will steadily improve with increasing sensitivity of temperature and polarization maps, coupled with control over systematic effects, leading to improved large-scale structure characterization, dark matter and dark energy constraints,  and neutrino mass limits. 
The thermal and kinematic SZ effects will become more powerful probes of both structure formation and astrophysical processes in galaxies and galaxy clusters. Measurements of the polarization power spectrum to cosmic variance limits down to arcminute angular scales will provide a nearly-independent determination of cosmological parameters \citep{galli/etal/2014,calabrese/etal/2016}, and hence provide  important consistency checks \citep{2016ApJ...818..132A,2017A&A...607A..95P}. The polarization signature of a primordial gravitational wave signal still beckons. 

The Simons Observatory (SO) is a project designed to target these goals. We are a collaboration of over 200 scientists from around 40 institutions, constituted in 2016. The collaboration is building a Large Aperture Telescope (LAT) with a 6-meter primary mirror similar in size to ACT, and three 0.5-meter refracting Small Aperture Telescopes (SATs) similar in size to BICEP3. New optical designs \citep{niemack:2016,parshley/etal:2018} will provide much larger focal planes for the LAT than current experiments. We initially plan to deploy a total of 60,000 detectors, approximately evenly split between the LAT and the set of SATs. Each detector pixel will be sensitive to two orthogonal linear polarizations and two frequency bands \citep{henderson/etal:2016,posada/etal:2016}. This number of detectors represents an order of magnitude increase over the size of current microwave detector arrays, and is more total detectors than have been deployed by all previous microwave background experiments combined. 

SO will be located in the Atacama Desert at an altitude of 5,200 meters in Chile's Parque Astronomico. It will share the same site on Cerro Toco as ACT, Simons Array, and the Cosmology Large Angular Scale Surveyor (CLASS\footnote{\url{http://sites.krieger.jhu.edu/class/}}), overlooking the Atacama Large Millimeter Array (ALMA\footnote{\url{http://www.almaobservatory.org/en/home/}}) on the Chajnantor Plateau. The site is also one of those planned for the future CMB-S4 experiment \citep{CMBS42016}. Nearly two decades of observations from this site inform the project. SO will cover a sky region which overlaps many astronomical surveys at other wavelengths; particularly important will be the Large Synoptic Survey Telescope (LSST\footnote{\url{https://www.lsst.org/}}), as well as the Dark Energy Survey (DES\footnote{\url{https://www.darkenergysurvey.org/}}), the Dark Energy Spectroscopic Instrument (DESI\footnote{\url{https://www.desi.lbl.gov/}}), and the {\it Euclid} satellite\footnote{\url{http://sci.esa.int/euclid/}}. A full description of the experiment design will be presented in a companion paper. 

This paper presents a baseline model for the SO instrument performance, including detector noise, frequency bands, and angular resolution.  We also present a set of simple but realistic assumptions for atmospheric  and galactic foreground emission. We translate these assumptions into anticipated properties of temperature and polarization maps, given nominal sky survey coverage and duration of observation.  We then summarize the constraints on  various cosmological signals that can be obtained from such maps. The results of this process have served as the basis for optimizing experimental design choices, particularly aperture sizes and angular resolutions,
division of detectors between large and small aperture telescopes, the range of frequency bands, and the division of detectors between frequency bands. 

In Sec.~\ref{sec:method} we give specifications for the Simons Observatory instruments, and our baseline assumptions for the atmosphere and for foreground sources of microwave emission.  We then describe science goals, design considerations and forecasts for each major science probe: $B$-mode polarization at large angular scales (Sec.~\ref{sec:bmodes}), the damping tail  of the power spectra at small angular scales (Sec.~\ref{sec:highell}), gravitational lensing (Sec.~\ref{sec:lensing}), probes of  non-Gaussian perturbation statistics, particularly the primordial bispectrum (Sec.~\ref{sec:bispec}), the thermal and kinematic SZ effects (Sec.~\ref{sec:sz}), and extragalactic sources (Sec.~\ref{sec:source}). We conclude in Sec.~\ref{sec:summary} with a summary of the forecasts and a discussion of the practical challenges of these measurements.

All science forecasts in this paper assume the standard $\Lambda$CDM cosmological model with parameters given by the temperature best-fit \planck{} values~\citep{Planck2015params} and an optical depth to reionization of 0.06~\citep{planck_tau:2016}, as fully specified in Sec.~\ref{sec:add_data}. However, the standard cosmology is now so well constrained that our analyses are essentially independent of assumed values of the cosmological parameters if they are consistent with current data. 
 
\section{Forecasting methods}
\label{sec:methods}
\label{sec:method}
Here we summarize common assumptions for all science projections, including our instrument model (Sec. \ref{subsec:instrument_summary}), atmospheric noise model (Sec.~\ref{subsec:sensitivity}), sky coverage (Sec.~\ref{subsec:coverage}), foreground emission model (Sec.~\ref{subsec:fg_model}), foreground cleaning for the Large Aperture Telescope (Sec.~\ref{sec:FG_LAT}), choice of cosmological parameters, and assumptions about external datasets (Sec. \ref{sec:add_data}). Details specific to the analysis of individual statistics are described in Secs.~\ref{sec:bmodes}--\ref{sec:source}. In particular, the substantially different approach taken for large-scale $B$-modes is described in Sec.~\ref{sec:bmodes}.

\subsection{Instrument summary}
\label{subsec:instrument_summary}
SO will consist of one 6-m Large Aperture Telescope (LAT) and three 0.5-m Small Aperture Telescopes (SATs). Early in the SO design process we concluded that both large and small telescopes were needed, consistent with \citet{CMBS42016}, to optimally measure the CMB anisotropy from few-degree scales down to arcminute scales. The large telescope provides high angular resolution; the small telescopes have a larger field of view and are better able to control atmospheric contamination in order to measure larger angular scales.

The LAT receiver will have 30,000 TES bolometric detectors distributed among seven optics tubes that span six frequency bands from 27 to 280~GHz. Each LAT tube will contain three arrays, each on a 150~mm detector wafer, each measuring two frequency bands and in two linear polarizations. One `low-frequency' (LF) tube will make measurements in two bands centered at 27 and 39~GHz, four `mid-frequency' (MF) tubes will have bands centered at 93 and 145~GHz, and two `high-frequency' (HF) tubes will have bands at 225 and 280~GHz. These seven tubes will fill half of the LAT receiver's focal plane. The LAT will attain arcminute angular resolution, as shown in Table \ref{tab:noise}.  The field of view of each LAT optics tube will be approximately $1.3^\circ$ in diameter, and the total field of view will be approximately $7.8^\circ$ in diameter.

The three SATs, each with a single optics tube, will together also contain 30,000 detectors. The SAT optics tubes will each house seven arrays, and will each have a continuously rotating half-wave plate to modulate the large-scale atmospheric signal. Two SATs will observe at 93 and 145~GHz (MF) and one will measure at 225 and 280~GHz (HF); an additional low-frequency optics tube at 27 and 39~GHz will be deployed in one of the MF SATs for a single year of observations. The SATs will have $0.5^\circ$ angular resolution at 93~GHz. Further details can be found in the companion instrument paper \citep{so_instrument}.

\subsection{Noise model}
\label{subsec:sensitivity}
\newcommand{\ellknee}{\ell_{\rm knee}}
\newcommand{\alphaknee}{\alpha_{\rm knee}}
\newcommand{\muK}{\mu{\rm K}}

\begin{table*}[]
\centering
\caption[Simons Observatory Surveys]{Properties of the planned SO surveys$^a$. } \small
\begin{tabular}{ c | c c c | c c c}
 & & SATs ($f_{\rm sky} = 0.1$) & & & LAT ($f_{\rm sky} = 0.4$) &\\
\hline
\hline
 Freq.~[GHz] & FWHM ($^\prime$) & Noise (baseline)& Noise (goal) & FWHM ($^\prime$) & Noise (baseline)& Noise (goal)\\
 & & [$\mu$K-arcmin] & [$\mu$K-arcmin] & & [$\mu$K-arcmin] & [$\mu$K-arcmin] \\
\hline
27 & 91 & 35 & 25 & 7.4 & 71 & 52\\
39 & 63 & 21 & 17  & 5.1 & 36 & 27\\
93 & 30 & 2.6 & 1.9 & 2.2 & 8.0 & 5.8\\
145 & 17 & 3.3 & 2.1  & 1.4 & 10 & 6.3\\
225 & 11 & 6.3 & 4.2 & 1.0 & 22 & 15\\
280 & 9 & 16 & 10 & 0.9 & 54 & 37\\
\hline
\hline
\end{tabular}
\begin{tablenotes}
\item \textsuperscript{a} The detector passbands are being optimized (see \citealp{so_instrument}) and are subject to variations in fabrication. For these reasons we expect the SO band centers to differ slightly from the frequencies presented here. `Noise' columns give anticipated white noise levels for temperature, with polarization noise $\sqrt{2}$ higher as both $Q$ and $U$ Stokes parameters are measured. Noise levels are quoted as appropriate for a homogeneous hits map.
\end{tablenotes}
\label{tab:noise}
\end{table*}

\begin{table}[]
\centering
\caption[Simons Observatory Surveys]{Band-dependent parameters for the large-angular-scale noise model described in Eq.~\ref{eq:actual_Nell}.  Parameters that do not vary with frequency are in the text.
} \small
\begin{tabular}{ c | c c c | c }

 & \multicolumn{3}{c|}{SAT Polarization} & LAT Temperature\\
\hline
\hline
 Freq.~[GHz] & $\ellknee{}^{\rm a}$  & $\ellknee{}^{\rm b}$ &  $\alphaknee{}$ & $N_{\rm red} [\muK{}^2s]$\\
\hline
 27 &  30 & 15 & -2.4 &     100\\
 39 &  30 & 15 & -2.4 &      39\\
 93 &  50 & 25 & -2.5 &     230\\
145 &  50 & 25 & -3.0 &   1,500\\
225 &  70 & 35 & -3.0 &  17,000\\
280 & 100 & 40 & -3.0 &  31,000\\
\hline
\hline
\end{tabular}
\begin{tablenotes}
\item \textsuperscript{a} Pessimistic case. \textsuperscript{b} Optimistic case.
\end{tablenotes}
\label{tab:oneoverf}
\end{table}

We consider two cases for the \so{} performance: a nominal `baseline' level which requires only a modest amount of technical development over currently deployed experiments, and a more aggressive `goal' level.  We assume a 5-year survey with 20\% of the total observing time used for science analysis, consistent with the realized performance after accounting for data quality cuts of both the \polarbear{} and \act{} experiments. For the LAT, we additionally increase the noise levels to mimic the effects of discarding 15\% of the maps at the edges, where the noise properties are expected to be non-uniform. This is consistent with the sky cuts applied in, e.g., \citet{2017JCAP...06..031L}. The expected white noise levels are shown in Table~\ref{tab:noise}. These are computed from the estimated detector array noise-equivalent temperatures (NETs), which include the impact of imperfect detector yield, for the given survey areas and effective observing time. The detector NETs and details of this calculation are in the companion paper \citep{so_instrument}.

These noise levels are expected to be appropriate for small angular scales, but large angular scales are contaminated by $1/f$ noise at low frequencies in the detector time stream, which arises primarily from the atmosphere and electronic noise. We model the overall expected SO noise spectrum in each telescope and band to have the form
\begin{equation}
 \centering
  N_\ell  = N_{\rm red}\left( \frac{\ell}{\ellknee{}}\right)^{\alphaknee{}} + N_{\rm white},
  \label{eq:actual_Nell}
\end{equation}
where $N_{\rm white}$ is the white noise component and $N_{\rm red}$, $\ellknee{}$, and $\alphaknee$ describe the contribution from $1/f$ noise.
We adopt values for these parameters using data from previous and on-going ground-based CMB experiments. We do not model the $1/f$ noise for the SATs in temperature: we do not anticipate using the SAT temperature measurements for scientific analysis, as the CMB signal is already well measured by {\it WMAP} and {\it Planck} on these scales. \\

\begin{figure}[!t]
 \centering
 \includegraphics[width=\columnwidth]{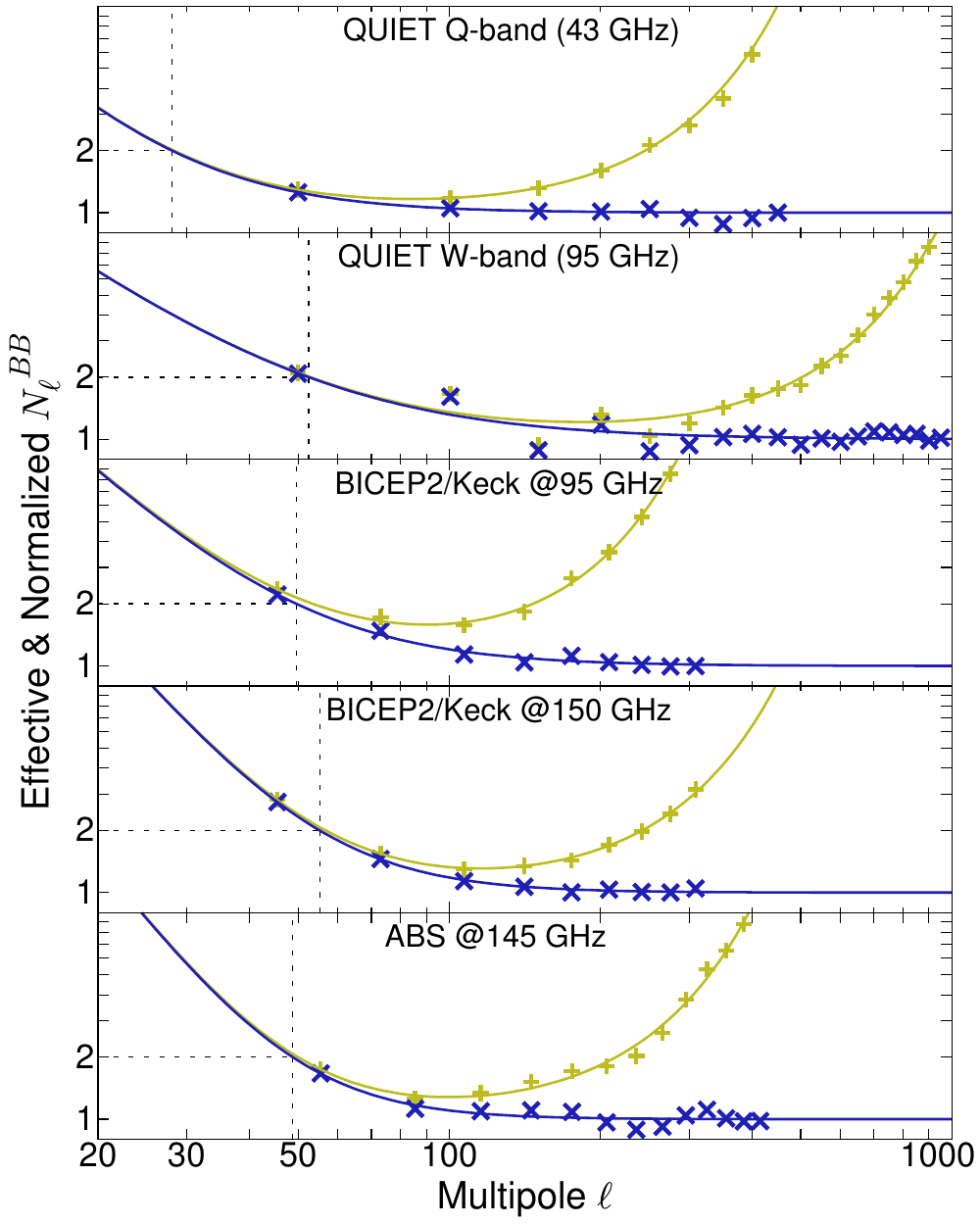} 
 \caption{
 The normalized uncertainties on the $C_\ell^{BB}$ power spectrum achieved
 by QUIET~\citep{2011ApJ...741..111Q,2012ApJ...760..145Q}, BICEP2 and Keck Array~\citep{2016PhRvL.116c1302B}, and ABS~\citep{2018arXiv180101218K}.
 The yellow data points are $\Delta C_\ell^{BB} / \sqrt{2/[(2\ell+1)\Delta\ell]} \propto N_\ell^{BB}$; the blue points have the beam divided out and are normalized to unity at high $\ell$. Solid lines show the modeled curves with Eq.~\ref{eq:actual_Nell}.
 Dashed horizontal lines indicate the location of $\ell_{\rm knee}$ and are at $\ell \approx 50$ or below.
 \label{fig:sat_noise_curves}}
\end{figure}

{\it SAT polarization:} In this case we normalize the model such that $N_{\rm red} = N_{\rm white}$. At a reference frequency of 93~GHz, we find that a noise model with $\ellknee{} \approx 50$ and $\alphaknee{}$ in the range of $-3.0$ to $-2.4$ describes the uncertainty on the $B$-mode power spectrum, $C_\ell^{BB}$,
achieved by QUIET~\citep{2011ApJ...741..111Q,2012ApJ...760..145Q}, BICEP2 and Keck Array~\citep{2016PhRvL.116c1302B}, and ABS~\citep{2018arXiv180101218K} as shown in Fig.~\ref{fig:sat_noise_curves}. QUIET and ABS were both near the SO site in Chile, and used fast polarization modulation techniques, while BICEP2 and Keck Array are at the South Pole. This $\ell_\mathrm{knee}$ accounts for both the $1/f$ noise and the loss of modes  due to filtering, and is also consistent with data taken by \pb{} in Chile with a continuously rotating half-wave plate~\citep{2017JCAP...05..008T}. We adopt $\ell_\mathrm{knee}=50$ for a pessimistic case and $\ell_\mathrm{knee}=25$ for an optimistic case. Here we assume a scan speed twice as fast as that adopted by ABS and QUIET. 
We scale the $1/f$ noise to each of the SO bands by evaluating the deviation of the brightness temperature due to expected changes in Precipitable Water Vapor~(PWV) level using the AM model~\citep{am_paine.scott_2018} and the Atmospheric Transmission at Microwaves (ATM) code~\citep{atm982447}. The parameters we adopt for each band are given in Table \ref{tab:oneoverf}. In forecasting parameters derived from the SATs, we consider these pessimistic and optimistic $1/f$ cases in combination with the SO baseline and goal white noise levels.\\

{\it LAT polarization:} Again we fix $N_{\rm red} = N_{\rm white}$.  We find that $\ellknee{} = 700$ and $\alphaknee{}=-1.4$ approximates the $\ell$-dependence of the uncertainties on the polarization power spectrum achieved by ACTPol~\citep{2017JCAP...06..031L} in Chile at 150~GHz, and is consistent with data from \pb{} without a continuously rotating half-wave plate~\citep{2014ApJ...794..171P,2017ApJ...848..121P}. We use these parameters at all frequencies, although in practice we expect the emission to be frequency-dependent. Upcoming data from ACTPol and \pb{} will inform a future refinement to this model.\\

{\it LAT temperature:} The intensity noise is primarily due to brightness variation in the atmosphere. We model the intensity noise by first measuring the contamination in time-ordered-data (TODs) and sky maps from ACTPol's 90 and 145 GHz bands, assuming that the SO detector passbands will be similar to those of ACTPol.  We then extrapolate this result to the full set of LAT bands, and account for the LAT's large field of view.

To characterize the intensity noise, we fix $\ellknee{} = 1000$.  Using data from ACTPol \citep{2017JCAP...06..031L}, we estimate a noise parameter $N_{\rm red} = 1800\,\muK{}^2s$ at 90 GHz and $N_{\rm red} = 12000\,\muK{}^2s$ at 145 GHz, with $\alphaknee{} = -3.5$ in both cases.  The dominant contribution to atmospheric contamination at 90 and 145 GHz is due to PWV. We use the 145~GHz noise power measured by ACTPol to fix the overall scaling of the contamination. To extrapolate to other frequency bands, the brightness temperature variance due to changes in PWV level is computed for each of the SO bands using the ATM code. 

\begin{figure}
  \includegraphics[width=\columnwidth]{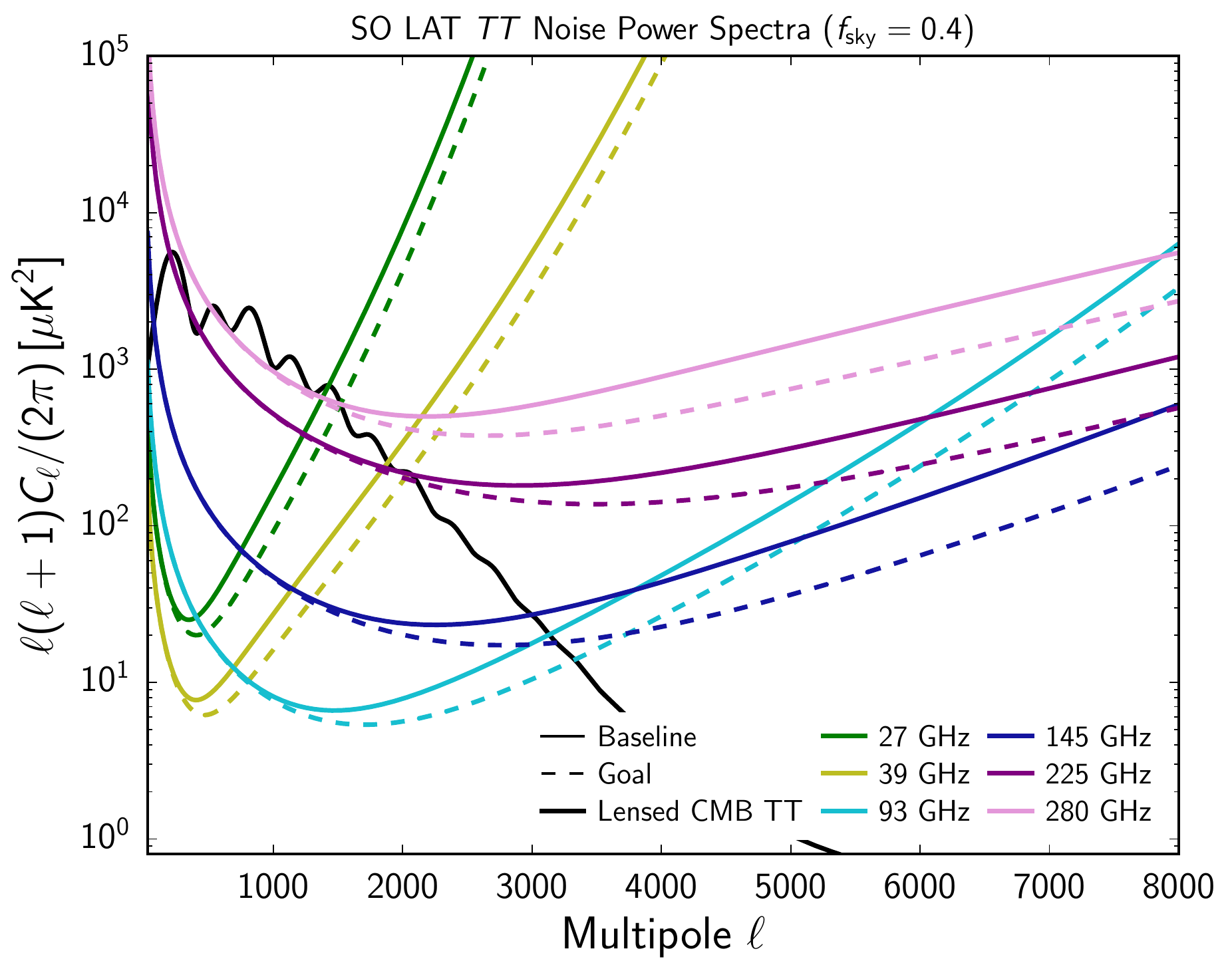} 
  \includegraphics[width=\columnwidth]{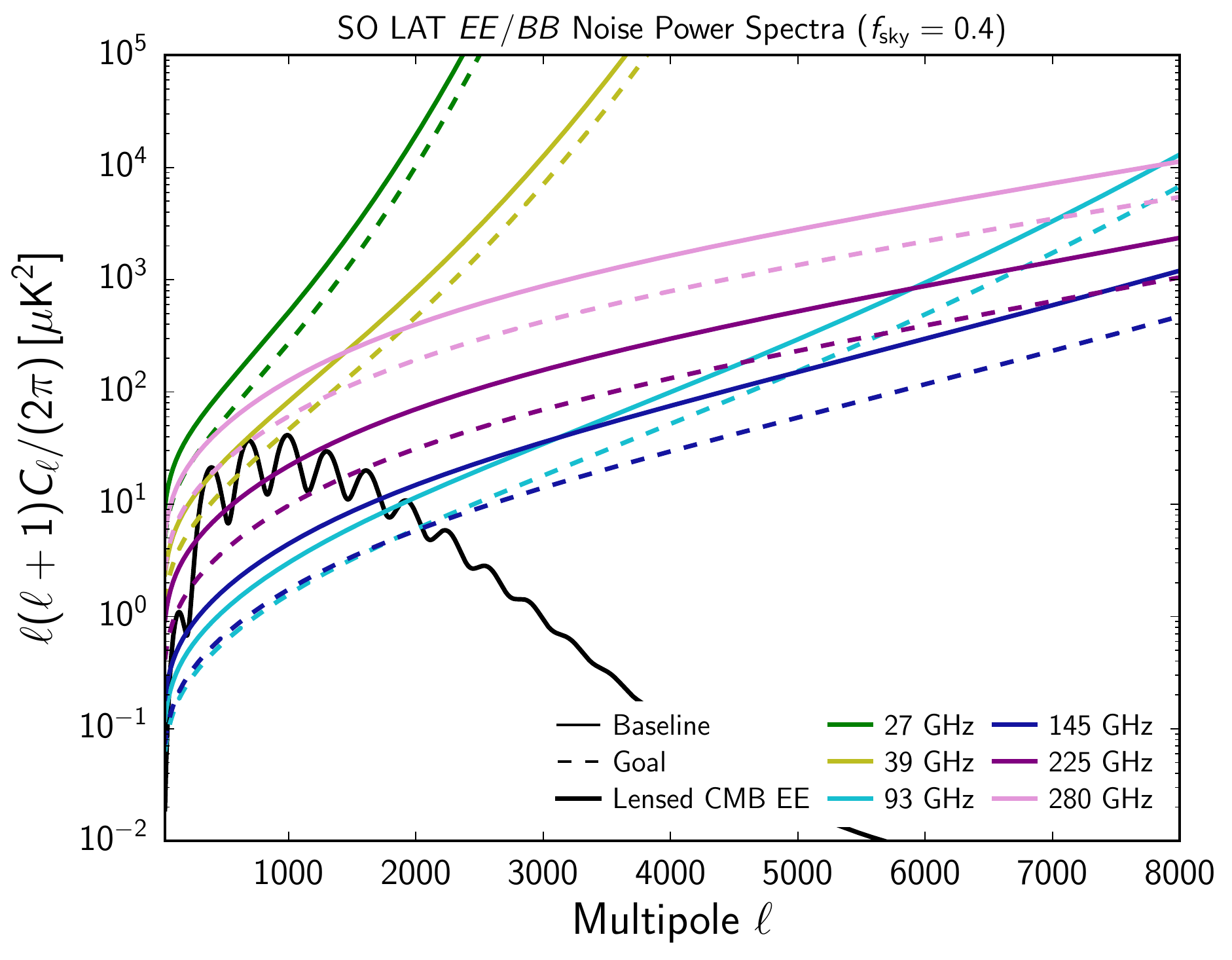}
    \includegraphics[width=\columnwidth]{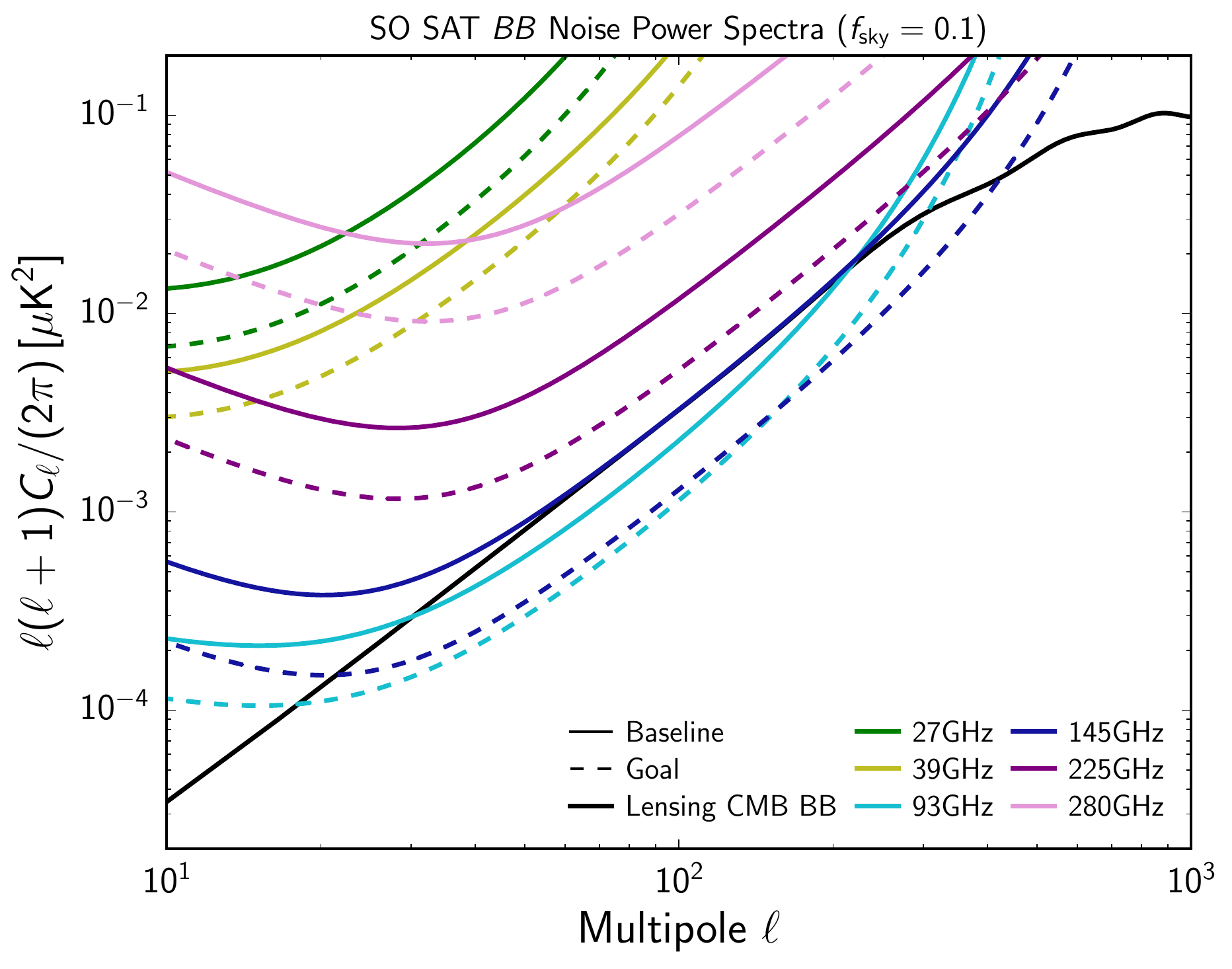}
  \caption{Per-frequency, beam-corrected noise power spectra as in Sec.~\ref{subsec:sensitivity} for the LAT temperature (top) and polarization (middle), and the SATs in polarization for the optimistic $\ell_{\rm knee}$ case of Table~\ref{tab:oneoverf} (bottom). Baseline (goal) sensitivity levels are shown with solid (dashed) lines, as well as the $\Lambda$CDM signal power spectra (assuming $r=0$).  The noise curves include instrumental and atmospheric contributions. Atmospheric noise correlated between frequency channels in the same optics tube is not shown for clarity, but is included in calculations.\label{fig:LAT_SAT_noises}}
\end{figure}

Atmospheric noise has strong spatial correlations and thus does not scale simply with the number of detectors.  We account for the increased field of view of the SO LAT relative to ACTPol using the following arguments.  First, each optics tube is assumed to provide an independent realization of the atmospheric noise, for angular scales smaller than the separation between the optics tubes.  The distance between optics tubes corresponds to $\ell$~$\approx$~50, well below the effective knee scale in the central frequencies.  Since the ACTPol noise spectra are based on measurements for arrays of diameter $\approx$ 0.1$\degree$, we thus divide the extrapolated ACTPol noise power by the number of SO optics tubes carrying a given band (one for LF, four for MF, and two for HF).  Second, each optics tube covers 3 to 4 times the sky area of an ACTPol array; we reduce the noise power by an additional factor of 2 to account for this. The $N_{\rm red}$ factors we adopt are given in Table \ref{tab:oneoverf}\footnote{In all forecasts we mistakenly used $N_{\rm red}=4\,\muK{}^2s$ instead of $39\,\muK{}^2s$ at 39~GHz. We corrected this error in Fig.~\ref{fig:LAT_SAT_noises}, and it has negligible effect on forecasts as the {\it Planck} temperature noise is below the SO noise at these scales.}. Because the same water vapor provides the $1/f$ noise at all frequencies, contamination of the two bands within a single optics tube is assumed to be highly correlated and we assign it a correlation coefficient of 0.9.

Figure~\ref{fig:LAT_SAT_noises} shows the instrumental and atmospheric noise power spectra for the SO LAT and SAT frequency channels, in temperature and polarization for the LAT, and polarization for the SATs. Correlated $1/f$ noise between channels in the same optics tube (due to the atmosphere) is not shown for clarity, but is included in all calculations in this paper according to the prescription described above.

\subsection{Sky coverage}
\label{subsec:coverage}
The SATs will primarily be used to constrain the tensor-to-scalar ratio through measuring large-scale $B$-modes. The LAT will be used to measure the small-scale temperature and polarization power spectra, the lensing of the CMB, the primordial bispectrum, the SZ effects, and to detect extragalactic sources. The LAT's measurement of CMB lensing may also be used to delens the large-scale $B$-mode signal measured by the SATs. Our nominal plan for sky coverage is to observe $\approx 40$\% of the sky with the LAT, and $\approx 10$\% with the SATs. As we show in this paper, we find this to be the optimal configuration to achieve our science goals given the observing location in Chile and the anticipated noise levels of SO. We do not plan to conduct a dedicated `delensing' survey with the LAT. 

\begin{figure*}
  \centering
  \includegraphics[width=\textwidth]{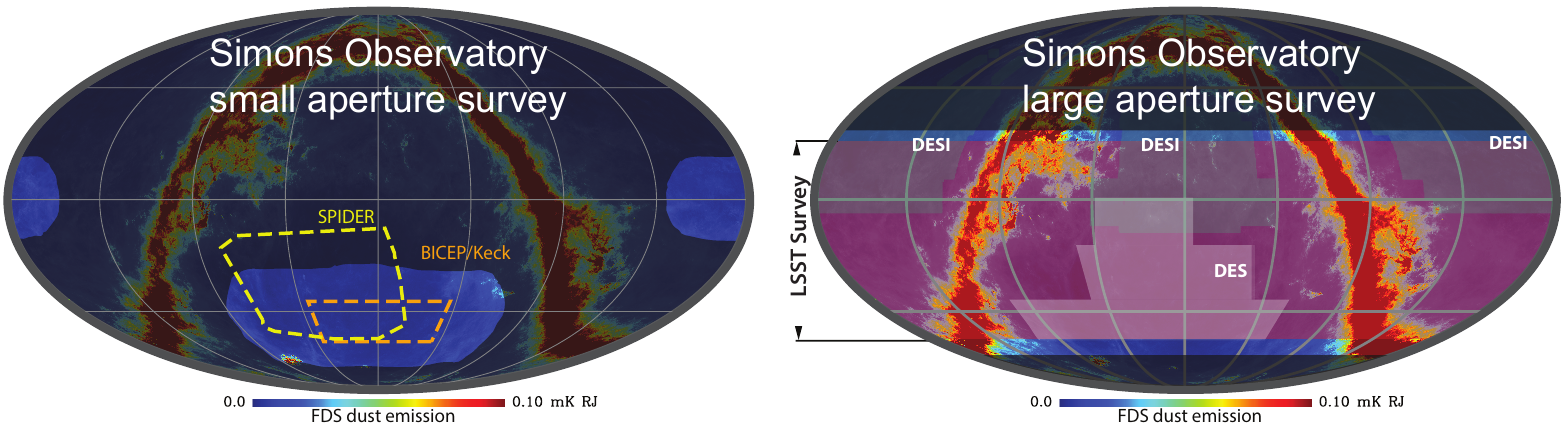}
  \caption{Anticipated coverage (lighter region) of the SATs (left) and LAT (right) in Equatorial coordinates, overlaid on a map of Galactic dust emission. For the SATs we consider a non-uniform coverage shown in Sec.~\ref{sec:bmodes}. For the LAT, we currently assume uniform coverage over $40\%$ of the sky, avoiding observations where the Galactic emission is high (red), and maximally overlapping with LSST and the available DESI region. This coverage will be refined with future scanning simulations following, e.g., \citet{debernardis/etal:2016}. The survey regions of other experiments are also indicated. The LSST coverage shown here represents the maximal possible overlap with the proposed SO LAT area; while this requires LSST to observe significantly further to the North than originally planned, such modifications to the LSST survey design are under active consideration~\citep{2018arXiv181200515L,2018arXiv181202204O}.} \label{fig:cover}
\end{figure*}

Key elements of our SO design study involved assessing the area that can realistically be observed from Chile with the SATs, and determining the optimal area to be surveyed by the LAT. The anticipated SAT coverage, motivated in~\cite{stevens/etal:prep}, is indicated in Fig.~\ref{fig:cover} in Equatorial coordinates, and is shown in more detail in Sec.~\ref{sec:bmodes}. The coverage is non-uniform, represents an effective sky fraction of $\approx$10-20\% accounting for the non-uniform weighting, and has the majority of weight in the Southern sky. This coverage arises from the wide field of view of the SATs, the desire to avoid regions of high Galactic emission, and the need to observe at a limited range of elevations from to achieve lower atmospheric loading. The exact coverage will be refined in future studies, but the sky area is unlikely to change significantly.

The LAT is anticipated to cover the 40\% of sky that optimally overlaps with LSST, avoids the brightest part of the Galaxy, and optimally overlaps with DESI given the sky overlap possible from Chile. In our design study we considered 10\%, 20\%, and 40\% sky fractions for the LAT, to determine the optimal coverage for our science goals. 

A limited sky fraction of 10\% would, for example, provide maximal overlap between the SATs and LAT, which would be optimal for removing the contaminating lensing signal from the large-scale $B$-mode polarization, as discussed in Sec.~\ref{sec:lensing}. However, we find in Sec.~\ref{sec:bmodes} that the impact of limiting the LAT sky coverage on our measurement of the tensor-to-scalar ratio is not significant, which is why we do not anticipate performing a deep LAT survey. In Secs.~\ref{sec:highell}--\ref{sec:sz} we show how our science forecasts depend on the LAT area, and conclude that SO science is optimized for maximum LAT sky coverage, and maximum overlap with LSST and DESI. We show a possible choice of sky coverage in Fig.~\ref{fig:cover}, which will be refined in further studies.

\subsection{Foreground model}
\label{subsec:fg_model}

Our forecasts all include models for the intensity and polarization of the sky emission, for both extragalactic and Galactic components, and unless stated otherwise we use the common models described in this section.  In intensity, our main targets of interest are the higher-resolution primary and secondary CMB signals measured by the LAT. In polarization our primary concern is Galactic emission as a contaminant of large-scale $B$-modes for the SATs. We also consider Galactic emission as a contaminant for the smaller-scale signal that will be measured by the LAT. We use map-based \citep[{\tt HEALPix}\footnote{\url{http://healpix.sf.net/}}]{2005ApJ...622..759G} sky simulations in all cases, except for small-scale extragalactic and Galactic polarization for which we use simulated power spectra. 

\subsubsection{Extragalactic intensity}

We simulate maps of the extragalactic components using the \citet{Sehgaletal2010} model, with modifications to more closely match recent measurements.  The extragalactic contributions arise from CMB lensing, the thermal and kinematic SZ effects (tSZ and kSZ, respectively), the cosmic infrared background (CIB), and radio point source emission. The components are partially correlated; the sources of emission are generated by post-processing the output of an $N$-body simulation.\\

{\it Lensed CMB}: We use the lensed CMB $T$ map from \citet{FerraroHill2018}, generated by applying the {\tt LensPix}\footnote{\url{http://cosmologist.info/lenspix/}} code to an unlensed CMB temperature map (generated at $N_{\rm side} = 4096$ from a CMB power spectrum extending to $\ell =  10000$ computed with {\tt camb}\footnote{\url{http://camb.info}}) and a deflection field computed from the $\kappa_{\rm CMB}$ map derived from the \citet{Sehgaletal2010} simulation.\\

{\it CIB}: We rescale the~\citet{Sehgaletal2010} CIB maps at all frequencies by a factor of 0.75, consistent with the~\citet{Dunkleyetal2013} constraint on the 148 GHz CIB power at $\ell=3000$. These simulations fall short of the actual CIB sky in some ways. The resulting CIB power spectrum at 353 GHz is low compared to the~\citet{Maketal2017} constraints at lower $\ell$. The spectral energy distribution (SED) of the simulated CIB power spectra is also too shallow compared to recent measurements~\citep[e.g.,][]{2012ApJ...756..142V}, in the sense that the model over-predicts the true CIB foreground at frequencies below 143~GHz. The CIB fluctuations in the simulation are correlated more strongly across frequencies than indicated by {\it Planck} measurements on moderate to large angular scales~\citep{Planck2013CIB,Maketal2017}. However, few constraints currently exist on cross-frequency CIB decorrelation on the small scales relevant for tSZ and kSZ component separation. The tSZ--CIB  correlation \citep[][]{Addison2012} has a coefficient ($35$\% at $\ell=3000$) a factor of two higher in the simulation than the SPT constraint~\citep{Georgeetal2015} and {\it Planck}~\citep{Planck2015tSZCIB}.

While not perfect, this CIB model is plausible and has realistic correlation properties with other fields in the microwave sky. The original simulated CIB maps are provided at 30, 90, 148, 219, 277, and 350 GHz; to construct maps at the SO and {\it Planck} frequencies, we perform a pixel-by-pixel interpolation of the flux as a function of frequency using a piecewise linear spline in log-log space.\\

{\it tSZ}: We rescale the \citet{Sehgaletal2010} tSZ map by a factor of 0.75 to approximately match measurements from {\it Planck}~\citep{Planck2013ymap,Planck2015ymap},  ACT~\citep{Sieversetal2013}, and SPT~\citep{Georgeetal2015}. The resulting tSZ power spectrum is in good agreement with the 2013 {\it Planck} $y$-map power spectrum. 
From the tSZ template map, we construct the tSZ field at all SO and {\it Planck} frequencies using the standard non-relativistic tSZ spectral function.\\

{\it kSZ}: The power spectrum of the~\citet{Sehgaletal2010} kSZ map is consistent with current upper limits from ACT~\citep{Sieversetal2013} and SPT~\citep{Georgeetal2015}. The kSZ map is approximately frequency independent in the blackbody temperature units that we use here.\\

{\it Radio point sources}: We apply a flux cut to the source population in the ~\citet{Sehgaletal2010} simulations, removing those with flux density greater than 7 mJy at 148 GHz. This models the effect of applying a point source mask constructed from sources detected in the maps.
We construct the maps by populating the true density field in the simulation with sources, and interpolate to the SO and {\it Planck} frequencies.

\subsubsection{Extragalactic source polarization}

We adopt a Poissonian power spectrum of radio point sources, with amplitude  $\ell(\ell+1)C_{\ell}/(2\pi) = 0.009 \,\, \mu {\rm K}^2$ at $\ell=3000$ at 150 GHz (for both $EE$ and $BB$).  This value is consistent with upper limits from ACTPol~\citep{2017JCAP...06..031L} and SPTpol~\citep{2018ApJ...852...97H}, and is computed by assuming a Poisson amplitude of $\ell(\ell+1)C_{\ell}/(2\pi) = 3 \, \mu {\rm K}^2$ in intensity and a polarization fraction of $0.05$.  The SED follows~\citet{Dunkleyetal2013}, with a spectral index of $-0.5$ in flux units. We assume that the polarization of the CIB, tSZ, and kSZ signals are negligible. 
For the lensed CMB, we generate the polarization power spectra using {\tt camb}.

\subsubsection{Galactic intensity}

Our model for Galactic emission intensity includes thermal dust, synchrotron, bremsstrahlung (free--free), and anomalous microwave emission (AME).
We use simulated maps that give forecast results consistent with the {\tt PySM} model~\citep{PySMpaper}, but were generated using an alternative code. For thermal dust we use `model 8' of \citet{FDS1999} in the \citet{Sehgaletal2010} simulation, with maps interpolated to the SO and {\it Planck} frequencies. Due to the SO LAT sky mask and large-scale atmospheric noise, our results are not particularly sensitive to the choice of Galactic thermal dust model. For synchrotron, free--free, and AME we use the \planck\ {\tt Commander} models~\citep{Planck2015Commander} generated at $N_{\rm side} = 256$ resolution.
We refine these maps to $N_{\rm side} = 4096$ to match the pixelization of the other maps in our analysis (with additional smoothing applied to suppress spurious numerical artifacts on small scales), but no additional information on sub-degree scales is added. 
\vfill\null

\subsubsection{Galactic polarization}\label{sssec:4cast.fgs.pol}

The dominant emission in Galactic polarization is from synchrotron and thermal dust, which we generate in map space for the SAT forecasts using the {\tt {PySM}} model~\citep{PySMpaper}, which extrapolates template Galactic emission maps estimated from \planck{} and \map{} data. They are scaled in frequency for both $Q$ and $U$ Stokes parameters assuming a curved power law and a modified blackbody spectrum, respectively
\be 
\label{eq:scalings}
 S_\nu^{\rm synch}=\left(\frac{\nu}{\nu_0}\right)^{\beta_s+C\log(\nu/\nu_0)},\hspace{12pt}
 S_\nu^{\rm dust}=\frac{\nu^{\beta_d}B_{\nu}(T_d)}{\nu_0^{\beta_d}B_{\nu_0}(T_d)},
\ee 
where $\beta_s$ is the synchrotron spectral index, $C$ is the curvature of the synchrotron index, $\beta_d$ the dust emissivity, $T_d$ the dust temperature, and $\nu_0$ a pivot frequency. 
All spectral parameters, except for the synchrotron curvature, vary across the sky on degree scales. We make the following choices to generate {\tt PySM} simulations with three different levels of complexity: \\

\begin{figure} [t!]
\centering
\includegraphics[width=\columnwidth]{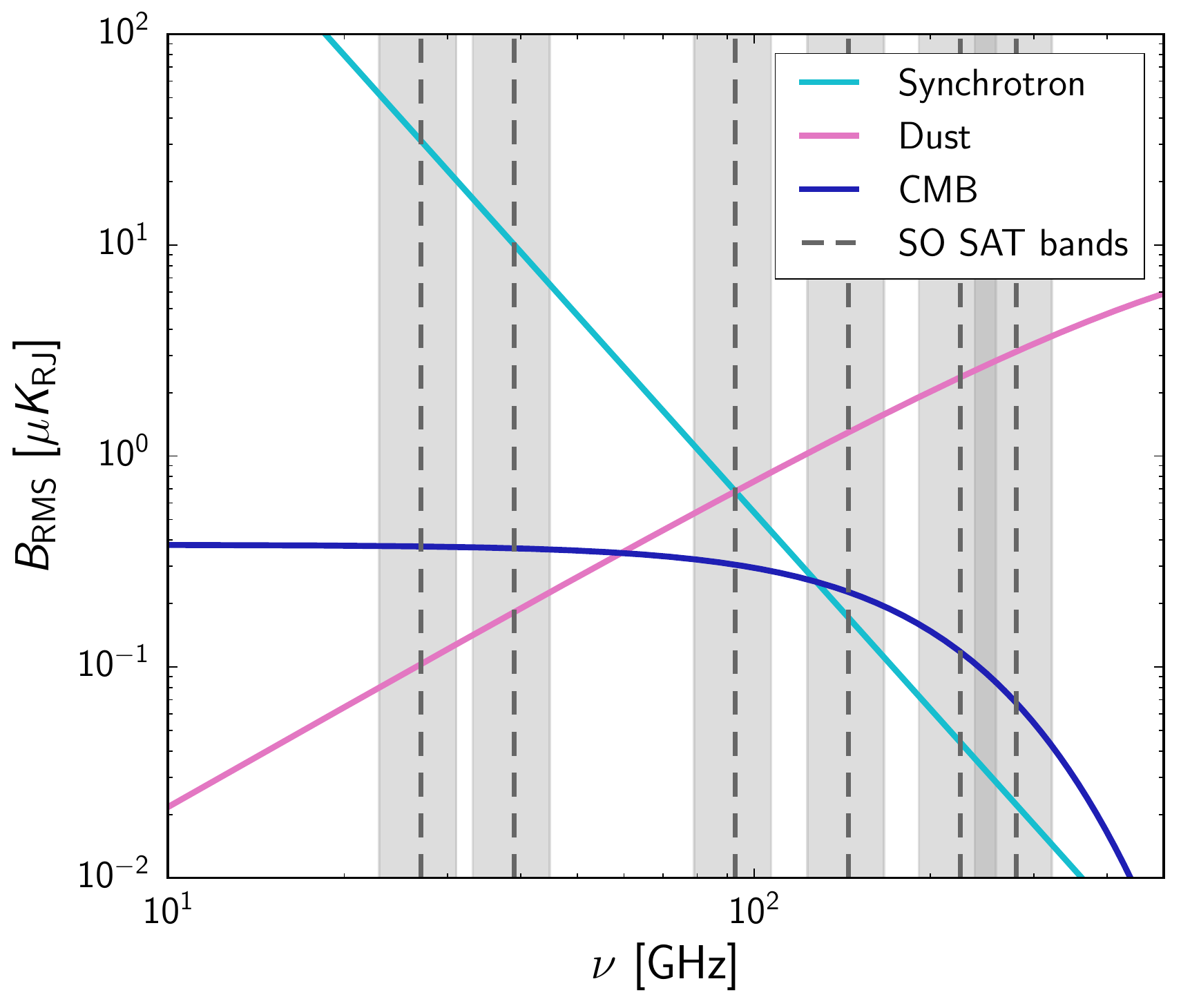}
\caption{Frequency dependence, in RJ brightness temperature, of the synchrotron and thermal dust emission at degree scales within the proposed footprint for the SATs, compared to the CMB lensing $B$-mode signal. The turnover of the modified blackbody law for the dust lies above this frequency range.}\label{fig:fg_freq}
\end{figure}

{\it  `Standard'}: this corresponds to the {\tt {PySM}} `a1d1f1s1'  simulation~\citep{PySMpaper}, assuming a single modified-blackbody polarized dust and a single power-law synchrotron component. These use spatially varying spectral indices derived from the intensity measurements from \planck{}~\citep{Planck_X_2015}.
We compare the $B$-mode amplitude and frequency dependence of these foregrounds to the expected cosmological $B$-mode signal from lensing in Fig.~\ref{fig:fg_freq}.\\

{\it `2 dust + AME'}: this corresponds to the `a2d7f1s3' {\tt {PySM}} model described in \citet{PySMpaper}, i.e., a power law with a curved index for synchrotron, dust that is decorrelated between frequencies, and an additional polarized AME component with 2\% polarization fraction.\\

{\it `High-res. $\beta_s$'}: this includes small scale (sub-degree) variations of the synchrotron spectral index $\beta_s$ simulated as a Gaussian realization of a power law angular spectrum with $C_{\ell}\propto\ell^{-2.6}$~\citep{Krachmalnicoff18}.\\

For the LAT forecasts, we model the power spectra of the polarized components instead of using simulated maps.
The power spectra of thermal dust and synchrotron are taken as power laws, with $C_{\ell}^{\rm dust} \propto \ell^{-2.42}$, following measurements by {\it Planck}~\citep{Planck2015dust}, and $C_{\ell}^{\rm synch} \propto \ell^{-2.3}$~\citep{Choi-Page2015}. The dust spectral parameters defined in Eq.~\ref{eq:scalings} are $\beta_d = 1.59$ and dust temperature 19.6 K~\citep{Planck2015dust}, and the synchrotron spectral index is fixed to $\beta_s = -3.1$~\citep{Choi-Page2015}. The $EE$ and $BB$ dust power spectrum amplitudes are normalized to the amplitudes at $\ell=80$ for the {\tt {PySM}} default sky model evaluated at 353 GHz for dust and 27 GHz for synchrotron (using the appropriate SO LAT sky mask). We include the cross-power due to dust--synchrotron correlations via a correlation coefficient (see, e.g., Eq.~6 of~\citealp{Choi-Page2015}), determined by normalizing the model to the {\tt PySM} model evaluated at 143 GHz in the SO LAT sky mask, after accounting for the dust and synchrotron auto-power.\\

\subsection{Foreground cleaning for the LAT}\label{sec:FG_LAT}
Here we describe the foreground removal method used for the LAT, which is then propagated to Fisher forecasts for parameters derived from the temperature and $E$-mode power spectrum (Sec.~\ref{sec:highell}), the lensing spectrum (Sec.~\ref{sec:lensing}), the primordial bispectrum (Sec.~\ref{sec:bispec}) and the SZ effects (Sec.~\ref{sec:sz}). Our forecasts for primordial large-scale $B$-modes, the main science case for the SAT, are based on the comparison of a number of component separation methods run on map-level foreground and noise simulations, and are described in Sec.~\ref{sec:bmodes}.

We generate signal-only simulations at the SO LAT frequencies and the {\it Planck} frequencies at 30, 44, 70, 100, 143, 217, and 353 GHz.  The noise properties of the SO LAT channels are described in Sec.~\ref{subsec:sensitivity}.  The white noise levels and beams for the {\it Planck} frequencies are drawn from~\citet{Planck2015LFImap} (30, 44, and 70 GHz) and~\citet{Planck2015HFImap} (70, 100, 143, 217, and 353 GHz).\footnote{After these calculations were performed, the \planck{} products were updated with the final LFI and HFI mission processing~\citep{planck2018:lfi_processing,planck2018:hfi_processing}. Since the temperature noise levels have not changed compared to 2015, and as the \so{} LAT polarization noise is lower than that of \planck{} at all relevant scales, we do not anticipate our forecasts to change with the updated \planck{} noise levels.}  This yields thirteen frequency channels, which we take to have $\delta$-function bandpasses for simplicity. We consider three SO LAT survey regions covering 10\%, 20\%, and 40\% of the sky.\footnote{The region retaining a sky fraction of 40\% was originally selected to minimize the polarized Galactic contamination and does not precisely match the planned sky area for the LAT shown in Fig.~\ref{fig:cover}, which has since been tuned to have improved overlap with LSST and DESI. However, we estimate the impact on forecasts of choosing between these two different sky masks to be small.}  We measure the auto- and cross-power spectra at all frequencies, considering scales up to $\ell_{\rm max} = 8000$. We correct for the mask window function using a simple $f_{\rm sky}$ factor, given the large sky area and lack of small-scale structure in the mask. We then add the instrumental and atmospheric noise power spectra for SO described in Sec.~\ref{subsec:sensitivity}, and white noise power for \planck, to form a model of the total observed auto- and cross-power spectra. 

\subsubsection{Component separation method}
\label{sec:LAT_ILC}
We implement a harmonic-space Internal Linear Combination (ILC) code to compute post-component-separation noise curves for various LAT observables \citep[e.g.,][]{2003ApJS..148....1B,Eriksen2004}.  While a more sophisticated method \citep[e.g., Needlet ILC]{Delabrouille2009} will likely be used in actual analyses, the harmonic-space ILC is rapid enough to enable calculations for many experimental scenarios and sky models. This is essential for optimization of the SO LAT frequency channels, while still being representative of the likely outcome using other methods.  Moreover, harmonic-space ILC forecasts can be evaluated given only models for the power spectra of the sky components (i.e., maps are not explicitly required).  Although we simulate maps for the temperature sky, our limited knowledge of small-scale polarized foregrounds forces us to rely on power-spectrum-level modeling for polarization (however, see~\citet{
Hervias-Caimapo2016} for steps toward simulating such maps), and thus harmonic-space ILC is necessary in this case.

In the following, we consider forecasts for `standard ILC', in which the only constraints imposed on the ILC weights are: (i) unbiased response to the known spectral energy distribution (SED) of the component of interest (e.g., CMB) and (ii) minimum variance (in our case, at each $\ell$).  We also consider `constrained ILC'~(e.g.,~\citealp{Remazeilles2011}), in which an additional constraint is imposed: (iii) zero response to some other component(s) with specified SED(s).  We refer to this additional constraint as `deprojection'.  Since this constraint uses a degree of freedom, i.e., one of the frequency channel maps, the residual noise after component separation is higher for constrained ILC than for standard ILC. We use the deprojection method as a conservative choice that reflects the need to explicitly remove contaminating components that may bias some analysis, even at the cost of increased noise (e.g., tSZ biases in CMB lensing reconstruction). Deprojection will impose more stringent requirements than standard ILC on the ability of the experiment's frequency coverage to remove foregrounds.

For temperature forecasts, we consider deprojection of the thermal SZ spectral function and/or a fiducial CIB SED (or, in the case of tSZ reconstruction, deprojection of CMB and/or CIB). For polarization forecasts, we consider deprojection of a fiducial polarized dust SED and/or of a fiducial polarized synchrotron SED.  Clearly, for components with SEDs that are not known \emph{a priori} from first principles, deprojection could leave residual biases; these can be minimized in practice by sampling over families of SEDs~\citep{Hillinprep}.  
For later reference, we provide a dictionary of the relevant deprojection choices here:\\

\noindent
CMB Temperature Cleaning:\\
Deproj-0: Standard ILC\\ 
Deproj-1: tSZ deprojection\\
Deproj-2: Fiducial CIB SED deprojection\\
Deproj-3: tSZ and fiducial CIB SED deprojection\\

\noindent
Thermal SZ Cleaning:\\
Deproj-0: Standard ILC \\
Deproj-1: CMB deprojection\\
Deproj-2: Fiducial CIB SED deprojection\\
Deproj-3: CMB and fiducial CIB SED deprojection\\

\noindent
CMB Polarization Cleaning:\\
Deproj-0: Standard ILC \\
Deproj-1: Fiducial polarized dust SED deprojection\\
Deproj-2: Fiducial polarized synchrotron SED deprojection\\
Deproj-3: Fiducial polarized dust and synchrotron SED deprojection\\

\subsubsection{Post-component-separation noise}
\label{subsubsec:compsepnoise}

\begin{figure*}[t!]
\centering
\begin{tabular}{cc}
  \includegraphics[width=\columnwidth]{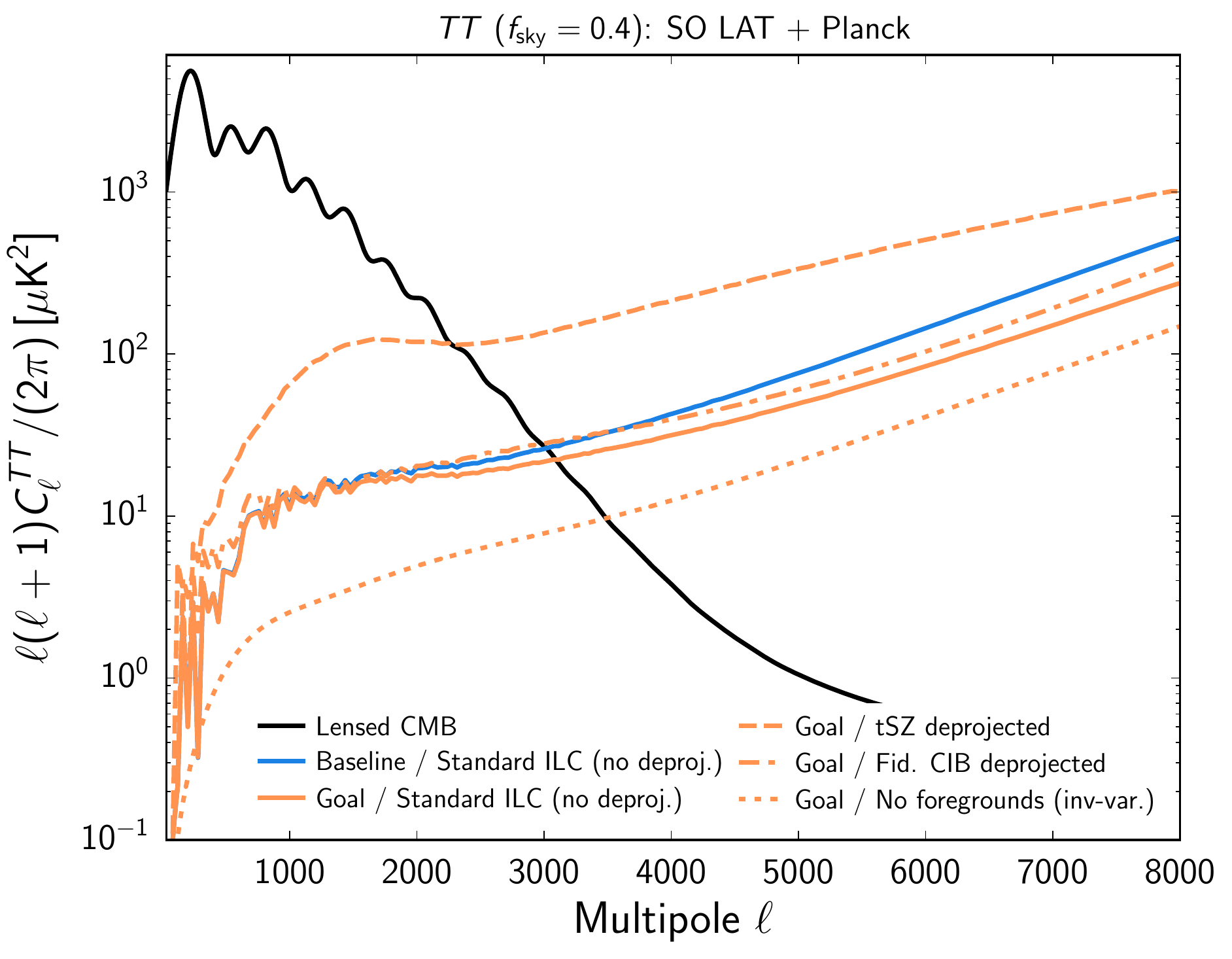} & 
  \includegraphics[width=\columnwidth]{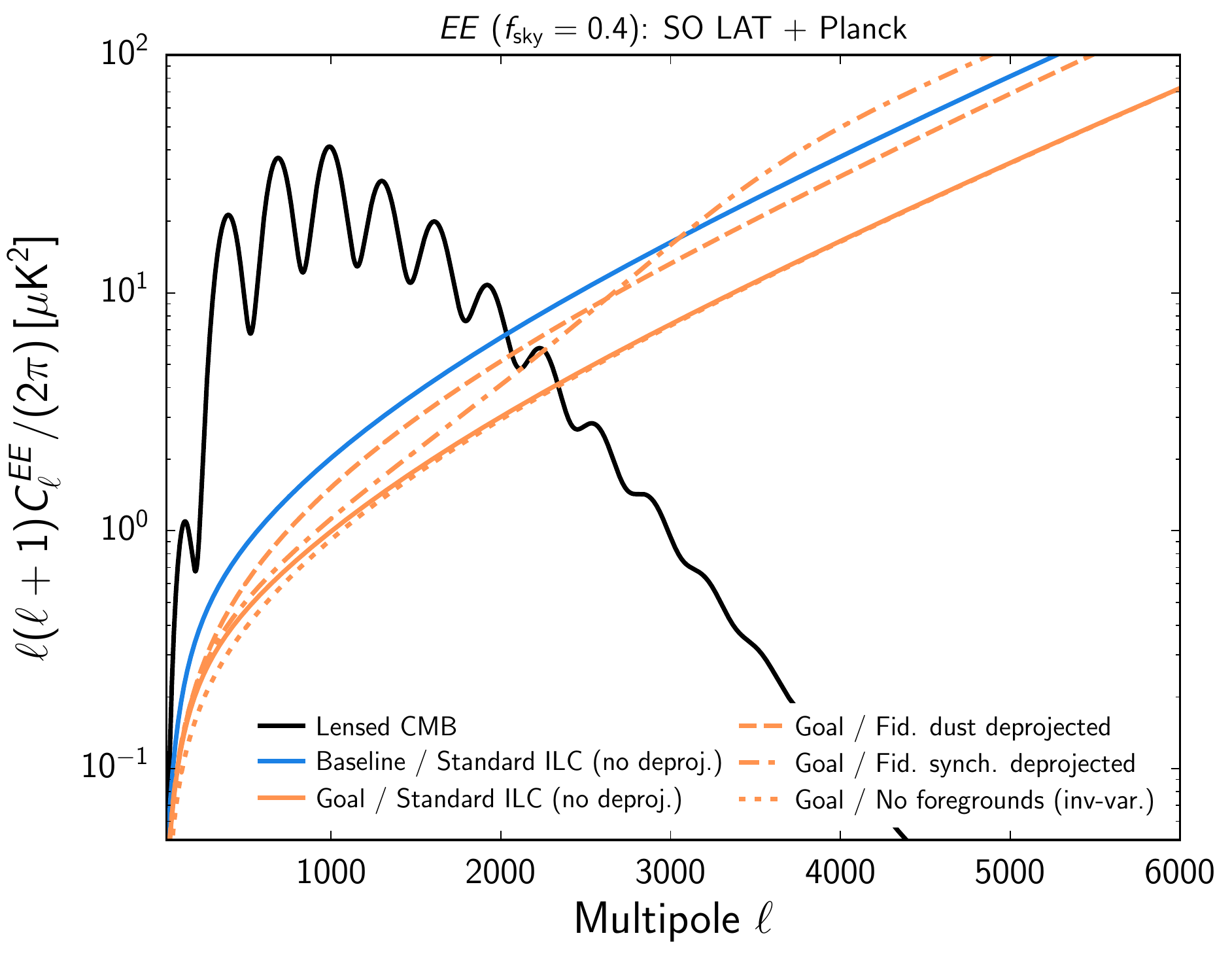}
\end{tabular}
\caption{Post-component-separation noise curves for the combination of six SO LAT (27--280 GHz) and seven {\it Planck} (30--353 GHz) frequency channels, assuming a wide SO survey with $f_{\rm sky} = 0.4$, compared to the expected signal (black).  The left (right) panel shows CMB temperature ($E$-mode polarization). Foregrounds and component separation are implemented as in Sec.~\ref{subsec:fg_model} and Sec.~\ref{sec:LAT_ILC}, considering multipoles up to $\ell_{\rm max}=8000$. The blue (orange) curves show the component-separated noise for the SO baseline (goal) noise levels, assuming standard ILC cleaning. The dashed and dash-dotted curves show various ILC foreground deprojection options, described in Sec.~\ref{sec:LAT_ILC}.  The tSZ deprojection penalty is larger than that for CIB deprojection because of (i) the relatively high noise at 225 GHz compared to 93 and 145 GHz and (ii) the lack of a steep frequency lever arm for the tSZ signal as compared to the CIB. The dotted orange curves show the no-foreground goal noise, i.e., when SO LAT and {\it Planck} channels are combined via inverse-noise weighting. This is the minimal possible noise that could be achieved. The temperature noise curves fluctuate at low-$\ell$ due to the use of actual sky map realizations, as opposed to the analytic power-spectrum models in  polarization.}
\label{fig:comp_sep_noise}
\end{figure*}
We compute SO LAT post-component separation noise for CMB temperature maps, thermal SZ maps, and CMB polarization maps ($E$- and $B$-mode).  We consider both the baseline and goal \so{} noise levels, as well as three sky fraction options (10\%, 20\%, and 40\%) and the four foreground deprojection methods.
These are the final LAT noise curves used throughout later sections of the paper for forecasting.
As an illustration, we show noise curves for the LAT in Fig.~\ref{fig:comp_sep_noise} for CMB temperature and CMB $E$-mode polarization, for a wide survey ($f_{\rm sky} = 0.4$).  A similar figure for tSZ reconstruction can be found in Sec.~\ref{sec:tSZ_PS}. The figure shows the post-component-separation noise for various foreground cleaning methods and assumed noise levels.  It also shows the pure inverse-noise-weighted (i.e., zero-foreground) channel-combined noise for the goal scenario, which allows a straightforward assessment of the level to which the foregrounds inflate the noise. In temperature, the foregrounds have a large effect; in contrast, in $E$-mode polarization at high-$\ell$, the foregrounds are expected to have little effect, making this a prime region for cosmological parameter extraction from the primary CMB.

\begin{figure}[t!]
\centering
\includegraphics[width=\columnwidth]{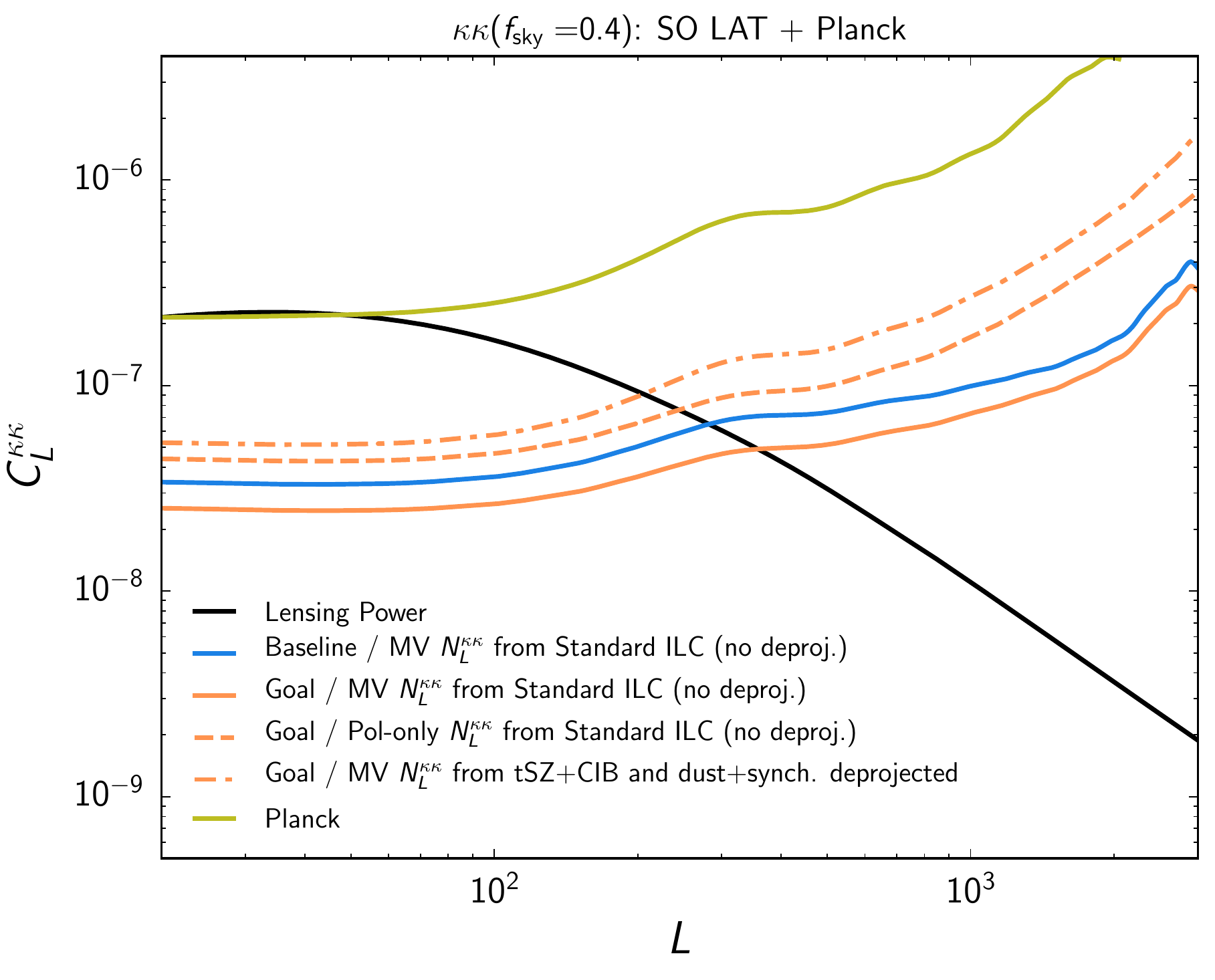}
\caption{$\Lambda$CDM CMB lensing power spectrum (black) compared to \so{} LAT lensing noise curves, $N_L^{\kappa\kappa}$, reconstructed assuming a polarization only (Pol-only) or  minimum variance (MV) combination of estimators in the case of standard ILC for both CMB temperature and polarization cleaning (solid and dashed curves), and tSZ and fiducial CIB SED deprojection for CMB temperature cleaning and fiducial polarized dust and synchrotron SED deprojection for CMB polarization cleaning (dot-dashed curve). \so{} baseline and goal scenarios are shown in blue and orange, respectively, and compared to the \planck\ lensing noise (\citealp{planck2018:lensing}, yellow). SO will be able to map lensing modes with $S/N>1$ to $L>200$.}
\label{fig:lensing_noise}
\end{figure}

We use the temperature and polarization noise curves to obtain the lensing noise $N_L^{\kappa\kappa}$ assuming quadratic estimators are used to reconstruct the lensing field, as described in~\cite{HDV2007}. We calculate the noise from five estimators ($TT, ET, TB, EE, EB$), and we combine the last two to obtain `polarization only' noise curves and combine all of them to obtain `minimum variance' noise curves. We show example lensing noise curves in Fig.~\ref{fig:lensing_noise} for a wide survey with SO LAT ($f_{\rm sky} = 0.4$) and two foreground cleaning cases: (i) standard ILC for both CMB temperature and polarization cleaning, and (ii) tSZ and fiducial CIB SED deprojection for CMB temperature cleaning and fiducial polarized dust and synchrotron SED deprojection for CMB polarization cleaning.

\begin{figure*}
\includegraphics[width=\textwidth]{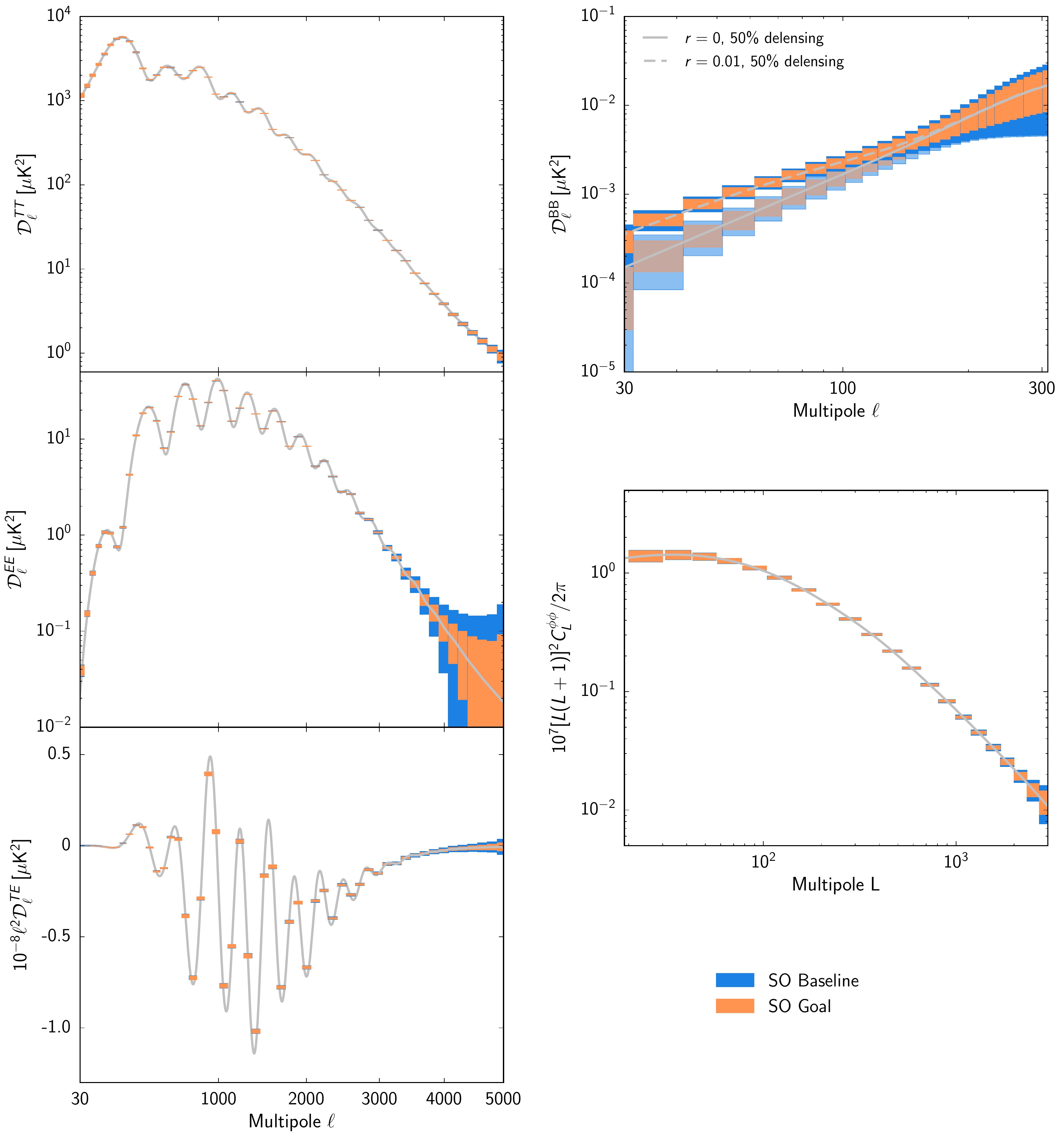}
\caption{Forecast \so{} baseline (blue) and goal (orange) errors on CMB temperature ($TT$), polarization ($EE$, $BB$), cross-correlation ($TE$), and lensing ($\phi \phi$) power spectra, with $\mathcal{D}_\ell \equiv \ell(\ell+1)C_\ell/(2\pi)$. The errors are cosmic-variance limited at multipoles $\ell \ltsim 3000$ in $T$ and $\ell \ltsim 2000$ in $E$. The $B$-mode errors include observations from both SAT and LAT surveys, and incorporate the uncertainty associated with foreground removal using {\tt BFoRe} (see Sec.~\ref{ssec:mr.tools}) for the optimistic $\ell_{\rm knee}$ given in Table \ref{tab:oneoverf}. The CMB signals for a fiducial $\Lambda$CDM cosmology ($\Lambda$CDM+tensor modes in the case of $BB$) are shown with gray solid (dashed) lines.}
\label{fig:allspecs}
\end{figure*}

Using these noise curves and anticipated sky coverage (40\% for the LAT, and 10\% for the SATs), we show the forecast errors on the temperature, polarization, and lensing power spectra in Fig.~\ref{fig:allspecs}. These include the anticipated instrument noise and foreground uncertainty, but do not include any additional systematic error budget. Fig.~\ref{fig:allspecs} also shows projected errors for the $B$-mode power spectrum described in Sec.~\ref{sec:bmodes}.

\vfill\null

\subsubsection{Optimization}
Our nominal noise curves correspond to the SO LAT frequency distribution given in Table~\ref{tab:noise}.  However, to determine this frequency distribution, we performed a full end-to-end optimization for various LAT observables.  This study will be described elsewhere, but we provide a summary here for reference.  We considered a range of sky areas (from $f_{\rm sky} = 0.03$ to $0.4$) and all configurations of LAT optics tubes, with the constraint that there are a total of seven tubes, and they can each have 27/39 GHz, 93/145 GHz, or 225/280 GHz.  

Using the noise calculator described in~\citet{so_instrument} and~\cite{hill/etal:prep}, we computed the SO LAT noise properties for each choice of survey region and experimental configuration, and then processed these noise curves through the foreground modeling and component separation methodology described in the previous subsections.  We then used the post-component-separation noise curves to determine the $S/N$ of various SO LAT observables: the CMB $TT$ power spectrum, the CMB lensing power spectrum reconstructed via the $TT$ estimator, the tSZ power spectrum, the kSZ power spectrum, the CMB $EE$ power spectrum, the CMB $BB$ power spectrum (lensing-only), and the CMB lensing power spectrum reconstructed via the $EB$ estimator.  
We repeated this analysis for the set of deprojection assumptions in the ILC foreground cleaning, which impose different constraints on the frequency channel distribution.  We found a set of configurations that was near-optimal for all observables (i.e., maximized their $S/N$) when using the simplest foreground cleaning method, and then we identified a near-optimal configuration that was also robust to varying the foreground cleaning method.  This process yielded the final choice of the SO LAT optics tube distribution and survey area: one low-frequency tube, four mid-frequency tubes, and two high-frequency tubes, with the widest possible survey ($f_{\rm sky} = 0.4$).

\subsection{Parameter estimation and external data}
\label{sec:add_data}
The majority of our parameter forecasts use Fisher matrix approaches, with the foreground-marginalized noise curves described above as inputs. An important exception to this is the measurement of large-scale $B$-modes, described in Sec.~\ref{sec:bmodes}. In this case the foregrounds are a more significant contaminant so we perform end-to-end parameter estimates on a suite of simulations: foreground-cleaning the maps, estimating $B$-mode power spectra, and then estimating parameters. 

 Unless stated otherwise we assume a six-parameter $\Lambda$CDM model as nominal (baryon density, $\Omega_bh^2$, cold dark matter density, $\Omega_ch^2$, acoustic peak scale, $\theta$, amplitude and spectral index of primordial fluctuations, $A_s$ and $n_s$, and optical depth, $\tau$), and add additional parameters as described in each of the following sections. As reference cosmology we assume the parameters from the \cite{Planck2015params} $\Lambda$CDM temperature fit, except for $\tau$ which is assumed to be 0.06 in agreement with~\citet{planck_tau:2016}. We use the Boltzmann codes {\tt camb} and {\tt class\footnote{\url{http://www.class-code.net/}}} to generate theoretical predictions, using updated recombination models, with additional numerical codes to generate statistics including cluster number counts and cross-power spectra between CMB lensing and galaxy clustering. 

We combine \so{} data with additional sky and frequency coverage provided by \planck{}. For the SATs this is done by assuming the \planck\ intensity data will be used at large scales. We also assume a prior on the optical depth of $\tau=0.06\pm 0.01$ (\citealp{planck_tau:2016}; neglecting the small change in the mean value and the improvement to $\sigma(\tau)=0.007$ with the 2018 results;~\citealp{planck2018:parameters}). For the LAT, the \planck\ data are included in the co-added noise curves over the sky common to both experiments. Additionally, for the largest angular ranges not probed by SO, we include $TT$, $TE$ and $EE$ from \planck\ over 80\% of the sky at $2\leq \ell \leq 29$. For the sky area not accessible to SO, we add an additional 20\% of sky from \planck\ in the angular range $30\leq \ell \leq 2500$. This produces an overall sky area of 60\% which is compatible with the area used by \planck{} after masking the Galaxy. For the \planck{} specifications we follow the procedure described in~\cite{allison/etal/2015} and ~\citet{calabrese/etal/2016}, scaling the overall white noise levels to reproduce the full mission parameter constraints. For reference, we give forecast constraints on the \LCDM\ parameters in Table \ref{tab:lcdm} for SO combined with {\it Planck}, compared to the published results from {\it Planck} alone (TT,TE,EE+lowE+lensing, ~\citealp{planck2018:parameters}). Both cases use temperature, polarization, and lensing data. In this paper we will refer to `SO Baseline' and `SO Goal' forecasts; these all implicitly include {\it Planck}.

\begin{table}[]
\centering
\caption[Simons Observatory Surveys]{Forecasts of \LCDM\ parameter uncertainties for SO compared to {\it Planck$^a$}}
\small
\begin{tabular}{ l | c | c}
\hline
\hline
Parameter & {\it Planck} & {\bf{SO-Baseline}+\it Planck} \\
\hline
$\Omega_bh^2$  &0.0001 &{\bf $0.00005$} \\
$\Omega_ch^2$  &0.001 &{\bf $0.0007$} \\
$H_0{\rm [km/s/Mpc]}$  & 0.5 & {\bf $0.3$}\\
$10^{9} A_s$  & 0.03& {\bf $0.03$}\\
$n_s$  & 0.004 & {\bf $0.002$}\\
$\tau$  & $0.007$& {\bf $0.007$}\\
\hline
\hline
\end{tabular}
\footnotetext[1]{The `\planck'-only constraints reported here are from the final 2018 \planck{} data \citep{planck2018:parameters}. We check that our \planck\ forecast code (using $T/E$ at $2\leq \ell \leq 29$ with $f_{\rm sky}=0.8$, $TT/TE/EE$ at $30 \leq \ell \leq 2500$ with $f_{\rm sky}=0.6$, and $\kappa\kappa$ at $8\leq L \leq 400$ with $f_{\rm sky}=0.6$) yields similar results, except for small differences: we find $\sigma(H_0)=0.6~{\rm km/s/Mpc}, \sigma(10^9 A_s)=0.04, \sigma(\tau)=0.009$.\\}
\label{tab:lcdm}
\end{table}

In many cases we combine SO forecasts with DESI and LSST. For LSST we consider an overlap area of $f_{\rm sky}=0.4$ and two possible galaxy samples. First is the `gold' sample, which has galaxies with a dust-corrected $i<24.5$ magnitude cut after three years of LSST observations.  This corresponds to $29.4$ galaxies $\mathrm{arcmin}^{-2}$ and $n(z) \propto z^2\exp[-(z/0.27)^{0.92}]$ following \citet{0912.0201} and \citet{1305.0793}. Second, we consider a more optimistic LSST galaxy sample with dust-corrected $i<27$ and a $S/N>5$ cut with ten years of LSST observation, following \cite{1301.3010}. In that sample we include a possible sample of Lyman-break galaxies at~$z$=4--7, identified using the dropout technique (see \citealp{Dunlop1205} for a review), with a number density estimated by extrapolating recent Hyper Suprime-Cam (HSC) results (\citealp{1704.06004}, \citealp{1704.06535}, following \citealp{1710.09465}). 

For DESI we include projected measurements of the baryon acoustic oscillation (BAO) scale, by imposing a prior on $r_s/D_V$ at multiple redshifts, as described in \citet{DESI}. Here, $r_s$ is the sound horizon at decoupling and $D_V$ is the volume distance. We consider the DESI Luminous Red Galaxy (LRG) catalog as providing the target galaxies for the SZ studies described in Sec.~\ref{sec:sz}. In these forecasts we assume an overlap area of 9000 square degrees between \so{} and DESI ($f_{\rm sky}=0.23$).

Throughout the paper we will retain two significant figures in many of our forecast errors to enable comparison of different experimental configurations. In the summary table we restrict errors to one significant figure.

\section{Large-scale $B$-modes}
\label{sec:bmodes}
In this section we describe the motivation for measuring large-scale $B$-modes (Sec.~\ref{ssec:bb_motivate}), the challenges for measuring them in practice (Sec.~\ref{ssec:tiny_b_modes}), our forecasting machinery (Sec.~\ref{ssec:mr.tools}), and our forecast constraints (Sec.~\ref{ssec:bb_results}), including a discussion of limitations in Sec.~\ref{ssec:bblimits}. 

\subsection{Motivation}
\label{ssec:bb_motivate}
Large-scale $B$-modes offer a unique window into the early universe and the physics taking place at very high energies. Primordial tensor perturbations, propagating as gravitational waves, would polarize the CMB with this particular pattern (\cite{Kamionkowski:1996zd,1997PhRvD..55.1830Z}). Since scalar perturbations generate only primary $E$-mode polarization, the amplitude of the $B$-mode signal provides an estimate of the tensor-to-scalar ratio, $r$. While the scalar perturbations have been well characterized by, e.g.,~\citet{planck2018:parameters}, we so far have only upper limits on the amplitude of tensor perturbations, with $r<0.07$ at 95\% confidence~\citep{bkp,planck2018:parameters} at a pivot scale of $k=0.05$/Mpc.

Theories for the early universe can be tested via their predictions for the tensor-to-scalar ratio, in addition to other properties of the primordial perturbations including the shape of the primordial scalar power spectrum and the scalar spectral index (Sec.~\ref{sec:pk} and Table~\ref{tab:lcdm}), degree of non-Gaussianity (Sec.~\ref{sec:bispec}), and degree of adiabaticity. While late-time probes of structure formation can be used to further characterize the scalar perturbations, the CMB is likely to be the most powerful probe to constrain the tensor-to-scalar ratio on large scales. 

Simple inflationary models can generate tensor perturbations at a measurable level. 
The current upper limit on $r$ already excludes a set of single-field slow-roll inflation models, as illustrated in, e.g.,~\citet{planck2018:parameters,planck2018:inflation}. As discussed in, e.g., \citet{CMBS42016}, there is a strong motivation to further lower the current limits, with certain large-field plateau models predicting an $r\approx0.003$ \citep[e.g.,][]{1980PhLB...91...99S}. 
SO, through its sensitivity, frequency coverage and angular resolution, is designed to be able to measure a signal at the $r=0.01$ level at a few $\sigma$ significance, or to exclude it at similar significance, using the $B$-mode amplitude around the recombination bump at $\ell\approx90$.
A detection of a signal at this level or higher would constitute evidence against classes of inflationary models~\citep{Martin_2014a,Martin_2014b}, e.g., $r\propto1/N^2$ models\footnote{Here $N$ is the number of $e$-folds of inflation.} such as Higgs or $R^2$ inflation \citep{1979JETPL..30..682S,2008PhLB..659..703B}. 
Measurably large tensor perturbations can also be generated by additional time-varying fields during inflation which contribute negligibly to inflation dynamics \citep{2016JCAP...01..041N}.

On the other hand, alternative non-inflationary cosmologies include scenarios in which the big bang singularity is replaced by a bounce -- a smooth transition from contraction to expansion \citep{ijjas/steinhardt:2018}. During the slow contraction phase that precedes a bounce, the universe is smoothed and flattened and nearly scale-invariant super-horizon perturbations of quantum origin are generated that seed structure in the post-bounce universe \citep{levy/etal:2015}. These models are not expected to produce detectable tensor perturbations, and therefore a detection of primordial $B$-modes would allow us to rule them out.

Beyond the tensor-to-scalar ratio, measurements of the large scale polarization signal could be used to explore constraints on the tensor tilt, $n_T$. Although these contraints would be weak even in a case with $r\sim 0.01$ and ~50\% delensing (with projected statistical uncertainty $\sigma(n_T) \approx 0.6$), SO would be able to test large deviations from the consistency relation for simple inflationary models predicting $r = - 8n_T$. 

The data could also be used to test for the presence of non-standard correlations such as non-zero $TB$ and $EB$, generated by early- or late-time phenomena, e.g., chiral gravitational waves and Faraday rotation \citep[see e.g.,][]{polarbear:2015,contaldi:2016,CMBS42016,2016A&A...596A.110P,2017PhRvD..96j2003B}.

\subsection{Measuring $B$-modes}\label{ssec:tiny_b_modes}

Figure~\ref{fig:fg_freq} shows the amplitude of the CMB lensing $B$-modes compared to the two main polarized Galactic contaminants: synchrotron and dust emission. Our goal is to search for a primordial $B$-mode signal that is of the same order or smaller than this lensing signal. Although the level of contamination depends on the sky region, polarized foregrounds are known to have amplitudes corresponding roughly to $r$=$0.01$--$0.1$~\citep{Krach2016} at their minimum frequency ($70$--$90$~GHz). As a function of scale, foreground contamination can be up to two orders of magnitude higher than the CMB $B$-mode power spectrum (see Fig. \ref{fig:clBB_summary}), which itself is dominated by non-primordial lensing $B$-modes over the scales of interest for SO \citep{bkp}. Our forecasts will therefore focus both on $\sigma(r)$, the 1$\sigma$ statistical uncertainty on $r$ after foreground cleaning, as well as on the possible bias on $r$ caused by foreground contamination.

To discriminate between the primordial CMB signal and other sources of $B$-modes, SO will use multi-frequency observations to characterize the spectral and spatial properties of the different components and remove the foreground contribution to the sky's $B$-mode signal. Lensing $B$-modes can be viewed as an additional source of stochastic noise, with an almost-white power spectrum with a $\sim5\,\mu$K-arcmin amplitude on the largest angular scales \citep{hu/okamoto:2002}. Fortunately, this contamination can be partially removed at the map level through a reconstruction of the lensing potential. For SO this reconstruction will be based on external large-scale structure datasets, such as maps of the CIB, as well as on internal CMB lensing maps estimated from the LAT observations. Further details can be found in Sec.~\ref{sec:lensing.delensing}.

While our forecasts in this paper focus on using data from the SATs to clean and estimate the large-scale $B$-modes, the complementarity between the SATs and the LAT will allow us to perform component characterization and subtraction over a wide range of angular scales, adding to the robustness of our foreground cleaning. SO will also provide the community with new high-resolution templates of the Galactic polarized emission (both on its own, and in combination with high-frequency data from balloon-borne experiments or low frequency measurements from other ground-based facilities). Complementary to {\it Planck} data, this will provide valuable information on the characteristics of the dust populations and the properties of the Galactic magnetic field.

\subsection{Forecasting tools}\label{ssec:mr.tools}
Our $B$-mode forecasts are based on a set of foreground cleaning and power spectrum estimation tools. We use the following suite of four foreground cleaning codes:

\begin{itemize}
  \item {\bf Cross-spectrum (`$\bm C_\ell$-MCMC'):} is a method that models the $BB$ cross-spectra between the six frequencies similarly to \cite{Cardoso2008} and \cite{bkp}. The free parameters of the foreground contribution are the dust and synchrotron spectral indices $\{\beta_d, \beta_s\}$, and the amplitude and power-law tilt of their power spectra ($\{A^d_{BB}, A^s_{BB}\}$ and $\{\alpha_d, \alpha_s\}$, respectively). The method fixes the dust temperature to $T_d=19.6\,{\rm K}$, the synchrotron curvature to $C=0$, and explicitly ignores the spatial variation of the other spectral parameters. Finally, the noise bias is modeled by averaging the spectra of noise-only simulations that reproduce both the inhomogeneous sky coverage and the effect of $1/f$ noise. Fifty simulations were used for each case explored.
    
  A Fisher-matrix version of this method ({\bf `$\bm C_\ell$-Fisher'}) was also used to obtain fast estimates of $\sigma(r)$ for a large number of different instrumental configurations and survey strategies. The code marginalizes over a larger set of 11 cosmological and foreground parameters, including $E$-mode and $B$-mode amplitudes $\{A^d_{EE},A^d_{BB},A^s_{EE},A^s_{BB}\}$, and tilts $\{\alpha_s,\alpha_d\}$, spectral parameters $\{\beta_s,\beta_d\}$, a dust-synchrotron correlation parameter $\rho_{\rm ds}$, the lensing amplitude $A_{\rm lens}$ and the tensor-to-scalar index $r$. The results of this method were verified against {\tt xForecast}, and informed some of the main decisions taken during the experiment design stage. 
  \item {\bf xForecast:}~\citep{Stompor2016} is a forecast code that uses a parametric pixel-based component separation method, explicitly propagating systematic and statistical foregrounds residuals into a cosmological likelihood. For this paper, the algorithm has been adapted to handle inhomogeneous noise and delensing. By default, the code fits for a single set of spectral indices $\{\beta_d,\beta_s\}$ over the whole sky region, fixing the dust temperature to $19.6\,$K. The inherent spatial variability of the spectral parameters in the input sky simulations naturally leads to the presence of systematic foreground residuals in the cleaned CMB map, and therefore can produce a non-zero bias in the estimation of $r$. We explore extensions to this method that marginalize over residual foregrounds as described in Sec.~\ref{ssec:results_fid}.
  \item {\bf BFoRe:}~\citep{Alonso2017} is a map-based foreground removal tool that fits for independent spectral parameters for the synchrotron and dust in separate patches of the sky. We use {\tt BFoRe} as an alternative to {\tt xForecast} to explore certain scenarios with a higher degree of realism. These imply running an ensemble of simulations with independent CMB and noise realizations in order to account for the impact of foreground residuals in the mean and standard deviation of the recovered tensor-to-scalar ratio. For these forecasts, 20 simulations were generated for each combination of sky and instrument model.
  \item {\bf Internal Linear Combination:} We also implemented a $B$-mode foreground-cleaning pipeline based on the Internal Linear Combination (ILC) method~\citep{2003ApJS..148....1B}. Our implementation calculates the ILC weights in harmonic space in a number of $\ell$ bands and marginalizes over the residual foregrounds at the power spectrum level in the cosmological likelihood, see Sec.~\ref{ssec:results_fid}. The method is similar to that used in Appendix A of the CMB-S4 CDT report\footnote{\url{https://www.nsf.gov/mps/ast/aaac/cmb_s4/report/CMBS4_final_report_NL.pdf}}. This pipeline was run on 100 sky simulations for each of the cases explored here.
\end{itemize}
All of these methods are based on the same signal, noise and foregrounds models and, except for the $C_\ell$-Fisher method, all use the same foreground simulations. All methods make extensive use of {\tt HEALPix} for the manipulation of sky maps.

For power spectrum estimation we use the following code:
\begin{itemize}
\item{\bf NaMaster:}\footnote{\url{https://github.com/damonge/NaMaster}} a software library that implements a variety of methods to compute angular power spectra of arbitrary spin fields defined on the sphere. We use the code to estimate pure-$B$ power spectra from our simulations. Pure-$B$ estimators \citep{2006PhRvD..74h3002S,Grain2009} minimize the additional sample variance from the leakage of $E$ modes due to the survey footprint and data weighting.
\end{itemize}

Finally, we note that, for simplicity, our forecasts assume a Gaussian likelihood for the $B$-mode power spectrum over the scales probed ($30\leq\ell\leq300$). This is not guaranteed to be a valid assumption given the reduced sky fraction and possible filtering of the data, and should be replaced in the future by, for example, the more accurate likelihood of \cite{2008PhRvD..77j3013H}.

\subsection{Forecast constraints}
\label{ssec:bb_results}

This section discusses the impact of the large-area scanning strategy pursued by the SATs, and our fiducial forecasts for $\sigma(r)$ after component separation. We also explore departures from the fiducial hypotheses, including different assumptions about delensing, the fiducial $r$ model, and the foreground models.

\subsubsection{Impact of large-area scanning strategy}
\begin{figure}[t!]
  \centering
  \includegraphics[width=\columnwidth]{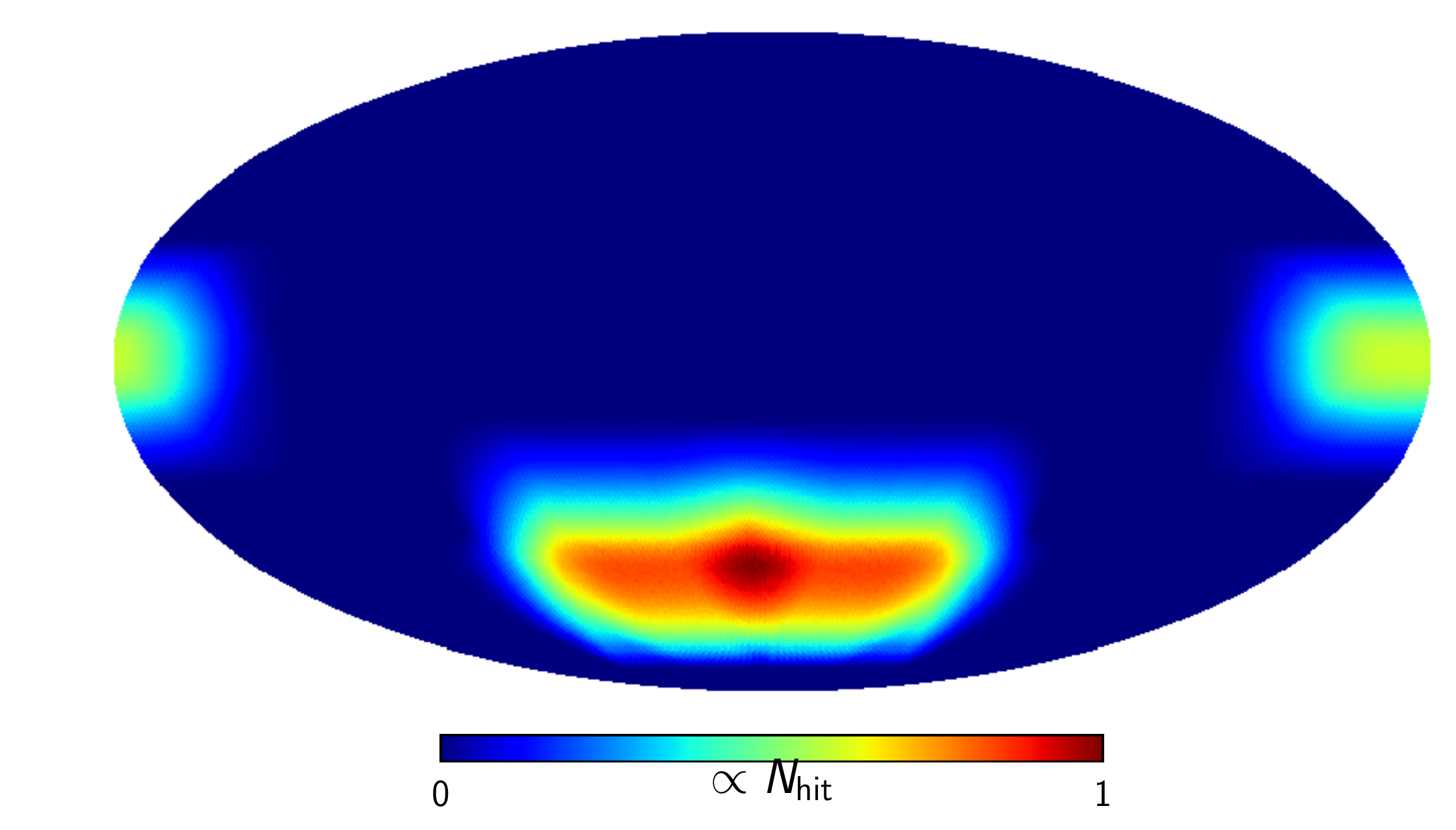}
  \caption{Simulated map of hit counts in Equatorial coordinates for the SATs, resulting from the preliminary scan strategy considered in this study. The $N_{\rm hit}$ is proportional to the amount of time anticipated to be spent observing each pixel. Regions of high Galactic emission are avoided.\\} \label{fig:nhits}
\end{figure}
The combination of the anticipated scanning strategy with the large field-of-view of the SATs gives rise to a sky coverage with broad depth gradients around a small effective area of $A\simeq 4000$ square degrees, as shown in Fig. \ref{fig:nhits}. A more compact sky coverage would arguably be more optimal for $B$-mode searches. However, $B$-modes cannot be measured locally, and on a cut sky some signal is lost near the patch boundaries, which could have a significant impact on the signal-to-noise obtained from a compact sky mask. This is probably negligible, however, for the broad area covered by this scanning strategy, given that the shallower regions must also be down-weighted in an inverse-variance way (i.e., with a window function proportional to the local hit counts in the simplest case). It is therefore important to assess the level to which this choice of survey strategy and field of view could degrade the achievable constraints on the tensor-to-scalar ratio with SO.

To do this we have considered two different types of survey window functions: the one described above and a window function with homogeneous depth over a circular patch with the same hit counts. For each of them, we generate 1000 sky simulations containing only the CMB signal with $r=0$ and lensing amplitude $A_{\rm lens}=1,\,0.5$ and $0.25$ (see definition in Eq.  \ref{eq:mr.rlike}), and non-white noise corresponding to both the baseline and goal noise models described in Sec.~\ref{subsec:sensitivity}. For all simulations using the SO small-aperture window function, the homogeneous noise realizations are scaled by the local $1/\sqrt{N_{\rm hits}}$ to account for the inhomogeneous coverage. In all cases we explore the impact of the additional apodization required by the pure-$B$ $C_\ell$ estimator \citep{2006PhRvD..74h3002S} for different choices of apodization scale. In particular, we use the $C^2$ apodization scheme described in \cite{Grain2009} with $5^\circ$, $10^\circ$, and $20^\circ$ apodization scales. For each combination of $A_{\rm lens}$, noise level, window function and apodization we estimate the associated uncertainty in the $B$-mode power spectrum as the scatter of the estimated power spectrum in the 1000 realizations.

\begin{figure}
  \centering
  \includegraphics[width=\columnwidth]{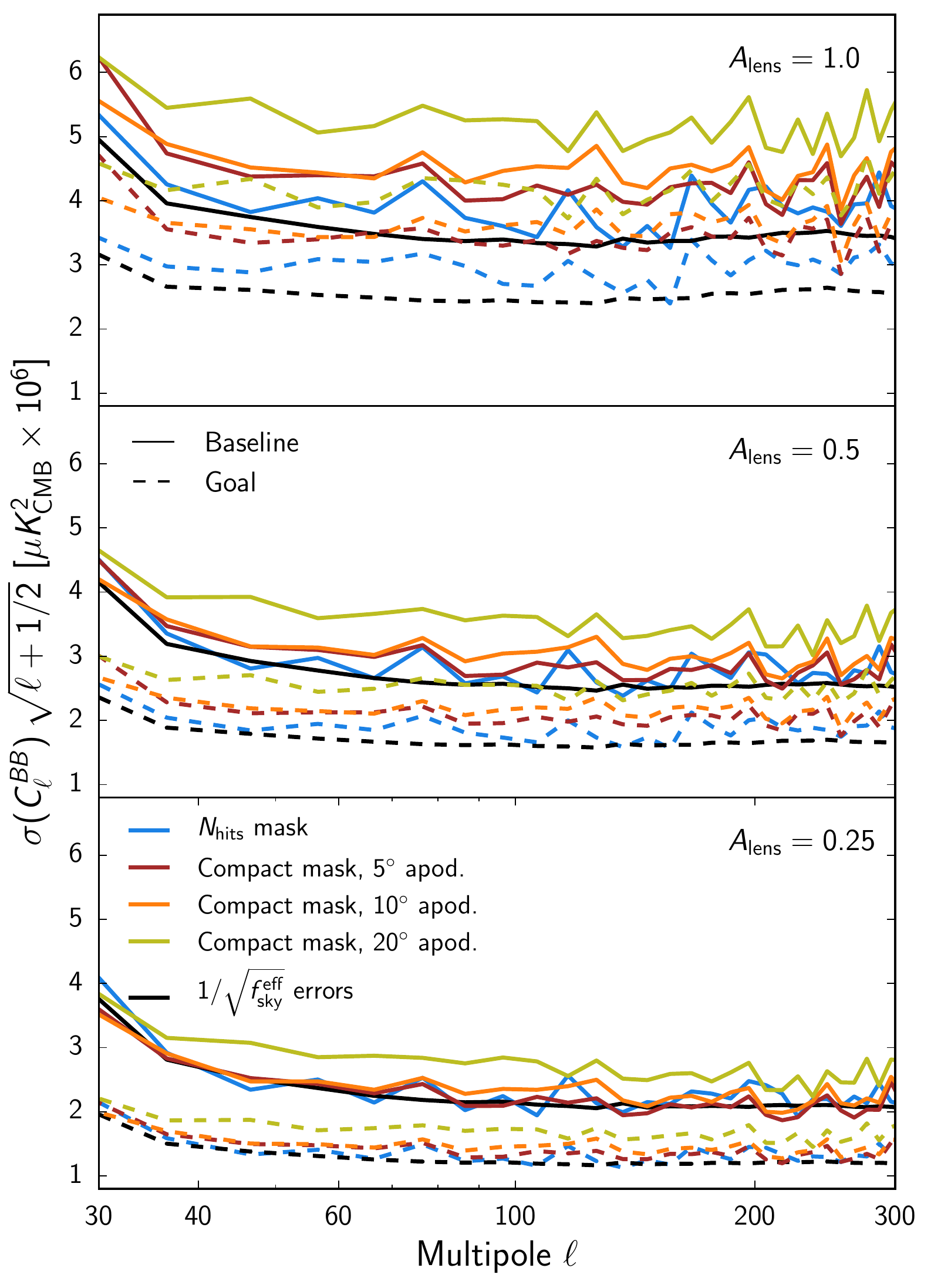}
  \caption{Impact of field of view and scanning strategy on the $B$-mode power spectrum ($C^{BB}_\ell$) uncertainties. The panels show increasing levels of delensing (top to bottom), parametrized by the lensing amplitude $A_{\rm lens}$, for the SO baseline and goal noise levels described in Table \ref{tab:noise}. We compare two different sky masks: the fiducial SO footprint in Fig.~\ref{fig:nhits} with $1/N_{\rm hits}$ weighting and additional $10^\circ$ tapering, compared to a compact circular mask with the same effective sky area and three different levels of apodization. Errors using the fiducial footprint closely match the mode-counting errors (labeled $1/\sqrt{f_{\rm sky}^{\rm eff}}$) achievable with this sky fraction.}\label{fig:r_fov}
\end{figure}
The results are shown in Fig.~\ref{fig:r_fov}: in all cases we find that, within the range of multipoles relevant to the SO SATs ($30\lesssim\ell\lesssim 300$), the uncertainties associated with our fiducial window function are equal or smaller than the uncertainties corresponding to the compact mask, regardless of the apodization scale, and that the apodization scale is mostly irrelevant for this fiducial window function. Furthermore, the uncertainties obtained for the fiducial window function are remarkably close to the `mode-counting' error bars $\sigma(C_\ell)=(C_\ell+N_\ell)/\sqrt{(\ell+1/2)f_{\rm sky}^{\rm eff}\Delta \ell}$~\citep{Knox:1995dq}, where $C_\ell$ and $N_\ell$ are the signal and noise power spectra, $\Delta \ell$ is the bandpower bin width and $f^{\rm eff}_{\rm sky}$ is the effective sky fraction associated with the window function, given by $f^{\rm eff}_{\rm sky}=\langle N_{\rm hits}\rangle^2/\langle N_{\rm hits}^2\rangle$\footnote{Note that this definition of $f_{\rm sky}^{\rm eff}$ is appropriate to quantify the variance of the power spectra computed from noise-dominated maps. For the hit counts map shown in Fig.~\ref{fig:nhits}, this is $f_{\rm sky}^{\rm eff}=0.19$. For signal-dominated maps this hit counts maps has $f_{\rm sky}^{\rm eff}=0.1$.}.

We therefore conclude that, in the absence of foreground systematics, assuming $r=0$ and given the achievable sensitivity, the sky coverage that results from the scanning strategy and instrumental field-of-view has a sub-dominant effect on the achievable constraints on primordial $B$-modes for SO. Throughout the rest of this section we use this sky coverage in our simulations to assess the impact of foregrounds.

Before we move on, we should also note that some of the results shown here are specific of the power spectrum estimator used (pseudo-$C_\ell$ with $B$-mode purification), and more optimal results could be obtained with brute-force $E/B$ projection or quadratic estimators (e.g., \citealp{2001PhRvD..64f3001T,2003PhRvD..68h3509L}).

\subsubsection{Fiducial forecasts for $r$}\label{ssec:results_fid}

\begin{table*}[htp!]
\centering
{
\renewcommand{\arraystretch}{1.2}
\caption{Forecasts for $r=0$ model using six different cleaning methods, for fiducial foreground model$^a$}
\newcolumntype{C}{ @{}>{${}}c<{{}$}@{} }
\begin{tabular}{ll| *4{rCl}}
\hline \hline
& & \multicolumn{6}{c}{SO Baseline} & \multicolumn{6}{c}{SO Goal}\\
& Method & \multicolumn{3}{c}{\quad pess-$1/f$} & \multicolumn{3}{c}{\quad opt-$1/f$} & \multicolumn{3}{c}{\quad pess-$1/f$} & \multicolumn{3}{c}{\quad opt-$1/f$} \\
\hline
$A_{\rm lens}=1$ & $C_\ell$-Fisher & $\sigma$&$=$&$2.4$ & $\sigma$&$=$&$1.9$ & $\sigma$&$=$&$1.7$ & $\sigma$&$=$&$1.5$ \\
& $C_\ell$-MCMC & $1.9$ & \pm & $2.6$ & $2.3$ & \pm & $2.3$ & $2.2$ & \pm & $2.1$& $2.4$ & \pm & $2.1$\\
&{\tt xForecast} & $1.3$ & \pm & $2.7$ & $1.6$ & \pm & $2.1$ & $1.4$ & \pm & $1.9$ & $1.6$ & \pm & $1.6$\\
&{\tt xForecast}\tablenotemark{b} & $0.0$ & \pm & $4.0$ & $0.0$ & \pm & $3.5$ & $0.0$ & \pm & $3.3$ & $0.2$ & \pm & $2.8$\\
&{\tt BFoRe}\tablenotemark{b} & $-0.5$ & \pm & $5.8$ & $-0.5$ & \pm & $3.6$ & $-0.6$ & \pm & $4.3$ & $-0.5$ & \pm & $3.4$\\
&ILC\tablenotemark{b} & $-0.4$ & \pm & $3.9$ & $-0.3$ & \pm & $3.1$ & $-0.2$ & \pm & $3.9$ & $-0.3$ & \pm & $3.0$\\
\hline
$A_{\rm lens}=0.5$ &$C_\ell$-Fisher & $\sigma$ & $=$ & $1.8$ & \bm{$\sigma$} & \bm{=} & \bm{$1.4$} & $\sigma$ & $=$ & $1.2$ &$\sigma$ & $=$ & $0.9$\\
&$C_\ell$-MCMC & $1.7$ & \pm & $2.1$ & \bm{$2.2$} & \bm{\pm} & \bm{$2.0$} & $2.0$ & \pm &$1.7$ & $2.2$& \pm &$1.7$\\
&{\tt xForecast} & $1.3$ &  \pm & $2.1$ & \bm{$1.6$} & \bm{\pm} & \bm{$1.5$} & $1.3$& \pm &$1.3$ & $1.5$& \pm &$1.0$\\
&{\tt xForecast}\tablenotemark{b} & $0.1$ & \pm & $3.2$ & $\bm{0.1}$& \bm{\pm} & \bm{$2.6$} & $0.0$ & \pm & $2.5$ & $0.3$& \pm &$1.8$\\
&{\tt BFoRe}\tablenotemark{b}  & $-0.2$ & \pm & $5.0$ &  $\bm{-0.4}$& \bm{\pm} &\bm{$2.6$} & $-0.6$ & \pm & $3.2$ & $-0.5$ & \pm & $2.0$\\
&ILC\tablenotemark{b}  & $-0.3$ & \pm & $3.0$ & \bm{$-0.3$} & \bm{\pm} & \bm{$2.4$} & $-0.1$ & \pm & $2.8$ & $-0.2$ & \pm & $2.3$\\
\hline
\hline
\label{table:r0}

\end{tabular}
}
\begin{tablenotes}
\item \textsuperscript{a} 
Table gives $(r\pm\sigma(r))\times10^3$ for different analysis pipelines (different rows), and different noise configurations (different columns, see Sec.~\ref{subsec:sensitivity} for details.). The results for the fiducial lensing and noise combination are highlighted in boldface.
\item \textsuperscript{b} In these cases a foreground residual is additionally marginalized over after map-based cleaning.
\end{tablenotes}
\end{table*}

We generate forecasts for the different SO instrument configurations described in Sec.~\ref{subsec:sensitivity} and shown in Table~\ref{tab:noise}. At the likelihood level, all our calculations assume that the $B$-mode power spectrum can be described as a sum of two contributions, corresponding to primordial tensor perturbations and lensing $B$-modes respectively
\begin{equation}\label{eq:mr.rlike}
  C^{BB}_\ell=r\,C^{\rm tensor}_{\ell}+A_{\rm lens}C_{\ell}^{\rm lensing},
\end{equation}
where $C_{\ell}^{\rm tensor}$ and $C_{\ell}^{\rm lensing}$ are templates for the unlensed, $r=1$ $B$-mode power spectrum and the lensed scalar $B$-mode power spectrum respectively\footnote{We assume a scale-invariant spectrum of tensor modes, and define $r$ at the pivot scale $k=0.005\,{\rm Mpc}^{-1}$. Our fiducial cosmology has $A_s=2.4\times10^{-9}$ at that pivot scale, which is $\sim4\%$ higher than the \planck{} 2018 cosmology~\citep{planck2018:parameters}. Our lensing template is $\sim2\%$ higher at low $\ell$ than predicted by \planck{} 2018 with power $C_{\ell=100}^{\rm lensing}=2.03\times10^{-6}\,\mu{\rm K}^2$. We expect these differences in fiducial cosmology to have a smaller impact on our results than other assumptions made here (e.g., on foreground modeling).}. Thus we assume that the effects of delensing can be encapsulated into a single parameter $A_{\rm lens}$ parametrizing the amplitude of the lensing contribution relative to the expected one without delensing for the fiducial cosmological parameters. In the nominal case we present results for $A_{\rm lens}=1$ (no delensing) and $A_{\rm lens}=0.5$, corresponding to the delensing efficiency that is expected to be achievable with external maps of the large-scale structure or the CIB~\citep{2017PhRvD..96l3511Y}. $A_{\rm lens}=0.5$ could also optimistically be achieved by internally delensing with SO data, as described in Sec.~\ref{sec:lensing}.

The achievable constraints on the tensor-to-scalar ratio depend, to some extent, on the choice of analysis method.\\

\begin{figure}
\centering
\includegraphics[width=\columnwidth]{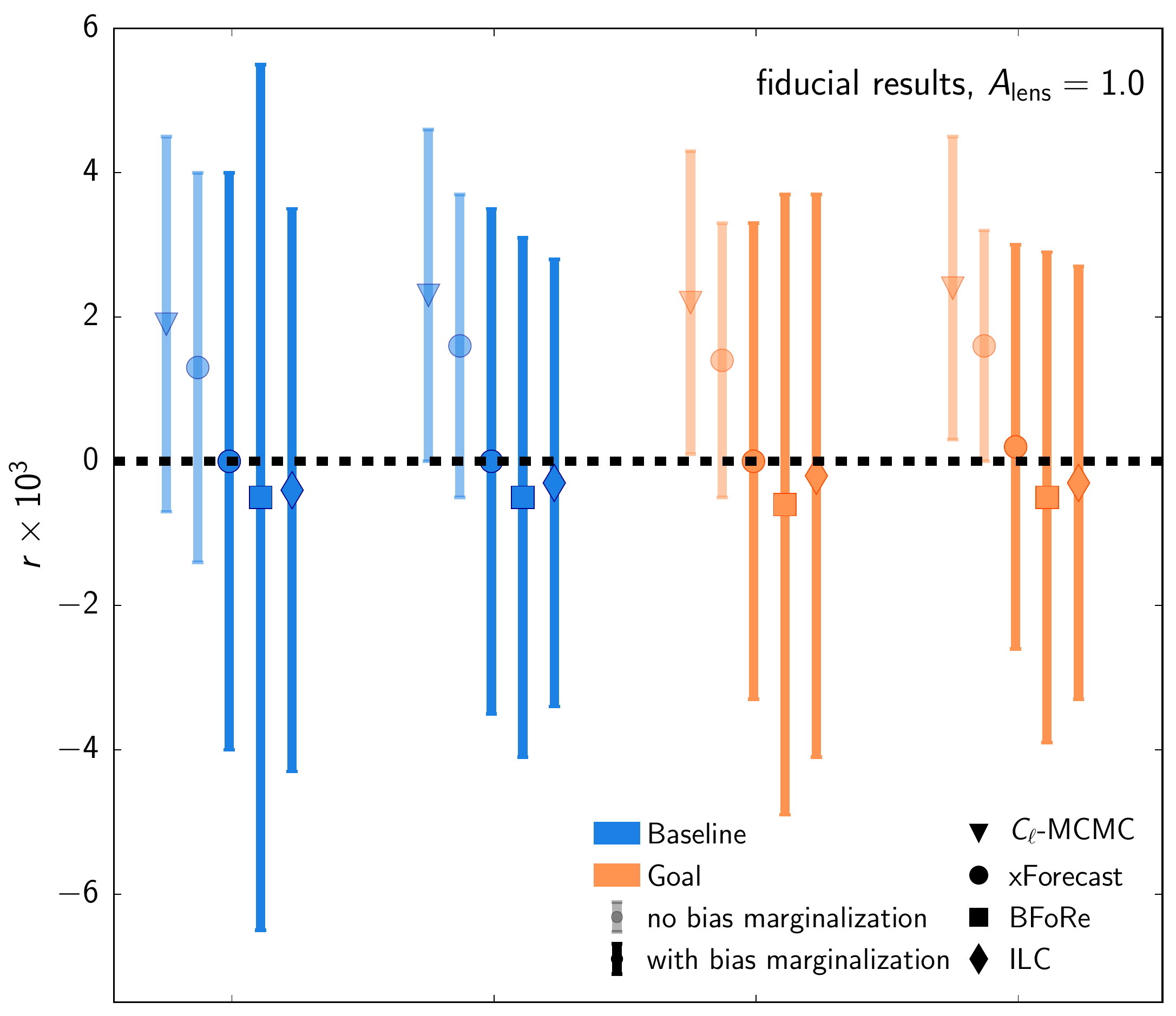}
\includegraphics[width=\columnwidth]{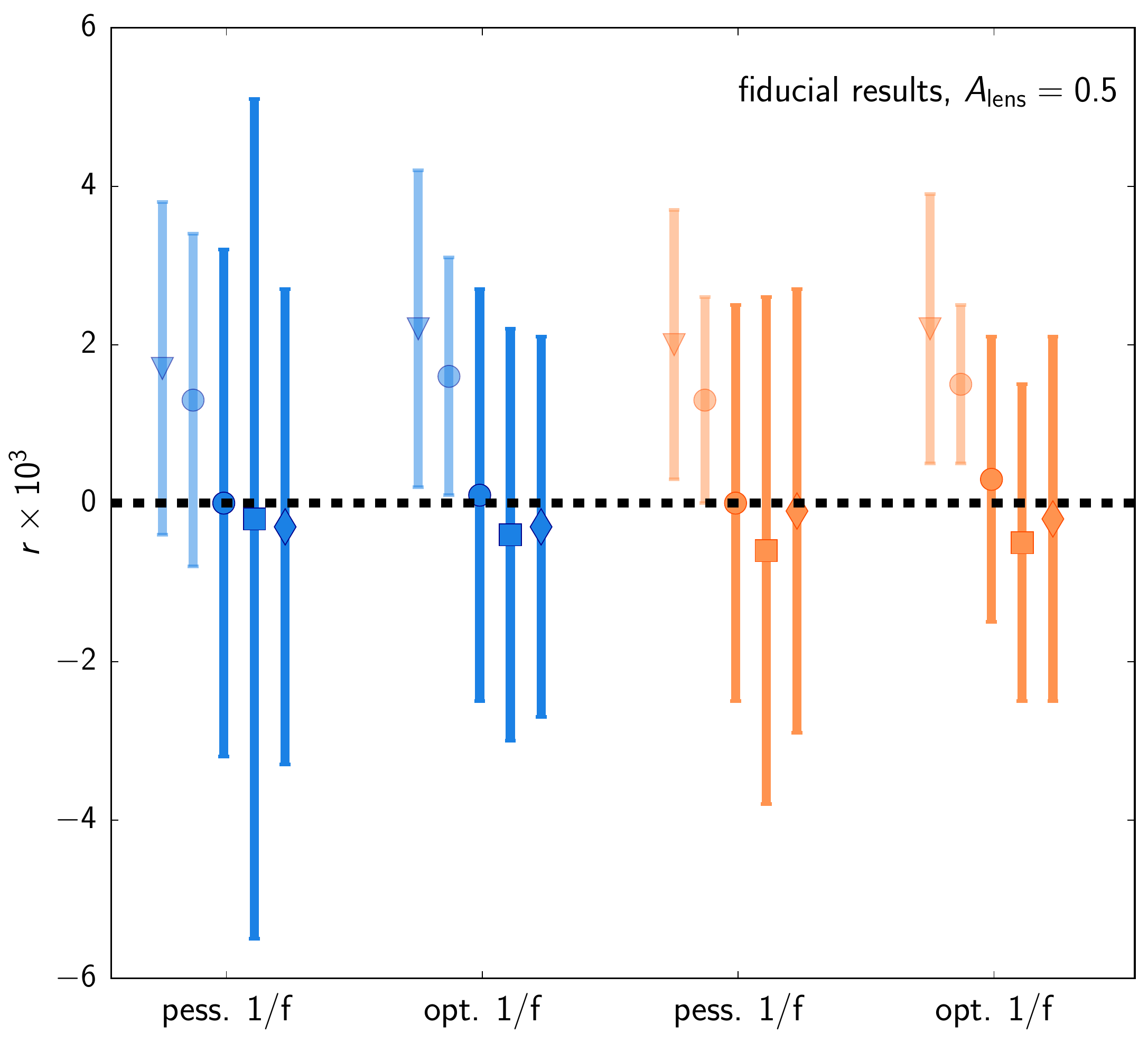}
\caption{Estimated $r$ and 1$\sigma$ uncertainties for simulations with $r=0$ and $A_{\rm lens}=1$ (top) and $0.5$ (bottom). The fiducial sky model has single dust and synchrotron components with spatially varying spectral indices, described in Sec~\ref{sssec:4cast.fgs.pol}. We show baseline and goal noise levels, with either pessimistic or optimistic $1/f$ scenarios, for the $C_\ell$-MCMC and {\tt xForecast} methods, without residual marginalization (translucent), and for the {\tt xForecast}, {\tt BFoRe} and ILC methods that additionally marginalize over foreground residuals at the power spectrum level (solid).}
\label{fig:fiducial_results_on_r}
\end{figure}

{\bf $\bm C_\ell$ method:} The simplest scenario corresponds to the $C_\ell$ cross-spectrum approach. These forecasts are shown in Table \ref{table:r0} for the $C_\ell$-Fisher method. This optimistic method predicts a sensitivity to primordial $B$-modes with $\sigma(r)\simeq 1$--$2\times10^{-3}$, depending on the different noise and delensing assumptions. We run the same pipeline on the 50 simulations and compute the bias and uncertainty on $r$ as its mean and standard deviation over the simulations set ($C_\ell$-MCMC of Table \ref{table:r0} and Fig.~\ref{fig:fiducial_results_on_r}). This case yields similar, although slightly higher values of $\sigma(r)$. However, the simple parametric model assumed by this method, with constant spectral indices across the full footprint (see Fig.~\ref{fig:nhits}), gives rise to foreground contamination and an associated bias on $r$. We find that this bias is of the order of the 1$\sigma$ error ($\sim2\times10^{-3}$) for our standard foreground configuration, and significantly larger for simulations with more highly-varying synchrotron spectral indices (described in the next subsection). \\

{\bf xForecast, BFoRe and ILC methods:} Motivated by these results, we explore more sophisticated, map-based cleaning methods with {\tt xForecast}, {\tt BFoRe} and the ILC pipeline. We analyze the maps in two steps:
\begin{enumerate}
 \item After foreground cleaning, an estimate of the $B$-mode power spectrum is obtained from the foreground-cleaned map. This will contain contamination from statistical and systematic foreground residuals.
 \item We then marginalize over residual foreground contamination at the likelihood level. To do so, we add a term of the form $A_{\rm FG}^TC^{\rm FG}_\ell$ to the model in Eq.~\ref{eq:mr.rlike}, where $C^{\rm FG}_\ell$ is a vector of templates for the expected foreground contamination from different components and $A_{\rm FG}$ is a vector of free amplitudes\footnote{{\tt xForecast} uses residual templates built as $C_\ell^{\rm FG}\equiv  \mathbf{W}^T \mathbf{F}_\ell \mathbf{W}$, where $ \mathbf{W}$ is the operator projecting sky templates onto frequency maps, and $\mathbf{F}_\ell$ the matrix containing all frequency cross spectra for that component. The ILC pipeline uses a similar approach, where $\mathbf{W}$ is constructed from the ILC weights. {\tt BFoRe} uses a single template built from the power spectrum of the foreground residual map obtained from the simulations.}.
\end{enumerate}
The marginalization over foreground residuals at the power spectrum level will reduce the foreground bias at the cost of degrading the final uncertainty on $r$. We therefore also compare the results before and after this additional marginalization, to provide a rough estimate of the level of foreground contamination in the cleaned maps.

These forecasts are summarized in Table \ref{table:r0}. For {\tt xForecast}, the map-based pipeline is able to recover values of $\sigma(r)$ similar to those predicted by the Fisher matrix, although foreground residuals cause a bias at the level of $r=2\times10^{-3}$, or equivalently at the $1\sigma$ level.

\begin{figure}
\centering
\includegraphics[width=\columnwidth]{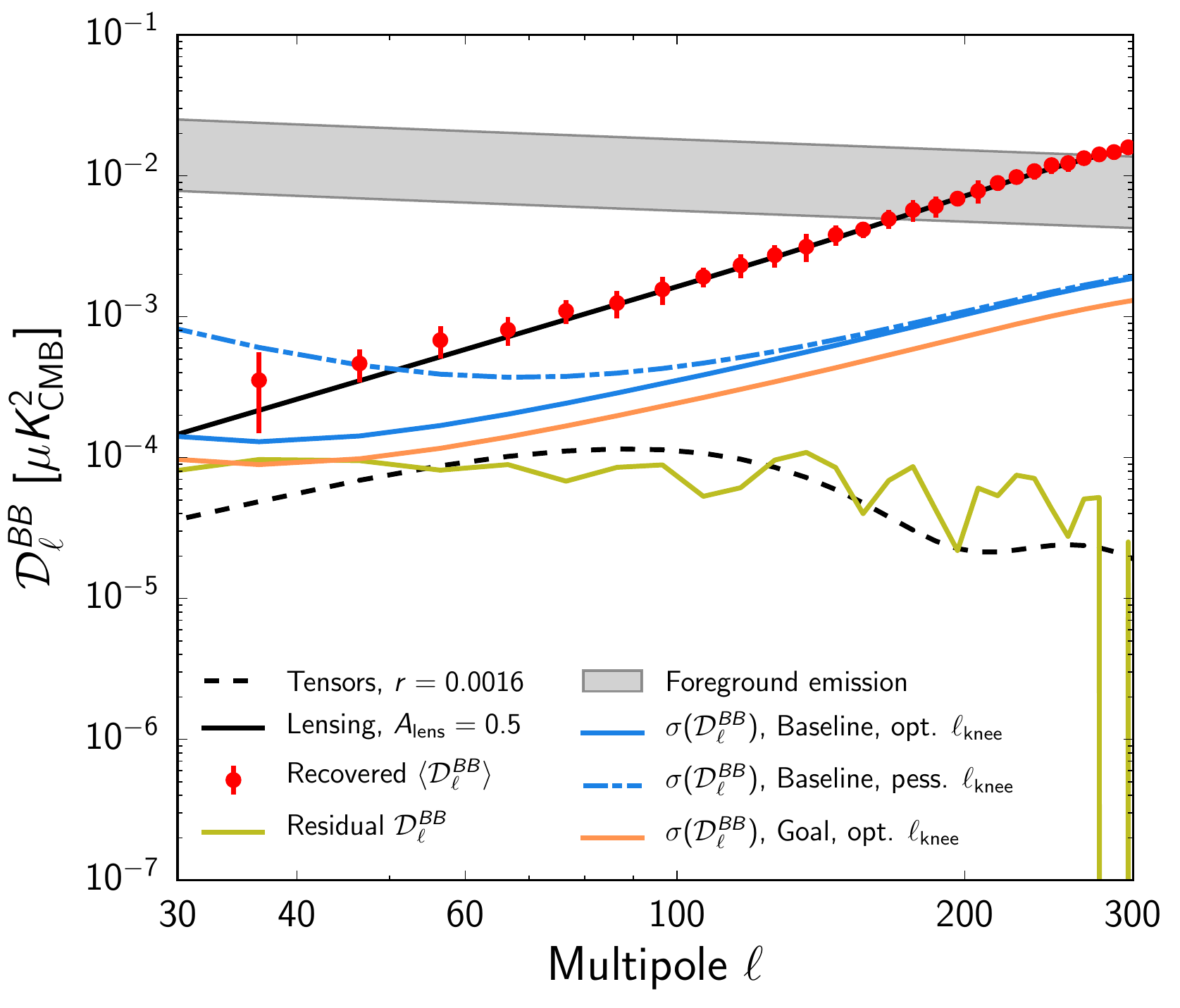}
\caption{Mean $B$-mode power spectrum, $D^{BB}_\ell$, estimated from simulations using map-level component separation (red) with errors from 100 realizations. The input spectrum (black solid) has $r=0$ and assumes 50\% delensing. The power of the total foreground emission is shown shaded (between 93~GHz, lower, and 145 GHz, upper). Power spectrum uncertainties for $\Delta\ell=10$ bandpowers are shown (blue, orange) for different SO noise configurations. The contribution of foreground residuals to the recovered $D^{BB}_\ell$ (yellow) biases the red circles above the input and is comparable to a signal with tensor-to-scalar ratio $r=0.0016$ (dashed). This bias can be suppressed by marginalizing over the foreground residuals in the likelihood.}
\label{fig:clBB_summary}
\end{figure}

We show the $B$-mode power spectrum of these foreground residuals, as computed by {\tt BFoRe}, in Fig.~\ref{fig:clBB_summary}, together with the average power spectrum measured from the simulated and foreground-cleaned maps before residual foreground marginalization. The foreground residuals are comparable to a primordial signal with $r=0.0016$, also shown for reference. We find that the foreground bias can be substantially reduced by the final marginalization over foreground residuals, at the cost of a noticeable increase in the final uncertainties. In this more conservative scenario, our forecasts predict that SO will be able to measure $r$ to the level of $2$--$3\times10^{-3}$, with the potential to rule out an $r=0.01$ model at the $3\sigma$ level. The statistical error bars on $\mathcal{D}_\ell^{BB}$ are shown in Fig.~\ref{fig:allspecs}.

To verify these results, we repeat these forecasts using the more realistic approaches of {\tt BFoRe} and the ILC method, that include the effects of $E$-to-$B$ leakage. The results, after residual foreground marginalization, are also reported in Table \ref{table:r0} and shown in Fig.~\ref{fig:fiducial_results_on_r}. These results agree well in most cases with the values predicted by {\tt xForecast}. For example we forecast
\ba
r &=& (+0.1 \pm 2.6) \times 10^{-3}  \quad {\tt xForecast}, \nonumber\\
r &=& (-0.4 \pm 2.6)  \times 10^{-3}\quad  {\tt BFoRE}, \nonumber\\
r &=& (-0.3 \pm 2.4)  \times 10^{-3} \quad  {\rm ILC}, 
\ea
for the baseline noise case, with $A_{\rm lens}=0.5$.
This agreement of methods increases our confidence in the ability of SO to separate the components to reach this level.

The approach described here of map-based foreground removal, combined with additional residual marginalization in the $B$-mode spectrum, provides an estimate of the anticipated errors due to foreground uncertainty. In practice, though, we will not know what the foreground residuals look like. We therefore anticipate using other strategies to mitigate the impact of foreground residuals. This can be done by increasing the complexity of the foreground model, either fitting for individual spectral parameters over smaller pixels in the sky \citep{Alonso2017}, or using a moment expansion to account for spatial variations of foreground SEDs, both at the map and power-spectrum levels \citep{2017MNRAS.472.1195C}.

Another approach would be to avoid the areas of larger foreground contamination. Using the same simulations, but focusing the analysis on the cleanest 5\% of the sky, we are able to limit the residual bias to the level of $r<10^{-4}$ without additional marginalization, at the cost of larger uncertainties ($\sigma(r)=0.003$ for baseline noise and $A_{\rm lens}=0.5$) due to the area loss. Finally, the robustness of any constraints on $r$ will ultimately be verified by exploring the dependence of the measured cosmological signal on different data cuts, both spatial and in frequency.

\begin{table}
\centering
{
\renewcommand{\arraystretch}{1.2}
\caption{Forecasts exploring departures from the fiducial case$^a$ \label{table:depart}}
\newcolumntype{C}{ @{}>{${}}c<{{}$}@{} }
\begin{tabular}{l| *4{rCl}}
\hline \hline
Method & \multicolumn{3}{c}{Fiducial} & \multicolumn{3}{c}{$r=0.01$} & \multicolumn{3}{c}{2-dust} & \multicolumn{3}{c}{High-res} \\
 & \multicolumn{3}{c}{} & \multicolumn{3}{c}{} & \multicolumn{3}{c}{$+$AME} & \multicolumn{3}{c}{$\beta_s$}\\
\hline
$C_\ell$-Fisher & $\sigma$ &$=$&$1.4$  & $\sigma$&$=$&$1.8$ & $\sigma$&$=$&$1.4$ & $\sigma$&$=$&$1.4$\\
$C_\ell$-MCMC & $2.2$&\pm&$2.0$ & $12$&\pm&$2.3$ & $1.5$&\pm&$1.9$ & $5.3$&\pm&$2.0$\\
{\tt xForecast}\tablenotemark{b} & $0.1$&\pm&$2.6$ & $9.9$&\pm&$3.3$ & $0.5$&\pm&$3.1$& $0.0$&\pm&$2.7$\\
{\tt BFoRe}\tablenotemark{b} & $-0.4$&\pm&$2.6$ & $9.5$&\pm&$3.2$ & $-0.3$&\pm&$2.4$ & $-0.4$&\pm&$2.7$\\
ILC\tablenotemark{b} & $-0.3$&\pm&$2.4$ & $9.7$&\pm&$2.8$& $-0.3$&\pm&$2.5$ &$-0.5$&\pm&$3.5$\\
\hline
\hline
\end{tabular}
}
\begin{tablenotes}
\item \hspace{-0.4cm} \textsuperscript{a} 
Table gives $(r\pm\sigma(r))\times10^3$ for different analysis pipelines (different rows), and different departures from the fiducial case (an $r=0.01$ model and two alternative foreground models).
\item \hspace{-0.4cm} \textsuperscript{b} Marginalized over foreground residual after cleaning.
\end{tablenotes}
\end{table}

\vspace{3pt}
\subsubsection{Departures from the fiducial case}\label{ssec:mr.non_fiducial}
Fixing the instrumental noise properties to the `optimistic' $\ell_{\rm knee}$ and baseline white noise case, we explore a set of departures from the fiducial scenario: increasing the foreground complexity, exploring the impact of delensing, and testing the impact of varying the low-frequency angular resolution.

\begin{figure}
\centering
\includegraphics[width=\columnwidth]{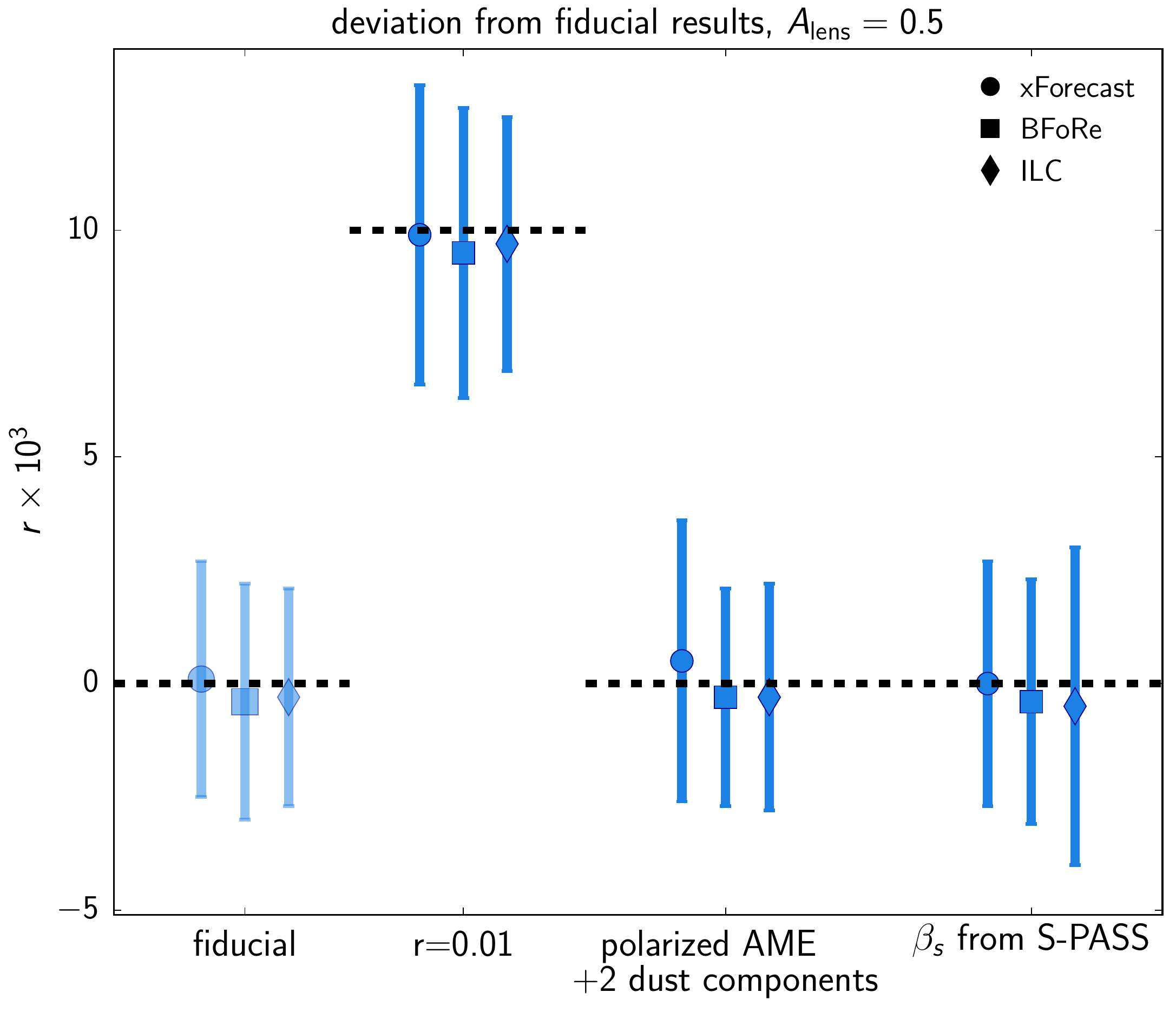}
\caption{As in Fig.~\ref{fig:fiducial_results_on_r}, for the cases deviating from the fiducial forecasts. The `fiducial' points match the second panel of Fig.~\ref{fig:fiducial_results_on_r} for baseline sensitivity and optimistic $1/f$. The three other cases assume $r=0.01$ in the input sky simulations (left), $r=0.0$ with 2 dust components and polarized AME (middle), and $r=0.0$ with synchrotron scaling based on a high-resolution $\beta_s$ template (right). These models are described in Sec.~\ref{sssec:4cast.fgs.pol} with forecasts  in Table \ref{table:depart}.}
\label{fig:departure_from_fiducial_results_on_r}
\end{figure}

\paragraph{Foreground complexity.} Beyond the `standard' foreground simulations used in the previous section, we repeat our forecasts using two more complex foreground models (the `2 dust $+$ AME' and `High-res. $\beta_s$' models described in Sec.~\ref{sssec:4cast.fgs.pol}). The projections for these two cases are reported in Table \ref{table:depart} and shown in Fig.~\ref{fig:departure_from_fiducial_results_on_r}. In both cases we observe that, while the more complex models lead to a larger $r$ bias before residual marginalization, particularly in the case of high-resolution $\beta_s$, marginalization over residuals  mitigates this bias without a significant loss of sensitivity when compared with our fiducial foreground simulations. 

\begin{figure}
\centering
\includegraphics[width=\columnwidth]{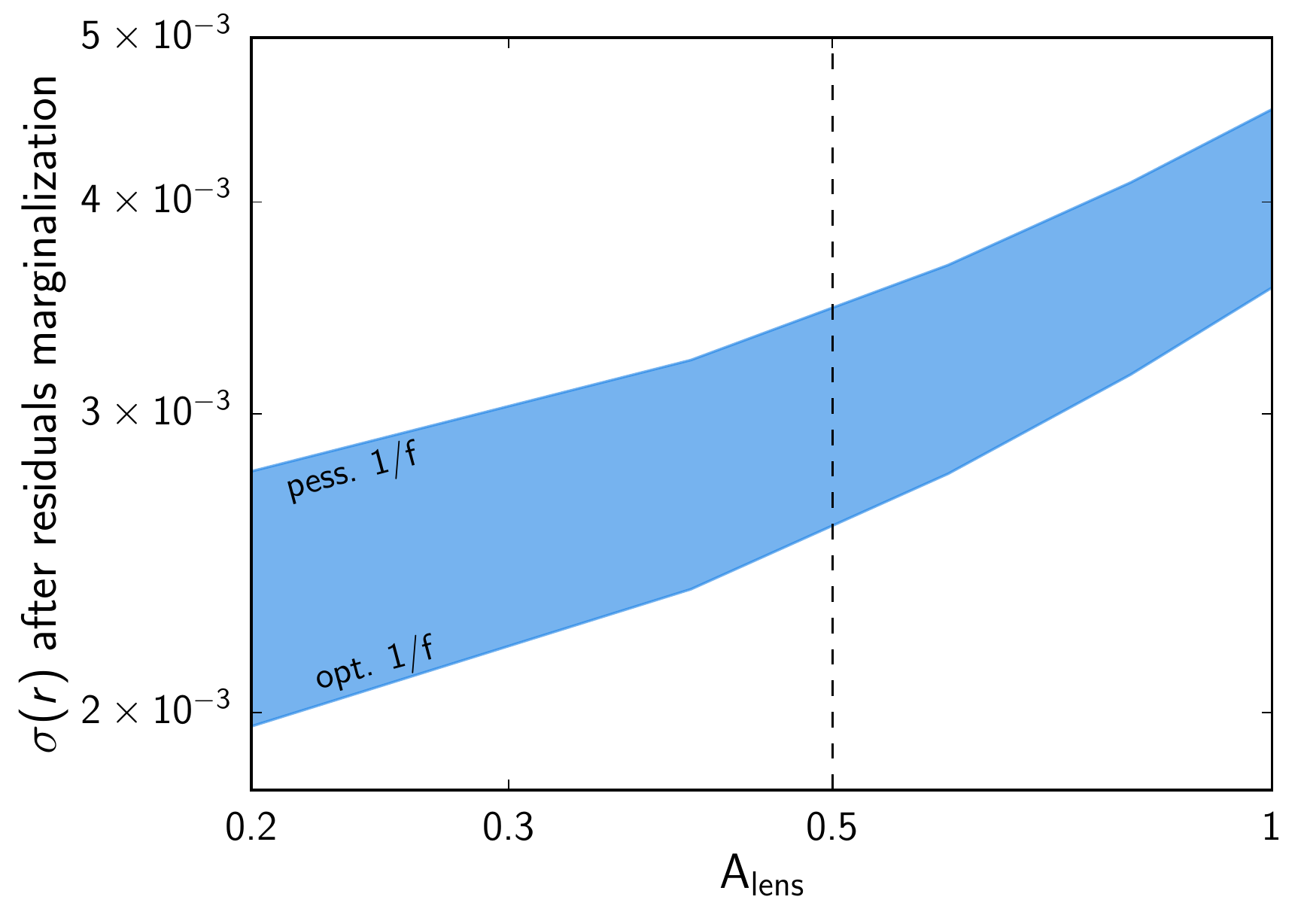}
\includegraphics[width=\columnwidth]{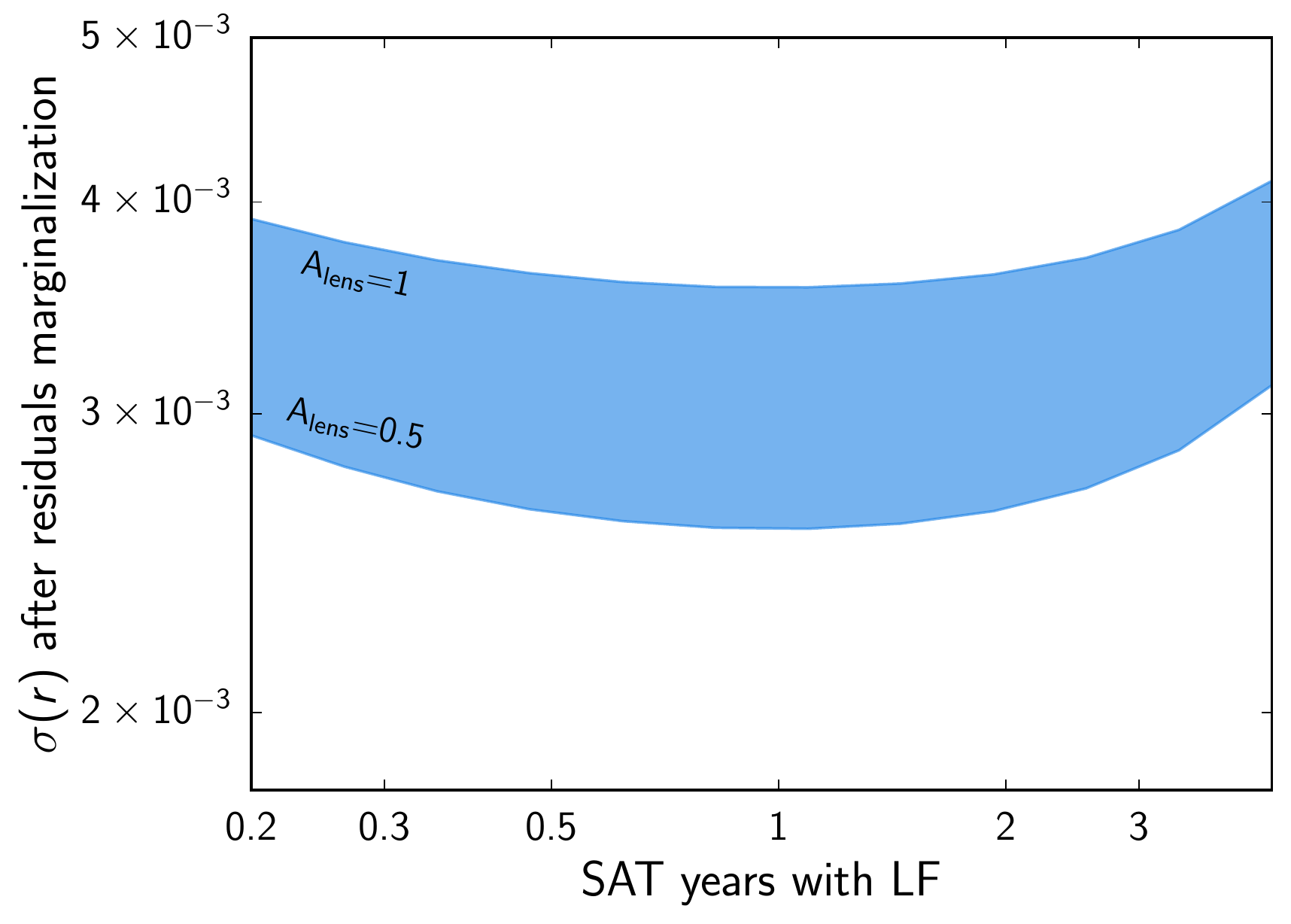}
\caption{{\sl Top:} forecast $\sigma(r)$  as a function of the residual lensing amplitude $A_{\rm lens}$, for an $r=0$ signal. The band corresponds to SO baseline between the optimistic and pessimistic $1/f$ noise models. We expect $A_{\rm lens}=0.5$ from external delensing and $0.7$ from internal delensing (this assumes the LAT covers $f_{\rm sky}=0.4$; focusing all of the LAT time on the $f_{\rm sky}=0.1$ SAT region would give $A_{\rm lens}=0.5$ from internal delensing -- not a significant gain). {\sl Bottom:} forecast $\sigma(r)$ as a function of the time devoted to low-frequency (LF) observations, at the expense of the mid-frequency channels. The band corresponds to $0.5\leq A_{\rm lens}\leq 1.0$. We find one year of LF observations to be optimal.}
\label{fig:r_sigma_r_vs_SACLF_and_Alens}
\end{figure}

\paragraph{Delensing.} Our fiducial forecasts only explore two possible scenarios regarding delensing: no delensing or 50\% delensing, achievable with external CIB and large-scale structure data. In the upper panel of Fig.~\ref{fig:r_sigma_r_vs_SACLF_and_Alens} we show how $\sigma(r)$ is expected to improve as a function of the lensing amplitude $A_{\rm lens}$. We find that the expected reduction of $\sigma(r)$ in lowering $A_{\rm lens}$ below $0.5$ is not substantial for the SO noise levels considered here. Since $A_{\rm lens}=0.5$ is expected to be achievable using external tracers of the lensing potential, we do not find strong motivation to optimize the LAT survey for internal iterative delensing (described in Sec.~\ref{sec:lensing}).

Specifically, we find that if we were to focus all of the LAT survey time on the 10\% of sky overlapping with the SATs, we could achieve $A_{\rm lens}=0.52~(0.42)$ for baseline (goal) noise level from internal delensing. This would give a negligible improvement in $\sigma(r)$ compared to the $A_{\rm lens}=0.5$ achievable from external delensing. Our nominal case, with the LAT surveying $f_{\rm sky}=0.4$, projects $A_{\rm lens}=0.71~(0.62)$. This is comparable to external delensing levels, and will provide a useful cross-check. Furthermore, the projected SO measurement of $r$ is not strongly lensing-limited. With no delensing, we project that errors would increase from $\sigma(r)=0.0026$ to $0.0035$ for the {\tt xForecast} method; this is non-negligible but, in practice, systematic and foreground effects are likely to be at least as impactful.

Our choice to keep the LAT survey as wide as possible was therefore simple, because the degradation in $\sigma(r)$ from a wide $f_{\rm sky}=0.4$ LAT survey, compared to a deeper $f_{\rm sky}=0.1$ LAT survey, appears negligible, and all of the other SO science cases benefit from surveying more sky with the LAT. Of course, this conclusion depends on the SO noise levels and the SAT survey area. If we were to improve the noise levels significantly, or reduce the SAT survey area significantly, these conclusions would change, and would be relevant for future survey design with SO or future telescopes. The impact of delensing is further discussed in Sec.~\ref{sec:lensing.delensing}.

\paragraph{LF sensitivity and resolution.} The bottom panel of Fig.~\ref{fig:r_sigma_r_vs_SACLF_and_Alens} shows how $\sigma(r)$ depend on the observation time for the low-frequency channels ($27$ and $39$ GHz), at the expense of the mid-frequency channels ($93$ and $145$ GHz, see description in Sec.~\ref{subsec:instrument_summary}), after component separation. We find a broad optimal duration around $T_{\rm LF}=1$ year, which has been used as the default in the results presented so far.

\begin{figure}
  \centering
  \includegraphics[width=\columnwidth]{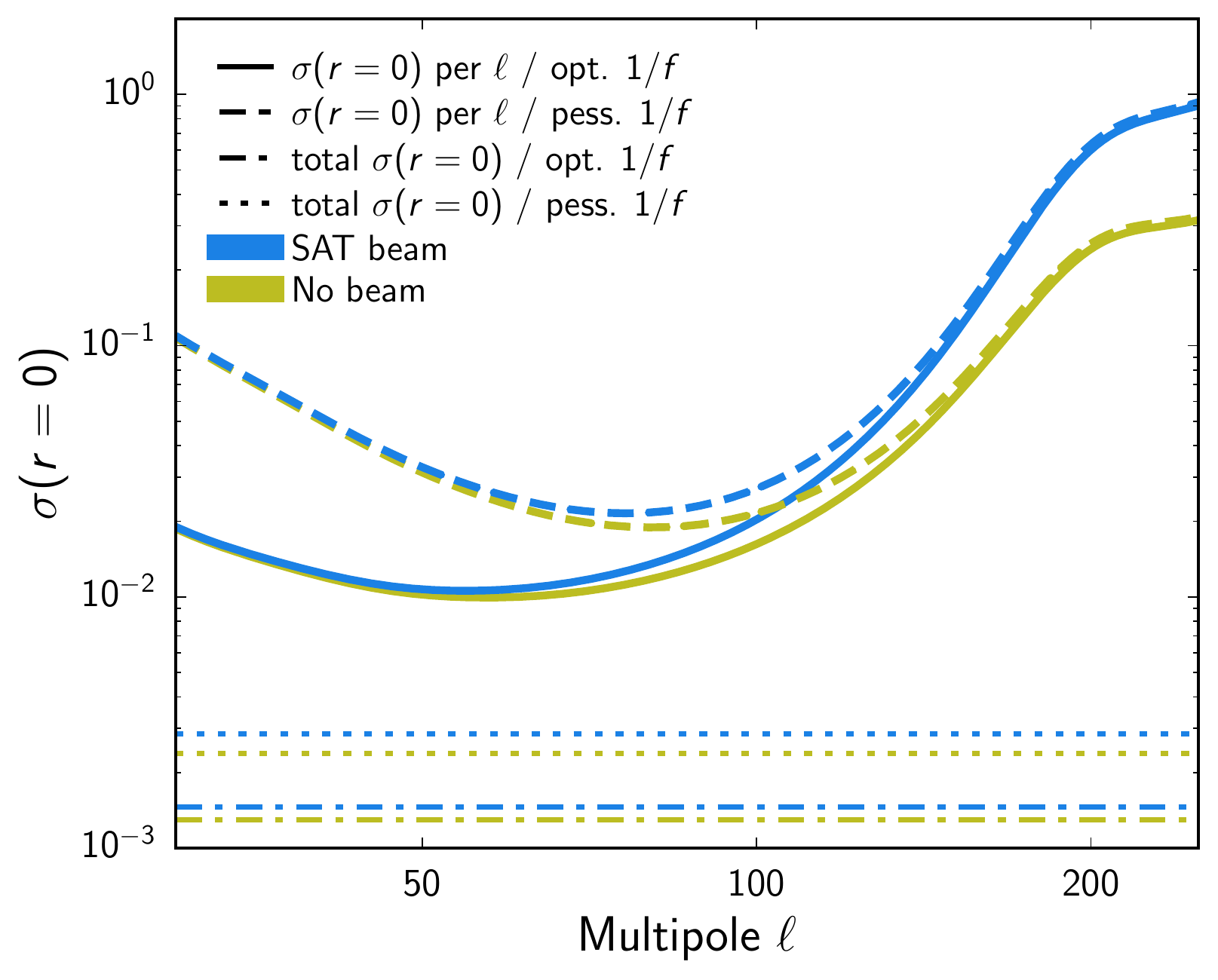}
  \caption{Contribution to the total error on the tensor-to-scalar ratio $r$ for each multipole $\ell$ ($r=0$ is used as fiducial). Results are shown assuming baseline noise levels for an instrument with the fiducial SAT resolution (blue) and for infinite resolution (light green), and for optimistic (solid) and pessimistic (dashed) $1/f$ noise. The horizontal dot-dashed and dotted lines show the final $\sigma(r)$ obtained by combining all multipoles for the optimistic and pessimistic $1/f$ noise cases, respectively. The finite resolution of the SATs, including the LF bands, does not impact the projections for our nominal modeling assumptions. These results were derived using power spectrum errors obtained from {\tt BFoRe} after component separation, accounting for inhomogeneous noise coverage and $E/B$ leakage, and assuming $A_{\rm lens}=0.5$. The curves assume an optimal combination of sensitivity from the SATs and the LAT.} \label{fig:LF_resolution_study_with_xForecast}
\end{figure}

Figure~\ref{fig:LF_resolution_study_with_xForecast} shows the impact of the SATs' finite resolution on the achievable constraints. Since most of the constraining power on $r$ is concentrated on angular scales $\ell\leq 100-150$, for angular resolutions $\leq 90'$, the constraints are mainly driven by scales larger than the beam FWHM. Thus, for SO, the relatively low resolution of the SATs does not appear to limit the instrument's performance, including the ability of the low-frequency channels to map the synchrotron emission. 

\subsection{Limitations of current forecasts}
\label{ssec:bblimits}
Although we have attempted to produce realistic forecasts in terms of foreground complexity and other technical sources of uncertainty, such as $E/B$ leakage, one of the main shortcomings of these forecasts has to do with the impact of instrumental systematics. Imperfect knowledge of the instrument bandpasses, relative gains, polarization angle, ground pickup, half-wave plate imperfections, crosstalk, etc. are challenges to achieve the promised performance of these next generation CMB polarization observatories. Although many mitigation solutions, based on both hardware and software, are currently designed and tested on real data, developing these methods in the context of high-sensitivity experiments will be critical for the success of SO.  As an example of this, in the context of bandpass and relative gain calibration, it is shown in~\cite{Ward2018} that post-calibration knowledge of bandpass shifts and relative gains at the percent level may be necessary in order to reduce instrumental systematics below the statistical uncertainties, unless the bandpass uncertainties can be successfully self-calibrated at the likelihood level. An assessment of systematic effects is now the subject of a new SO study. Future forecasts will also need to account for the effects of filtering in a realistic way, since this can be one of the major challenges for $B$-mode power spectrum analyses \citep{bkp}.

Our forecasts have also not assessed the possible benefit of combining the SO observations with external datasets in terms of characterizing the dust and synchrotron SEDs. At high frequencies, and particularly on large angular scales, {\it Planck}'s 353 GHz map will help in constraining the spatial variation of $\beta_d$ and setting tighter priors on $T_d$. At low frequencies, observations from C-BASS~\citep{2018arXiv180504490J}, S-PASS~\citep{Krachmalnicoff18} and QUIJOTE~\citep{Poidevin_2018} will help constrain the spatial variation of $\beta_s$ as well as set upper bounds on the non-zero synchrotron curvature or the presence of polarized AME. Our forecasts have also neglected the possible spatial variation of the synchrotron curvature, which could be informed by these experiments. Future forecasts and analysis pipelines will take this into account.

\section{Small-scale damping tail}
\label{sec:highell}
At small angular scales, the main SO science targets are the number of relativistic species in the early universe, the spectrum of primordial perturbations, the expansion rate of the universe, the masses of neutrinos, the abundances of primordial elements, and the particle nature of dark matter and its interactions.  These phenomena affect the damping tail (high-$\ell$ region) of the CMB power spectra in temperature and polarization, as well as the growth of structure revealed by CMB lensing.
In this section we describe our methodology, present forecasts for \neff\  (Sec.~\ref{ssec:neff}), for the primordial scalar power (Sec.~\ref{sec:pk}), the Hubble constant (Sec.~\ref{ssec:hubble}), and additional high-$\ell$ science parameters (Sec.~\ref{ssec:highl}).

We use the effective number of relativistic species, \neff, as our main proxy for the information encoded in the CMB at small scales, and use it to investigate the \so{} experimental requirements 
(the CMB-S4 science book followed a similar approach;~\citealp{CMBS42016}). The \neff\ parameter tracks the qualitative effect of early universe physics on the CMB, and also has precise theoretical predictions. In this section we then report forecast for \neff\ and how these are affected by the LAT survey specifications, and also report projections for other \so{} high-$\ell$ science targets.

We forecast the science performance with a Fisher matrix based on \so{} noise levels, with details described in Sec.~\ref{sec:method}.  We explore the experimental parameter space (including resolution, sensitivity, observed sky fraction) and systematic and contaminating effects (including beam errors, atmospheric noise, and polarized source contamination). Our observables are the lensed $TT, TE, EE$ power spectra and the lensing convergence power spectrum, $\kappa \kappa$. We include lensing-induced covariances~\citep{2012PhRvD..86l3008B}. To mitigate residual foreground contamination (dominating in temperature and leaking to lensing at $\ell \sim 3000$), we retain only the angular scales in the range $30\leq \ell \leq 3000$ for $TT$ and $\kappa \kappa$, while we use the full range of CMB scales, $30\leq \ell \leq 5000$, for the cleaner $TE$ and $EE$ correlations. We impose Big Bang Nucleosynthesis (BBN) consistency relations~\citep{Pisanti:2007hk} on the Helium abundance, $Y_p$, by assuming the value \neff\ was the same during BBN and recombination. In all cases we combine \so{} data with \planck{} as described in Sec.~\ref{sec:add_data}.  

We simplify all the damping tail forecasts in this section by assuming perfect foreground removal from the power spectra and a single, co-added noise level for the LAT. For this we use the CMB temperature and CMB polarization Deproj-0 $N_{\ell}^{TT}$ and $N_{\ell}^{EE}$ described in Sec.~\ref{sec:FG_LAT}, and the minimum-variance noise curves for $\kappa\kappa$. 
For a further suite of tests on individual components of the noise, including the atmosphere or point source terms, and tests of the foreground cleaning, we co-add the 93 GHz and 145 GHz white noise levels\footnote{We checked that using the full range of LAT frequencies does not improve on the 93 and 145 GHz co-added signal.} in Table~\ref{tab:noise}. We demonstrate the validity of these assumptions in Sec.~\ref{sec:highell_fg}. 

\subsection{Forecasts for \neff}
\label{ssec:neff}

At its baseline sensitivity, \so{} will be sensitive to well-motivated, non-standard scenarios for relativistic species in the early universe. The usual parametrization of the neutrino contribution to radiation density at early times is (see, e.g.,~\citealp{Kolb&Turner})
\begin{equation}
\rho_{\rm rad} = \left[1+\frac{7}{8} \left(\frac{4}{11}\right)^\frac{4}{3}N_{\rm eff}\right]\rho_{\gamma} \,,
\end{equation}
where $\rho_\gamma$ is the CMB photon energy density, and \neff\ is the effective number of relativistic species.  The Standard Model of particle physics predictions for \neff\ is 3.046, assuming standard electroweak interactions, three active light neutrinos, small effects from the non-instantaneous neutrino decoupling from the primordial photon--baryon plasma, and corrections due to energy-dependent neutrino interactions \citep{Dolgov:1997mb,Dolgov:1998sf,2002PhLB..534....8M,2016JCAP...07..051D}. 
Via its impact on the expansion rate, primordial element abundances, and radiation perturbations, \neff\ affects the damping and the position of the acoustic peaks in the temperature, $E$-mode of polarization, and $TE$ power spectra \citep[see, e.g.,][]{2004PhRvD..69h3002B,2013PhRvD..87h3008H,2015APh....63...66A}.

For the nominal \so{} LAT survey covering 40\% of the sky, our Fisher forecast yields errors:
\begin{eqnarray}
\sigma(N_{\rm eff})&=& 0.055 \quad {\rm SO \ Baseline,} \nonumber\\
\sigma(N_{\rm eff})&=& 0.050 \quad {\rm SO \ Goal.}
\end{eqnarray}
The current limit from \planck{} is $\sigma$(\neff)=0.19 (TT,TE,EE+lowE+lensing,~\citealp{planck2018:parameters}). 

\begin{figure}[t!]
\includegraphics[width=\columnwidth]{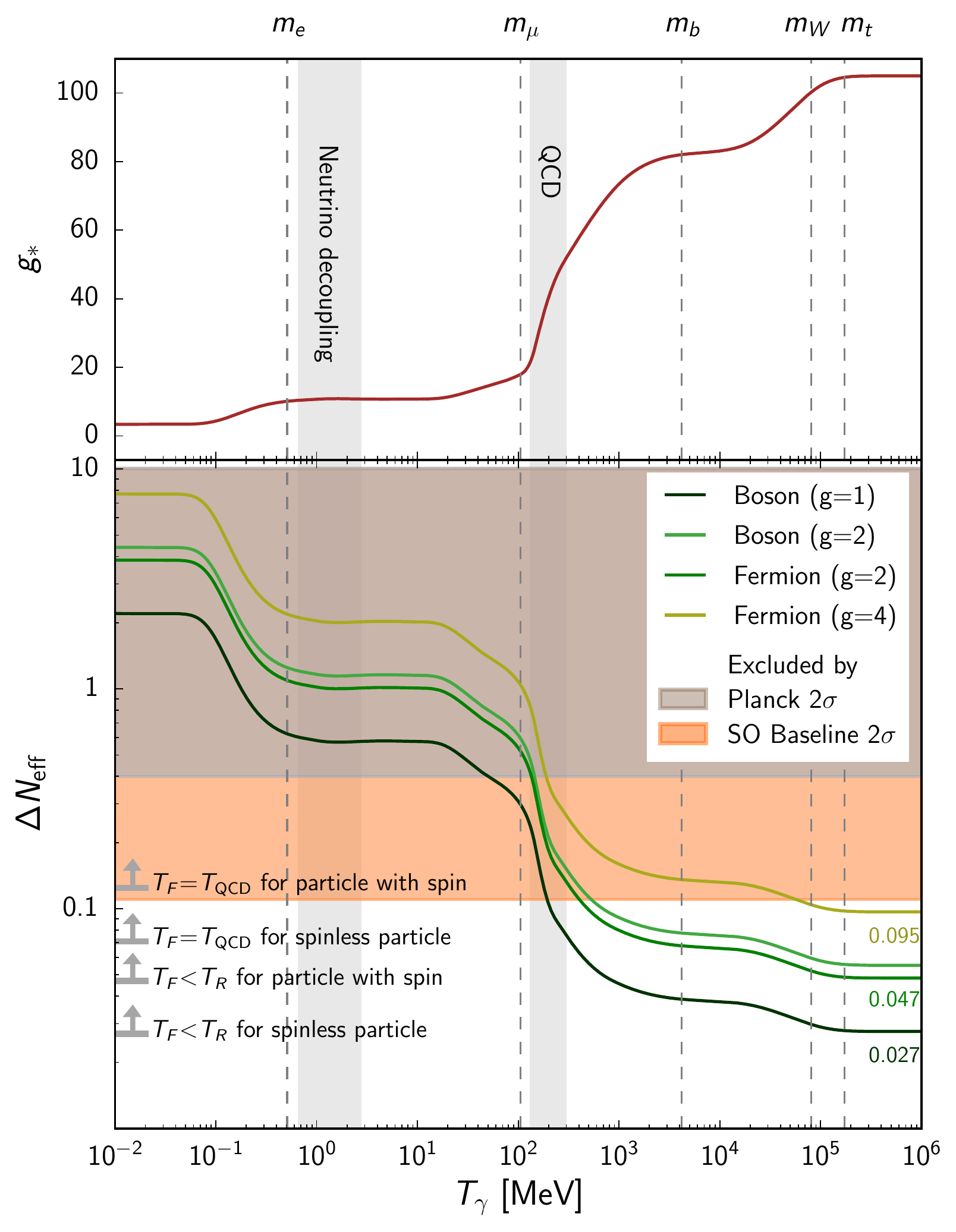}
\caption{Theoretical targets for \neff, assuming the existence of an additional Beyond-the-Standard-Model relativistic particle that was in thermal equilibrium with the Standard Model at temperatures $T>T_{F={\rm freeze-out}}$. 
{\it Top}:
evolution of the effective degrees of freedom for Standard Model particle density, $g_*$, as a function of photon temperature in the early universe, $T_\gamma$. Vertical bands show the approximate temperature of neutrino decoupling and the QCD phase transition, and dashed vertical lines denote some mass scales at which corresponding particles annihilate with their antiparticles, reducing $g_*$. The solid line shows the fit of \citet{Borsanyi:2016ksw} plus standard evolution at $T_\gamma<1~ {\rm MeV}$. 
{\it Bottom}:
expected $\Delta N_{\rm eff}$ today for species decoupling from thermal equilibrium as a function of the decoupling temperature, where lines show the prediction from the \citet{Borsanyi:2016ksw} fit assuming a single scalar boson with $g=1$ (dark green), bosons with $g=2$ (e.g., a gauge vector boson, light green), a Weyl fermion with $g=2$ (green), or fermions with $g=4$ (yellow).
Shaded regions show the exclusion regions for \planck{} and \so{} (baseline).
Arrows on the left show the lower limits for specific cases, for example any particle with spin that decoupled after the start of the QCD phase transition would typically be measurable by \so{} at $2\sigma$. The lower two arrows and corresponding numbers on the right give lower bounds for particles produced any time after reheating (at $T_R$). }
\label{fig:Neff_targets}
\end{figure}

Figure ~\ref{fig:Neff_targets} compares the \so{} baseline sensitivity to interesting science targets for new particles that were at some point in thermal equilibrium with Standard Model particles and then decoupled.  The predicted \neff\ depends on the number and spin of the particles which decouple (determining the degrees of freedom, $g$, of the new particles), and effective degrees of freedom for entropy $g_s$, or equivalently the effective
degrees of freedom for density $g_*$, of the Standard Model plasma at the time of decoupling.

A single particle with spin $1/2$ that decouples just before or after the QCD phase transition contributes $\Delta N_{\rm eff} \geq 0.12$, and a similar particle with spin 1 contributes $\Delta N_{\rm eff} \geq 0.14$; these cases should be measurable at the 2$\sigma$ significance level for \so{}.   Models predict deviations  at $\Delta$\neff$ \geq 0.047$ for additional light particles of spin 1/2, 1 and/or 3/2 that were in thermal equilibrium with the particles of the Standard Model at any point back to the time of reheating \citep[see, e.g.,][]{Brust:2013xpv,Chacko:2015noa,Baumann:2016wac}; these could be measurable by \so{} if multiple species are involved. 
Four or more scalar bosons would also be measurable at 2$\sigma$ significance by \so{}, as would various scenarios involving multiple non-thermal axions if they contribute sufficiently to the energy density. 

In the following we assess how the forecasts for \neff\ depend on survey strategy, foreground residuals, and instrument beams. These studies were used in practice to guide the nominal choice of the SO experimental specifications.

\subsubsection{Survey requirements}

\begin{figure}[t!]
\includegraphics[width=\columnwidth]{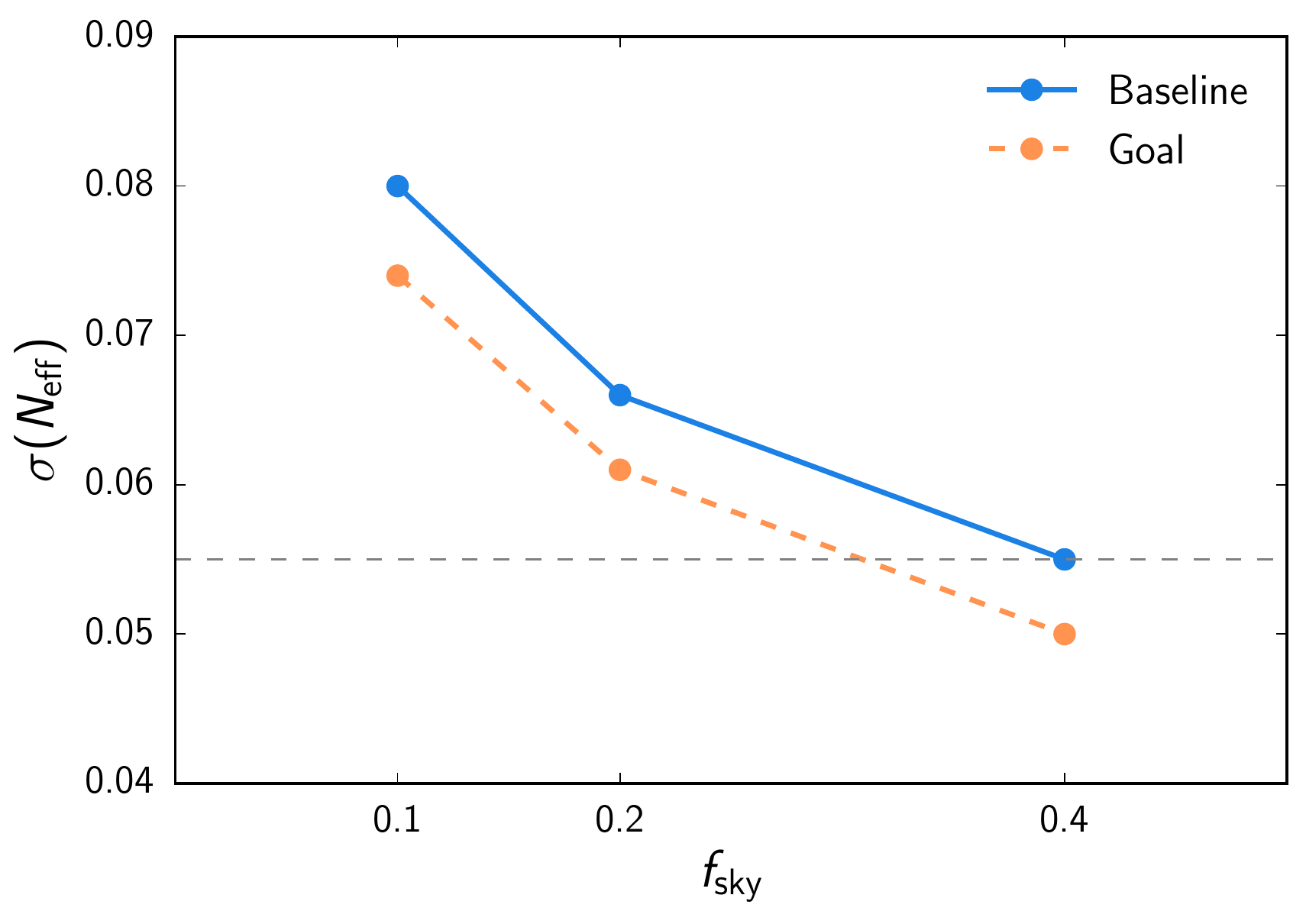}
\caption{
Dependence of the forecast $\sigma(N_{\rm eff})$ on the fractional sky coverage for SO baseline  and goal scenarios. The measurement degrades by 40\% when reducing $f_{\rm sky}=0.4$ to $f_{\rm sky}=0.1$. The dashed gray line gives a reference for the baseline forecast.\\}\label{fig:Neff_fsky}
\end{figure}

\begin{figure*}[tp!]
\centering
\includegraphics[width=\textwidth]{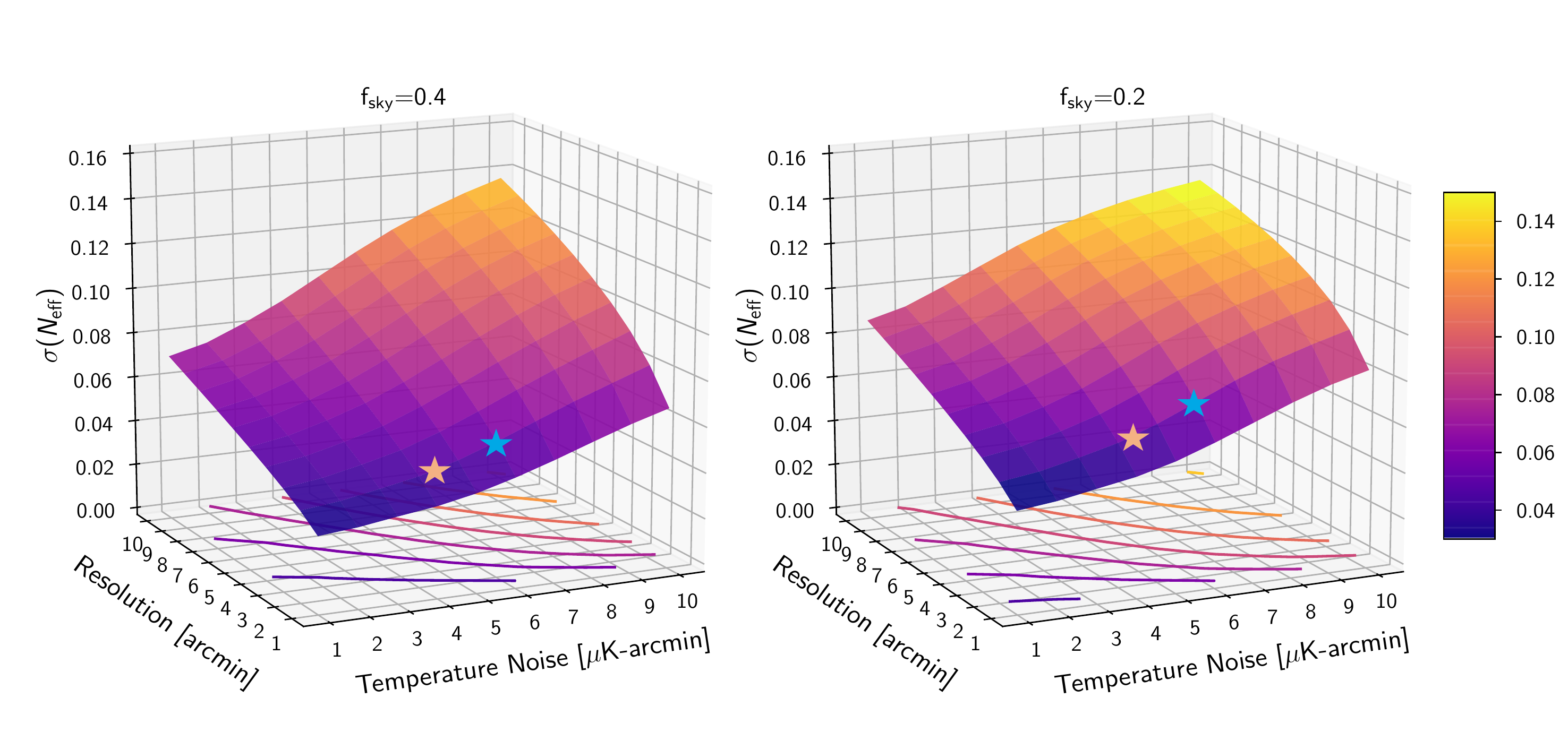}
\caption{Dependence of the \neff\ estimate on resolution, sensitivity and sky coverage (with wide coverage shown on the left and a smaller, deeper survey on the right). In order to reach the science target of $\sigma$(\neff)$=0.06$, a reduced noise of 5 $\mu$K-arcmin on $40\%$ of the sky and resolution $<3'$ at 145 GHz are needed. Stars show the expected \so{} measurements in the case of goal (orange) and baseline (blue) configurations. \label{fig:Neff3D}}
\end{figure*}  

Although different experimental configurations can trade sensitivity, resolution and observed sky area to achieve the same errors for \neff, the errors depend most strongly on sky area.  We show the dependence on survey and instrument design in Figs.~\ref{fig:Neff_fsky} and \ref{fig:Neff3D}. 
\paragraph{Sky area} At fixed-cost effort, and for maps with temperature noise exceeding 4 $\mu$K-arcmin, damping tail science depends strongly on $f_{\rm sky}$. In particular, $f_{\rm sky} \geq 0.4$  is needed to reach $\sigma$(\neff)$\leq 0.06$, as shown in Fig.~\ref{fig:Neff_fsky}. The \neff\ constraint worsens for smaller sky coverage, with a 40\% degradation going from $f_{\rm sky}=0.4$ to $f_{\rm sky}=0.1$. This is also visible in the two scenarios shown in Fig.~\ref{fig:Neff3D}, where the amplitude of the likelihood surface increases for lower $f_{\rm sky}$ (right hand side). 
\paragraph{Resolution} The uncertainty in \neff\ does not improve significantly with resolution finer than  $2'$--$3'$ if other specifications are held fixed (see Fig.~\ref{fig:Neff3D}).
\paragraph{Sensitivity} Assuming a wide survey ($f_{\rm sky}=0.4$) and resolution better than $3'$ at 145 GHz, a white noise level of 5 $\mu$K-arcmin is needed to reach $\sigma (N_{\rm eff})=0.06$ (see Fig.~\ref{fig:Neff3D}). 
\paragraph{Frequency coverage} In the next subsection we show that cosmological parameters from the damping tail are not strongly reliant on having broad frequency coverage to mitigate foreground contamination.

\subsubsection{Testing foreground contamination}\label{sec:highell_fg}

\begin{figure}[t!]
\centering
\includegraphics[width=8cm]{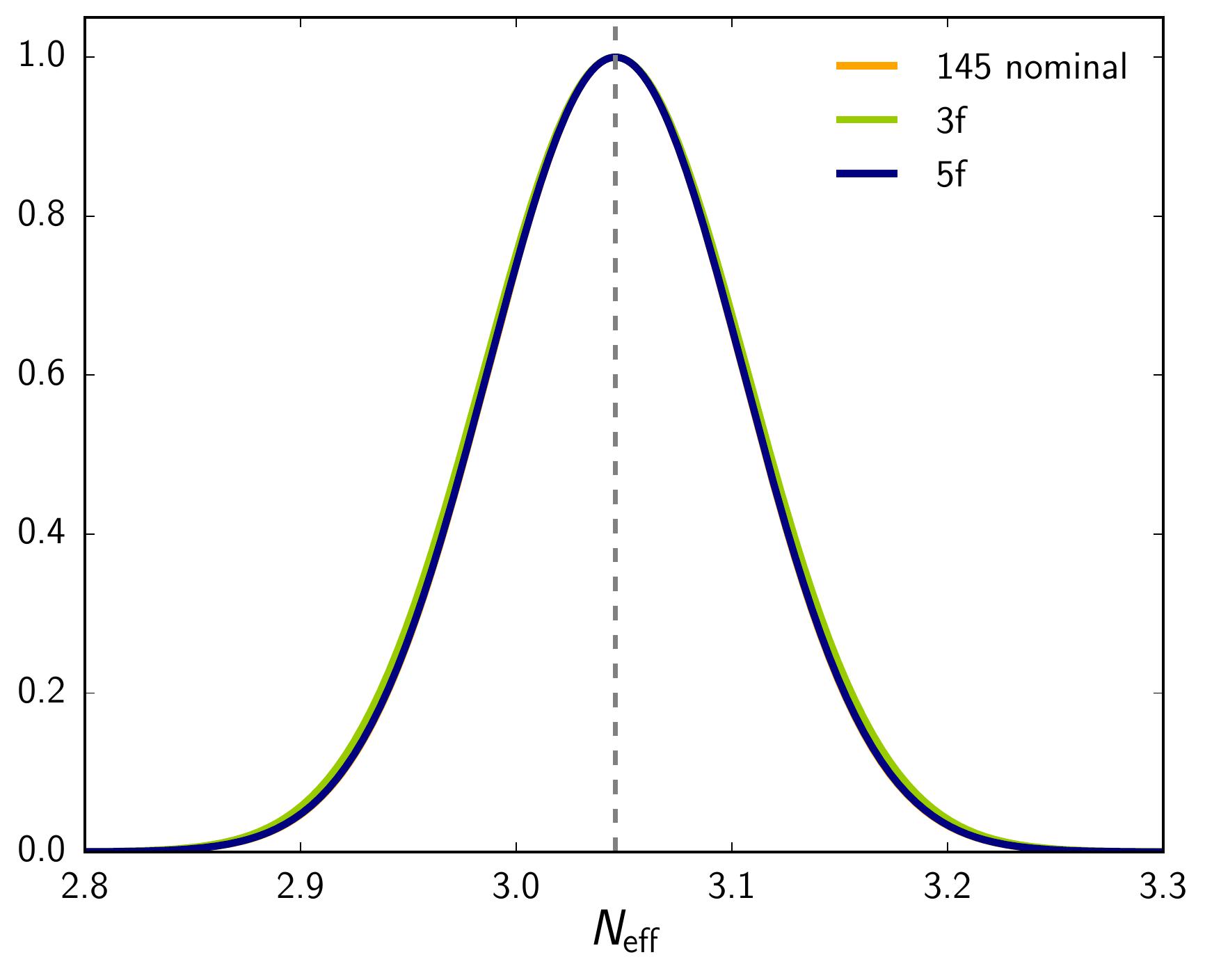}
\includegraphics[width=8cm]{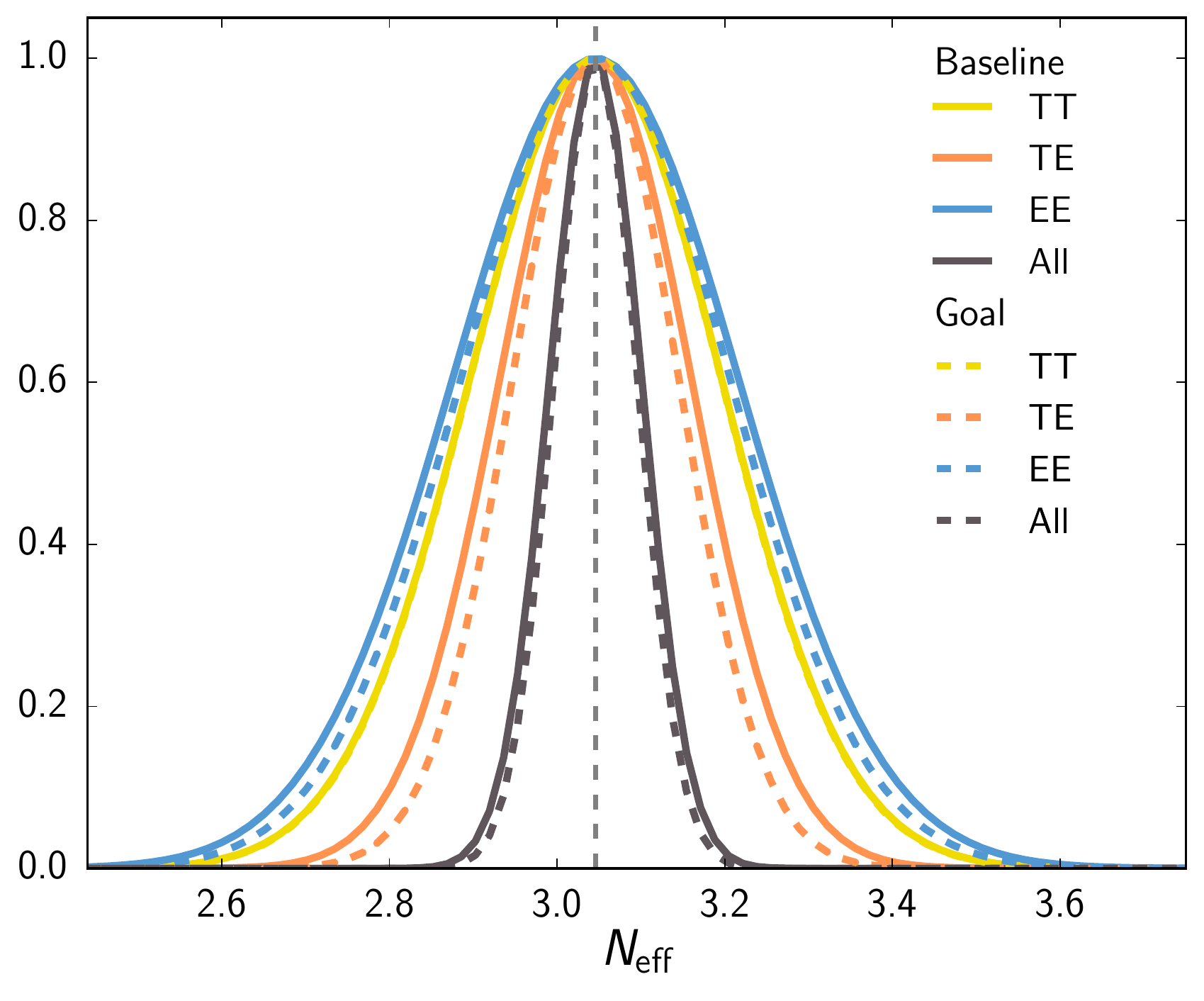}
\caption{\emph{Top:} Impact of foreground cleaning the LAT using 5 (blue) or 3 (green) frequencies on the \neff\ estimate. The 145 nominal orange curve shows the result assuming the ideal case of having no foregrounds in the data. The gain from retaining frequency information is completely negligible. \emph{Bottom}: Constraints on \neff\ from individual small-scale spectra ($TT, TE, EE$) and the total, shown for baseline (solid)  and goal (dashed) scenarios. The most constraining channel is the foreground-clean $TE$ spectrum.
\label{fig:Neff_f}}
\end{figure} 

We find that foreground contamination, at the level that exists in our simulations, has little impact on these forecasts. At the \so{} level of sensitivity and resolution (and for all beyond-Stage-3  CMB experiments), we expect polarization observations to dominate cosmological constraints \citep[see, e.g.,][]{galli/etal/2014,calabrese/etal/2016}. At small scales, this comes with the advantage of reducing the impact of foreground contaminants.
In the damping tail, the $TT$ power spectrum is more contaminated by Galactic and extragalactic foreground emission than the $TE$ and $EE$ spectra, which are mainly affected by residual power in unresolved radio sources. However, the shape of the power spectrum for radio source emission is well known and therefore easy to separate from the CMB.  Polarized thermal dust and synchrotron emission from the Galaxy, in our current simulations and after masking, have a negligible impact on these small angular scales.  

All our forecasts for parameters derived from the damping tail therefore assume that the LAT information has been compressed in a single channel, co-adding the foreground-marginalized noise curves of Sec.~\ref{sec:FG_LAT} and without retaining multi-frequency information. To demonstrate the validity of this assumption, we perform a full multi-frequency power spectrum analysis, running simulations to test the impact on the \neff\ constraints of marginalizing over Galactic and extragalactic emission in $TT$, $TE$ and $EE$.

We do not use the map-based simulations described in Sec.~\ref{sec:method}, but instead simulate LAT multi-frequency spectra with a model for the sky. The lensed CMB signal is added to the following foreground components: (i) emission in temperature from the tSZ and kSZ effects; Poisson-like and clustered dusty star-forming galaxies for the CIB; radio galaxies; a cross-correlation term between the clustered component of the CIB and the thermal SZ; thermal dust emission from the Galaxy; and (ii) emission in polarization from unresolved Poisson sources. The modeling follows methods used for ACT, SPT, and \planck\ analyses \citep{Dunkleyetal2013,reichardt/etal:2012,Georgeetal2015,planck_like:2015,planck_like:2013}, and is consistent with the map-based modeling used in Sec.~\ref{sec:FG_LAT}.
We implement the temperature model described in \cite{Dunkleyetal2013}, and include polarized radio sources at the level measured in \cite{2017JCAP...06..031L}. The level of foregrounds present in the simulated data has the following ${\cal D}_\ell$ power at $\ell=3000$ at 150 GHz: 4 and 2 $\mu$K$^2$ for the tSZ and kSZ components; 7, 3 and 6 $\mu$K$^2$ for CIB-Poisson, radio, and CIB-clustered sources; a 10\% correlation between the tSZ and CIB clustered component; 0.5 $\mu$K$^2$ of Galactic dust emission; 1 $\mu$K$^2$ for radio sources in $EE$ (this is based on a very conservative upper level, as no polarized sources were masked in the analysis of \citealp{2017JCAP...06..031L}), and no emission in $TE$ (consistently with what was found in \citealp{2017JCAP...06..031L}). We model the \so{} LAT noise following the white-noise levels and atmospheric contributions described in Sec.~\ref{sec:methods}.

The simulations are used to form a multi-frequency likelihood and processed in two steps: first, the foreground power is marginalized over and the uncertainty due to foreground marginalization is propagated to marginalized CMB bandpowers. Second, parameter estimation is then carried out from the marginalized CMB bandpowers using standard MCMC techniques. The results are shown in the top panel of Fig.~\ref{fig:Neff_f}: when using 5 frequencies (39, 93, 145, 225, 280) to marginalize over the foregrounds present in the LAT data we recover the nominal foreground-free estimate, and with 3 frequencies (93, 145, 225) the degradation in the \neff\ estimate is negligible. 

We find that component separation has a relatively small impact on damping-tail forecasts because, at the noise level and resolution of \so{}, constraints on \neff\ are mainly driven by the $TE$+$EE$ constraining power, as shown in the bottom panel of Fig.~\ref{fig:Neff_f}. We also find that \planck{} $TT$ data at $\ell < 1500$ could be used in place of \so{} 
$TT$ information, and still achieve the full constraining power of \so{} without needing to model and clean the SO $TT$ spectrum.
Of course, since we have not included Galactic contamination in the $TE$ and $EE$ statistics in these simulations, and we included only $EE$ point source power described by a Poisson term, it is unsurprising that we find no impact of foregrounds on the forecast constraints from $TE$ and $EE$. Future studies will be useful to validate these assumptions. 

\subsubsection{Impact of point source and  atmospheric noise}

In addition to component separation, atmospheric noise and noise from polarized point sources can both impact the forecast constraints\footnote{Here we study the impact of point sources as an additional source of small-scale noise, rather than as a foreground contaminant.}. In our nominal forecasts, we include atmospheric noise as described in Sec.~\ref{sec:methods}. This effectively increases the noise levels above the white noise level on scales $\ell < \ell_{\rm knee}$. For temperature noise, $\ell^{TT}_{\rm knee} \approx 3000$ and so this can be an important effect for $TT$ and $TE$. Some of this impact is mitigated by adding \planck\ data to minimize $N^{TT}_\ell$ at large scales, as described in Sec.~\ref{sec:methods}.

\begin{figure}[t!]
\centering
\includegraphics[width=\columnwidth]{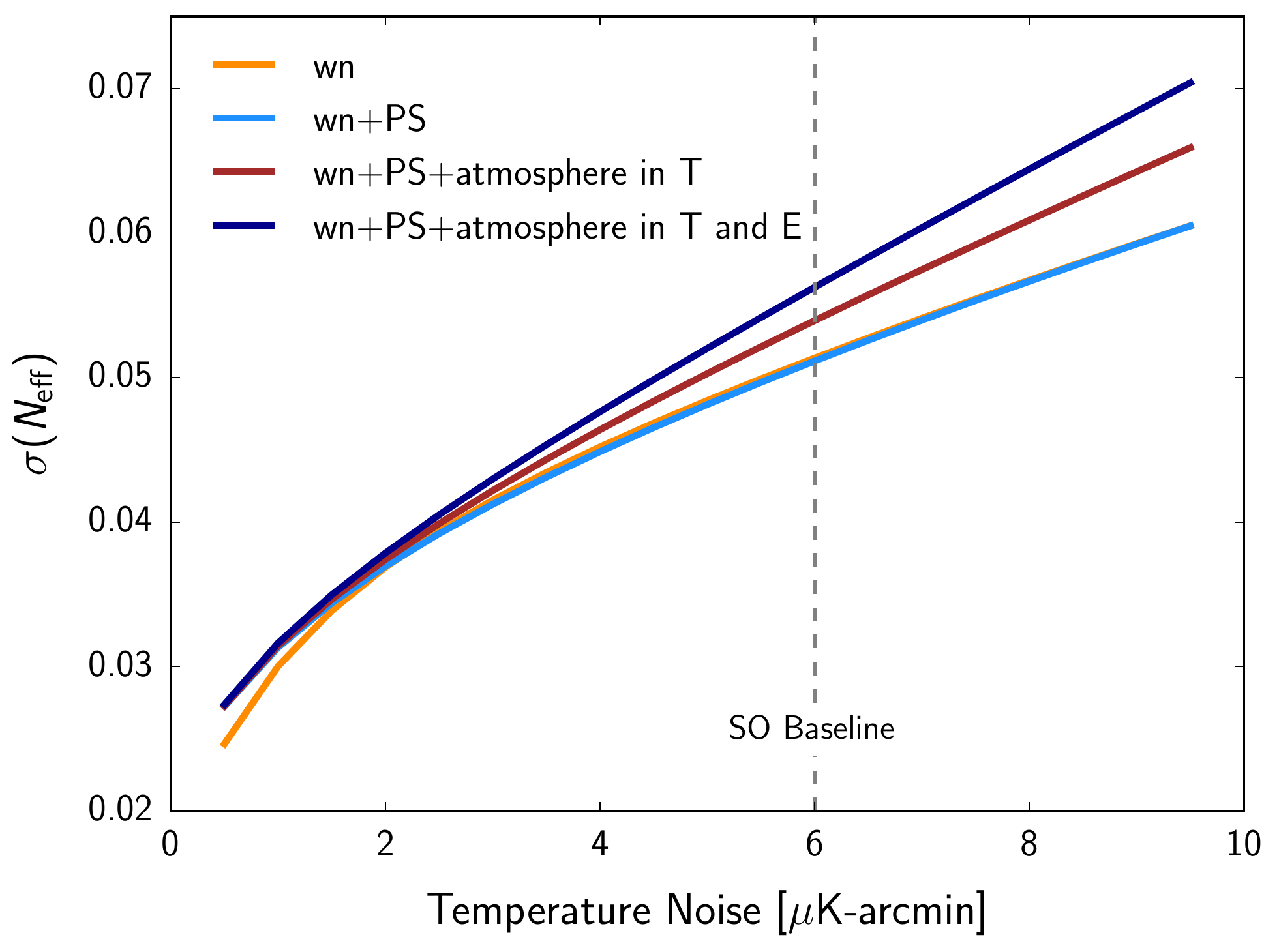}
\caption{Impact of atmospheric and point source (PS) noise contamination on \neff\ forecasts. For temperature noise (white noise levels, wn) above 1.5 $\mu$K-arcmin, the dominant effect is the atmosphere. This is due to the increase in the temperature noise at $\ell<3000$. \planck{} data on the same \so{} sky coverage are included to minimize the impact of the atmosphere as much as possible. The \so{} baseline noise case is highlighted with a dashed vertical line.}
 \label{fig:ps_nw}
\end{figure}

We test the impact of atmospheric noise on the \neff\ forecasts by removing the atmospheric noise component, as shown in Fig.~\ref{fig:ps_nw}.  For the combined \planck\ and \so{} surveys, it shows that the atmospheric contribution to $N^{TT}_\ell$ has a small impact on the \neff\ forecasts. The SO baseline temperature noise level is 6~$\mu$K-arcmin co-added over 93 and 145 GHz, so the effect of removing atmospheric noise would be to reduce the \neff\ error by $\approx$ 0.005. For $EE$, atmospheric noise impacts scales at $\ell < 700$, which play an important role in breaking degeneracies with other cosmological parameters (and in particular $n_s$).  We find that the atmospheric contribution to $N_\ell^{EE}$ impacts \neff\ forecasts at the same level as temperature, as shown in Fig.~\ref{fig:ps_nw}, even though $\ell_{\rm knee}^{TT} \gg \ell_{\rm knee}^{EE}$.  This reflects the role the \planck\ data play in $N_{\ell}^{TT}$.

Unresolved polarized emission is expected from extragalactic radio sources as shown in \citet{2004MNRAS.349.1267T,2011MNRAS.413..132B,Tucci2012,2017arXiv171209639P} and discussed in detail in Sec.~\ref{sec:source}. This emission effectively contributes as an additional source of noise for small-scale polarization data. 
The point source (PS) power enters $D_\ell^{TT}$ and $D^{EE}_\ell$ with a white spectrum, characterized by a single amplitude parameter at a pivot scale, $D_{\ell}=A (\ell/\ell_0)^2$.  We study the impact of this additional noise term by including power at 150~GHz from point sources with an amplitude of
\begin{eqnarray}
D^{TT}_{\rm PS, \ell = 3000} &=& 6 \, \mu{\rm K}^2 \,, \nonumber\\
D^{EE}_{\rm PS, \ell = 3000} &=& (3\times 10^{-3}) \times 6 \,   \mu{\rm K}^2\, \label{eq:EEps}.
\end{eqnarray}
These values are based on \planck, ACT, and SPT measurements of dusty and radio emission in temperature and  on the measured polarization fraction of the point sources~\citep{dunkley/etal/2011,planck_params/2013,Georgeetal2015}. Here we assume that all dusty-star-forming and radio sources present in temperature are polarized, but in practice, at frequencies below $150$ GHz, we expect  only polarized radio sources to contribute to the signal, as described in Sec.~\ref{sec:source}.

A more realistic estimate of the polarized source level comes from~\citet{2017arXiv171209639P}, who combine state-of-the-art catalogs of polarized radio sources at frequencies ranging from $1.4$ to $ 217$ GHz, to derive the statistical properties of fractional polarization, $\Pi = P/S$. They find  $\langle \Pi^2 \rangle$ to have a roughly constant value, $2.5\times 10^{-3} $. The amplitude of $D_{\rm PS}^{EE}$ depends on the differential number counts expected at a given frequency, the detection flux, commonly set at $5\sigma$ the sensitivity flux, and the average fractional polarization of radio sources. We use these results to estimate $D^{EE}_{\rm PS, \ell = 3000} =0.015\, \mu{\rm K}^2$ for SO, consistent with the value considered above. 

Even with the over-estimate of the $EE$ point source power assumed in Eq.~\ref{eq:EEps}, the contribution from $EE$ point sources would have to be two orders of magnitude larger to meaningfully impact the forecasts. This is illustrated in Fig.~\ref{fig:ps_nw}, which shows no significant impact of $EE$ point sources on the SO \neff\ forecasts.

\subsubsection{Beam requirements}

\begin{figure}[t!]
\centering
\includegraphics[width=\columnwidth]{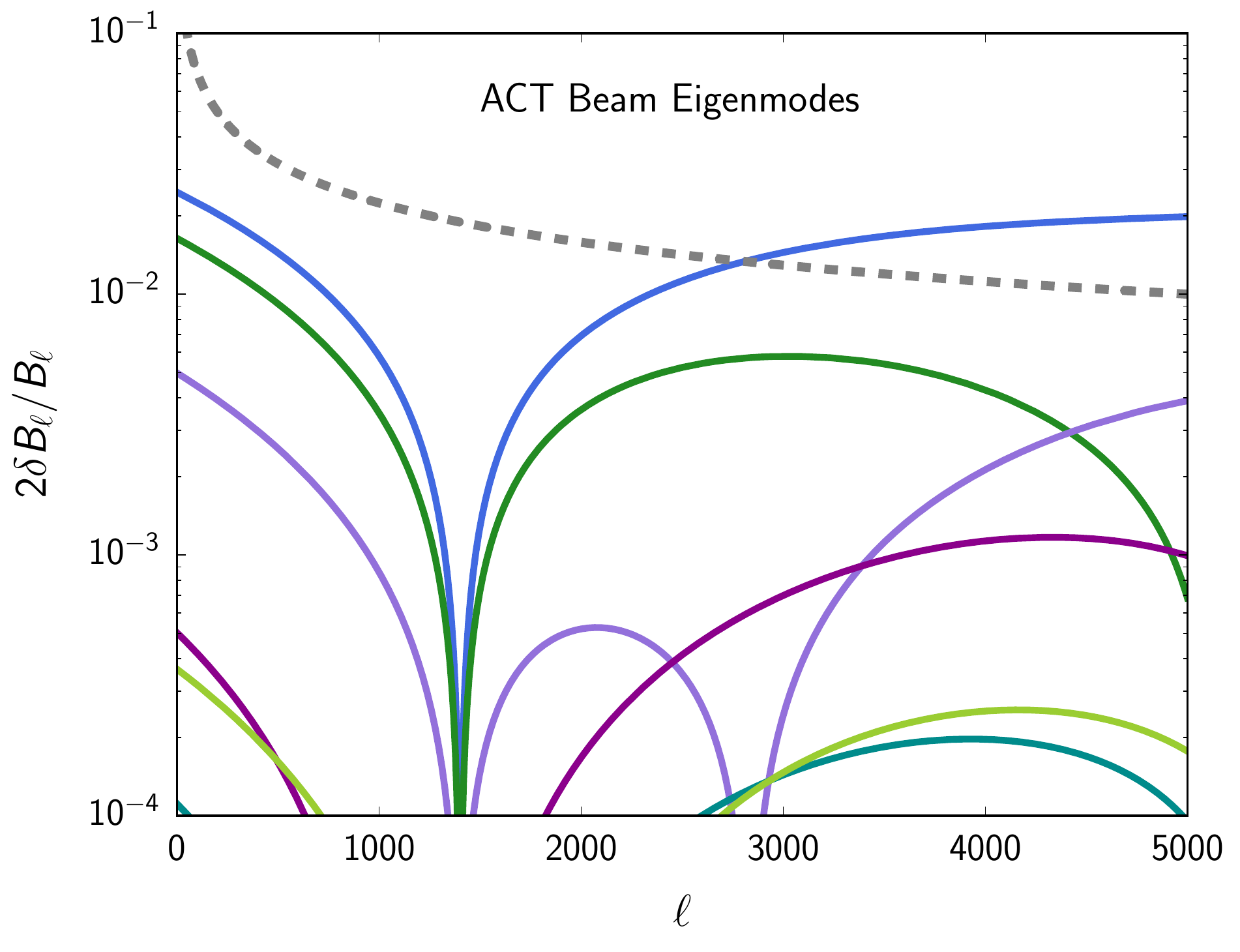}
\includegraphics[width=\columnwidth]{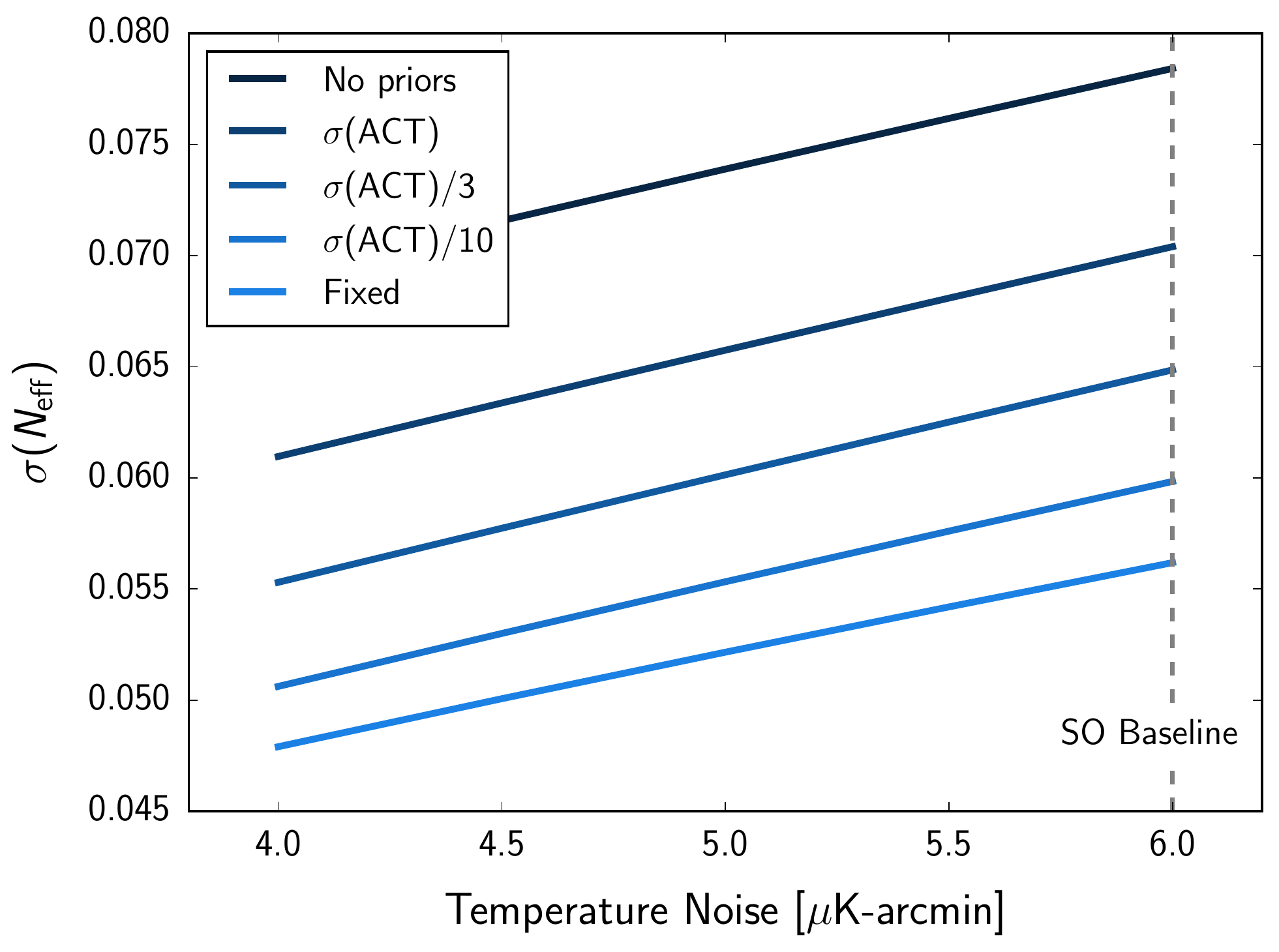}
\caption{{\it Top:} The first five ACT beam eigenmodes described in Eq.~\ref{eq:beam} (solid colored lines) compared to the cosmic variance term (gray dashed line). Each eigenmode is normalized to have unit variance with ACT calibration.  {\it Bottom:} \neff\ forecasts for different assumptions in the beam calibration uncertainty, ranging from no marginalization (`Fixed', lower light-blue line) to marginalization without any prior (`No priors', top dark blue line). The \so{} baseline noise case is highlighted with a dashed vertical line.} \label{fig:beams}
\end{figure} 

High-$\ell$ science goals can be adversely affected by beam imperfections, as the beam alters the shape of the observed power spectrum.  We can recover the primordial power spectrum by measuring and removing the effect of the beam, but marginalizing over uncertainties in this calibration can potentially weaken the constraining power for cosmology.  In addition, a bias in the beam calibration can also bias the cosmological parameters.  

The observed $C_\ell^{\rm obs}$ is a product of the beam shape, $B_\ell$, times the true power spectrum, $C_\ell$, or $C_\ell^{\rm obs} = B_\ell^2 C_\ell$.
For a Gaussian beam, $B_\ell \approx \exp({-\ell(\ell+1) \theta_{\rm FWHM}^2 /8 \log 2})$.  The primordial spectra at high $\ell$ are similarly damped due to diffusion, $C_\ell \propto e^{-\ell^2/\ell_d^2}$; high-$\ell$ parameters like \neff\ and $Y_p$ alter the amount of damping and are measured primarily through effects on $\ell_d$.  In the observed spectra, a change to $\theta_{\rm FWHM}^2$ is indistinguishable from a change to $\ell_d$.  Of course, beam calibration allows us to distinguish these two effects but only if the uncertainty in the beam shape is smaller than cosmic variance on the scales of interest.

It is conventional to decompose $B_\ell$ into eigenmodes that diagonalize the beam covariance matrix.  To model the calibration uncertainty, we use the ACT temperature\footnote{For the purpose of this analysis we assume that beams are equal in temperature and polarization.} beam eigenmodes from~\cite{2013ApJS..209...17H} and project out the average $B_\ell$ such that
\begin{eqnarray}\label{eq:beam}
C_\ell^{\rm eff} &=& \left(1+\sum_{i=1}^5 b_i \frac{\delta B^{(i)}_\ell}{B_\ell}\right)^2 C_\ell \ ,
\end{eqnarray}
where $\delta B^{(i)}$ are the beam eigenmodes shown in the top panel of Fig.~\ref{fig:beams}. The eigenmodes are normalized such that the coefficients $b_i$ have unit variance for the calibration achieved by ACT. The beam factor multiplying $C_\ell$ is also compared to the cosmic variance term, $1/\sqrt{2\ell+1}$, shown with a gray dashed line. This highlights that the uncertainty in the power spectrum due to the variance of the largest beam eigenmode is larger than cosmic variance for $\ell > 2500$ and therefore degrades the measurement of \neff.
A quantitative estimate of this degradation is obtained by marginalizing over the amplitudes of each mode, $b_i$, assuming a mean of $b_i =0$. In this normalization, $\sigma(b_i) = 1$ corresponds to the size of beam errors from ACT in temperature.

The resulting \neff\ forecasts are shown in the bottom panel of Fig.~\ref{fig:beams} for choices of prior on $b_i$ with variances of $\sigma(b_i) = 0,1/10, 1/3, 1, \infty$. (In the figure, these correspond to a perfect beam with no error (`Fixed'), $\sigma({\rm ACT})/10$,  $\sigma({\rm ACT})/3$, $\sigma({\rm ACT})$ and `No priors'.)  We see that $\sigma(b_i)< 1/3$ is needed to avoid inflating the errors, and thus would require significant improvements over current ACT beam calibration. From the normalization of the eigenmodes in Fig.~\ref{fig:beams}, this corresponds to sub-percent calibration of the beam shape. However, as will be described in Sec.~\ref{sec:summary}, we allow for a systematic error inflation term in our baseline science forecasts, that could include this beam uncertainty and other effects, and will be the subject of future SO studies.

\subsection{Primordial scalar power}\label{sec:pk}

\so{} will make substantial improvements to our knowledge of the primordial density perturbations.  The standard approach is to model their power spectrum with a power law with the scalar spectral index $n_s$ and an amplitude of perturbations $A_s$ as $\mathcal{P}(k) = A_s\left(k/k_*\right)^{n_s-1}.$ The power $\mathcal{P}(k)$ is then mapped onto the angular power spectrum $C_\ell$ through the transfer function.

Deviations from this power law prescription can take many forms, including adding a running of the spectral index with wavenumber.  Instead, we parameterize the power spectrum as a series of band powers binned in 20 $k$ bins between wave numbers $k=0.001$~Mpc$^{-1}$ and $k=0.35$~Mpc$^{-1}$, with $k_{i+1} = 0.756k_i.$ We perform a cubic spline over the band powers to ensure smoothness and we normalize the power spectrum in units of $10^{-9}.$ Similar approaches at reconstructing the primordial power are followed by, e.g., \citet{2003MNRAS.342L..72B, 2011JCAP...08..031G, 2012ApJ...749...90H, 2013PhRvD..87h3526A, 2014JCAP...01..025H, 2014PhRvD..89j3502D, 2016PhRvD..93b3504M, 2016JCAP...09..009H, 2017PhRvD..96h3526O}. The \planck\ inflation papers (see~\citealp{2016A&A...594A..20P,planck2018:inflation} and references therein) performs an exhaustive search over model space, but we restrict ourselves to one model of perturbations.
\begin{figure}[t!]
\centering
\includegraphics[width=\columnwidth]{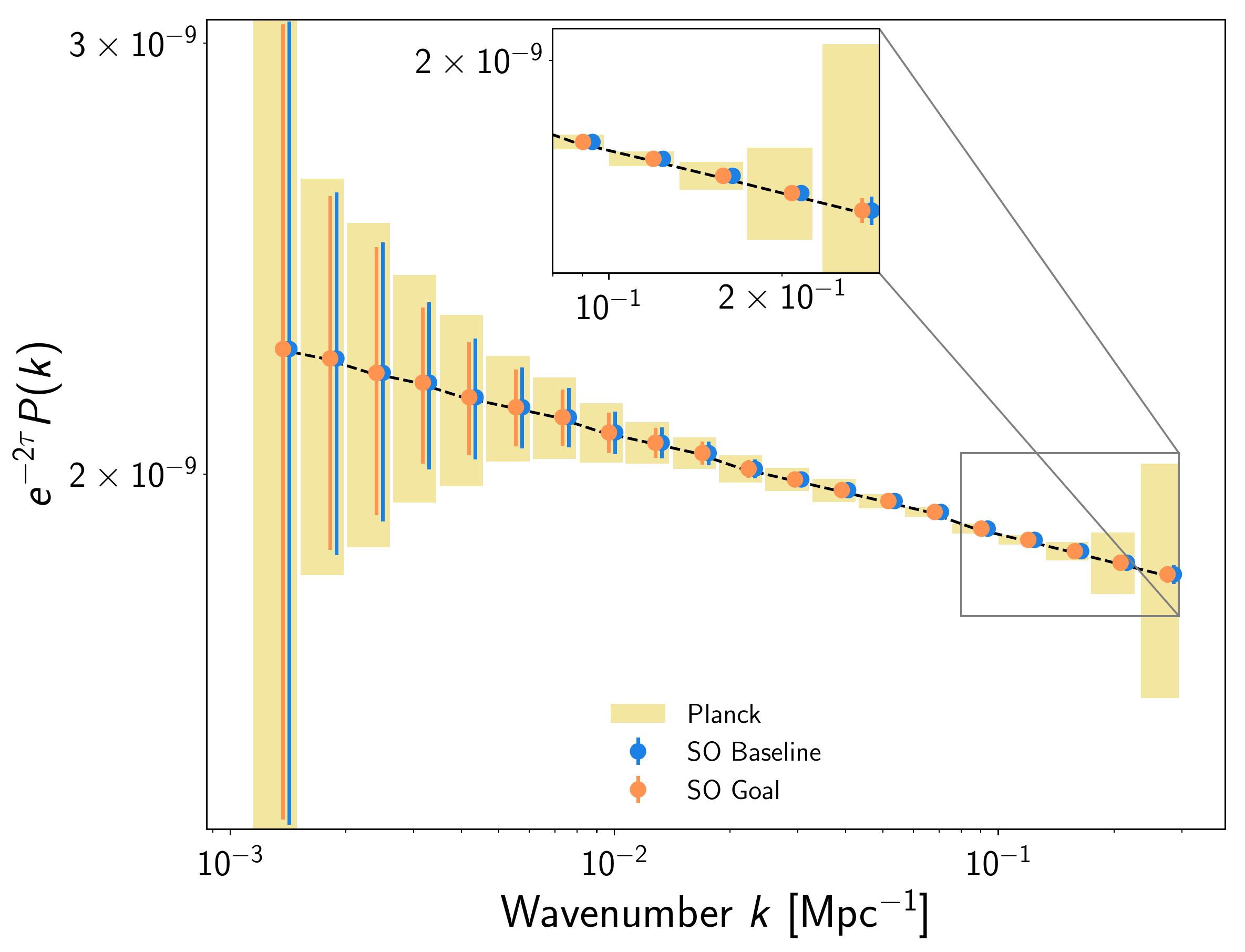}
\caption{Constraints on the primordial power $e^{-2\tau}P(k)$ from \so{} baseline (blue) and goal (orange) configurations, compared to estimated constraints from \planck\ temperature and polarization (yellow boxes). The large-scale constraints come from the combined \so{}+\planck\ temperature and polarization, with \planck\ significantly contributing to the constraint. 
The largest improvement in the spectra is seen on small scales, where the error on the primordial power spectrum at $k=0.2$ Mpc$^{-1}$ improves by an order of magnitude thanks to the \so{} polarization data.
\label{fig:SO_pk}}
\end{figure}

We compute the power deviation in bins for $0.001<k/\mathrm{Mpc}^{-1}<0.35$. This roughly corresponds to a range in multipoles of $14 < \ell < 5000, $ given that $\ell= k D(z=1089),$ where $D(z=1089) \simeq 14000$ Mpc is the comoving distance to the last scattering surface. Given the degeneracy between the power spectrum amplitude and the optical depth, $\tau,$ we present the results in terms of marginalized errors on the quantity $e^{-2\tau}P(k)$. The forecasts are shown in Fig.~\ref{fig:SO_pk}. 
Picking out the $k^*=0.2$~Mpc$^{-1}$ scale, which is the scale that shows the largest improvement over current results, we forecast
\ba
\sigma(e^{-2\tau}\mathcal{P}(k^*)) &=& 6.5 \times 10^{-12} \quad {\rm SO \ Baseline}, \nonumber \\
\sigma(e^{-2\tau}\mathcal{P}(k^*)) &=& 5.9 \times 10^{-12} \quad {\rm SO \ Goal},
\ea
for $e^{-2\tau}\mathcal{P}(k^*)=1.8\times10^{-9},$ a 0.4\% measurement. For comparison, 
$\sigma(e^{-2\tau}\mathcal{P}(k^*))=5.3\times10^{-11}$ currently for \planck, a 3\% measurement of the power at these scales.  
The improvement in the polarization of \so{} over the current \planck\ constraints (which include high-$\ell$ temperature and polarization, and low-$\ell$ polarization with the same prior on $\tau$) leads to a reduction in the error on the primordial power over a wide range of scales $0.001<k/\mathrm{Mpc}^{-1}<0.3$. 
The errors are somewhat sensitive to the choice of binning in the primordial power, future work will investigate the optimal number of bins given the noise properties of the survey. Finally we note that the constraints are not strongly dependent on the lensing reconstruction $\kappa\kappa$: the \so{} improvement over \planck{} is dominated by high-precision polarization.

\subsection{The Hubble constant}
\label{ssec:hubble}
The local Hubble constant, $H_0$, can be estimated from the CMB power spectrum for a given cosmological model. 

Since 2013, with the first \planck{} cosmological release, the Hubble constant derived from the CMB and its local measurement from Cepheids in Type Ia Supernovae host galaxies are in increasing tension if one assumes the \LCDM\ model. They are now discrepant at the level of $3.6\sigma$ (see e.g.,~\citealp{planck2018:parameters,Riess:2018byc} for recent discussions). Recent works have demonstrated that this tension with the local measurement exists even independently of \planck{}, considering for example \map{} or BBN results combined with BAO and SNe~\citep{2018ApJ...853..119A,2017arXiv171100403D,2018arXiv180606781L}. The explanation for this discrepancy is currently unknown, and simple extensions to the \LCDM\ model cannot explain all current cosmological data. Residual systematics in the data and/or new physics are possible explanations. 

The constraint on the power spectra from \so{} in its baseline configuration\footnote{We note that \so{} Goal leads to the same measurement, as the constraining power on $\Lambda$CDM basic parameters saturates at \so{} noise levels.} translates into a forecast measurement of $H_0$ in units of  km/s/Mpc, within \LCDM, with
\begin{equation}
\sigma(H_0)=0.3 \quad {\rm \so{} \ Baseline} \,.
\end{equation}
This has contributions from each high-$\ell$ spectrum, with
\begin{eqnarray}
\sigma(H_0)&=&0.6 \quad {\rm \so{}} \ TT \,, \nonumber \\
\sigma(H_0)&=&0.5 \quad {\rm \so{}} \ TE \,, \nonumber \\
\sigma(H_0)&=&0.5 \quad {\rm \so{}} \ EE \,.
\end{eqnarray}

\begin{figure}[tp!]
\centering
\includegraphics[width=\columnwidth]{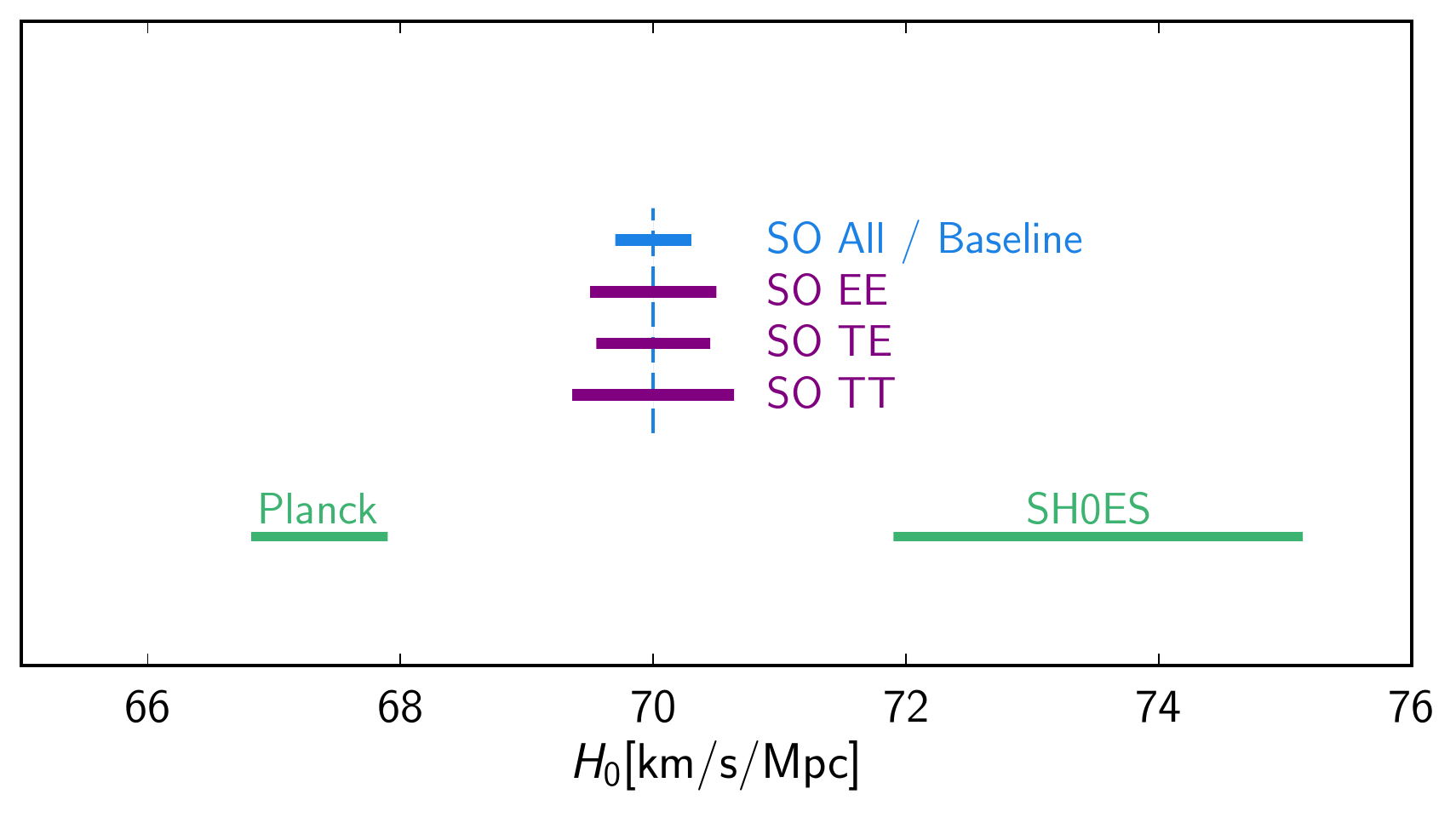}
\caption{
Constraints on the Hubble constant in a $\Lambda$CDM model from different \so{} high-$\ell$ channels and the full \so{} baseline dataset (purple and blue bars), compared to the current estimate provided by \planck{} (TT,TT,EE+lowE+lensing,~\citealp{planck2018:parameters}) and the most recent local measurement from the SH0ES project~\citep{Riess:2018byc}. The \so{} forecasts are centered on a fiducial value of $H_0=70$~km/s/Mpc, i.e, the mean value between the \planck{} and the SH0ES measurements. \\ \label{fig:H0}}
\end{figure}

The total uncertainty expected from \so{} is about a factor of two better than current estimates from \planck{} and five times better than determinations of $H_0$ from Cepheids in the local universe (see~\citealp{Riess:2018byc} for the most recent estimates), as shown in Fig.~\ref{fig:H0}. 

\so{} is therefore expected to provide a new compelling CMB estimate of $H_0$ which will test the \planck{} and other early universe results. 
Beyond improvement at the statistical level, the sensitivity of the different spectra ($TT, TE, EE$) will provide three nearly-independent and similarly powerful routes for SO to estimate $H_0$, checking for internal consistency and ensuring robustness against systematics. Although the forecast \so{} numbers include \planck{} data, we note that the significant improvement expected from \so{}, and a \planck-independent new estimate of $H_0$, is coming from the $TE$+$EE$ combination at scales not well measured by \planck{}. 

\subsection{Additional high-$\ell$ science}
\label{ssec:highl}
\subsubsection{Neutrino mass}

One of the ways in which \so{} will measure the neutrino mass is indirectly via precise measurements of the CMB lensing signal (see Sec.~\ref{sec:lensing} for details). The lensing signal is best measured in the convergence power spectrum computed from temperature and polarization four-point correlations (described in Sec.~\ref{sec:lensing}), but it also affects the two-point correlation function, through lensing-induced peak smearing in the temperature and polarization power spectrum. For the LAT baseline configuration, and for CMB power spectrum data alone, we forecast a constraint on the total sum of the masses of
\be
\sigma(\Sigma m_\nu)^{\rm 2pt} =110~{\rm meV}.
\ee

This constraint from the two-point function is a somewhat weaker constraint than can be derived from the CMB alone using the 4-point function, described in Sec.~\ref{sec:lensing}, where $\sigma(\Sigma m_\nu)^{\rm 4pt} =90~{\rm meV}$\footnote{We emphasize that, differently from the numbers reported in Sec.~\ref{sec:lensing}, these estimates are CMB-only constraints, i.e., without adding large-scale-structure or improved optical depth data.}.
From either method, the error improves significantly when including BAO information from DESI. As discussed in Sec.~\ref{sec:lensing}, the error would also improve if improved optical depth measurements become available. Both the two-point and four-point estimators yield neutrino mass constraints of the same order of magnitude. Considering that at SO noise levels, correlations between two-point and four-point statistics are at the level of only a few percent~\citep{Peloton2017}, these neutrino mass measurements will provide two nearly-independent constraints and therefore an important robustness check on each other.

\subsubsection{Big bang nucleosynthesis}

Big bang nucleosynthesis is the process by which light elements were formed in the early universe \citep[for reviews see][]{Patrignani:2016xqp, 2016RvMP...88a5004C}.  In standard cosmology, the predicted primordial light element abundances are determined by the baryon-to-photon ratio (which is fixed by $\Omega_bh^2$ and the current CMB temperature) and radiation energy density as measured by $N_\mathrm{eff}$.

We focus here on the primordial abundances of deuterium, parameterized by $y_D \equiv 10^5 n_\mathrm{D} / n_\mathrm{H}$, and \ce{^4 He}, parametrized by $Y_p \equiv 4n_\mathrm{He} / n_b$.  Theoretical predictions exist for the abundances of \ce{^3 He}, \ce{^6 Li}, and \ce{^7 Li}, but the observational status of these isotopes is much less certain~\citep{2016RvMP...88a5004C}.  
We use an interpolation of the results of the PRIMAT code~\citep{2018arXiv180108023P} to predict $y_D$ and $Y_p$. The uncertainty on the predicted deuterium abundance is $\sigma(y_\mathrm{D}) = 0.06$ and is dominated by the nuclear reaction rate d(p,$\gamma$)\ce{^3 He}~\citep{2011RvMP...83..195A}.  The theoretical uncertainty of $\sigma(Y_p) = 0.0003$ on the primordial helium abundance is dominated by the neutron lifetime~\citep{Patrignani:2016xqp}.

\begin{figure}[t!]
\centering
\includegraphics[width=\columnwidth]{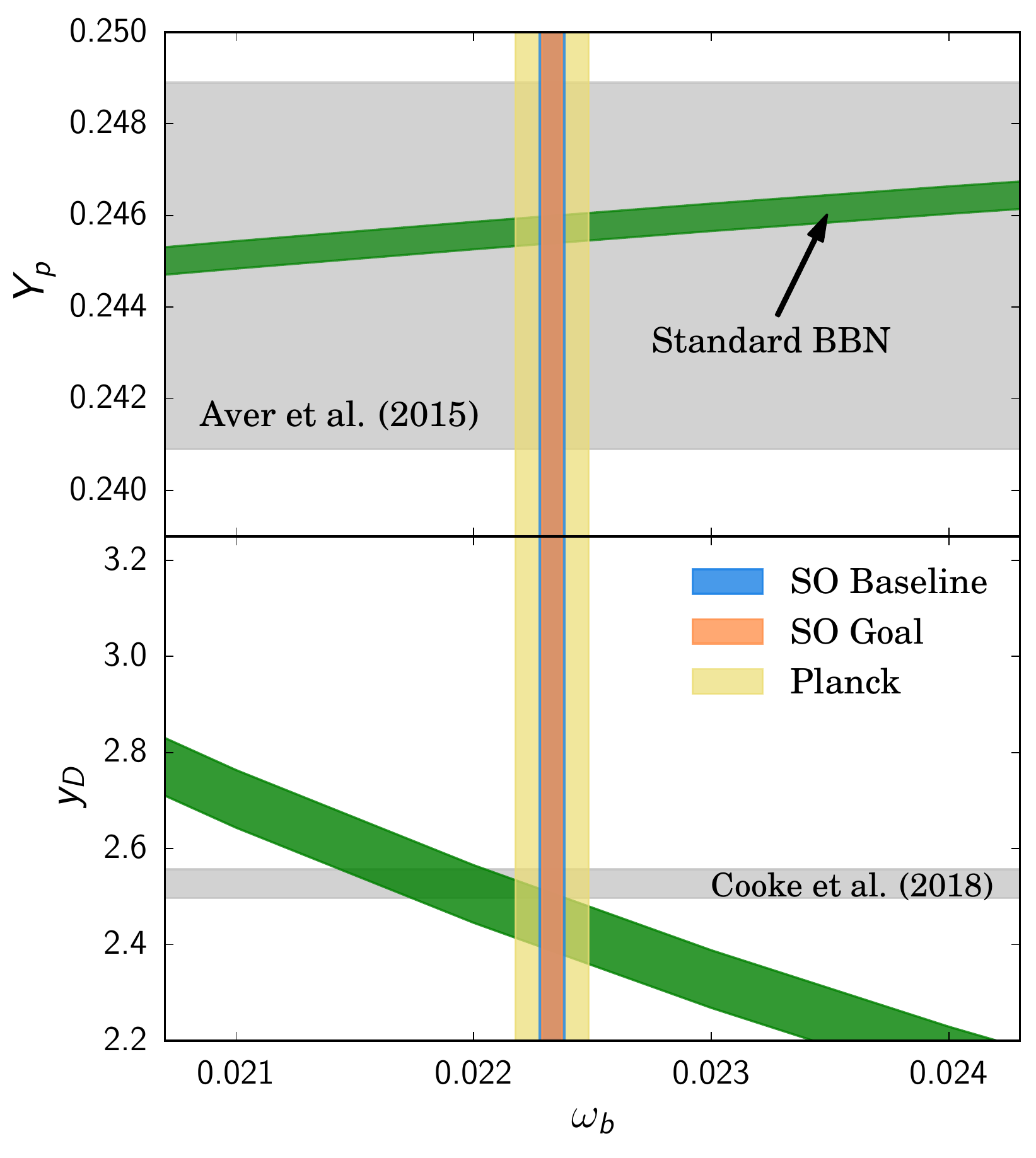}
\caption{Predictions of the primordial \ce{^4 He} and deuterium abundances assuming standard cosmology and $N_\mathrm{eff} = 3.046$ for the current \planck\ constraint and the forecast \so{} constraint on $\omega_b\equiv \Omega_bh^2$, compared to current astrophysical measurements of the primordial abundances.}\label{fig:BBN}
\end{figure}

Under the assumption of a standard cosmology with $N_\mathrm{eff}=3.046$, BBN is a one-parameter model depending only on the baryon density, $\omega_b\equiv\Omega_bh^2$, which is precisely measured with the CMB. Astrophysical measurements of primordial light element abundances~\citep{2015JCAP...07..011A,2018ApJ...855..102C} can then be used as a consistency check on the standard BBN model.  The results of this comparison with current CMB constraints on $\Omega_bh^2$ from \planck{}~\citep{planck2018:parameters} and forecast constraints from \so{} are shown in Fig.~\ref{fig:BBN}.  For the precision of \so{}, the consistency check on the primordial abundance of deuterium will be limited by uncertainties on the nuclear reaction rates used for the theoretical prediction from BBN. These uncertainties should be reduced by forthcoming results from low-energy accelerator experiments~\citep{Gustavino:2016qlm}.

\begin{figure}[t!]
\centering
\includegraphics[width=\columnwidth]{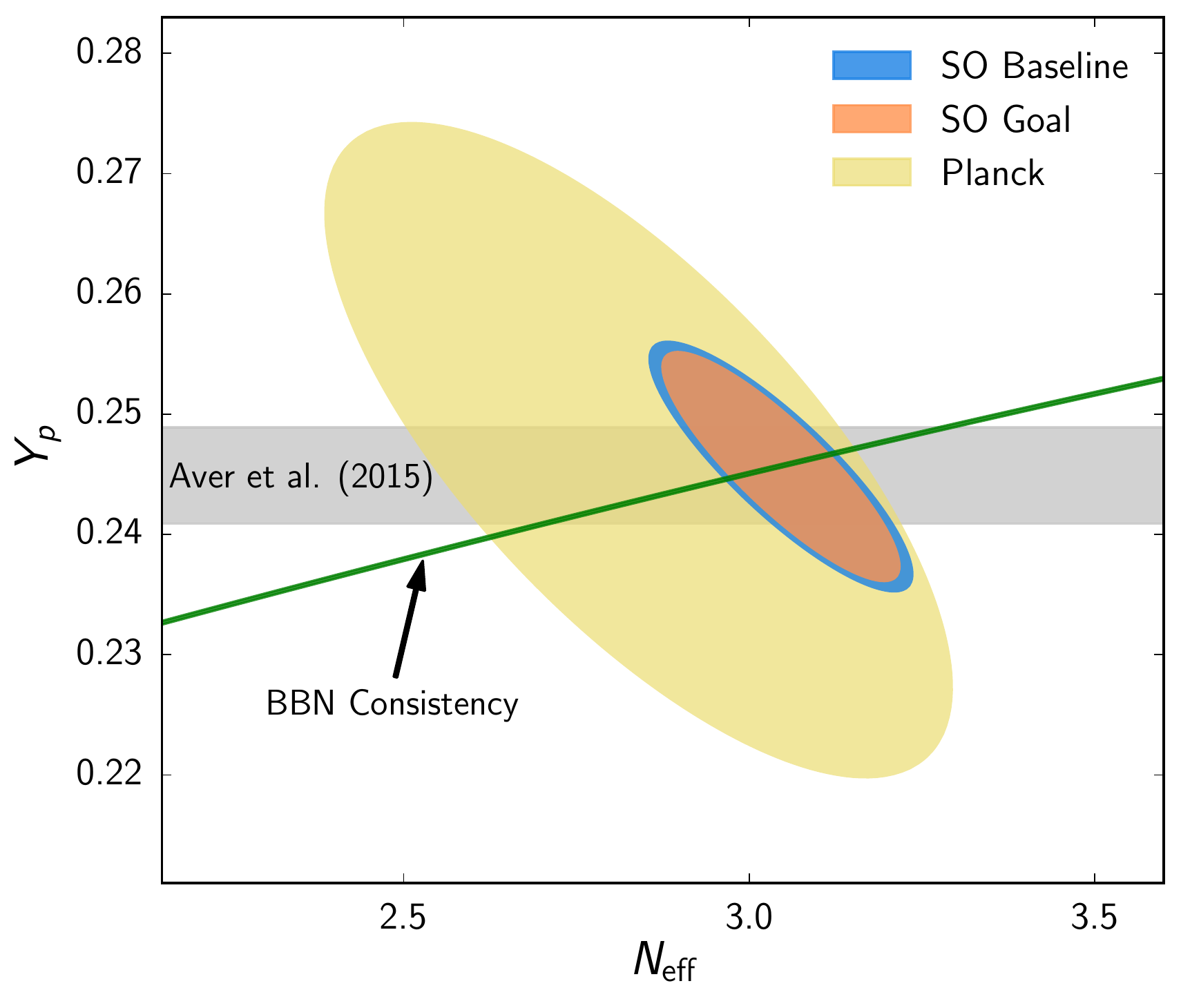}
\caption{Simultaneous CMB constraints, at 68\% confidence level, on the primordial helium abundance and light relic density from \planck\ and forecast constraints from \so{}. We also show the region predicted by BBN consistency, assuming a constant value for $N_\mathrm{eff}$ after neutrino freeze-out. }\label{fig:yp}
\end{figure} 

Motivated by models with non-standard thermal histories, we should seek simultaneous constraints on the primordial helium abundance and the light relic energy density \citep[see e.g.,][]{2011PhRvD..83f3520F,2015PhRvD..92b3010M}.  Both the primordial helium abundance and the light relic density impact the damping tail of the CMB power spectrum, and so $Y_p$ and $N_\mathrm{eff}$ are partially degenerate in CMB constraints when both are allowed to vary. However, this degeneracy is mitigated by effects on the polarization and on the phase of the acoustic oscillations~\citep{2004PhRvD..69h3002B,2013PhRvD..87h3008H,2016JCAP...01..007B}.  In Fig.~\ref{fig:yp} we show the current constraints from \planck~\citep{planck2018:parameters} along with the error forecasts for \so{}.  We forecast
\be
\sigma(Y_P)=0.007 \quad {\rm SO \ Baseline,~varying}~N_{\rm eff}\footnote{Delensing the temperature and polarization sharpens acoustic peaks and further helps to reduce the degree of this degeneracy~\citep{2017JCAP...12..005G}. The improvement of delensing is however negligible at \so{} noise levels -- only 5\% -- and therefore not included here.}.
\ee
These are plotted with the prediction from BBN assuming that the value of $N_\mathrm{eff}$ remains constant after neutrino freeze-out. 

\subsubsection{Dark Matter nature and interactions}

Evidence for non-gravitational interactions of dark matter particles would provide insight into the unknown physical nature of dark matter.
To date, the CMB has opened a valuable window into some of these interactions through the limits on dark matter annihilation \citep{Padmanabhan:2005es}.  
However, {\it Planck} has largely saturated this measurement~\citep{Madhavacheril:2013cna,Planck2015params,planck2018:parameters,Green:2018pmd}, and so SO is expected to improve it by only $\sim 30$ percent over {\it Planck}. 
Below we focus on two scenarios in which SO can substantially improve the sensitivity of CMB searches: elastic scattering of sub-GeV dark matter with Standard Model baryons and ultra-light axion dark matter.
We also note that galaxy cluster measurements discussed in Sec.~\ref{sec:sz} will also provide information on the dark-matter density profiles and thus may prove valuable in constraining other types of DM interactions (see, e.g.,~\citealp{More:2016vgs,Berezhiani:2017tth,Adhikari:2018izo}), but those scenarios are not directly discussed in this paper.

\paragraph{Dark matter--baryon interactions} 
The dark matter--baryon scattering processes sought by traditional direct-detection experiments can also leave imprints on cosmological observables. 
They transfer momentum and heat between the dark matter and photon-baryon fluids, damp the acoustic oscillations, and suppress power on small scales in the primary CMB, the linear matter power spectrum, and the CMB lensing anisotropy. 

CMB measurements have been used to search for interactions of dark matter particles with masses down to 1 keV \citep{Gluscevic:2017ywp,Boddy:2018kfv,Slatyer:2018aqg,Xu:2018efh} -- far beyond the reach of current nuclear-recoil based experiments that are optimized to detect weakly interacting massive particles (WIMPs) much heavier than the proton \citep{2013arXiv1310.8327C}. 
Furthermore, cosmological searches for dark matter are conducted in the context of a wide variety of interaction theories (including the most general non-relativistic effective theory of dark matter--proton elastic scattering) and need not be restricted to a particular dark matter model \citep{Sigurdson:2004zp,Boddy:2018kfv,Xu:2018efh,Slatyer:2018aqg}. Finally, they probe large (nuclear-scale) interaction cross sections which are inaccessible to traditional dark-matter direct searches, due to the extensive shielding of those experiments
\citep{Chen:2002yh,Dvorkin:2013cea,Emken:2017qmp}. For these reasons, they present a unique avenue for testing dark matter theory, complementary to laboratory searches.

\begin{figure}[t!]
\centering
\includegraphics[width=\columnwidth]{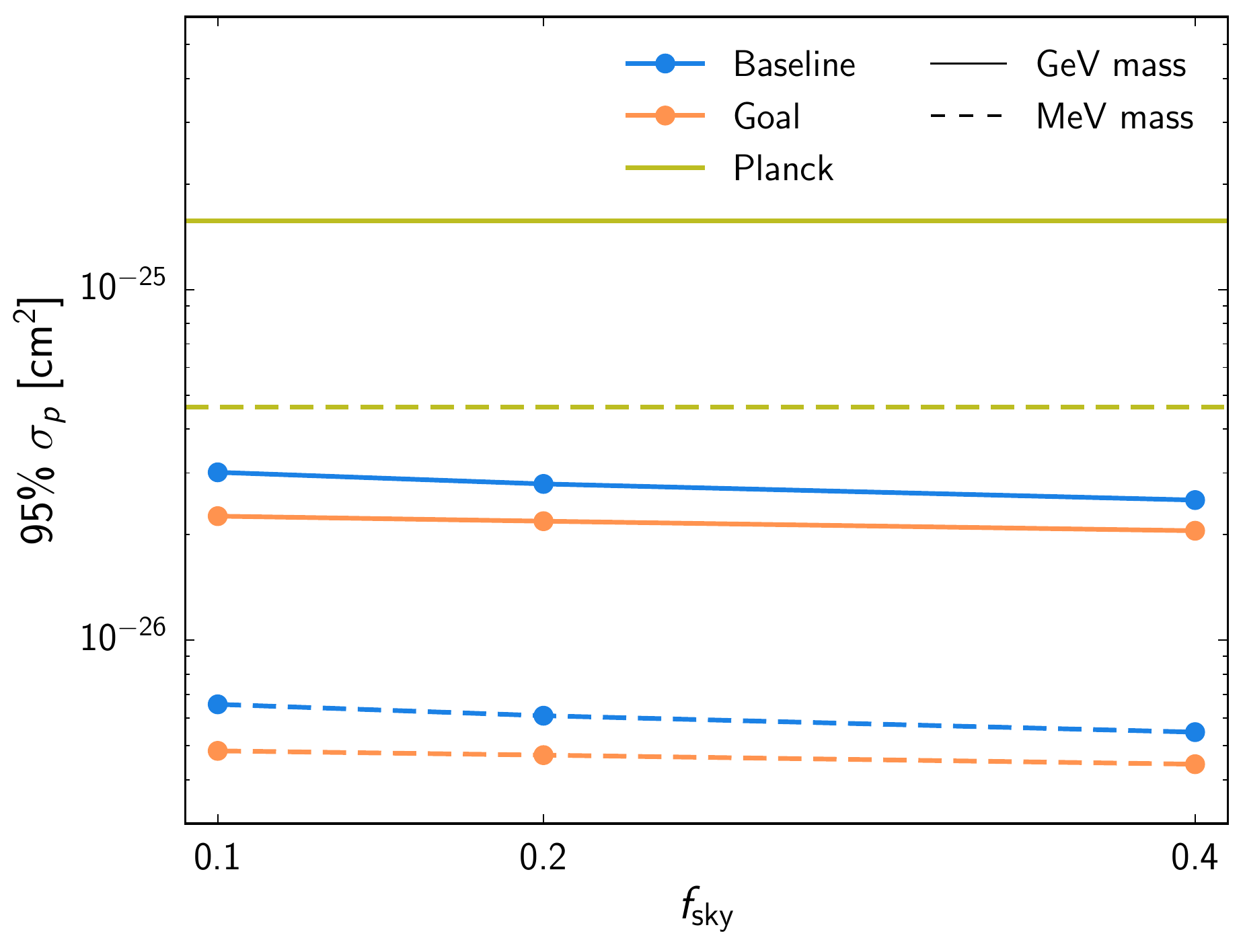}
\caption{We show current 95\% confidence-level upper limits on the cross section for dark matter--proton elastic scattering, derived using \planck\ temperature, polarization, and lensing anisotropy measurements (\citealp{Gluscevic:2017ywp}, yellow lines) assuming a velocity-independent spin-independent interaction. Limits for dark matter masses of 1 GeV and 1 MeV are shown as horizontal solid and dashed line, respectively. We also show the projected limits for the baseline and goal \so{} LAT configurations, as a function of the fractional sky coverage, for the same two masses. We see an improvement of a factor of 7.6 in constraining power for \so{} with $f_\text{sky}=0.4$, over \planck. 
\label{fig:SO_dmeff}}
\end{figure}

In Fig.~\ref{fig:SO_dmeff}, we show the current 95\% confidence-level upper limits on the cross section for elastic scattering of dark matter and protons, for two dark matter particle masses: 1 GeV and 1 MeV, and assuming velocity-independent scattering, from~\cite{Gluscevic:2017ywp}. In the same plot, we also show the projected \so{} upper limits for two configurations (baseline and goal noise levels), for a range of sky areas, $f_\text{sky}$, given a fixed observing time. \so{} is expected to improve the sensitivity to dark matter--proton scattering cross section by a factor of $\sim 8$, for a survey that covers 40\% of the sky, when temperature, polarization, and lensing are all included in the analysis. We forecast
\ba
\sigma_p ({\rm GeV}) &<& 3 \times 10^{-26}~{\rm cm}^2~({\rm 95\%})  \quad {\rm \so{} \ Baseline,}  \nonumber\\
\sigma_p ({\rm MeV}) &<& 5 \times 10^{-27}~{\rm cm}^2~({\rm 95\%})  \quad {\rm \so{} \ Baseline},
\ea
compared to current limits of $2\times 10^{-25}$ and $5\times 10^{-26}$ cm$^2$ respectively from {\it Planck}~\citep{Gluscevic:2017ywp}.

We find that reducing the noise from the baseline to goal level would improve the limit by about 25\%, and narrowing the sky coverage to $f_\text{sky}=0.1$ would degrade the limit by about 15\%. We note that the limit is driven by both the measurement of the primary CMB and intermediate-$\ell$ lensing power spectrum (inclusion of the SO lensing signal, which is discussed in Sec.~\ref{sec:lensing}, improves the limits by a factor of 3).

\paragraph{Ultra-light axions}
\begin{figure}[t!]
\centering
\includegraphics[width=\columnwidth]{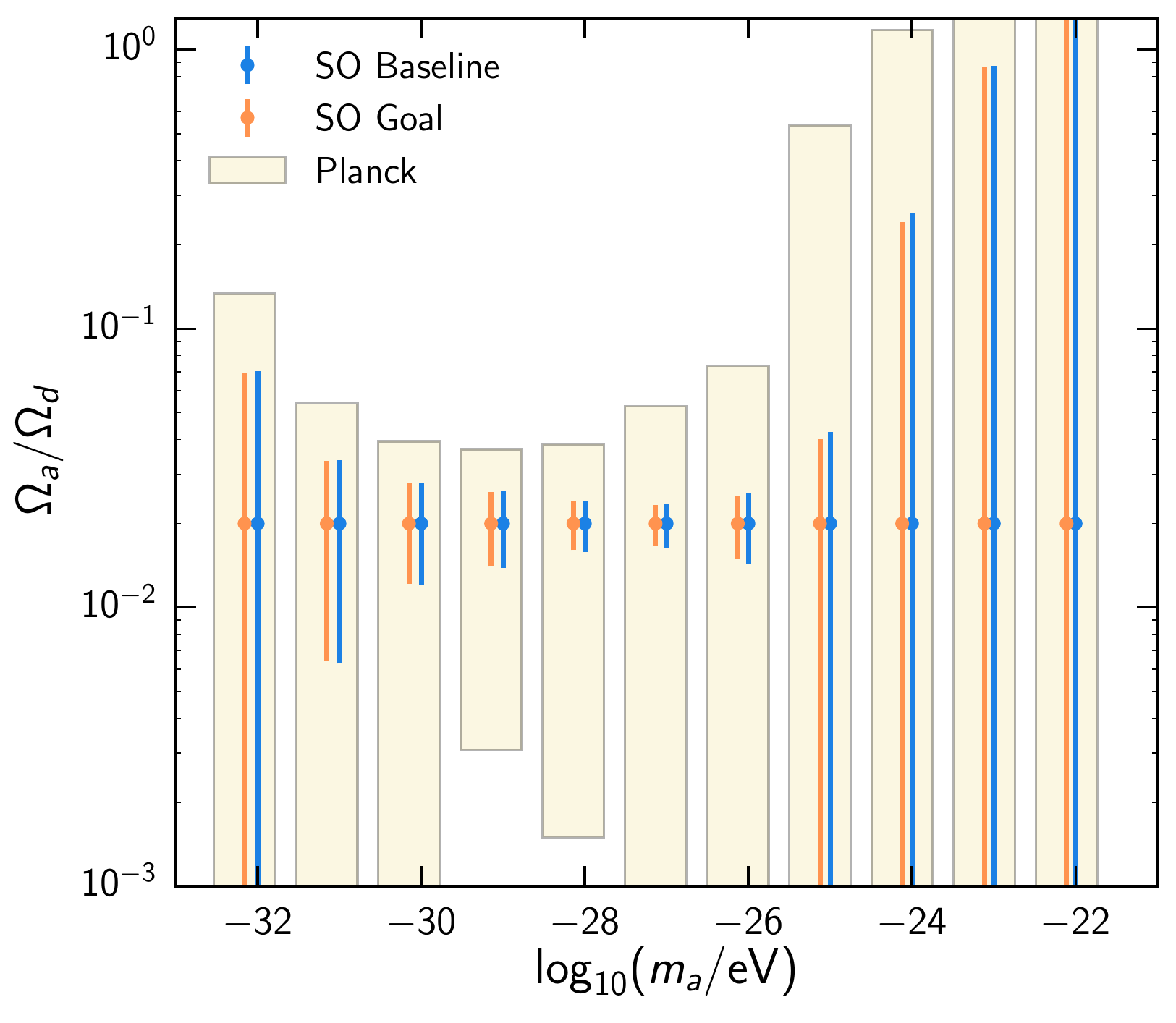}
\caption{
Forecast constraints on the axion fraction, assuming an axion fraction of $2\%$ of the total dark matter content for the baseline (blue) and goal (orange) \so{} LAT configurations, at a fixed neutrino mass of 0.06 eV. The boxes and error bars represent the 1$\sigma$ errors from a Fisher matrix analysis. The \planck\ errors match the current constraints in~\cite{Hlozek:2014lca}. \so{} will greatly improve the chance of detecting even a small amount of axions in the intermediate mass regime, which is bounded at $2\sigma$ by current \planck\ constraints.\\
\label{fig:SO_axions}}
\end{figure}
In addition to interactions from massive dark matter candidates, SO will open up a window into non-thermal ultra-light particle dark matter. One example of this candidate dark matter are ultra-light axions (ULAs). Axions were initially proposed to solve the strong CP problem ~\citep{pecceiquinn1977,weinberg1978, wilczek1978,Abbott:1982af,Ipser:1983mw,Moody:1984ba,2010RvMP...82..557K}, however we consider here not the QCD axion with mass $\sim 10^{-7}~\mathrm{eV}$ but its much lighter cousin, the ULA, with masses between $10^{-33}<m_a/\mathrm{eV}<10^{-22}$ \citep{ 2006PhLB..642..192A,axiverse, 2017PhRvD..95d3541H} The light end of the mass range are axions that exhibit dark-energy-like behavior, while the axions with masses $m_a \geq 10^{-27}~\mathrm{eV}$ are dark-matter-like.

The ULAs we consider here are created through vacuum realignment and as such are non-thermal dark matter candidates~\citep{2006PhLB..642..192A,Hwang:2009js,axiverse,Noh:2013coa,2017PhRvD..95d3541H}. A recent review of the cosmology of axions is found in~\cite{2016PhR...643....1M}. 

At a redshift specified by their mass, the axion field coherently oscillates, and the ULAs transition from behaving more like scalar field dark energy to being a dark-matter-like cosmological component, which has strong cosmological implications for structure formation~\citep{Schive:2015kza, 2017MNRAS.471.4606A,2018arXiv180610608P, 2018PhRvD..97h3502F}. Hence for the lightest axions masses ($m_a \simeq 10^{-33}~\mathrm{eV},$ the value of $H_0$ in eV today), the field is frozen and never begins oscillating and the axions appear dark-energy-like today. 

Axions affect the primordial power spectrum by smoothing out structures on small scales through their scale-dependent sound speed~\citep{Marsh:2010wq,Hlozek:2014lca}. In addition to these matter-power-spectrum effects, they impart signatures on the CMB temperature and polarization spectra depending on whether the mass transition was before or after decoupling. The lightest axions have a signature on the CMB power spectra similar to dark energy; they generate an Integrated Sachs--Wolfe signal due to the fact that the gravitational potential wells are shallower when the $10^{-33}\mathrm{eV}$ axions dominate the total energy of the universe. We consider the full range of axion masses, to test the ability of \so{} to constrain both dark-matter-like and dark-energy-like axions.

For our constraints we make use of the publicly available code {\tt axionCAMB}\footnote{Available at \url{https://github.com/dgrin1/axionCAMB}.}, integrated into a modified version of the Fisher forecast code {\tt OxFish}~\citep{allison/etal/2015}.
We perform a Fisher matrix forecast at fixed mass and vary the allowed axion fraction, where we assume a fiducial value set by the upper limit of current experiments $\Omega_a/\Omega_d = 0.02$~\citep{Hlozek:2014lca}. We compute the error on the axion fraction for the baseline and goal LAT configurations and using a sky fraction of $f_{\rm sky}=0.4$. These are shown in Fig.~\ref{fig:SO_axions}. The improvement in the constraints with respect to current data is significant, with for example the uncertainty on the axion fraction reducing current limits by more than a factor of $5$ for $m_a=10^{-26}\mathrm{eV}$, as summarized in Table~\ref{tab:surveys}.

\section{Gravitational lensing}
\label{sec:lensing}
Measurements of CMB lensing underlie many key areas of SO science, including constraints on the tensor-to-scalar ratio, neutrino mass, dark energy, and high-redshift astrophysics. In this section, we detail the calculations we use to guide the design of SO from the perspective of lensing science, focusing on the delensing efficiency (Sec. \ref{sec:lensing.delensing}), neutrino mass constraints (Sec. \ref{sec:mnuForecast}), and science from cross-correlations (Sec. \ref{sec:cross-corr}) including growth of structure, primordial non-Gaussianity, shear bias, and spatial curvature. We show how lensing can be used for halo mass calibration in Sec \ref{sec:HaloLens}, and discuss the impacts of foregrounds in Sec. \ref{sec:lensingforegrounds}.

Our nominal lensing noise curves are described in Sec.~\ref{sec:method} and shown in Fig.~\ref{fig:lensing_noise}. SO lensing noise is derived based on LAT temperature and polarization data (the SATs have much lower resolution and hence are not useful for lensing measurements); in our derivation, we assume that lensing is reconstructed using the minimum-variance lensing quadratic estimator~\citep{HDV2007}, which is nearly optimal for our noise levels. The noise curves we obtain are used for most of the results of this paper (including Secs.~\ref{sec:lensing.delensing},~\ref{sec:mnuForecast}, and also Sec.~\ref{sec:highell}). Deviations from the nominal case are considered in some specific cases. In particular: in Sec.~\ref{sec:mnuForecast} we also explore polarization-only estimators; in Secs.~\ref{sec:cross-corr} and ~\ref{sec:HaloLens}, we apply a foreground cleaning prescription to the gradient leg of the estimator (i.e. the low-pass filtered field in the quadratic estimator), as described in Sec.~\ref{sec:lensingforegrounds}; in Sec.~\ref{sec:HaloLens} we in addition apply a maximum $\ell$ cut of $\ell=2000$ to the gradient leg of the estimator, as detailed in that section.  

\subsection{Delensing efficiency}\label{sec:lensing.delensing}

Lensing-induced $B$-modes limit the measurement of the tensor-to-scalar ratio, $r$. Fortunately, given an estimate of the particular realization of the CMB lensing potential in a given patch of sky, it is possible to partially undo the effect of lensing on CMB maps \citep{2002PhRvL..89a1303K,2004PhRvD..69d3005S}. This procedure is called `delensing'. Delensing reduces the lensing contribution to the $B$-mode polarization power spectrum,  reducing cosmic variance and allowing an improved measurement of any underlying primordial $B$-modes. The estimate of the lensing potential can be obtained through internal reconstruction, using the CMB maps themselves, or through appropriately weighted combinations of external tracers of the dark matter, such as the CIB~\citep{1502.05356} or high-redshift galaxies~\citep{2012JCAP...06..014S}.

\begin{figure}[t!]
\centering
\includegraphics[width=\columnwidth]{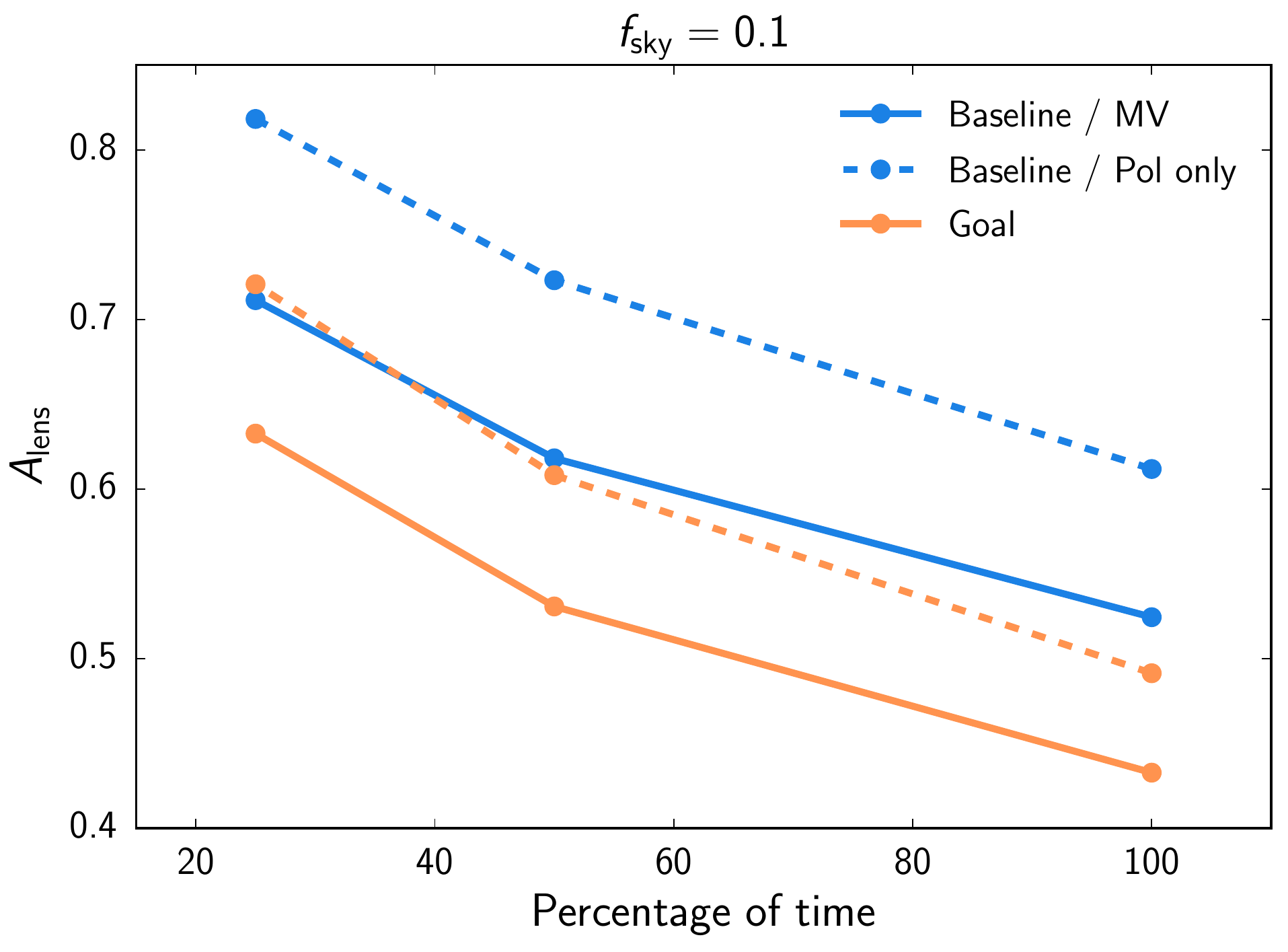}
\caption{Fraction of lensing $B$-mode power remaining after delensing, $A_\mathrm{lens}$, as a function of time spent observing the SAT survey footprint with the LAT.  We show SO baseline and goal sensitivities, using the minimum variance lensing estimator (`MV)', which combines temperature and polarization data, as well as the polarization lensing estimators alone (`Pol only'). Here we assume standard ILC foreground cleaning as in Sec.~\ref{sec:LAT_ILC}. We note that the nominal  plan for sky coverage, as discussed Sec.~\ref{subsec:coverage}, is to spend 25\% of the LAT time observing the SAT footprint.  For comparison, we anticipate that $A_{\rm lens}=0.5$ can be achieved using external delensing.}
\label{delensingEfficiency}
\end{figure} 
\vspace{2mm}

We forecast the fraction of lensing $B$-mode power remaining in the delensed CMB map, $A_{\rm{lens}}$, after delensing.  We assume iterative maximum likelihood lensing reconstruction in our forecasts, using the formalism of \citet{2012JCAP...06..014S}. We assume the lensing potential will be obtained using the minimum variance lensing estimator, using both temperature and polarization, though, where indicated, we also show polarization-only lensing reconstruction for comparison.

In Fig.~\ref{delensingEfficiency}, we show $A_{\rm{lens}}$ as a function of the fraction of time spent observing the SAT survey footprint with the LAT. Here, 100\% observing time corresponds to the LAT survey covering only the $f_{\rm sky}=0.1$ coincident with the SAT survey. Observing percentages of 50\% and 25\% correspond to a LAT survey covering $f_{\rm sky}=0.2$ and $0.4$ respectively.

The delensing performance is best for a LAT survey overlapping fully with the SAT. In this case, from Fig.~\ref{delensingEfficiency}, the fraction of $B$-mode lensing power remaining after delensing is optimistically of order 0.5 for baseline noise, which would reduce~$\sigma(r)$ by~$\approx 30\%$, as summarized in Table~\ref{table:r0}. Polarization-only reconstruction, which potentially has fewer systematic concerns, does not perform as well. Foreground cleaning causes a more significant loss of delensing performance in this case.

While this configuration gives optimal delensing, the smaller sky area is non-optimal for SO's broader science goals. In addition, we expect to achieve $A_{\rm lens}$ of order 0.5 from external delensing \citep{1502.05356} using CIB emission and other external matter tracers. Though practical delensing demonstrations have not yet achieved such low  $A_{\rm lens}$ directly \citep{Larsen,Manzotti,CarronDelens}, recent work has shown the construction of multi-tracer maps with sufficient correlation with lensing to achieve $A_{\rm lens}\approx 0.5$ \citep{2017PhRvD..96l3511Y,planck2018:lensing}.   

We therefore anticipate the primary delensing path for SO to come from external delensing. Though the delensing performance is expected to be fairly robust to astrophysical uncertainty \citep{1502.05356},  systematic errors in external delensing are only now being investigated.  The internal delensing approach will therefore provide an important cross-check. For the nominal configuration where the LAT surveys $f_{\rm sky}=0.4$, we forecast
\be
A_{\rm lens}= 0.71 \quad {\rm \so{} \ Baseline}.
\ee
Here we assume the standard ILC foreground cleaning (Deproj-0) described in Sec.~\ref{sec:method}. We find $A_{\rm{lens}}=0.80$ with more pessimistic foreground cleaning assumptions, using the Deproj-1 method.

\subsection{Neutrino mass and lensing spectra}
\label{sec:mnuForecast}

Non-zero neutrino masses suppress the amplitude of fluctuations measured at late times relative to the amplitude measured early on, at the CMB last scattering surface (see, e.g.,~\citealp{lesgourgues_mangano_miele_pastor_2013}). We can exploit this to measure the sum of the neutrino masses using the combination of the primary CMB power spectrum and the CMB lensing potential spectrum. We use a Fisher code to forecast constraints on the sum of neutrino masses using this combination, together with DESI BAO information as described in Sec.~\ref{sec:method}.
We use unlensed $C_{\ell}$s for the primary CMB and thus do not include any covariance between the primary CMB and the CMB lensing spectra.  As discussed in \citet{Peloton2017}, this is reasonable for the neutrino mass forecasts here since we include the BAO information which breaks parameter degeneracies (in particular, with the matter density). Here we assume a single-parameter extension to $\Lambda$CDM, varying the total sum of the neutrino masses around a fiducial value of 0.06~eV. We consider that only one of the three neutrino species is massive, and that it carries the total mass.  This is approximately correct if neutrino masses obey the `normal hierarchy' with the lowest mass zero and the larger mass splitting between the middle and largest-mass states.

\begin{figure}[t!]
\centering
\includegraphics[width=\columnwidth]{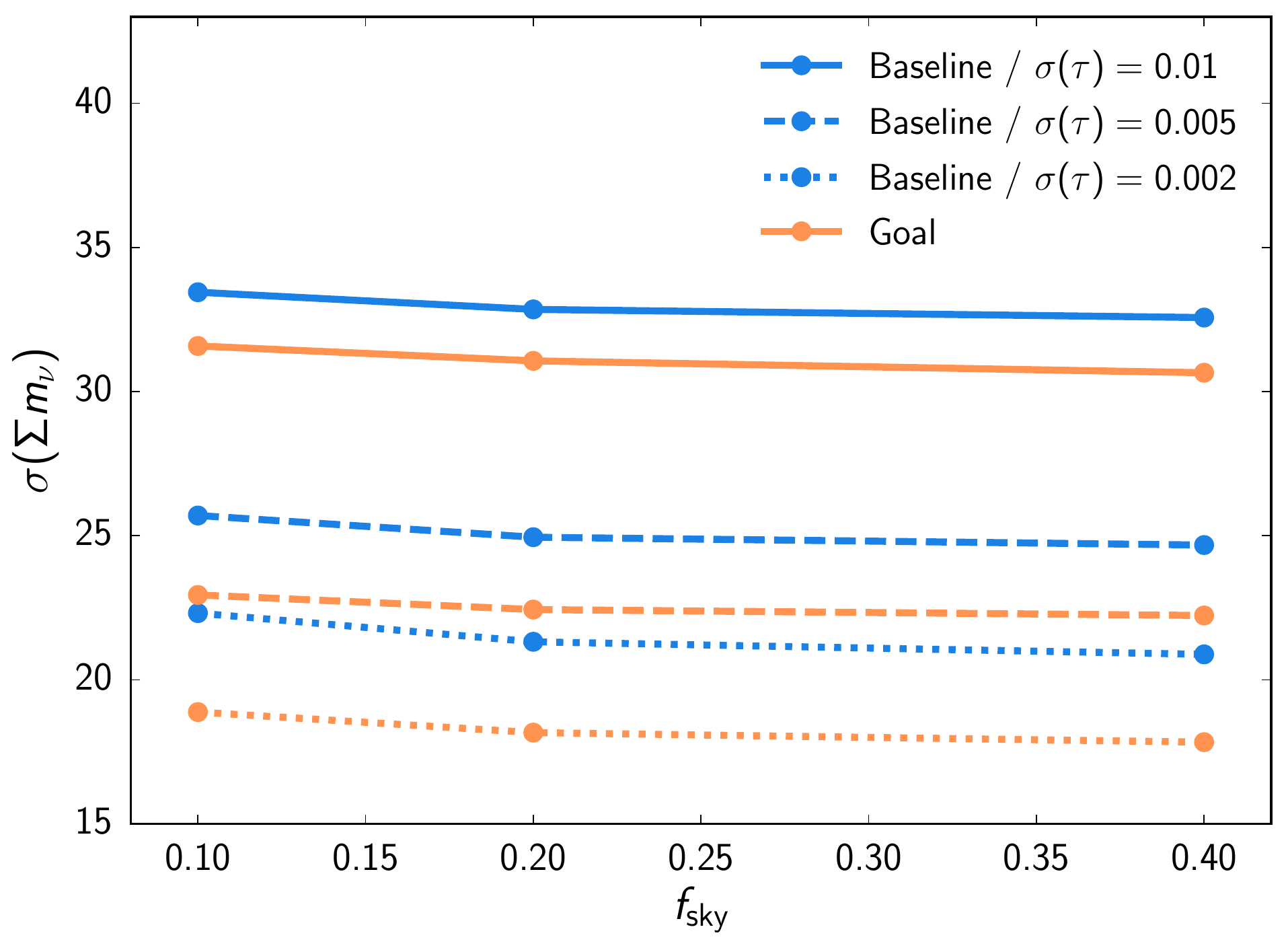}
\includegraphics[width=\columnwidth]{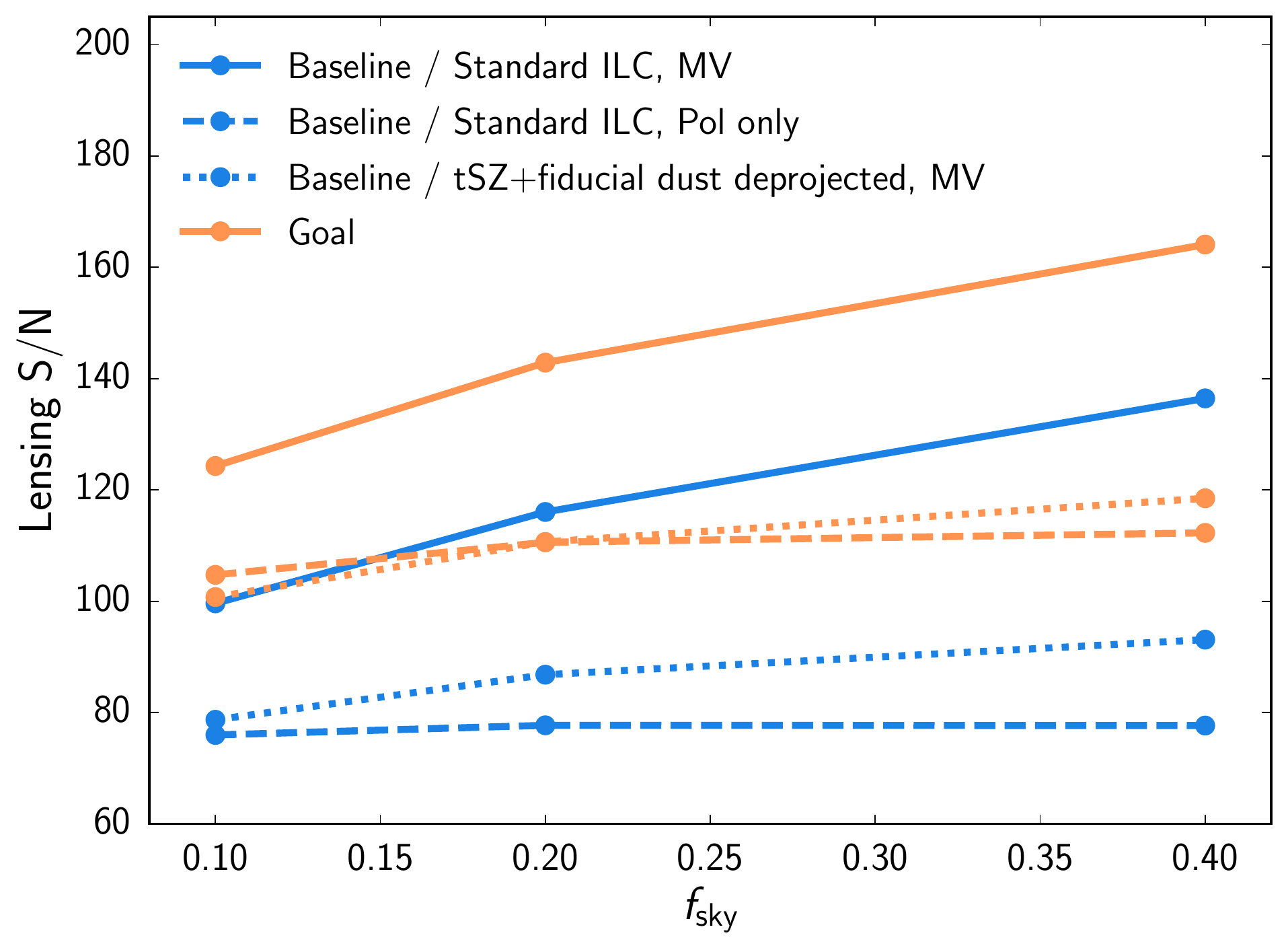}
\caption{\emph{Top:} Neutrino mass constraints from temperature and polarization CMB power spectra and lensing potential autospectrum, including foreground cleaning with explicit tSZ deprojection for temperature and dust deprojection for polarization, as a function of sky area. \emph{Bottom:} Lensing potential autospectrum signal-to-noise ratio from temperature and polarization CMB data as a function of sky area. Foreground cleaning is included with standard ILC without explicit deprojection (Deproj-0) or tSZ deprojection for temperature and dust deprojection for polarization (Deproj-1), as discussed in Sec.~\ref{sec:LAT_ILC}. }
\label{fig:mnu-SN}
\end{figure} 

We show the anticipated constraints as a function of sky area in Fig.~\ref{fig:mnu-SN}. Here we include foreground cleaning with explicit tSZ deprojection for temperature and dust deprojection for polarization (Deproj-1), since these foregrounds could potentially cause the largest systematic effects (see Sec.~\ref{sec:lensingforegrounds} below). We forecast
\ba
\sigma(\mnu)&=&33~{\rm meV}  \quad {\rm \so{} \ Baseline + DESI\mbox{-}BAO}\,, \nonumber\\
\sigma(\mnu)&=&31~{\rm meV}  \quad {\rm \so{} \ Goal + DESI\mbox{-}BAO}\,,
\ea
for $\sigma(\tau)=0.01$ -- i.e., for current measurements of $\tau$. The constraints depend only weakly on sky area, and moderately on sensitivity. This is due to  constraints being limited by the knowledge of the CMB optical depth $\tau$ \citep[e.g.,][]{allison/etal/2015}, which allows the high-redshift amplitude of structure to be determined.  With a different foreground cleaning assumption, namely Deproj-0, we find $\sigma(\mnu)=30$~meV for the baseline case.

The constraints on neutrino mass improve significantly with the addition of better $\tau$ data, as can be obtained from improved measurements of large-angular-scale ($\ell<30$, $\theta>6^\circ$) CMB $E$-modes. Current and upcoming ground-based, balloon, and satellite experiments show great promise to make these measurements. Challenges include dust and synchrotron foreground removal, as well atmospheric instability at $\sim$150 GHz and above for ground-based measurements. CLASS \citep{essinger-hileman14,2018arXiv181108287A} is a currently ongoing ground-based experiment aiming to measure large angular scale CMB from the ground at 40/90/150 GHz with a single dust channel at 220 GHz. If the data is sufficient for full foreground cleaning, it would provide a measurement of $\tau$ with $\sigma(\tau)=0.003$~\citep{watts2018}, which is near the half-sky cosmic variance (CV) limit.
BFORE \citep{bryan17,bryan18} aims to use the improved access to large angular scales and high frequencies offered by the balloon platform to measure the CMB at 150 GHz, and dust at 217/280/353 GHz.  The latter would yield a $\tau$ measurement, robust to dust contamination, also of $\sigma(\tau)=0.003$. BFORE has been selected by NASA for further study, and the current schedule has a launch in 2022. The LiteBIRD satellite \citep{Hazumi:2012gjy,matsumura14,Sekimoto:2018} aims to take all-sky CMB and foreground data at 40--400 GHz, and would be in a position to make an all-sky CV-limited measurement of reionization (i.e., $\sigma(\tau)=0.002$). LiteBIRD is under study by a large international team, and the current schedule has a launch in 2027.
Combining SO with CV-limited reionization data and DESI BAO data, we forecast
\ba
\sigma(\mnu)&=&22~{\rm meV}  \quad {\rm \so{} \ Baseline + DESI\mbox{-}BAO + \tau}\,,\nonumber\\
\sigma(\mnu)&=&17~{\rm meV}  \quad {\rm \so{} \ Goal + DESI\mbox{-}BAO + \tau}. 
\ea
The latter combination of data would enable a 3.5$\sigma$ detection of $\Sigma m_\nu=0.06$~eV (the minimal summed mass allowed in the case of normal mass hierarchy;~\citealp{Capozzi:2017ipn, Esteban:2016qun,2018PhLB..782..633D}) and a 5.9$\sigma$ detection of $\Sigma m_\nu=0.1$~eV (the minimal summed mass in the case of inverted mass hierarchy).  
  
Even without new measurements of large-scale $E$-modes, it may also be possible  to obtain improved  measures of $\tau$ using indirect methods with surveys operating from the ground.  On the timescale of SO, these methods involve using measurements of ionization perturbations, either of neutral hydrogen from an experiment such as the Hydrogen Epoch of Reionization Array \citep{2016PhRvD..93d3013L} or the kSZ effect from the SO itself (\citealp{2018arXiv180307036F}; see Sec.~\ref{sec:kszreionization} below), and to tie these measurements to the spatially averaged ionization state using modeling of reionization.  Alternative indirect methods with the CMB that do not rely on such modeling, such as reconstructing large-scale $E$-modes from the lensed small-scale data \citep{2017PhRvD..95l3538M} and using the polarized SZ effect from lower-redshift objects \citep{2018PhRvD..97j3505M}, require noise levels beyond the reach of SO.

Since the lensing potential power spectrum is a powerful cosmological probe, we also show forecasts for the signal-to-noise ratio of that observable in Fig.~\ref{fig:mnu-SN}. We forecast
\be
S/N_{\rm lensing} =140  \quad {\rm \so{} \ Baseline.}
\ee
We note that this signal-to-noise ratio only includes the lensing power spectrum derived from the four-point function. We find that the signal-to-noise ratio improves as sky area increases, although there is minimal improvement when we consider the potentially more systematic-error-free polarization-only lensing power spectrum. We find $S/N = 93$ for Deproj-1 foreground cleaning.

\subsection{Cross-correlations}
\label{sec:cross-corr}

Cross-correlations between maps of the reconstructed CMB lensing potential and low-redshift tracers of the matter distribution provide a handle on the growth of structure, primordial non-Gaussianity, curvature, and dark energy.  Here we forecast the science obtainable from cross-correlations of the SO reconstructed lensing potential with LSST, which is particularly suited for this type of analysis because it will observe a large number of galaxies out to high redshift, covering a large fraction of the CMB lensing redshift kernel. The LSST galaxy samples we consider are described in Sec.~\ref{sec:method}.

For the gold sample, we forecast the cross-correlation coefficient between the SO lensing field, $\kappa$ (using the minimum variance estimator), and the galaxy overdensity field, $g$, given by $\rho=C_L^{\kappa g}/(C_L^{\kappa\kappa}C_L^{gg})^{1/2}$, to be $\rho\ge 80\%$ for lensing scales $L\le 80$, with a maximum correlation of $\rho\simeq 86\%$ for $10\le L\le 20$, if the LSST galaxies are optimally weighted in redshift to match the CMB lensing kernel \citep{1502.05356,1710.09465}.
We forecast that the optimistic LSST sample will improve the cross-correlation coefficient to $\rho\ge 80\%$ for $L\le 150$, with a maximum correlation of $\rho\simeq 90\%$ for $15\le L\le 25$.

These high cross-correlation coefficients imply that sample variance cancellation can be useful \citep{0807.1770}, and thus that the naive signal-to-noise ratio calculation for a cross spectrum can understate the potential constraining power of combining CMB and large-scale structure (LSS).
We include this sample variance cancellation by forecasting joint constraints from SO lensing and LSST clustering power spectra, $C^{\kappa\kappa}_L$, $C^{\kappa g}_L$, and $C^{gg}_L$, accounting for (Gaussian) covariances between these power spectra.

Here and in Sec.~\ref{sec:HaloLens}, we assume a CMB lensing estimator where only the CMB gradient in the quadratic estimator pair has explicit deprojection of foregrounds~\citep{MMHill}, which is suitable for cross-correlation measurements.  We discuss this gradient foreground cleaning in Sec.~\ref{sec:lensingforegrounds}.

\subsubsection{Growth of structure: $\sigma_8(z)$}
\label{sec:growth}
We forecast how well the growth of structure as a function of redshift, parametrized by the amplitude of matter perturbations, $\sigma_8(z)$, could be constrained by combining LSST galaxies and SO CMB lensing data, for different sky fractions and sensitivities. We consider six tomographic redshift bins ($z =$ 0--0.5, 0.5--1, 1--2, 2--3, 3--4, 4--7) and marginalize over one linear galaxy bias parameter in each bin. As in Sec.~\ref{sec:cross-corr}, our observables are $C_L^{\kappa\kappa}$, $C_L^{\kappa g}$ and $C_L^{gg}$. We assume scale-cuts in both $L$ and $k$ (which the Limber approximation maps to an $L$ for each redshift bin) of $L_\mathrm{min}=50$ and $k_\mathrm{max}=0.3 h/\mathrm{Mpc}$. 
We follow the standard methods described in Sec.~\ref{sec:methods}, marginalizing over $\Lambda$CDM parameters and including DESI BAO information.
In this analysis, however, we allow $\sigma_8(z)$ to vary independent of the expansion history. We assume $\Lambda$CDM for the background expansion; we also neglect any effect on the CMB power spectra.

\begin{figure}[t!]
\centering
\includegraphics[width=\columnwidth]{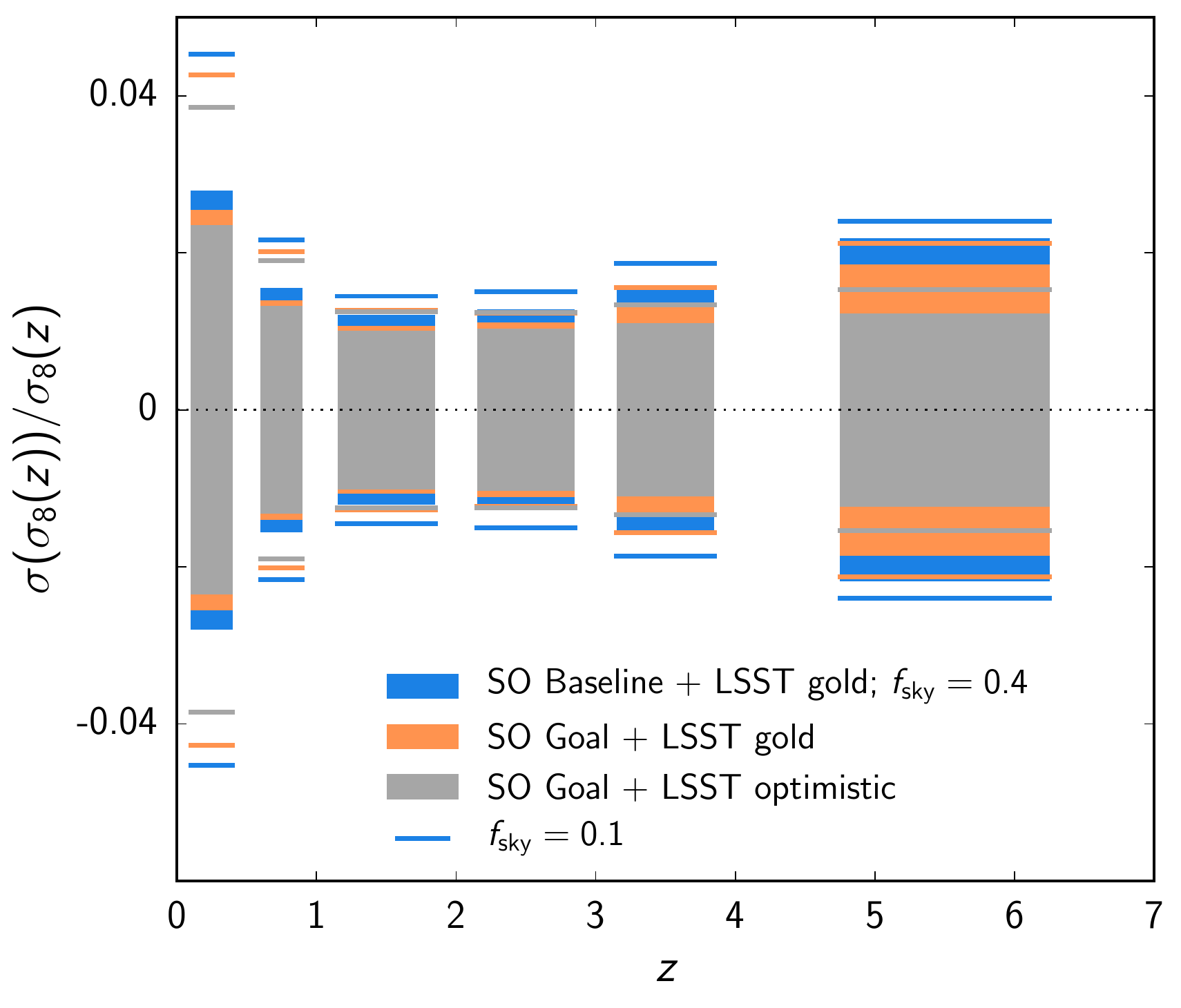}
\caption{The relative uncertainty on $\sigma_8$, defined in six tomographic redshift bins ($z$=0--0.5, 0.5--1, 1--2, 2--3, 3--4, 4--7), as a function of redshift.  The forecast assumes a joint analysis of $C_L^{\kappa\kappa}$, $C_L^{\kappa g}$ and $C_L^{gg}$, where $\kappa$ is \so{} CMB lensing and $g$ denotes LSST galaxies binned in tomographic redshift bins. We use multipoles from $L_{\mathrm{min}} = 30$ to $L_\mathrm{max}$ corresponding to $k_{\mathrm{max}} = 0.3h$Mpc$^{-1}$ in each tomographic bin. We marginalize over one linear galaxy bias parameter in each tomographic bin and over $\Lambda$CDM parameters.  We also include {\it Planck} CMB information -- with a $\tau$ prior, $\sigma(\tau) = 0.01$ -- and BAO measurements (DESI forecast). The filled bands and single lines distinguish between an \so{} survey over $f_{\mathrm{sky}} = 0.4$ and $0.1$, respectively. Different colors distinguish between the \so{} baseline and goal configurations, and for this latter case different LSST number density, with 29.4 and 66~$\mathrm{arcmin}^{-2}$ galaxies for the gold and optimistic sample, respectively. 
}
\label{fig:s8}
\end{figure} 
We show projected constraints in Fig.~\ref{fig:s8}, finding forecasts for $\sigma_8(z)$ that are competitive with cluster and weak lensing probes, with percent-level forecast constraints on the amplitude of structure for a significant number of redshift bins. We find that constraints improve moderately as sky area and sensitivity are increased, as shown in Fig.~\ref{fig:s8}. For the nominal SO survey we project
\ba
\sigma(\sigma_8)/\sigma_8 &=&0.015~(z=\textrm{1--2}), \, {\rm \so{} \ Baseline + LSST} \nonumber\\
 &=& 0.015~(z=\textrm{2--3}).
\ea
These measurements would provide unique constraints on deviations from the cosmological constant, $\Lambda$, at redshifts higher than typically accessible to optical cluster and weak lensing probes. These deviations could include modified gravity, non-standard dark energy models, as well as other deviations from $\Lambda$CDM. 
\vfill\null

\subsubsection{Local primordial non-Gaussianity} 
\label{sec:non-gaussianity}
Local primordial non-Gaussianity, parameterized by the amplitude $f_\mathrm{NL}$, can be generated by multi-field inflation models, and a measurement of non-zero  $|f_\mathrm{NL}|\gtrsim 1$ would robustly rule out all single-field inflation models with standard Bunch-Davies initial conditions \citep{Maldacena2003,Creminelli2004}.
Observationally, in addition to the effect on higher-point functions described in Sec.~\ref{sec:bispec}, this type of non-Gaussianity leads to a distinct scale-dependence of galaxy bias on large scales, scaling as a function of comoving wavenumber $k$ as $f_\mathrm{NL}/k^2$ \citep{0710.4560}.
A joint analysis of CMB lensing and galaxy clustering data can search for this effect by comparing the scale-dependence of lensing and clustering auto- and cross-spectra on the largest scales. 
The high cross-correlation coefficient between SO lensing and LSST clustering discussed above allows us to cancel part of the sample variance that usually limits constraints derived from  large scales. 

\begin{figure}[t!]
\centering
\includegraphics[ width=\columnwidth]{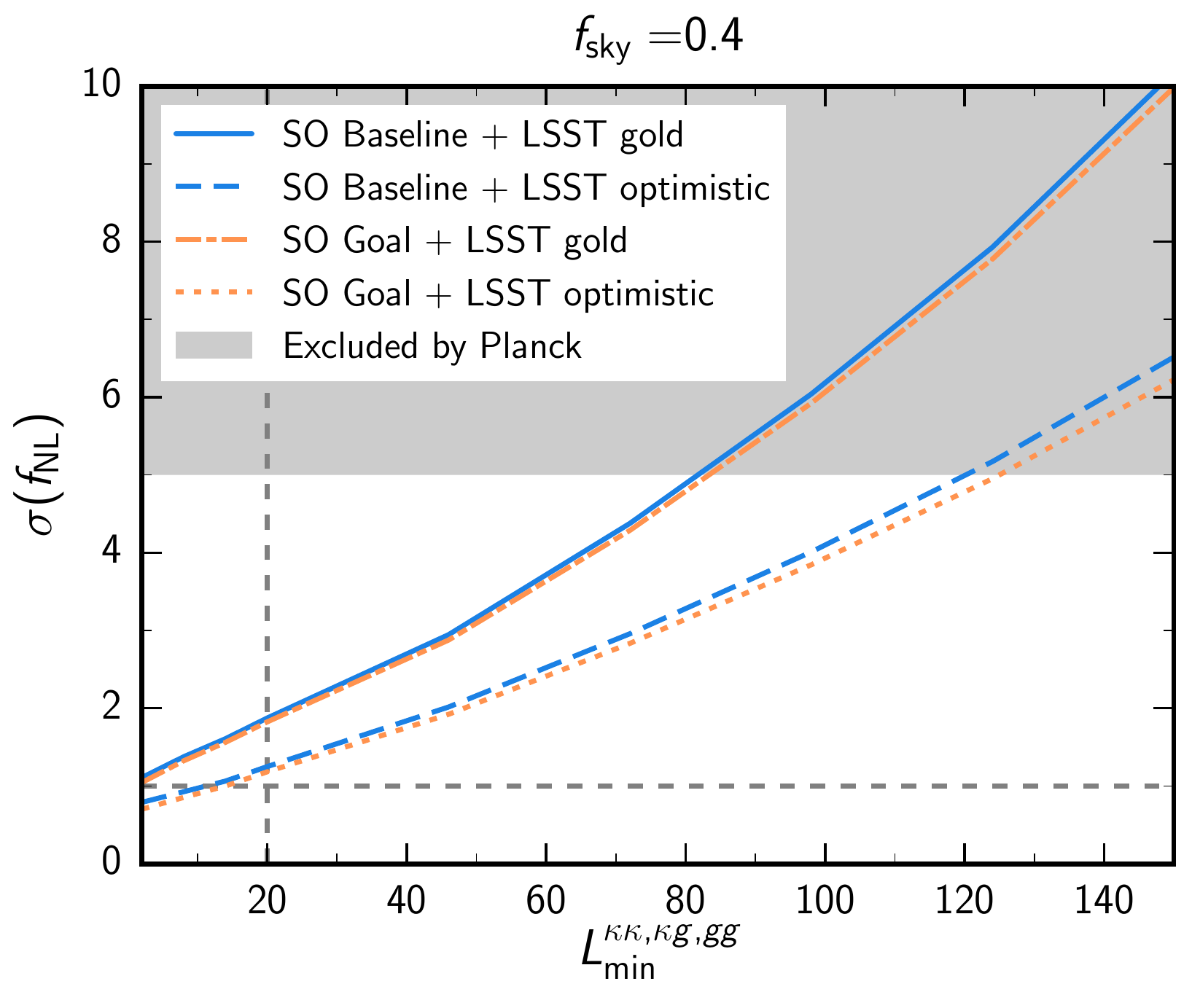}
\caption{Constraint on local primordial non-Gaussianity expected from \so{} CMB lensing (minimum variance combination, including foreground-deprojected gradient cleaning; ~\citealp{MMHill}) cross-correlated with LSST galaxy clustering, exploiting the large-scale scale-dependent bias effect from $f_\mathrm{NL}$~\citep{0710.4560}. Different lines show different assumptions about \so{} sensitivity and LSST number density and redshift distribution. The gray shaded region is excluded by \planck~\citep{1502.01592}.
We marginalize over the amplitude of one linear galaxy bias parameter per tomographic redshift bin that rescales the fiducial bias redshift dependence $b(z)=1+z$~\citep{0912.0201}.  We also marginalize over an artificial parameter that changes the matter power spectrum with the same shape as $f_\mathrm{NL}$ changes the galaxy bias (see~\citealp{1710.09465} for details). The gray dashed vertical line marks our primary forecasts in Eq.~\ref{eq:fnl_lmin20} with $L_{\rm min}$=20, the horizontal line marks the $f_{\rm NL}$=1 target.}
\label{fig:NG}
\end{figure}
The forecasts of local $f_\mathrm{NL}$ from cross-correlations between SO lensing and LSST clustering are shown in Fig.~\ref{fig:NG}, following a similar procedure as for $\sigma_8(z)$ (using the same six tomographic LSST redshift bins and marginalizing over the same linear bias parameters). 
We find that competitive constraints are possible with SO.
If clustering and CMB lensing data are used down to $L_\mathrm{min}=20$, we project
\ba
\sigma(f_\mathrm{NL}) &=& 1.8 \quad {\rm \so{} \ Baseline + LSST\mbox{-}gold}\,,\nonumber \\
\sigma(f_\mathrm{NL}) &=& 1.2 \quad {\rm \so{} \ Baseline + LSST\mbox{-}opt}\,.
\label{eq:fnl_lmin20}
\ea
This is close to the value of $f_\mathrm{NL}=1$ that is typically chosen as a target in efforts to separate multi-field from single-field inflation models.
These constraints improve as sky area is increased, but the strongest dependence is on the minimum lensing and galaxy density multipole used in our analysis. 
Indeed, we project that $\sigma(f_\mathrm{NL})=1$ could be achieved for lensing- and clustering-$L_\mathrm{min}=14$, assuming SO baseline sensitivity and optimistic LSST galaxies, and the error would be reduced even further with lower $L_\mathrm{min}$. This motivates preserving low multipoles when designing SO lensing and LSST clustering observations and analysis pipelines for this kind of measurement. 

Most of the $f_\mathrm{NL}$ constraint comes from comparing $\kappa\kappa$ and $gg$ power spectra on large scales. It is also possible to work only with $\kappa\kappa$ and $\kappa g$ on very large scales and include $gg$ only on smaller scales; however this degrades the $f_\mathrm{NL}$ forecast by a factor of two or more, since sample-variance cancellation is no longer fully effective. 

The above forecast assumes that $f_\mathrm{NL}$ is only measured from a scale-dependent difference between $\kappa\kappa$, $\kappa g$ and $gg$ that scales as $f_\mathrm{NL}/k^2$ on large scales.  This is implemented by marginalizing over an artificial $f_\mathrm{NL}$ parameter that rescales the matter power spectrum in the same way as $f_\mathrm{NL}$ rescales the galaxy bias, so that any information from $gg$ alone is not taken into account and only the relative bias between clustering and lensing is used~\citep{1710.09465}.

Alternatively,  $f_\mathrm{NL}$ can be constrained from LSST alone by measuring the shape of the large-scale $gg$ power spectrum assuming that the matter power spectrum and large-scale galaxy bias are known. 
This can also achieve $\sigma(f_\mathrm{NL})\simeq \textrm{1--2}$, but with different systematics. 
In particular, such a $gg$-only measurement is only sensitive to the absolute value of the large-scale $gg$ power spectrum rather than the relative ratio between lensing and clustering power spectra. 
Additionally, a cross-correlation analysis can help to reduce the impact of catastrophic photo-$z$ errors, because these errors affect $\kappa g$ and $gg$ power spectra differently, so that measuring these spectra can constrain catastrophic outlier rates and $f_\mathrm{NL}$ simultaneously. This is the case for a simple idealized parameterization of the distribution of catastrophic errors \citep[][]{1710.09465}, but requires further study in the case of more realistic LSST catastrophic photo-$z$ errors.
Using cross-correlations with SO can therefore help to increase the robustness of $f_\mathrm{NL}$ constraints from LSST alone. 

\subsubsection{Shear bias validation}
The upcoming optical lensing experiments, such as LSST, Euclid and WFIRST, will rely on shear measurements, achieved by estimating the correlated shapes of galaxies, to deliver percent-level precision measurements of the amplitude of structure as a function of redshift.
This task is complicated by various systematics which bias the shear. The resulting shear multiplicative bias is degenerate with the amplitude of structure, which makes it a critical systematic for optical lensing surveys.
State-of-the-art optical lensing analyses currently reach $\sim 1\%$ shear calibration \citep{2017arXiv170801533Z, 2017MNRAS.467.1627F, 2018PASJ...70S..25M}, relying in most -- but not all -- cases on image simulations, and are expected to achieve the stringent $\sim 0.5\%$ requirement for LSST, Euclid and WFIRST \citep{2006MNRAS.366..101H}.  Newer techniques using meta-calibration may even achieve $0.1\%$ shear calibration~\citep{Sheldon2017}.
However, because the shear multiplicative bias is such a crucial systematic, any independent method to calibrate or validate it will be valuable, and will add confidence to any potential dark energy discovery from LSST, Euclid and WFIRST.

\begin{figure}[t!]
\centering
\includegraphics[width=\columnwidth]{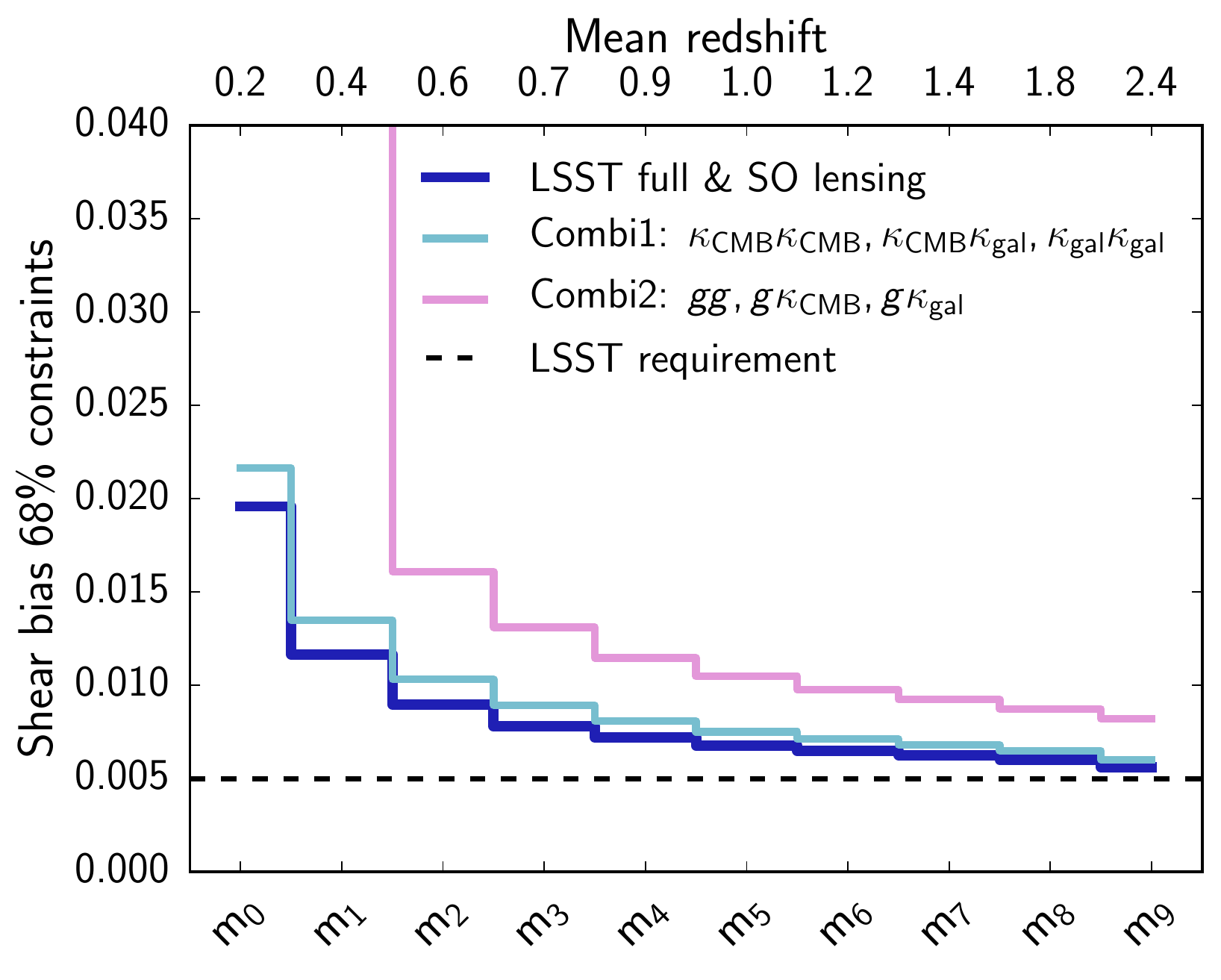}
\caption{
$68\%$ confidence-level constraints on the shear biases $m_i$ for the 10 tomographic bins of LSST, when self-calibrating them from the cross- and auto-correlations of CMB and galaxy lensing (combination 1, light blue), from clustering and the cross-correlation of galaxy density with CMB and galaxy lensing (combination 2, pink) and the full LSST \& SO lensing (dark blue). The self-calibration is very close to the level of the LSST requirements (black dashed line) for the highest redshift bins, where shear calibration is otherwise most difficult. We stress that all the solid lines correspond to self-calibration from the data alone, without relying on image simulations. CMB lensing will thus provide a valuable consistency check for building confidence in the results from LSST.\\
}
\label{shear_calibration_cmbs4}
\end{figure}

CMB lensing from \so{} can provide such an external validation \citep{2017PhRvD..95l3512S}. Indeed, jointly analyzing auto- and cross-correlations of CMB lensing with galaxy shear and galaxy number density gives information on the shear multiplicative bias.
This method has been proposed in \citet{2012ApJ...759...32V}, \citet{2013ApJ...778..108V}, and \citet{2013arXiv1311.2338D}
and subsequently applied to current datasets (e.g., in \citealt{2016PhRvD..93j3508L, 2016MNRAS.461.4099B,DistRatio1, DistRatio2}).
In Fig.~\ref{shear_calibration_cmbs4} we show how combining CMB lensing from \so{} with galaxy number density and shapes from LSST enables a shear calibration close to the LSST requirements in many redshift bins, while marginalizing over nuisance parameters (galaxy bias and photometric redshifts) and varying cosmological parameters. In this forecast, we assume a conservative LSST lens sample as in \citet{2017PhRvD..95l3512S}, to guarantee accurate photometric redshifts. This process is independent from any shear calibration internal to LSST, Euclid, and WFIRST. It is also robust to expected levels of intrinsic alignment, to uncertainties in non-linearities and baryonic effects, and to changes in the photo-$z$ accuracy, as verified in \citet{2017PhRvD..95l3512S}.
A similar level of shear calibration is achieved with WFIRST or Euclid instead of LSST.

\subsubsection{Lensing ratios: curvature}
Two-point correlation functions between galaxies and gravitational lensing are powerful cosmological probes.  However, fully exploiting the available signal is complicated by the challenges of modeling the small-scale regime, where non-linearities, baryons and galaxy bias all become important.  Several authors \citep{Jain:2003, Hu:2007,Das:2009} have pointed out that these complications can be circumvented by measuring ratios of lensing-galaxy two-point functions involving a single set of `lens' galaxies and multiple gravitational lensing source planes.  In the limit that the lens galaxies are narrowly distributed in redshift, such `lensing ratios' reduce to a purely geometrical ratio of angular diameter distances.  The sensitivity of lensing ratios to cosmological distances make them useful cosmological probes, while the fact that they are insensitive to the lens galaxy power spectrum means that all available signal at small scales can be exploited without difficult modeling.  

On account of the large distance to the last scattering surface, the cosmological sensitivity of the lensing ratios can be dramatically increased by using the CMB as one of the gravitational lensing source planes (see, for example, recent first measurements in \citealp{DistRatio1} and \citealp{DistRatio2}).  In this case, the lensing ratios are particularly sensitive to the curvature of the universe.   Purely geometric constraints on curvature from the primary CMB alone are significantly degraded by degeneracies between the matter density, curvature, and the equation of state of dark energy (i.e., the so-called geometric degeneracy).  The addition of lensing ratio measurements breaks this degeneracy and can yield significantly tighter constraints on the curvature density parameter, $\Omega_k$, while still relying on purely geometric information. 

The combination of CMB lensing maps from SO and galaxy imaging from LSST can be used to place tight constraints on lensing ratios.   In order for the galaxy power spectrum to cancel in the lensing ratio, the lens galaxies must be narrowly distributed in redshift.  We find that bins of width $\Delta z = 0.05$ are sufficiently narrow, given the projected uncertainties on the ratio measurements.  Such redshift accuracy can be achieved photometrically in LSST data using an algorithm like \code{redMaGiC} \citep{Rozo:2016}. A \code{redMaGiC}-like selection would sacrifice number density for the improved photometric accuracy necessary for using lensing ratios.

Given these considerations, for this forecast we do not use the same assumptions about LSST as described in Sec.~\ref{sec:method}. Instead we assume that LSST can provide a population of lens galaxies divided into redshift bins of width $\Delta z = 0.05$ between $z=0.2$ and $z = 0.7$ with a number density of 100 per square degree in each bin. These numbers are comparable to what is achieved currently from the Dark Energy Survey \citep{Cawthon:2017}.  We further assume that LSST provides a source galaxy population distributed between $z=1.2$ and $z=1.6$ with density of 25 galaxies per square arcminute; the source galaxies are assumed to have photometric redshifts that are accurate to $1\%$, and we assume that multiplicative shear bias can be controlled to the $0.1\%$ level \citep{Sheldon2017}, although our projections are not very sensitive to this assumption.

Assuming overlap between SO and LSST of $40\%$ of the sky, we find that the lensing ratios can be measured to roughly 3\% precision assuming SO goal sensitivity, with some variation depending on the lens galaxy redshift bin. We adopt SO projections for CMB lensing maps generated via the gradient field cleaning described in Sec.~\ref{sec:lensingforegrounds}.  When combined with geometric information from the primary CMB anisotropies (using the geometric CMB likelihood of \citealt{2015PhRvD..92l3516A}), we project that constraints of roughly $\sigma(\Omega_k) = 0.026$ can be achieved in a $\Lambda$CDM+$w_0$+$\Omega_k$ scenario, after marginalizing over the matter and baryon densities, $\Omega_{\rm m}h^2$ and $\Omega_b h^2$, the dark energy equation of state parameter, $w_0$, the Hubble constant, $h$, and systematics parameters describing photometric redshift and shear calibration uncertainty. It should be emphasized that a constraint obtained in this way is purely geometrical, depending only on the distances to the lens and source galaxies and to the last scattering surface. It does not make use of any information coming from the growth of structure or the matter power spectrum.

\subsection{Halo lensing: mass calibration}
\label{sec:HaloLens}

With arcminute resolution of the CMB, gravitational lensing on cluster and galaxy group scales can be resolved (see~\citealp{SZHLensing,HKHLensing}) and has been measured with data~\citep{2015PhRvL.114o1302M,2015ApJ...806..247B,Baxter2017,PlnkSZCos2015,geachpeacock}. Such `halo lensing' measurements of the 1-halo-dominated lensing contribution from dark matter halos allow for constraints on the total halo mass while being mostly independent of assumptions about baryonic physics. CMB lensing measurements of halos will enable mass calibration of the SZ cluster sample from SO (described in Sec.~\ref{sec:sz}) as well as cluster and galaxy group samples from external optical surveys. This will provide a complementary mass calibration to that obtained from optical weak lensing follow-up using LSST background galaxies. The high-redshift cluster sample ($z>0.7$) will rely on this measurement, since optical measurements at these high redshifts will suffer from photometric redshift uncertainties, systematics, and a general decline in the number of available background galaxies.

Polarization-derived lensing estimators can also be used to get complementary information on the mass calibration with minimal and differing systematics due to foregrounds. At SO sensitivities, polarization-only estimators have 2 to 3 times lower sensitivity than minimum variance estimators, but are still informative and can be used as a cross-check for systematics.

\begin{figure}[t!]
\centering
\includegraphics[width=\columnwidth]{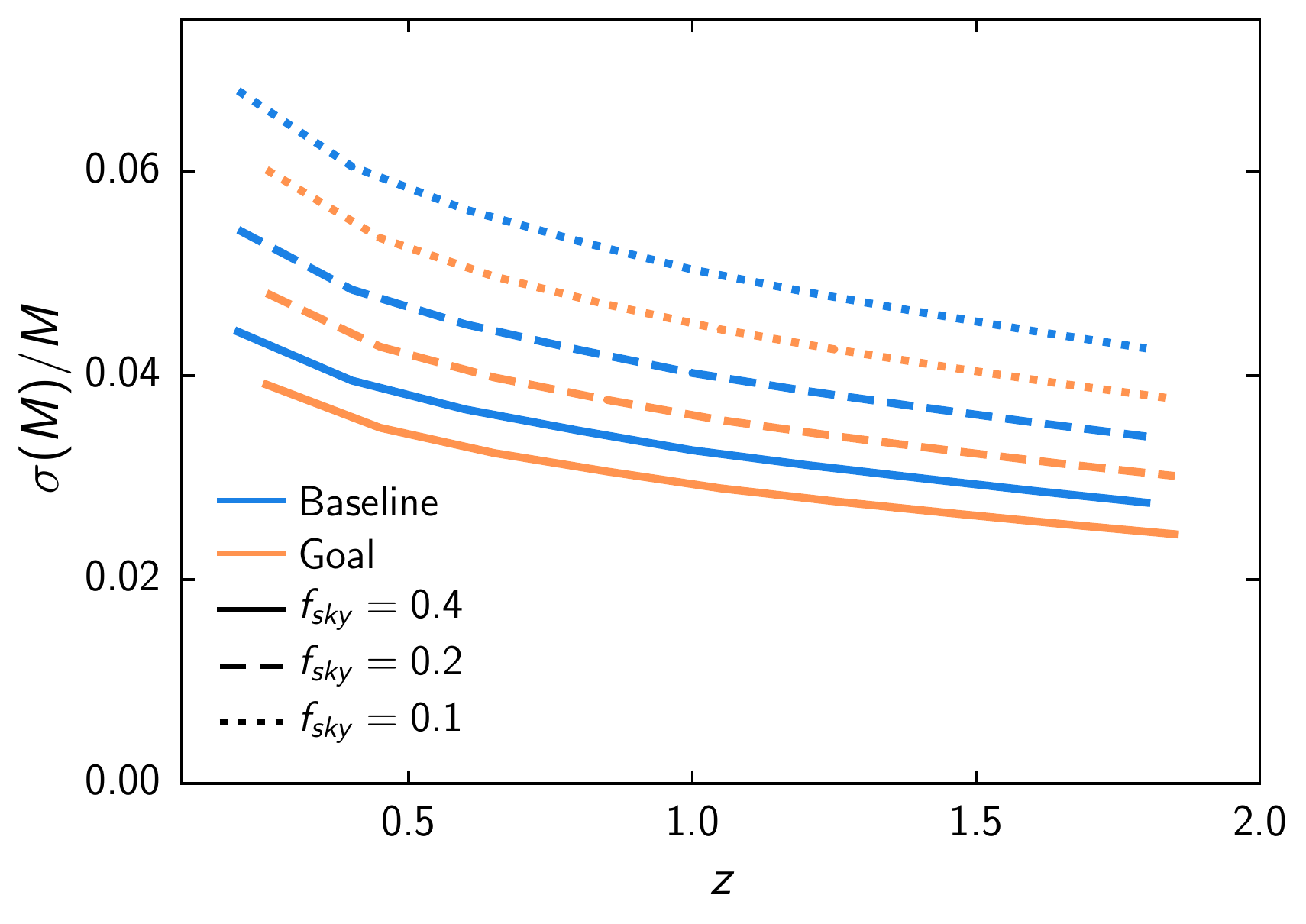}
\caption{
The relative mass calibration uncertainty from CMB halo lensing on a sample of clusters $N=\{1000,500,250\}$ for corresponding fixed-time observations over $f_{\mathrm{sky}}=\{0.4,0.2,0.1\}$  as a function of mean redshift of the sample. The sample is assumed to be of mean mass $M_{500}=2\times 10^{14} M_{\odot}/h$. Both temperature and polarization data are used and noise from temperature foregrounds is included.}
\label{fig:halo_lensing}
\end{figure}

In Fig.~\ref{fig:halo_lensing}, we show the relative mass calibration uncertainty, $\sigma(M)/M$, expected for SO as a function of redshift when using both temperature and polarization data (we follow the~\citealp{MBMSZ} method). This combination of foreground-cleaned temperature and polarization data is designed to be free from bias from the tSZ effect using the ideas described in~\cite{MMHill}, as discussed in Sec.~\ref{sec:lensingforegrounds}. We note that the quadratic estimator used for halo lensing uses a maximum CMB multipole of $\ell=2000$ for the gradient since this significantly reduces a mass-dependent bias when probing small scales. For the nominal $f_{\rm sky}=0.4$, we project that SO can calibrate the mass of a sample of 1000 clusters to
\be
\sigma(M)/M (z=1) = 0.03  \quad {\rm \so{} \ Baseline}.
\ee

Deeper measurements would improve the mass sensitivity on any given cluster, but since constraints are obtained on a stack, increasing the survey overlap with a fixed cluster sample improves the overall constraint. In Sec.~\ref{sec:sz}, we use an internal SZ-detected sample to forecast cosmological parameters, and discuss the optimal survey design.

A promising application of CMB halo lensing mass calibration is
applying it to measurements of the splashback radius (\citealp{More:2016vgs,2017ApJ...841...18B}), which has recently been established as a useful dynamical boundary of dark matter dominated halos. With the large SZ cluster sample provided by SO, high accuracy measurements of splashback which can be compared with simulation based predictions will be available, and this comparison will benefit from precise mass calibration from a combination of galaxy lensing and CMB halo lensing with SO. The splashback radius itself is also a test of the physics of dark matter and gravity, and useful for galaxy evolution studies.

\subsection{Impact of foregrounds on lensing}
\label{sec:lensingforegrounds}
Extragalactic and Galactic foregrounds can significantly affect measurements of both gravitational lensing and delensing \citep{2014ApJ...786...13V,2014JCAP...03..024O,FerraroHill2018}.  The trispectra of objects such as galaxy clusters and  galaxies and the bispectra from their cross-correlation with the lensing field will  give biases on measurements of the power spectrum of the lensing potential.  Though simulated estimates have been made in some cases, many of these higher order statistics have not yet been fully measured in CMB data.  

Several approaches exist for characterizing or avoiding these biases: One can attempt to reduce them using their known statistical properties (i.e., bias hardening); one can deproject them with multi-frequency observations; one can use a shear-only estimator; or one can use polarization-only data.\\

{\it Bias hardening:} One method to reduce the impact of non-Gaussian unresolved foregrounds is to form optimal estimates for these signals in the CMB maps.  One can then  project them out of lensing estimates, a procedure known as `bias hardening'~\citep{2013MNRAS.431..609N}. This has been shown to reduce the contamination from the tSZ effect and the CIB~\citep{2014JCAP...03..024O}, and was recently applied to the analysis of the {\it Planck} data~\citep{planck2018:lensing}. \\

{\it Gradient cleaning:} Another promising approach, particularly for cross-correlations and halo lensing, is to clean the CMB gradient part of the pair of maps used in the quadratic estimator using a combination of SO and existing {\it Planck} data \citep{MMHill}. At SO frequencies of 93 and 145 GHz where the CMB lensing signal is most informative, there is considerable contamination from tSZ, as well as emission from dusty sources, kSZ, and potential unmasked synchrotron sources. Foregrounds such as tSZ and CIB introduce a significant bias even if a standard ILC procedure is performed to obtain a minimum variance CMB map. This necessitates projecting out the foregrounds through constrained ILC. It is, however, sufficient to perform foreground deprojection on the low-resolution gradient estimate used in the quadratic estimator. Ensuring that the gradient alone is foreground-free eliminates much of the bias while ensuring that there is only a small loss of signal-to-noise ratio.

In our forecasts for cross-correlations (Sec.~\ref{sec:cross-corr}) and halo lensing (Sec.~\ref{sec:HaloLens}), we assume this gradient cleaning  when generating noise curves. Specifically, for the $TT$ estimator, the CMB gradient part of the quadratic estimator pair utilizes tSZ- and CIB-deprojected constrained ILC temperature maps for large-scale cross-correlations (Sec.~\ref{sec:cross-corr}) and tSZ-deprojected constrained ILC temperature maps for halo lensing (Sec.~\ref{sec:HaloLens}), where the foreground cleaning procedure is discussed in more detail in Sec.~\ref{sec:method}. We include CIB deprojection for the large-scale cross-correlations since a high-redshift LSST sample can have significant overlap with the CIB, which peaks at a redshift of around~2.

For CMB halo lensing, we only include tSZ deprojection since tSZ is expected to be the dominant foreground contaminant in an SZ selected sample of clusters for redshifts $z<2$. (However, CIB cleaning could also be included straightforwardly.) The non-gradient map of the quadratic estimator pair uses the standard ILC temperature map. All other estimators use corresponding standard ILC maps. This procedure does not remove the bias from kSZ, although that bias can be alleviated by lowering the maximum multipole used in the gradient at the cost of some signal-to-noise in the $TT$ estimator at large scales.\\

{\it Shear-only estimator:} It has been recently shown that we can use the different symmetries of the foregrounds and lensed CMB to obtain unbiased lensing reconstruction \citep{2018arXiv180406403S}. In particular, a `shear-only' estimator is approximately immune to foreground contamination. This method simultaneously eliminates biases from all of the extragalactic foregrounds, including those from the tSZ, kSZ, CIB, and radio point sources. In addition, it allows for reconstruction on a larger range of multipoles, therefore increasing the overall statistical power.\\

{\it Polarization-only estimators:} A final approach is to rely only on polarization data.  SO will be in a regime where roughly equal signal-to-noise ratio will be available with polarization and with temperature.  Using polarization data is beneficial as extragalactic emission is low in polarization. However, when using polarization data, sources of non-Gaussianity that can impact lensing measurements include Galactic dust at high frequencies and Galactic synchrotron at low frequencies \citep{2012JCAP...12..017F}.  Current data from the \textit{Planck} satellite do not have low enough noise on the small angular scales of interest to definitively determine the impact of the non-Gaussianity of the polarized dust.  A study by the CORE team \citep{2017arXiv170702259C} using simulations found a non-negligible bias on lensing from polarization data using part of the sky within the \textit{Planck} analysis mask, though it could  be mitigated with a modification of the lensing weights.  The situation is similar for small-scale synchrotron.  These potential challenges motivated the choice of a wide frequency coverage for the LAT, including frequencies to characterize both synchrotron and dust, to permit the removal of these possible contaminants if they are non-negligible, along with providing powerful null tests.
\vfill\null

\section{Primordial bispectrum}
\label{sec:bispec}
The CMB bispectrum is a powerful observable to constrain possible interactions in the early universe. In inflationary models, vacuum fluctuations in the early universe source both scalar and tensor perturbations, and through weak gravitational interactions one expects small corrections to Gaussianity \citep{Maldacena2003}. Predicted levels in the simplest inflationary scenarios will therefore be hard to observe, even with future CMB and LSS measurements. However, deviations from the simplest models can easily generate large non-Gaussianities, e.g., via non-canonical kinetic terms, the presence of multiple fields, or through particle production \citep{2009astro2010S.158K}.

The CMB has been at the forefront of constraining non-Gaussianities through their effect on the three- and four-point correlation functions. The most stringent bounds on any form of non-Gaussianity has been derived using the {\it Planck} data \citep{1502.01592}. The most interesting type of non-Gaussianities are the so-called local and equilateral type as they provide a fairly robust physical interpretation \citep[see, e.g.,][]{Alvarez:2014vva}. Observable levels of local non-Gaussianity are expected when more than one field is involved in the inflationary dynamics, while equilateral non-Gaussianities are expected if the inflaton is strongly coupled. As mentioned in Sec.~\ref{sec:non-gaussianity}, an interesting theoretical target widely shared within the community is $f_{\rm NL}\sim \mathcal{O}(1)$, both for local and equilateral non-Gaussianities \citep{Alvarez:2014vva}. {\it Planck} yields $\sigma(f_{\rm NL}^{\rm local})=5$ \citep{1502.01592}. 

Here we present the forecast constraints using SO combined with {\it Planck}. 
We combine errors as $\sigma^{-2}(f_{\rm NL}) = \sigma^{-2}(f^{\rm Planck}_{\rm NL}) + \sigma^{-2}(f^{\rm SO}_{\rm NL})$, which is non-optimal but approximates an analysis that combines the Fisher information before inverting. For the \planck{} constraints, we use only the fraction of the sky not observed by SO. The current \planck{} constraints are derived from $f_{\rm sky} \sim 0.75$, so combining with SO at $f_{\rm sky} = 0.4$, we will use only $f_{\rm sky} = 0.35$ for $\sigma(f^{\rm Planck}_{\rm NL})$.  Our results match the current constraints from \planck{} for $f_{\rm sky} = 0.75$ in the \planck{} only case. On scales below $\ell = 40$ in the SO LAT patch, we use \planck{} noise. 
We consider local, equilateral, and orthogonal non-Gaussianities.  
We forecast the scalar bispectra, $\langle \zeta \zeta \zeta \rangle$, using the SO LAT combined with {\it Planck}, using the LAT ILC noise curves described in Sec.~\ref{sec:method} (Deproj-0) and with tSZ and CIB SED deprojected out in temperature for $\langle TTT \rangle$ only (Deproj-3), as recently shown to be important for avoiding biases in \cite{Hill:2018ypf}.
For the tensor-scalar-scalar forecasts, $\langle \gamma \zeta \zeta \rangle$, we include the SO SATs, and constraints derived only use 10\% of the sky. Following Sec.~\ref{sec:method}, we use ILC noise curves at low multipoles ($30 \leq \ell \leq 260$). For the baseline forecasts we use $1/f$ noise parameterized by $\ell_{\mathrm{knee}} = 50$, and $\ell_{\mathrm{knee}} = 25$ for goal. Here we assume that the SATs have uniform coverage over 10\% of the sky, which is an approximation to the expected sky survey.
We assume that delensed $B$-modes have $A_{\rm lens}=0.75$ (baseline) and $0.5$ (goal). 

\begin{table}[t!]
\centering
\caption[Simons Observatory Surveys]{Forecast constraints on non-Gaussianity ($\sigma_{f_\mathrm{NL}}$).
  The scalar bispectra ($\langle \zeta \zeta \zeta \rangle$) use the SO LAT combined with \planck. The tensor-scalar-scalar ($\langle \gamma \zeta \zeta \rangle$) forecasts add the SO SATs.}
\small
\begin{tabular}{ c  c c c }
\hline
 Shape ($\langle \zeta \zeta \zeta \rangle $)& Current & \so{} Baseline & \so{} Goal\\
 $\langle TTT \rangle $,$\langle TTE \rangle $, &  ({\it Planck})   &    &  \\
  $\langle TEE \rangle $,$\langle EEE \rangle $ &     &    &  \\
\hline \hline
local & 5 &  4 & 3 \\
equilateral & 43 & 27 & 24 \\
orthogonal  & 21 &  14 & 13 \\
\hline
 Shape ($\langle \gamma \zeta \zeta \rangle $) &  &  &  \\
 $\langle BTT \rangle $,$\langle BTE \rangle $,$\langle BEE \rangle $ & (\map) &   & \\
\hline \hline
local & 28\footnote{Constraint derived from temperature only and shape considered is not exactly $f_{\rm NL}^{\rm local}$. There are currently no constraints on equilateral and orthogonal non-Gaussianities $\langle \gamma \zeta \zeta \rangle $.\\} &  2 & 1 \\
equilateral & - &  13 & 8 \\
orthogonal  & - &   6 & 3  \\
\hline
\hline
\end{tabular}
\label{tab:NGs}
\end{table}

Our forecasts are summarized in Table~\ref{tab:NGs}. In the most optimal case, combined forecasts with {\it Planck} cannot reach the threshold of $f_{\rm NL}^{\rm local} = 1$. Instead, constraints improve over {\it Planck} by $\approx~50\%$, which is anticipated by mode counting. In Secs.~\ref{sec:lensing} and \ref{sec:sz} we show that cross-correlation of LSST data with CMB gravitational lensing and the kSZ effect from SO can further improve constraints on $f_{\rm NL}^{\rm local}$. In principle, these three constraints can be combined.

Thus far we have only considered non-Gaussianities sourced by scalar perturbations in the early universe. Recently, a growing body of work \citep{Maldacena:2011nz,Lee:2016vti,Bordin:2016ruc,Baumann:2017jvh,Domenech:2017kno} has explored the possibility of non-Gaussianities sourced by tensors \citep{Meerburg:2016ecv}. Interestingly, tensor non-Gaussianities are hard to produce \citep{Bordin:2016ruc} and a detection would almost certainly signal new physics, such as the presence of higher-order massive spin particles \citep{Lee:2016vti,Baumann:2017jvh}, or extra gauge particles \citep{Agrawal:2017awz}. Unlike the scalar non-Gaussianities, observational constraints on tensor non-Gaussianities have only been considered in the CMB through their effect on the temperature bispectrum \citep{Shiraishi:2017yrq}.

In the scalar bispectra forecasts we use temperature, $T$, and $E$-mode polarization data. With tensors, we can add $B$-mode polarization, which has a smaller contribution from cosmic variance. Constraints on non-Gaussianities from a bispectrum containing at least one tensor will therefore improve significantly when adding $B$-mode data from SO \citep{Meerburg:2016ecv}. As an example, we show forecasts for a coupling between two scalars and a tensor derived from combining information from $\langle BTT\rangle$, $\langle BTE \rangle$, and $\langle BEE\rangle$ \citep{Shiraishi:2010kd} using a newly developed full sky framework \citep{Duivenvooorden:2018}. The forecasts assume $r=0$, i.e., the covariance contains only $B$-mode noise and no signal. For non-zero values of $r$, constraints will weaken due to the extra variance \citep[see, e.g.,][]{CMBS42016}. 
We find
\ba
&\sigma(f_{\rm NL}^{\rm local}, \langle \gamma \zeta \zeta \rangle)& = 2, \quad {\rm \so{} \ Baseline} \nonumber\\
&\sigma(f_{\rm NL}^{\rm equil}, \langle \gamma \zeta \zeta \rangle)& = 13,  \nonumber\\
&\sigma(f_{\rm NL}^{\rm orthog}, \langle \gamma \zeta \zeta \rangle)& = 6, 
\ea
for $r=0$. These would improve current constraints by more than an order of magnitude.  

\section{Sunyaev--Zel'dovich effects}
\label{sec:sz}
As CMB photons propagate through the universe, about 6\% of them are Thomson-scattered by free electrons in the intergalactic medium (IGM) and intracluster medium (ICM), leaving a measurable imprint on the CMB temperature fluctuations. These imprints are called secondary anisotropies in the CMB and they contain a wealth of information about how structure grows in the universe and about the thermodynamic history of baryons across cosmic time. Some fraction of the Thomson-scattered photons produce Sunyaev--Zel'dovich effects~\citep{SZ1969,SZ1972}, the focus of this section, in particular we will discuss the thermal and kinematic SZ effects.
The improvements in sensitivity and frequency coverage provided by \so{} will greatly contribute to our understanding of the growth of structure and baryonic physics in the late-time universe, through improved measurements of SZ. 

In this section we introduce the SZ effects, then forecast cosmological constraints from cluster counts (Sec.~\ref{sec:mnu-w-sz}) and from the thermal SZ power spectrum (Sec.~\ref{sec:tSZ_PS}). We show how well feedback efficiency and non-thermal pressure support can be constrained from both SZ effects (Sec.~\ref{sec:sz_astro}). We then forecast how the kinematic SZ effect can be used to constrain the growth of structure (Sec.~\ref{subsec:growth}), primordial non-Gaussianity (Sec.~\ref{ssec:nong_ksz}), and reionization properties (Sec.~\ref{sec:kszreionization}). 

The tSZ effect is the increase in energy of CMB photons due to scattering off hot electrons. This results in a spectral distortion of the CMB blackbody that corresponds to a decrement in CMB temperature at frequencies below 217 GHz and an increment at frequencies above. At a given frequency, $\nu$, the temperature shift due to the tSZ effect is given (in the non-relativistic limit) by
\begin{equation}
\left( \frac{\Delta T(\nu)}{T_{\rm CMB}} \right)_{\rm tSZ} = f_\nu y, \ \text{where } f_\nu = x\,{\rm coth}(x/2) - 4.
\end{equation}
Here, $x = h \nu / (k_{\rm B} T_{\rm CMB})$ is dimensionless, where $h$ is the Planck constant, and $k_{\rm B}$ is the Boltzmann constant. The tSZ amplitude is proportional to the Compton-$y$ parameter which is defined as an integral over the line of sight $l$ of the radial electron pressure $P_e$
\begin{equation}
y(\theta) = \frac{\sigma_T}{m_e c^2}  \int P_{e} \left(\sqrt{l^2 + d^2_A(z)|\theta|^2 } \right) d l.
\label{eq:yproj}
\end{equation} 
Here $r^2 = l^2+d^2_A(z)|\theta|^2$, $d_A(z)$ is the angular diameter distance to redshift $z$, $\theta$ is the angular coordinate on the sky, and the constants $c$, $m_e$, and $\sigma_T$ are the speed of light, electron mass, and Thomson cross-section respectively.

The kSZ is the Doppler shift of CMB photons Thomson-scattering off free electrons that have a non-zero peculiar velocity with respect to the CMB rest frame. This produces small shifts in the CMB temperature proportional to the radial velocity of the object, $v_r$, and to its optical depth, $\tau$, with
\begin{equation}
\left( \frac{\Delta T}{T_{\rm CMB}} \right)_{\rm kSZ} = - \tau(\theta) \ \frac{v_r}{c}.
\end{equation}
This shift preserves the blackbody spectrum of the CMB to first order, and therefore is independent of frequency in thermodynamic units. The optical depth is defined as an integral along the line of sight of the electron density $n_e$,
\begin{equation}
\tau(\theta) = \sigma_T \int n_e\left(\sqrt{l^2 + d^2_A(z)|\theta|^2 } \right) d l.
\label{eq:tproj}
\end{equation}
As shown in Eqs.~\ref{eq:yproj}--\ref{eq:tproj}, both tSZ and kSZ contain information about the thermodynamic properties of the IGM and ICM since their magnitudes are proportional to the integrated electron pressure (tSZ) and momentum (kSZ) along the line of sight. For ensemble statistics of clusters or galaxies the tSZ and kSZ effects contain cosmological information as they depend on the abundance of clusters or the velocity correlation function. In the following subsections we explore some of the information that we can extract from the anticipated SO SZ measurements:
\begin{itemize}
\item Cosmological parameters from the abundance of tSZ-detected clusters and statistics of component-separated tSZ maps.
\item Thermodynamic properties of galaxies, groups, and clusters from combined tSZ and kSZ cross-correlation measurements.
\item Measurements of peculiar velocities, which are powerful cosmological probes on large scales, through the kSZ effect.
\item Patchy reionization which imprints the CMB through higher order moments of the kSZ effect.

\end{itemize}

\subsection{Cosmology from tSZ cluster counts}
\label{sec:mnu-w-sz}

Galaxy clusters can be identified across the electromagnetic spectrum, from microwave to X-ray energies. The tSZ effect is emerging as a powerful tool to find and count galaxy clusters. Among the many ways to find clusters, the tSZ effect is unique because the detection efficiency is nearly independent of redshift as long as the beam size is about arcminute scale. Measuring cluster abundances as a function of redshift allows us to probe the physical parameters that govern the growth of structure, including the sum of neutrino masses and the dark energy equation of state. Tests of dark energy and gravity can be sharpened by distinguishing geometric information (via the distance-redshift relation) from growth of structure. Cluster abundances and power spectrum observables (see Sec.~\ref{sec:tSZ_PS}) contain a combination of both: these can be distinguished via forward modeling, or using techniques to separate information on geometry and growth relative to a fiducial model.

The ability to use cluster abundances to constrain cosmological parameters is limited by uncertainties in the observable-to-mass scaling relation \citep[e.g.,][]{2009ApJ...692.1060V,Vand2010,Sehgal2011,Benson_2013,Hass2013,Planckcounts,Mantz2014,PlnkSZCos2015,Mantz2015,dehaan2016}.
An accurate and precise calibration of the observable-to-mass relation is therefore essential for any future cosmological constraint from clusters. SZ-selected cluster samples have well-behaved selection functions that make it straightforward to calibrate observable-to-mass relations and constrain cosmological parameters.

We forecast cluster abundances, and associated cosmological parameters, following previous methods \citep{LA2017,MBMSZ}. The details of the methods and assumptions we use are described in \citet{MBMSZ}, which include a matched filter technique that exploits the unique spectral shape of the tSZ signal \citep{Herranz2002,JB2006} and empirical calibrations of the tSZ signal-to-mass relations via optical weak lensing and CMB halo lensing. We do not use the component-separated noise curves from Sec.~\ref{sec:LAT_ILC}, but instead apply our method directly to the per-frequency noise curves described in Sec.~\ref{sec:method}. We use these to compute the noise levels obtained by the matched filter technique, which provides the cluster selection function.

\begin{figure}[t!]
\centering
\includegraphics[width=\columnwidth]{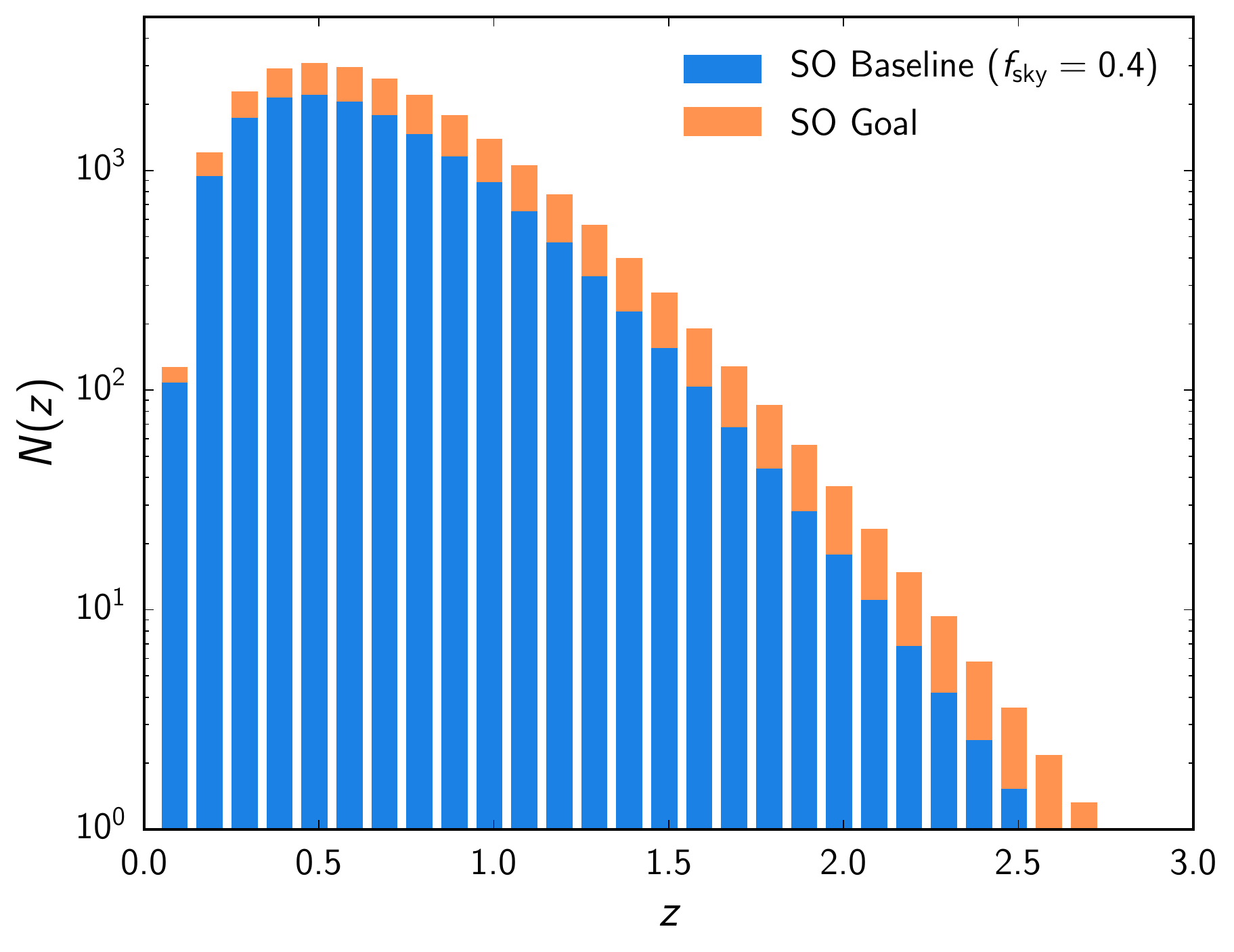}
\caption{The forecast SZ cluster abundances as a function of redshift for the SO baseline and goal configurations with $f_{\rm sky}=0.4$ in bins of redshift with width $\Delta z=0.1$ and a $S/N > 5$. We forecast approximately 16,000 clusters with baseline noise levels and approximately 24,000 clusters with the goal noise levels.}
\label{fig:cldndz}
\end{figure}

We include additional noise from the following CMB secondary anisotropies in the matched filter: a Poisson radio point source term, a Poisson and clustered term for the CIB, the kSZ signal, the unresolved tSZ signal, and the tSZ--CIB cross-correlation term. We estimate that half of the total tSZ auto-spectrum power is coming from clusters with masses $\sim 10^{14} M_\odot$ \citep[e.g.,][]{KS2002,Trac2011,BBPS2012}, which are expected to be detected by \so{} (see Fig.~\ref{fig:cldndz}). Therefore, when we model the unresolved tSZ contribution, we reduce the amplitude of the auto-spectrum power to account for the contribution from these clusters, for the purposes of additional secondary anisotropy noise. We use the functional forms and parameters for these secondary anisotropies presented in \citet{Dunkleyetal2013}. This foreground model is consistent with the model used in Sec.~\ref{sec:method}.

In Fig.~\ref{fig:cldndz} we show the number of clusters expected to be detected as a function of redshift for the SO baseline and goal configurations with the value of $f_{\rm sky} = 0.4$ in bins of redshift with width $\Delta z=0.1$ and a $S/N > 5$. 
With baseline noise levels and $f_{\rm sky} = 0.4$, we forecast approximately 16,000 clusters; with goal noise levels we forecast approximately 24,000 clusters. This is roughly an order of magnitude more SZ detected clusters than the current samples. We find that more clusters are found with larger $f_{\rm sky}$ for fixed total observing time. In addition to cosmological constraints, this sample of galaxy clusters obtained by \so{} will provide the greater astronomical community with a homogeneous and well-defined catalog for follow-up cluster studies out to high redshifts. Such a sample will be critical for studying the impact of over-dense environment on galaxy formation at the peak of the star-formation history in the universe.

\begin{figure}[t!]
\centering
\includegraphics[width=\columnwidth]{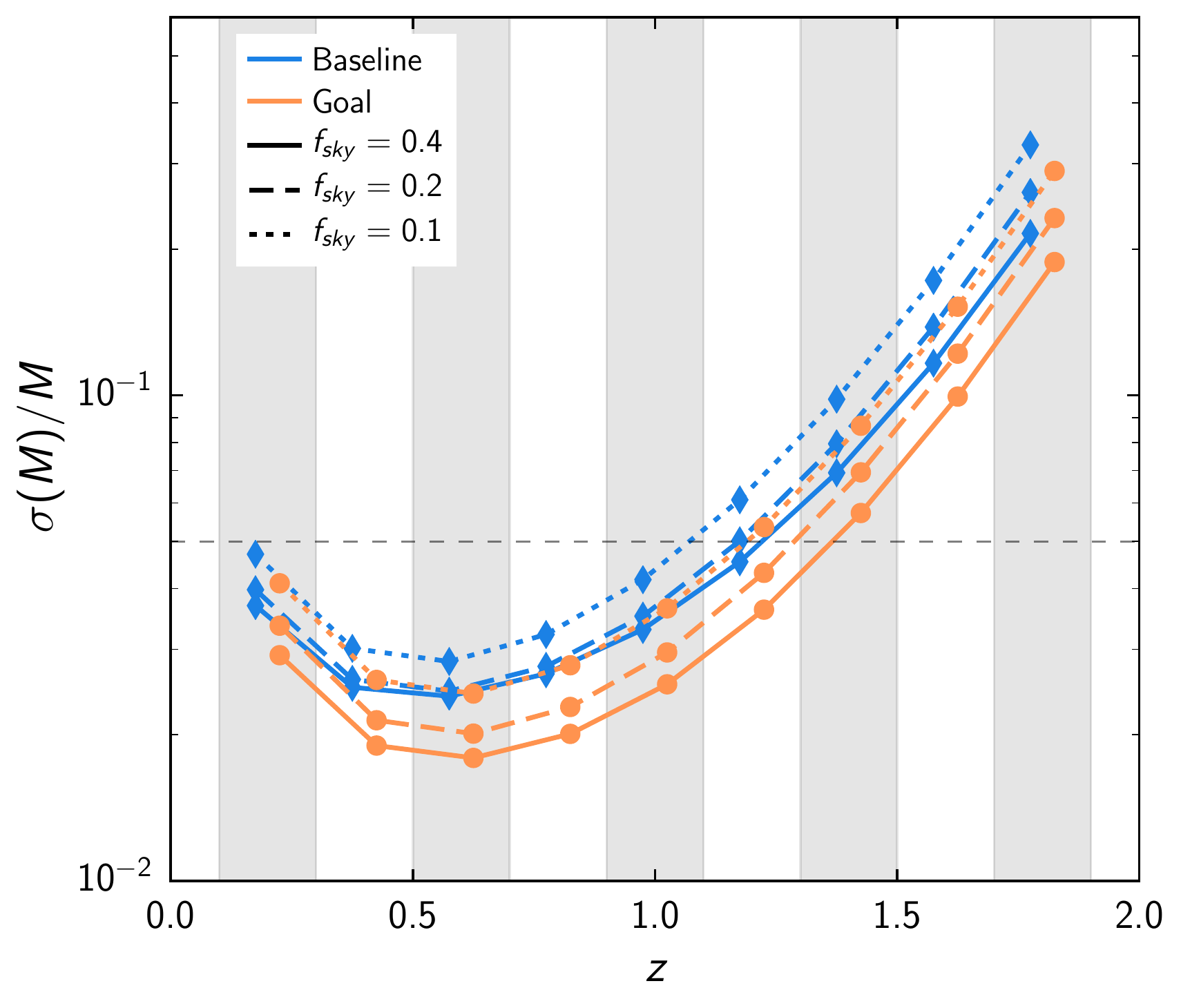}
\caption{The relative mass calibration uncertainty from CMB halo lensing for each redshift bin ($\Delta z=0.2$) of the SZ sample for various configurations of SO. Only the uncertainty on the stacked mass in the $1.5\times 10^{14} < M/M_{\odot} < 2.5\times 10^{14}$ bin is shown here. Constraints on individual clusters would be larger by a factor of $\sqrt{N}$ for $N$ clusters in a bin.\\}
\label{fig:halo_lensing_2}
\end{figure}

For the calibration of the SZ signal using optical weak lensing, we compute the shape noise assuming that a three-year LSST survey will cover the \so{} survey area, as described in Sec.~\ref{subsec:coverage}. In this forecast we make slightly different assumptions to the gold sample described in Sec.~\ref{sec:add_data}, by assuming 20 galaxies per square arcminute with the $d N_g / dz$ from \cite{Oguri:2011},
\begin{equation}
\frac{d N_g}{d z} = \frac{z^2}{2z_0^3}\exp\left(-\frac{z}{z_0}\right),
\end{equation}
where $z_0=1/3$ that corresponds to the mean redshift $z_m=1$. The constraining power of optical weak-lensing will be limited at cluster redshifts above $z \sim 2$ and conservatively $z \gtrsim 1.5$, due the lack of enough lensed galaxies behind the clusters and the large photometric uncertainties of source galaxies. For higher redshift halos we consider CMB halo lensing calibration and follow the methods outlined in Sec.~\ref{sec:HaloLens}. In Fig.~\ref{fig:halo_lensing_2} we show the relative error on the cluster mass for the $1.5\times 10^{14} < M/M_{\odot} < 2.5\times 10^{14}$ mass bin of the SO SZ sample as a function of redshift bin for the baseline and goal configurations with various $f_{\rm sky}$ we considered. We obtain the smallest errors with $f_{\rm sky} = 0.4$. Additionally, we assume that LSST will confirm and provide redshifts for all the clusters found by SO for $z \lesssim1.5$ clusters, the remaining clusters will require pointed, near-IR follow-up observations.

For the forecasts of cosmological parameters, our fiducial model includes the same nuisance parameters for the observable-to-mass scaling relation as \citet{MBMSZ}, $\Lambda$CDM cosmological parameters, and the specified extensions to this model, like the sum of neutrino masses. We include a prior on $\tau$ and {\it Planck} primary CMB observations as described in Sec.~\ref{sec:method}, in addition to SO primary lensed two-point CMB $TT/EE/TE$.
There are no priors on the 7 observable-to-mass scaling relation parameters and no systematic floors applied to the weak-lensing calibration of  observable-to-mass scaling relation parameters. A systematic that we will address in future work is how the uncertainty in the mass function will impact our forecasts \citep{Shimon2011}.

\subsubsection{Neutrino mass and dark energy}
In addition to \LCDM\ parameters, here we vary $\Sigma m_{\nu}$ when constraining neutrino mass, and either $w_0$ and $w_a$ or $w_0$, $w_a$ and $\Sigma m_{\nu}$ when constraining a time-dependent dark energy equation of state $w(a)=w_0+w_a(1-a)$ for scale factor $a$. For the dark energy equation of state parameters $w_0$ and $w_a$, we use the Parameterized post-Friedmann dark energy model implemented in CAMB \citep{ppf1,ppf2,ppf3,ppf4}.
For  $f_{\rm sky} = 0.4$ and SO Baseline we forecast for neutrino mass
\ba
\sigma(\Sigma m_\nu) &=& 27 ~{\rm meV} \quad {\rm \Lambda CDM}+\Sigma m_\nu
\ea
and for the dark energy equation of state
\ba
&\sigma(w_0)&= 0.06, \quad {\rm \Lambda CDM}+w_0+w_a \nonumber\\ &\sigma(w_a)&=0.20, \\ 
&\sigma(w_0)&=0.08, \quad {\rm \Lambda CDM}+w_0+w_a+\Sigma m_\nu \nonumber\\ &\sigma(w_a)&=0.32, 
\ea
with $\sigma(\Sigma m_\nu) = 25 ~{\rm meV}$, $\sigma(w_0) = 0.06$ and $\sigma(w_0) =0.07$ for Goal noise for the above respective cases. Larger $f_{\rm sky}$ provides better constraints. For example, the $\sigma(\Sigma m_\nu) = 27 ~{\rm meV}$ baseline constraint for $f_{\rm sky} = 0.4$ degrades to $\sigma(\Sigma m_\nu) = 35 ~{\rm meV}$ for $f_{\rm sky} = 0.1$. Excluding SO primary lensed two-point CMB $TT/EE/TE$ information degrades these constraints slightly, we get $\sigma(w_0) =0.08$ for $f_{\rm sky} = 0.4$ baseline and $\Lambda$CDM$+w_0+w_a$.

\subsubsection{Amplitude of structure: $\sigma_8(z)$}
\begin{figure}[t]
\centering
\includegraphics[width=\columnwidth]{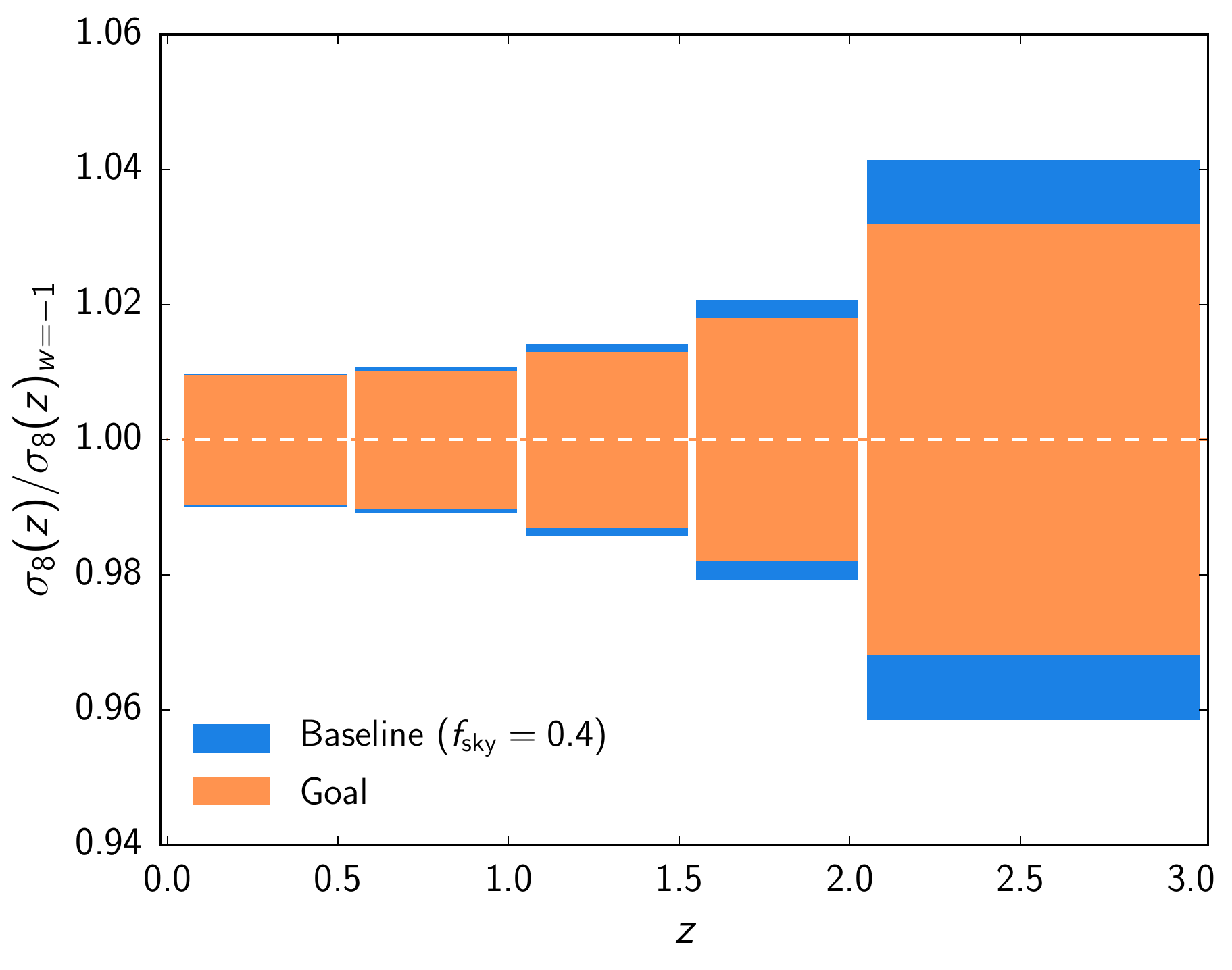}
\caption{The uncertainty on $\sigma_8(z)$ for various redshift bins obtained from the abundances of SO-detected SZ clusters calibrated using CMB halo lensing.}
\label{fig:sz_s8}
\end{figure}

In Fig.~\ref{fig:sz_s8}, we forecast constraints on the growth of structure as a function of redshift, parametrized by the amplitude of matter fluctuations, $\sigma_8(z)$. We consider five tomographic redshift bins ($z =$ 0--0.5, 0.5--1, 1--1.5, 1.5--2, 2--3) and marginalize over $\Lambda$CDM cosmological parameters and the same scaling relation nuisance parameters as above. For the mass calibration of SZ clusters, we use CMB halo lensing calibration, which allows us to push to higher redshift than the optical weak lensing calibration.
The forecast constraints on $\sigma_8 (z)$ for the redshift bins $z$=1--2 and $z$=2--3 are complementary to the constraints shown in Sec.~\ref{sec:growth} using the SO lensing signal in combination with galaxy clustering.

\subsection{Neutrino mass from tSZ power spectrum}
\label{sec:tSZ_PS}
While the number counts of tSZ-detected galaxy clusters contain substantial cosmological information, further constraints can also be extracted from statistical analyses of the tSZ signal in component-separated $y$-maps (or, in some cases, appropriately filtered single-frequency CMB maps).  While many statistics have been explored~\citep{KS2002,RBZ2003,Wilson2012,Bhattacharya2012,Hill-Sherwin2013,Crawford2014,Hill2014PDF,HS2014,Planck2013ymap,Planck2015ymap}, we focus for simplicity on the tSZ power spectrum here, leaving higher-order statistics to future work.\footnote{Also note that as the threshold for individual tSZ detections moves to lower masses and higher redshifts, the detected cluster catalog will contain progressively more of the information in higher-order statistics.} The tSZ power spectrum has long been recognized as a sensitive probe of cosmological and astrophysical parameters, particularly $\sigma_8$~\citep{KS2002,HP2013,Planck2013ymap,Planck2015ymap,Horowitz-Seljak2017,Bolliet2018}.\\

{\it Extracting the Compton-y map:} With sufficient multi-frequency coverage, it is possible to apply component separation methods to extract maps of the Compton-$y$ signal over large sky areas using the known tSZ spectral function, as has been possible recently for the first time with {\it Planck}~\citep{Planck2013ymap,HS2014,Planck2015ymap,Khatri2016}.  In Sec.~\ref{subsubsec:compsepnoise}, we used constrained harmonic-space ILC methods to obtain component-separated noise curves for Compton-$y$ reconstruction.

\begin{figure}[t!]
\centering
\includegraphics[width=\columnwidth]{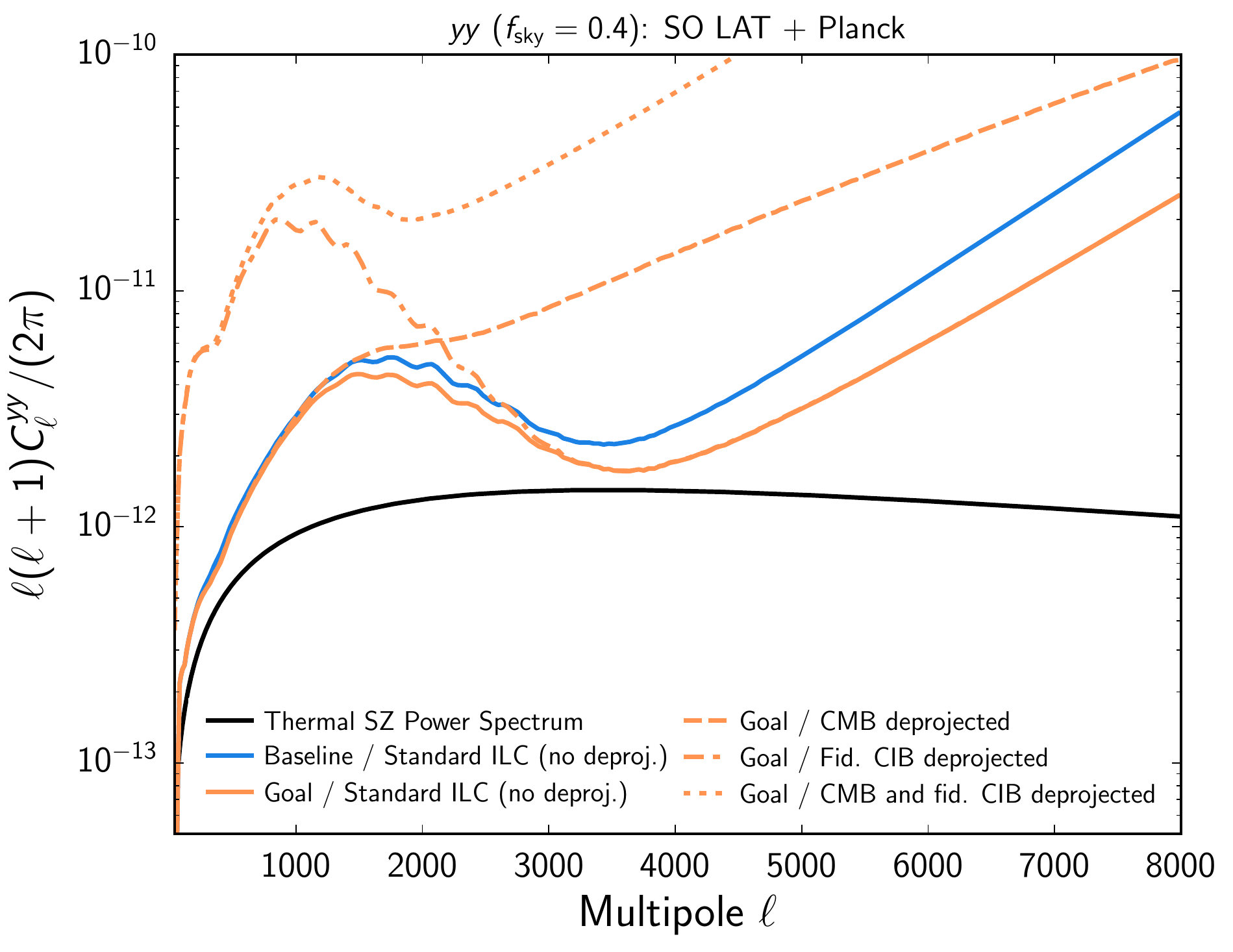}
\caption{Post-component-separation noise for tSZ reconstruction with SO LAT and \planck, for a wide survey with $f_{\rm sky} = 0.4$.  The noise curves are derived via the methodology described in Sec.~\ref{sec:LAT_ILC}.  The solid black curve shows the expected tSZ power spectrum signal. The solid blue (orange) curve shows the tSZ reconstruction noise for the baseline (goal) SO LAT noise levels. The other orange curves show the tSZ reconstruction noise for various assumptions about additional foreground deprojection in the constrained ILC formalism (see the text for further discussion).  The increase in the noise curves at $\ell \approx 1000$--$1500$ is due to the transition from the \planck{}-dominated to SO-dominated regimes (note that atmospheric noise is large for the SO LAT at low-$\ell$).}
\label{fig:comp_sep_noise_yy}
\end{figure}

In Fig.~\ref{fig:comp_sep_noise_yy}, we show the post-component-separation tSZ noise for the SO LAT in combination with {\it Planck} (30--353 GHz), analogous to the CMB temperature and $E$-mode polarization noise curves shown in Fig.~\ref{fig:comp_sep_noise}.  The figure shows the results for both the baseline and goal SO LAT noise levels, as well as various foreground-deprojection options in the constrained harmonic-space ILC used to obtain the noise curves.  The standard ILC case agrees precisely with the CMB-deprojection case at low-$\ell$, where the CMB is the dominant contaminant, and with the CIB-deprojection case at high-$\ell$, where the CIB is the dominant contaminant.  The CMB temperature acoustic oscillations can be seen in the tSZ noise for all cases in which the CMB is not explicitly deprojected.  Finally, one can also clearly see the transition between the regime in which the {\it Planck} channels dominate the reconstruction at low-$\ell$ and that in which the SO channels dominate at high-$\ell$ (note that low-$\ell$ modes in SO LAT temperature are noisy due to the atmosphere), with a bump in the effective noise at $\ell \approx 1000$--$1500$ between the two.\\

{\it Cosmological parameter forecast method:} We use these noise curves to forecast cosmological parameter constraints from the tSZ power spectrum, focusing on $\Sigma m_{\nu}$.  Following Sec.~\ref{sec:methods}, we also include primary CMB power spectrum information from {\it Planck} and SO, as well as information from anticipated DESI BAO measurements. Our fiducial model includes $\Lambda$CDM parameters (in addition to $\Sigma m_{\nu}$) and three parameters describing the gas physics of the intracluster medium: $P_0$, which represents the overall normalization of the $P(M,z)$ relation (where $P$ is gas pressure); $\beta_0$, which denotes the normalization of the relation between the outer slope of the pressure profile and the halo mass (and redshift); and $\alpha_M$, which is the power-law slope of the mass dependence in the $P(M,z)$ relation.  The exact form of the $P(M,z)$ relation used here can be found in~\citet{BBPS2012}.

We adopt values for the gas physics parameters drawn from cosmological hydrodynamics simulations incorporating AGN feedback, supernova feedback, and other sub-grid processes~\citep{Battaglia:2010tm,BBPS2012}.  We impose priors on the gas physics parameters to account for information obtained from studies of individually detected objects. These include a 1\% prior on $P_0$ (which is feasible when folding in all expected weak lensing constraints on the detected clusters -- see Sec.~\ref{sec:mnu-w-sz}), a 10\% prior on $\beta_0$, and a 10\% prior on $\alpha_M$.  We comment on the influence of these priors below.

In addition to the Gaussian errors due to cosmic variance and the component-separated noise curves described earlier, we also include the non-Gaussian error due to the tSZ trispectrum~\citep{KS2002,Shaw2009,HP2013,Horowitz-Seljak2017}.  The non-Gaussian error dominates the error budget at low- to moderate-$\ell$.  While this contribution can be suppressed by masking nearby, massive clusters, we do not consider such complexities here. In addition, we do not vary the trispectrum when varying parameters in the Fisher analysis below, i.e., we compute it only for the fiducial model. \\

{\it Forecast S/N of tSZ power spectrum:} We forecast that the tSZ power spectrum will be detected at high $S/N$ by the SO LAT survey combined with {\it Planck}, reaching $S/N \approx 250$ for a wide survey ($f_{\rm sky} = 0.4$) with goal noise levels.  Note that Fig.~\ref{fig:comp_sep_noise_yy} shows the noise per mode; the high $S/N$ is obtained due to measuring a large number of modes, particularly at high-$\ell$. We give the expected $S/N$ values for various survey choices and foreground cleaning methods in Table~\ref{tab:tSZ_PS}, with the Standard ILC method used unless stated. The alternative cleaning methods were described in Sec.~\ref{sec:method}, with Deproj-1 projecting out the CMB, Deproj-2 the CIB (assuming a fiducial CIB SED), and Deproj-3 both the CMB and CIB.
We also include the tSZ trispectrum contribution to the errors when computing the $S/N$.\\

{\it Neutrino mass:} Table~\ref{tab:tSZ_PS} also shows the marginalized constraints on $\Sigma m_{\nu}$.
The tSZ power spectrum is expected to yield constraints on $\Sigma m_{\nu}$ that are competitive with those from cluster counts, with $\sigma (\Sigma m_{\nu}) = 35$ meV in the best case.  The $\Sigma m_{\nu}$ results are not highly sensitive to the gas physics priors; if $\sigma(P_0)$ is increased to 0.03 (i.e., by a factor of three), then we find $\sigma (\Sigma m_{\nu}) = 38$ meV in the best case, i.e., a very small increase.

\begin{table}[t!]
\caption{Thermal SZ power spectrum (PS) forecasts for the SO LAT, in combination with \planck, for a variety of LAT configurations and foreground cleaning assumptions. Expected tSZ PS $S/N$ and constraints on the sum of the neutrino masses are given.
}
\centering
\begin{tabular}{ c | c c}
\hline
SO LAT configuration & tSZ PS \,\, & $\sigma(\Sigma m_{\nu})$  \\
 & $S/N$ \,\, & [meV] \\
\hline \hline
SO Baseline ($f_{\rm sky} = 0.1$) & 140 & 40 \\
SO Baseline ($f_{\rm sky} = 0.2$) & 180 & 38 \\
{\bf SO Baseline ($f_{\rm sky} = 0.4$)} & 220 & 36 \\
\\
SO Goal ($f_{\rm sky} = 0.1$) & 160 & 39 \\
SO Goal ($f_{\rm sky} = 0.2$) & 200 & 37 \\
{\bf SO Goal ($f_{\rm sky} = 0.4$)} & 250 & 35 \\
\\
SO Goal ($f_{\rm sky} = 0.4$, Deproj-1) & 140 & 36 \\
SO Goal ($f_{\rm sky} = 0.4$, Deproj-2) & 250 & 35 \\
SO Goal ($f_{\rm sky} = 0.4$, Deproj-3) & 60 & 43 \\
\hline
\hline
\end{tabular}
\label{tab:tSZ_PS}
\end{table}

We conclude that the tSZ power spectrum should be a useful source of cosmological information for SO, providing an additional independent route to estimating the neutrino mass and other cosmological parameters.  Nevertheless, it is possible that unforeseen difficulties could arise.  For example, the small-scale component separation could be more challenging than expected due to decorrelation in the CIB across frequencies at high-$\ell$, which is poorly constrained at present.  Additional gas physics parameters may also be needed, beyond the three-parameter model used here.  However, our analysis is conservative in that we have not attempted to combine the information in all tSZ probes, nor have we considered higher-order tSZ statistics.  Since the populations of objects sourcing these statistics are different (the tSZ power spectrum receives contributions from lower-mass clusters and other structures that are not part of the number count analysis of massive galaxy clusters), combining the statistics will yield improvements in cosmological parameters~(e.g.,~\citealp{Hill-Sherwin2013,Bhattacharya2012, Hill2014PDF,Salvati2017,Hurier-Lacasa2017}).  We leave detailed exploration of these avenues for future work.

\subsection{Feedback efficiency and non-thermal pressure support}
\label{sec:sz_astro}

Star formation in galaxies is known to be inefficient: less than 10\% of the available gas is ever turned into stars \citep[e.g.,][and references therein]{2004ApJ...616..643F,Gallazzi2008}. 
This implies that there must be powerful feedback processes that prevent gas from efficiently cooling and forming stars. The conventional theory is that feedback processes that arise from  star formation, supernovae, and active galactic nuclei (AGN), inject energy into the IGM and ICM arresting efficient star-formation and changing both the density and temperature of the gas. In particular it is thought such feedback can alter the gas density profile significantly, moving large amounts of it to the outskirts and causing the well-known `missing baryon problem'. Forms of non-thermal pressure support are also provided to the IGM and ICM through such processes like bulk motions within the gas and turbulence.

Cross-correlations of the thermal pressure of the gas, through the tSZ effect, with the gas density, through the kSZ effect, can trace the baryons out to the outskirts of galaxies and clusters, constrain the amount of non-thermal pressure support, and measure the amount of energy injected by feedback processes. The tSZ and kSZ measurements are supplemented with halo mass estimates obtained using weak lensing. Here we follow the formalism of \cite{2017JCAP...11..040B}, based on a semi-analytical model by \cite{OBB2005}, and show forecasts for the efficiency of energy injection by feedback processes, $\epsilon$, defined in terms of the stellar mass $M_\star$ as $E_{\rm feedback} = \epsilon M_{\star} c^2$, and the fraction of non-thermal pressure, $\alpha$. The kSZ analysis closely follows the method outlined in \cite{2016PhRvD..93h2002S}.

\begin{figure}[t!]
\centering
\includegraphics[width=\columnwidth]{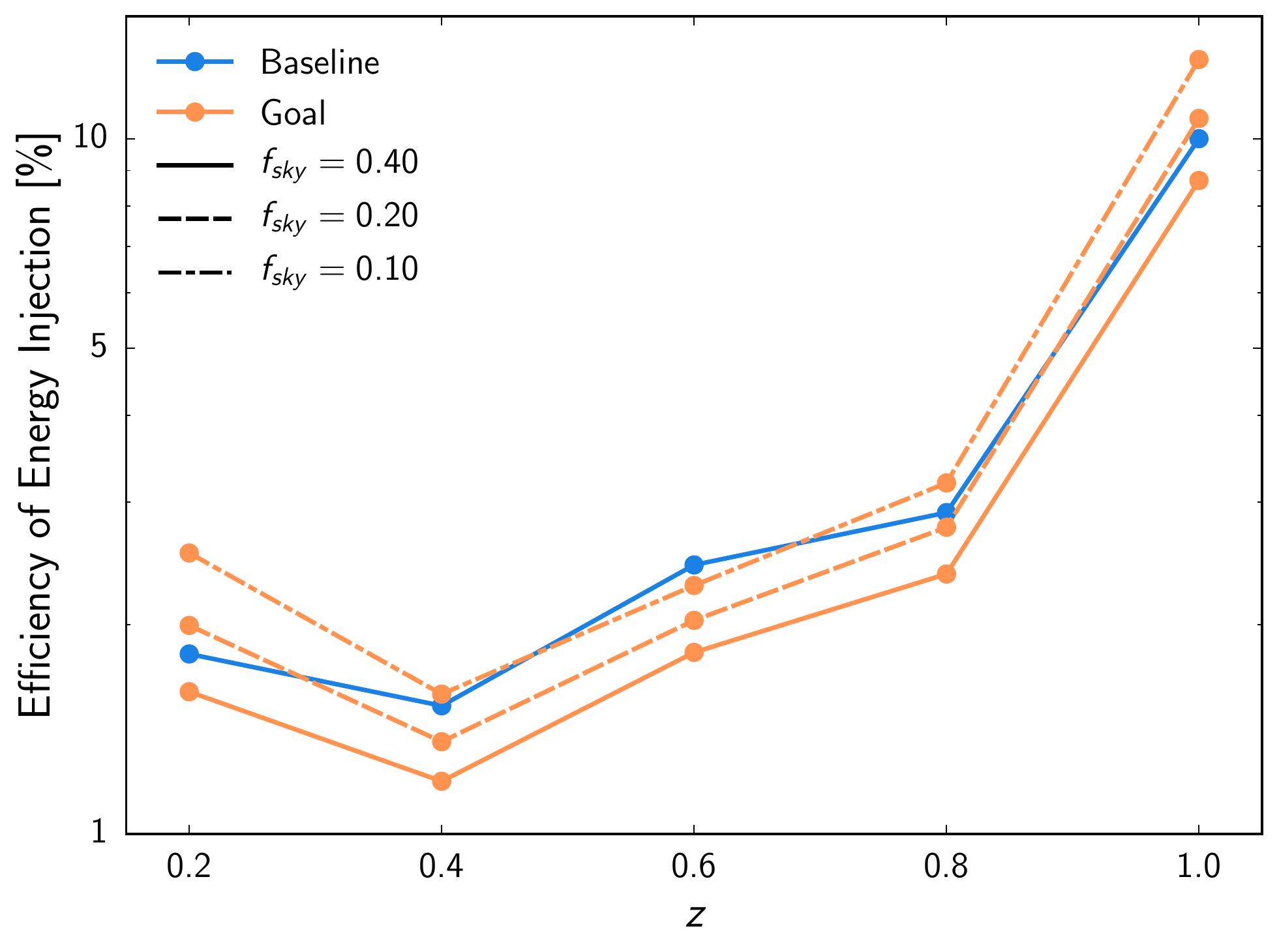}
\includegraphics[width=\columnwidth]{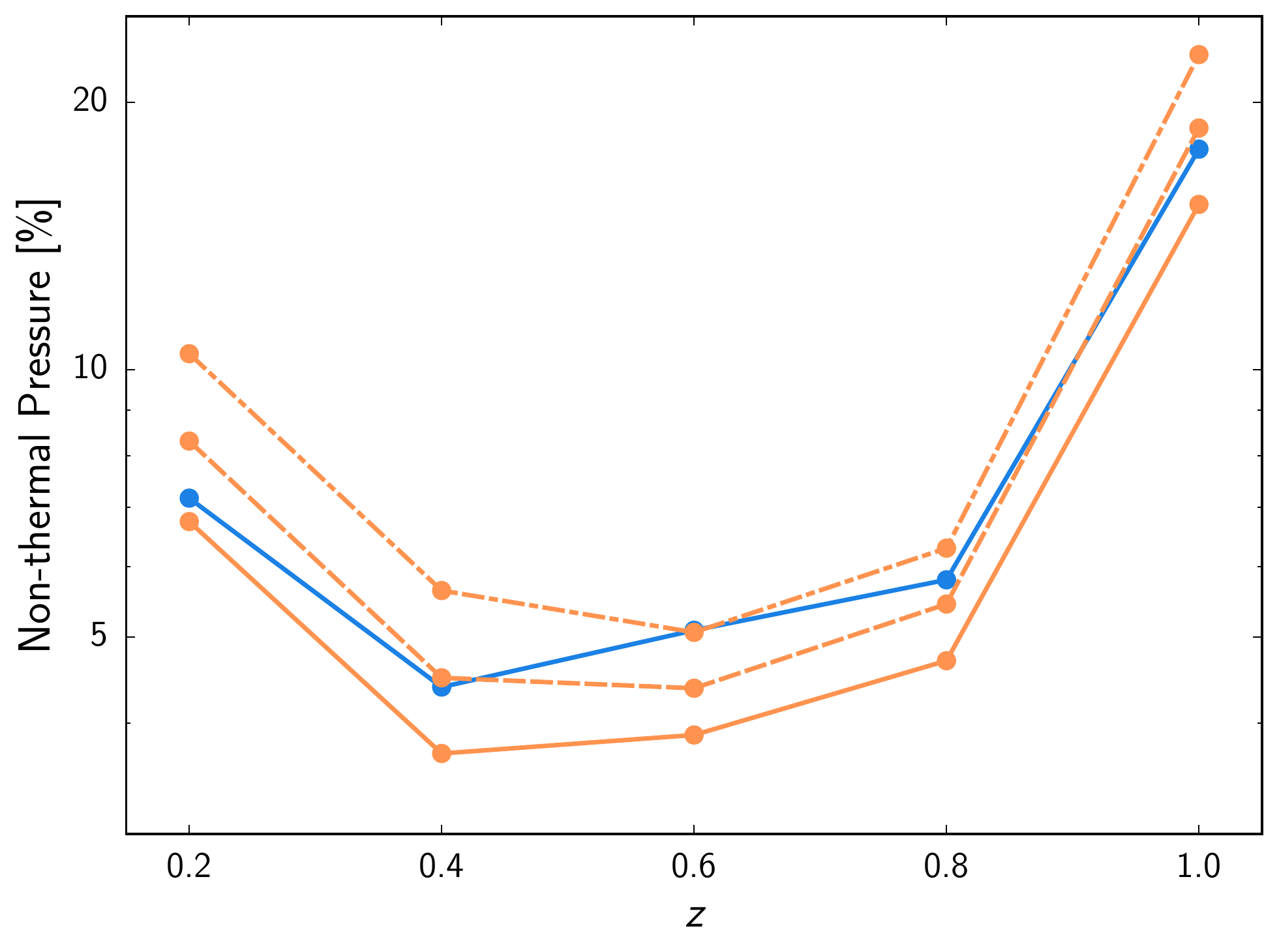}
\caption{Forecast 1$\sigma$ uncertainties on the feedback efficiency, $\epsilon$, (top) and the non-thermal pressure support, $\alpha$, (bottom) using SO combined with DESI Luminous Red Galaxies, through cross correlating the thermal and kinematic Sunyaev--Zel'dovich effects. Different colors distinguish between baseline and goal sensitivities and different line styles show the impact of different sky coverages.}
\label{fig:sz_energetics}
\end{figure}

We consider the LRG catalog of the upcoming DESI experiment \citep{Font-Ribera:2013rwa} as providing the target galaxies together with spectroscopic redshifts. In practice this analysis is not restricted to an LRG galaxy sample and can also use other extragalactic catalogs, including quasars and emission-line galaxies. Our forecasts are shown in Fig.~\ref{fig:sz_energetics}, and we find typical constraints
\ba
\sigma(\epsilon)/\epsilon &=& 2\%  \quad {\rm SO \ Baseline},  \nonumber \\
\sigma(\alpha)/\alpha &=& 6\% \quad {\rm SO \ Baseline},
\ea
from $z = 0.2$ to 0.8. 
These properties are currently only constrained at the 50--100\% level from existing SZ information.

The large SO signal-to-noise anticipated on these properties  will allow us to study the redshift and mass dependence of the signal, yielding important information about galaxy formation and evolution through cosmic time. Moreover it will help quantify the size of baryon effects in weak lensing, one of the main systematic effects of future lensing surveys \citep{2011MNRAS.417.2020S,2015MNRAS.454.2451E}. 

While working with a spectroscopic catalog is desirable, strong constraints on the baryon abundance and distribution can also be obtained in the absence of spectroscopic redshifts. As proposed in \cite{2004ApJ...606...46D} and demonstrated in \cite{2016PhRvL.117e1301H} and~\cite{2016PhRvD..94l3526F}, cross-correlating a foreground-reduced, filtered, and squared CMB temperature map with tracer galaxies is an effective probe of the baryons through the kSZ effect. Combining SO with LSST, we expect a statistical $S/N \gtrsim 100$ on the kSZ effect through this estimator~\citep{2016PhRvD..94l3526F}.
Imperfect foreground removal may ultimately be the limiting factor of this method, which nonetheless can produce tight constraints on the baryons around the galaxies in upcoming photometric catalogs.

\subsection{Growth of structure from kSZ}
\label{subsec:growth}

The kSZ effect has been identified by cross-correlating CMB surveys with the positions and redshifts of clusters, using LRGs as tracers for clusters \citep{2012PhRvL.109d1101H,DeBernardis:2016pdv}, or by using a photometrically selected cluster catalog such as redMaPPer \citep{Rykoff:2013ovv}. In this analysis we anticipate cross-correlating with LRG samples obtained from DESI, in nine redshift bins from $0.1$ to $1.0$. 

Forecast cosmological constraints from kSZ have so far come from calculating the correlation, across a full cluster sample, between the kSZ signatures of pairs of clusters as a function of their redshift and comoving separation, known as the pairwise velocity statistic, $V$.  To extract the pairwise velocity, $V$, rather than momentum, an independent measurement or estimate of each cluster's optical depth must be established. There are a number of ways this might happen, for example through calibration with hydrodynamical simulations \citep{Battaglia:2010tm} or estimation with complementary datasets, such as CMB polarization measurements \citep{Sazonov:1999zp, 2017PhRvD..96l3509L}. Uncertainties in the determination of the cluster optical depth are included in the forecasting by marginalizing over a nuisance parameter, independently in each redshift bin, as in \citet{Mueller:2014nsa}, that scales the amplitude of the pairwise velocity in each redshift bin, $\hat{V}(z)=b_{\tau_c}(z)V(z)$. In this forecast we consider two cases: a conservative one in which no knowledge of the optical depth is assumed, in which we fully marginalize over each $b_{\tau_c}(z)$, and a case in which we assume cluster optical depths can be measured to 10\% accuracy in each redshift bin, by imposing a prior on the nuisance parameters.

In the kSZ Fisher matrices, we marginalize over $\Lambda$CDM cosmological parameters, and include nuisance parameters, $b_{\tau_c}(z)$, and the logarithmic growth rate, $f_g(z)$, as independent parameters in each of the nine redshift bins. We use the same foreground cleaning as in Sec.~\ref{sec:sz_astro} for the baseline and goal SO configurations. 
This foreground cleaning yields a resultant noise level of 9.89~$\mu$K-arcmin, with only a small difference for the baseline and goal SO noise levels. We consider two survey areas of 4,000 and 9,000 square degrees and the achieved area will depend on the overlap between the SO and DESI surveys.

\begin{figure}[!t]
\begin{center}
\includegraphics[width=\columnwidth]{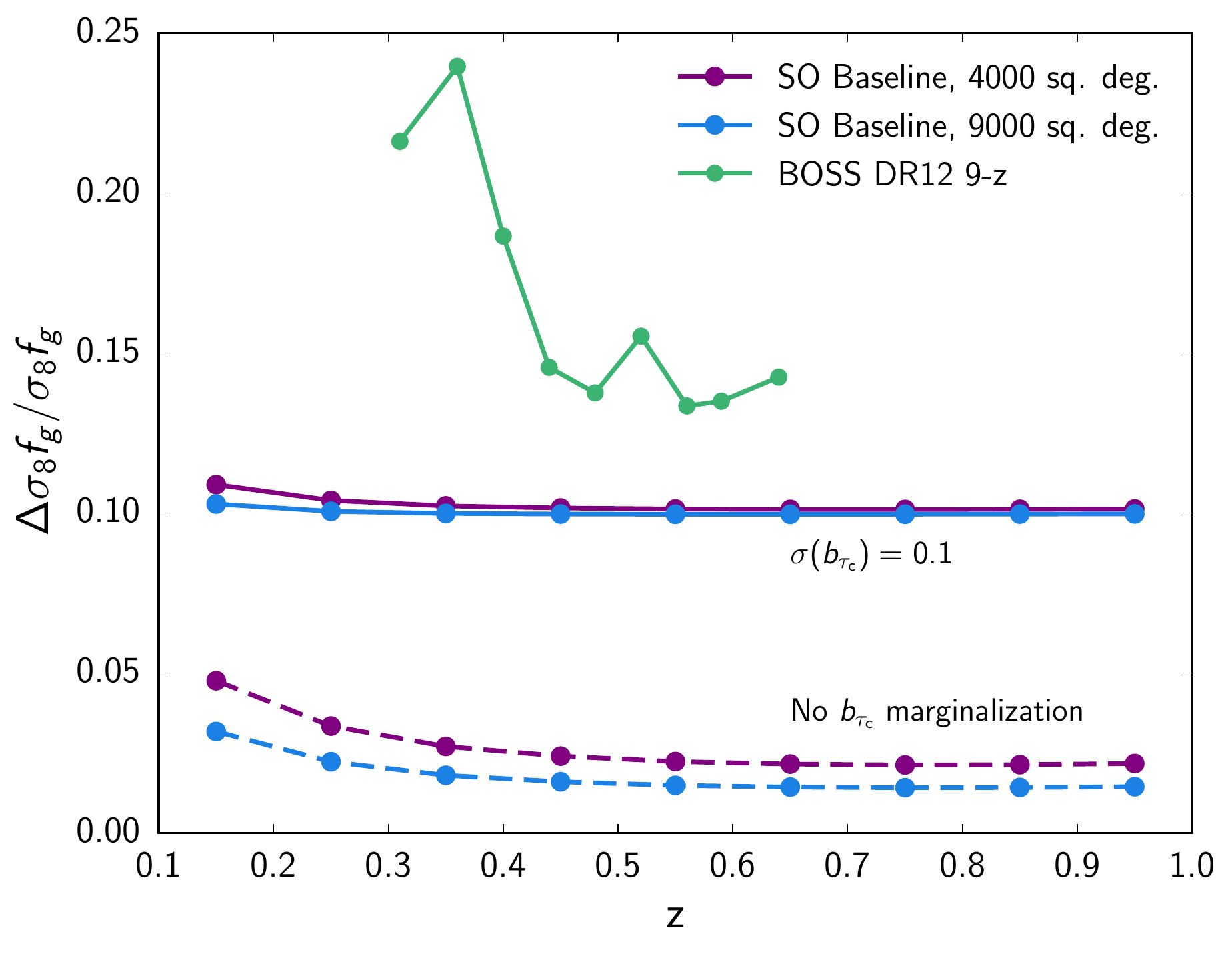}
\caption{Predictions for \so{} cluster pairwise kSZ measurements in combination with lensing: fractional 1$\sigma$ errors on the logarithmic growth rate, $f_g=d\ln\delta/d\ln a$, for large scale structure. We assume two different survey areas, 4,000 square degrees (purple) and 9,000 square degrees (blue, nominal), 
for SO baseline noise levels and foregrounds, and compared it with current limits from BOSS (green; \citealp{2017MNRAS.470.2617A}). Constraints are shown with two different assumptions about our knowledge of the cluster optical depth, either assuming the best-case scenario where $\tau_c$ is known perfectly (no $\tau_c$ marginalization, dashed lines) or assuming a 10\% prior on $\tau_c$ in each redshift bin (solid lines), that might come, for example, from combining with additional information such as CMB polarization data.}
\label{ksz_lens_fig} 
\end{center}
\end{figure}
We estimate the Fisher matrix for the kSZ signal, and combine it with the Fisher matrix that includes primary CMB constraints and the forecast SO lensing constraints, which provides complementary constraints on the $\Lambda$CDM parameters. Figure~\ref{ksz_lens_fig} shows the potential cosmological constraining power from combining pairwise velocity measurements of clusters identified in the SO survey with CMB lensing information. We project
\be
\frac{\Delta(\sigma_8 f_g)}{\sigma_8f_g} = 0.1  \quad {\rm SO \ Baseline}
\ee
over a broad redshift range between $z=0.1$ and $z=1$. If we can avoid marginalizing over the optical depth, using for example an external measurement on the optical depth, we would improve this measurement by a factor of about three. Increasing the overlap area with DESI would also be beneficial in this case. Even with marginalization over the optical depth, we forecast improvement over  current constraints for $\Delta(\sigma_8 f_g)/\sigma_8f_g$ (see Fig.~\ref{ksz_lens_fig}, green line). With a much better external constraint on the optical depth the $\Delta(\sigma_8 f_g)/\sigma_8f_g$ constraints becomes complementary to the forecast constraints from DESI.
 
\subsection{Primordial non-Gaussianity from kSZ}
\label{ssec:nong_ksz}
As described in Sec.~\ref{sec:non-gaussianity}, primordial non-Gaussianity can be constrained through its scale-dependent bias. Using a similar method to the approach described using lensing, we find that kSZ velocity reconstruction can be used as an alternative measurement of the unbiased large-scale density fluctuations. Small-scale kSZ fluctuations induced in the CMB temperature depend both on peculiar velocities of galaxies hosting free electrons as well as on astrophysically uncertain quantities such as the optical depth and electron pressure profile. However, the coherence of these fluctuations traces the large scale cosmic velocity field up to an overall scale-independent amplitude that encapsulates the small-scale astrophysics \citep{SmithEtAlInPrep}. In linear theory, valid on large scales, velocities directly trace the underlying matter density field and the growth rate. While the uncertain normalization of the velocity field makes direct inference of the growth rate or the amplitude of structure formation difficult, scale-dependent effects such as the effect of primordial non-Gaussianity on galaxy bias can be measured.

As was the case with lensing, the addition of kSZ to galaxy clustering improves the $f_{\mathrm{NL}}$ constraint when the galaxy clustering measurement is sample variance limited, since the scale-dependent galaxy bias can be directly measured (up to a constant) through cancellation of the sample variance by using a ratio of the measured galaxy and velocity fields \citep{MunchmeyerEtAl}.

We forecast how well SO temperature maps in combination with the LSST 3-year gold sample can constrain $f_{\mathrm{NL}}$ using galaxy clustering in combination with kSZ velocity reconstruction. We estimate the noise in the kSZ velocity reconstruction when either the SO baseline or goal standard ILC foreground cleaned temperature maps are combined in a quadratic estimator with the LSST galaxy positions \citep{Deutsch2017,SmithEtAlInPrep}. This allows us to construct a Fisher matrix for the auto-spectra and cross-spectrum of the velocity reconstruction and galaxy density fields in five redshift bins \citep{MunchmeyerEtAl}. We find that the combination of SO kSZ and LSST clustering allow us to constrain $f_{\mathrm{NL}}$ to an uncertainty of $\sigma(f_{\rm NL}) = 1$, reaching an interesting theoretical threshold for primordial non-Gaussianity.

\subsection{Epoch of Reionization from kSZ}
\label{sec:kszreionization}

The epoch of reionization leaves several imprints on the CMB. The most prominent on small scales is due to the `patchy' kSZ effect from the peculiar motion of ionized electron bubbles around the first stars and galaxies. The kSZ power spectrum has contributions from the epoch of reionization and from the `late-time' kSZ (i.e., from IGM, galaxies, and clusters considered in Sec.~\ref{subsec:growth}). The amplitude of these two contributions are predicted to be comparable \citep[e.g.,][]{Shaw2012,MMS2012,Battaglia2013}.

If the amplitude of the late-time contribution is known, and foregrounds are effectively removed at small scales using multi-frequency analysis, it is possible to use the power spectrum to extract information about reionization. If $C_\ell^{EE}$ is used to estimate the CMB power in $C_\ell^{TT}$, the remaining emission in the high-$\ell$ temperature spectrum is dominated by kSZ and can provide strong constraints on reionization~\citep{calabrese/etal/2014}.\\

{\it kSZ amplitude parameter:} We run a Fisher forecast for \so{} using baseline and goal noise levels, assuming that a patchy kSZ component from \cite{Battaglia2013} with $\mathcal{D}_{\ell=3000}=1.4~\mu$K$^2$ (corresponding to a reionization scenario with fiducial duration $\Delta z_{\rm re}=1$ and time $z_{\rm re}=8$) is added to the lensed primary CMB signal, and perfect knowledge of the power spectrum of noise and residual foregrounds in the ILC-cleaned map. We then forecast the \LCDM\ parameters plus a single additional patchy kSZ amplitude parameter. We impose a Gaussian prior on the late-time homogeneous contribution, bound to have $\mathcal{D}_{\ell=3000}=1.5 \pm 0.2~\mu$K$^2$. This term is known at the 10\% level, accounting for astrophysical and cosmological uncertainties \citep{Shaw2012,Park2018}. Additional uncertainty due to marginalization over non-kSZ foregrounds is incorporated with the Deproj-0 noise curves from Sec.~\ref{sec:method}.

\begin{figure}[t!]
\centering
\includegraphics[width=\columnwidth]{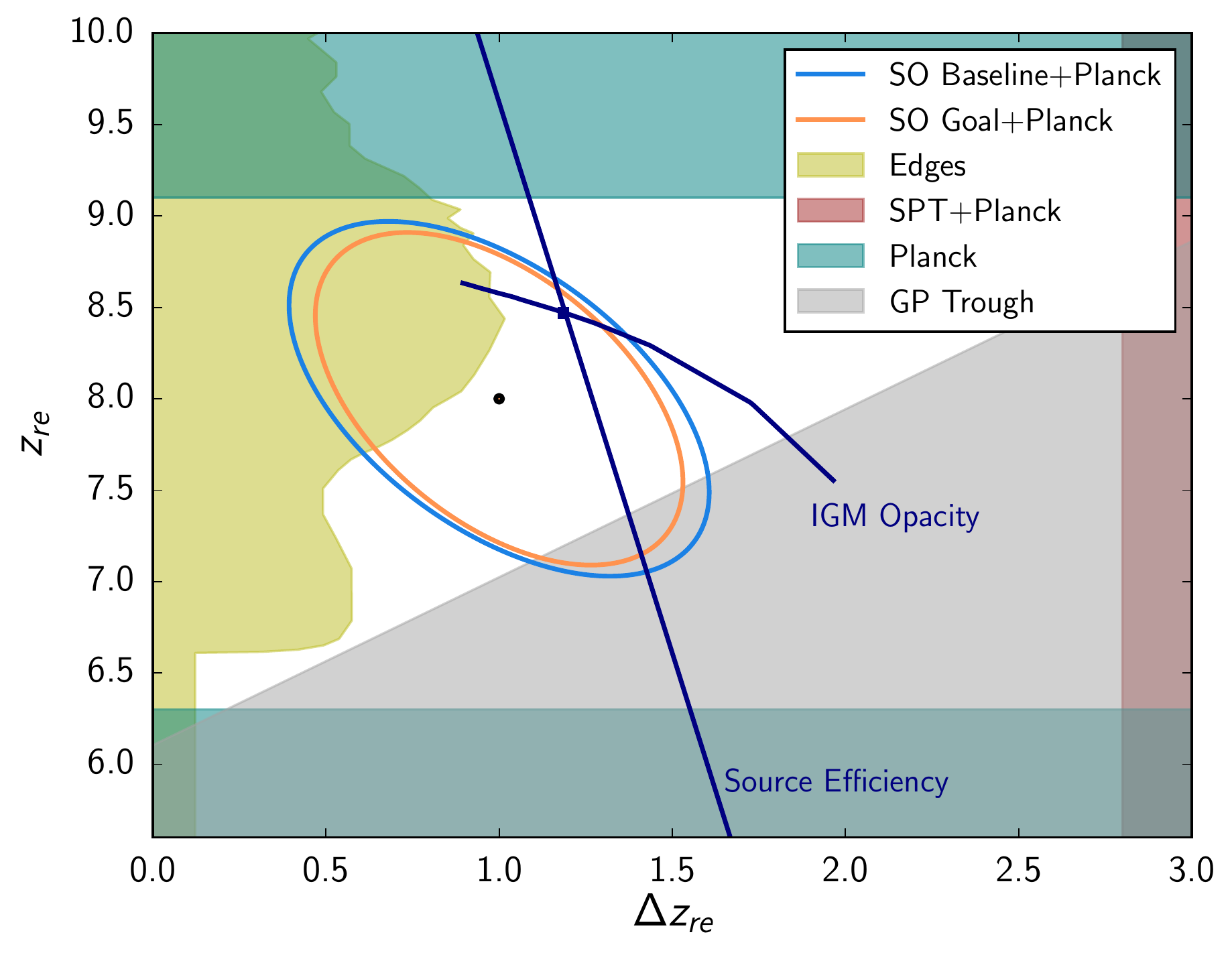}
\caption{Summary of constraints on the redshift and duration/width of reionization. The \so{} forecasts are reported with 68\% confidence-level contours from baseline/goal configurations in combination with \planck{} large-scale data (blue/orange). The solid navy lines show the redshift and width of reionization at constant values of the IGM opacity and source efficiency. The \so{} constraints on these parameters are shown in Fig.~\ref{fig:patchy_sz_ma}. The \so{} predictions are compared to current exclusion limits for the time of reionization from \planck{} (green band;~\citealp{planck2018:parameters}), recent measurements of the global 21 cm signal assuming standard thermal properties (i.e., spin temperature much larger than the CMB temperature) of the IGM (yellow band;~\citealp{edges2017}), and Gunn Peterson trough from fully absorbed Lyman alpha in quasar spectra (gray band;~\citealp{Fan2006}), and to current upper limits on the duration of reionization from \planck{} and SPT data (brown band;~\citealp{planck_reio:2016}).}
\label{fig:sz_reion}
\end{figure}

The measurement of the kSZ power and the constraint on the optical depth to reionization from the \planck{} large-scale measurements (\citealp{planck_tau:2016}; incorporated here as a $\sigma(\tau)=0.01$ measurement) are then converted into a bound on the time and the duration of reionization (following, e.g.,~\citealp{calabrese/etal/2014,calabrese/etal/2016}). We forecast
\begin{eqnarray}
\sigma(\Delta z_{\rm re}) &=& 0.40 \quad \text{\so{} Baseline}\,,\nonumber\\
\sigma(\Delta z_{\rm re}) &=& 0.35 \quad \text{\so{} Goal}\,, 
\end{eqnarray}
predicting a significant improvement over current upper limits~($\Delta z_{\rm re}<2.8$ at $95\%$ confidence when combining \planck{} and SPT data,~\citealp{planck_reio:2016}). We note that these constraints are somewhat conservative, due to the fact that the forecasts presented throughout this paper assume that the maximum multipole, $\ell$, in \so{} temperature is $\ell_{\rm max}=3000$. We anticipate that information from smaller scales, after foreground removal, could further increase the \so{} sensitivity to reionization parameters. These forecasts, alongside other current constraints on the reionization time and duration, are summarized in Fig.~\ref{fig:sz_reion}.\\

\begin{figure}[t!]
\includegraphics[width=\columnwidth]{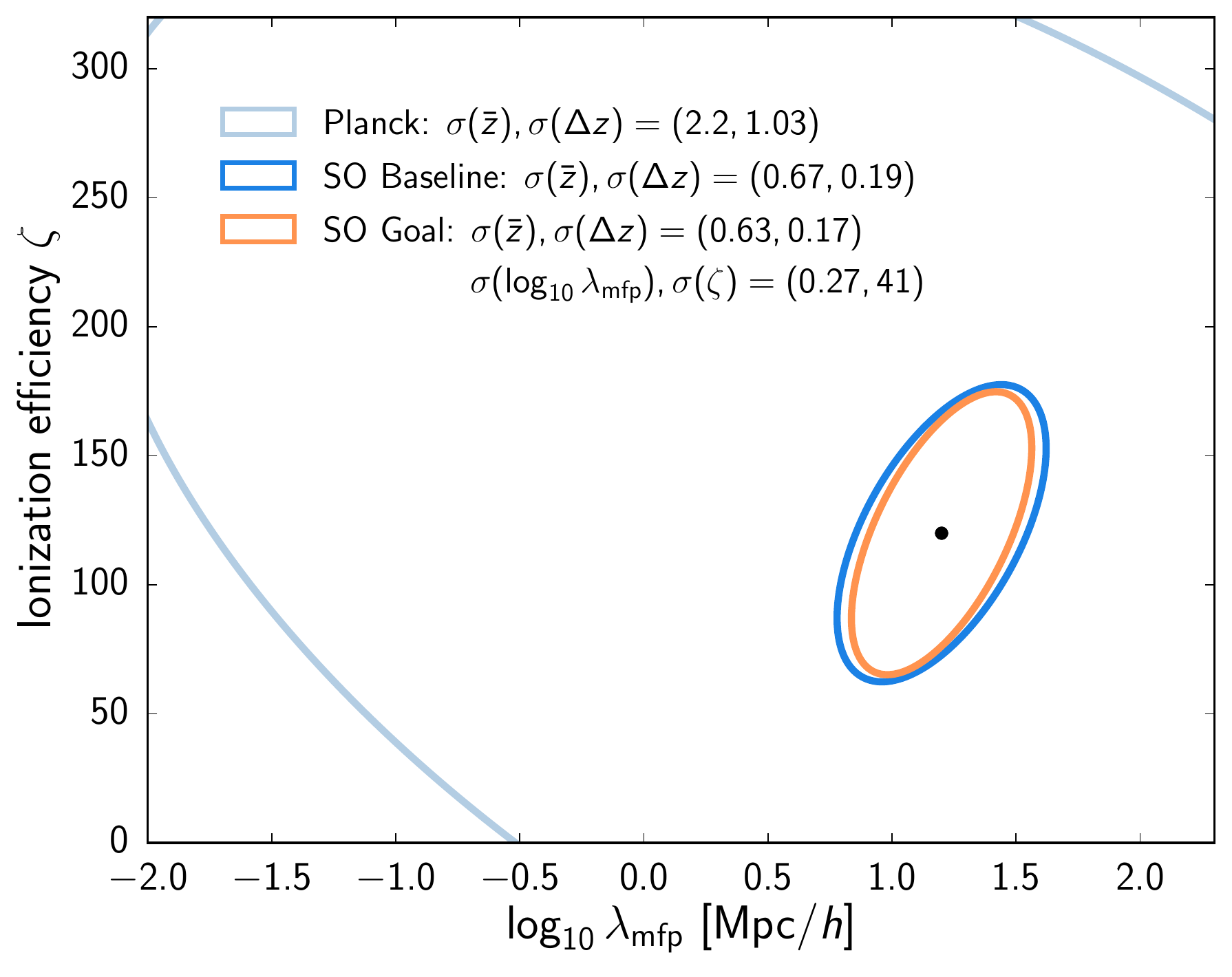}
\caption{Forecast $1\sigma$ constraints on the ionization efficiency and mean free path of reionization for a minimum galaxy halo mass of $M_{\rm min} = 10^9 M_{\odot}$, and around the fiducial model $(\lambda_{\rm mfp}, \zeta) =  (16~\mathrm{Mpc}/h, 100)$. We assume both baseline and goal sensitivities and foreground cleaning with an $f_{\rm sky}=0.4.$ The connection between these parameters and the redshift and duration of reionization is shown in Fig.~\ref{fig:sz_reion}.\\
\label{fig:patchy_sz_ma}}
\end{figure}

{\it Ionization efficiency and mean free path:} In addition to fitting for the kSZ amplitude, we can alternatively constrain physically motivated parameters directly: the ionization efficiency (or number of atoms ionized per atom in halos above the minimum mass), $\zeta$, and the mean free path of the ionizing photons, $\lambda_{\rm mfp}$. 
Since not all atoms are in these halos, the ionization efficiency by definition must be greater than unity. The rarer the halos (or equivalently the higher the minimum halo mass), the higher the required efficiency factor.
We show the constraints on these parameters in Fig.~\ref{fig:patchy_sz_ma}. These constraints were obtained by simulating kSZ maps using the method described in \cite{2016ApJ...824..118A} around a fiducial model of $(\log{\lambda_{\rm mfp}},\zeta) = (1.2, 100).$ The spectra are simulated at discrete `step sizes' away from the fiducial point, and finite differencing is used to obtain the derivative of the resulting temperature power spectrum with respect to these parameters. Figure~\ref{fig:patchy_sz_ma} also shows the corresponding errors on the width and duration of reionization, taken as the redshift at which the universe is $50\%$ ionized, and the time between ionization fractions of 25 and 75\%, respectively.  \\

{\it Separating kSZ components:} To improve the confidence in separating the kSZ components, there are promising prospects to use higher point functions of the temperature map to internally separate the reionization and late-time components \citep{2016arXiv160701769S, 2018arXiv180307036F}. We anticipate implementing these methods on the SO data.

\section{Extragalactic sources}
\label{sec:source}
In mapping the sky at frequencies between 30 and 280~GHz, SO will observe extragalactic sources, including active galactic nuclei (AGNs, Sec. \ref{AGNs}) and dusty star-forming galaxies (DSFGs, Sec. \ref{ssec:dsfg}), and transient sources such as Gamma Ray Burst (GRB) afterglows (Sec. \ref{ssec:transient}). 

\subsection{Active Galactic Nuclei}
\label{AGNs}
SO will enable statistical studies of AGN populations with its high-sensitivity, large-sky-area maps in temperature and polarization, measured in multiple frequencies simultaneously.  Such statistical studies of AGN \citep[e.g.,][]{2010MNRAS.402.2403M,mocanu2013,marsden2014,plancksources2016,gralla2018} reveal their abundance, spectral energy distributions, and global polarization properties \citep[e.g., ][]{2015ApJ...806..112H,bonavera2017, datta2018}. AGN samples that are selected at SO's frequencies will be dominated by blazars with flat or falling spectra, with an increasing fraction of steep spectrum AGN at lower frequencies. Depending on modeling assumptions, and based on \citet{Tucci2011} source counts, we expect to detect 10,000--15,000 sources at flux-densities exceeding $7$ mJy.  This is the $5\sigma$ detection limit for the baseline SO sensitivity at $93$ GHz, which is SO's most sensitive band for sources with an AGN-like SED.  A multi-frequency matched filter will find more sources.  The radio source counts found by \citet{mocanu2013} and \citet{gralla2018} somewhat exceed the \citet{Tucci2011} C2Ex model used here, which may raise the actual counts above our predicted level.

The SO frequency coverage will complement low-frequency radio continuum surveys (e.g., VLA/VLASS, ASKAP/EMU, MeerKAT/MIGHTEE).  In addition to finding bright objects, the SO can constrain the spectral energy distributions of AGN selected at lower frequencies due to the large area of overlap with those surveys \citep[e.g., ][]{gralla2014}. The blazars can also be variable over timescales of days to years \citep[e.g., ][]{chen2013,2014MNRAS.438.3058R}. Because of its rapid cadence, SO will provide one of the best continuous monitors of AGN variability at any wavelength, following many thousands of AGN continuously in the mm-cm band.  Studies of AGN variability can potentially constrain the evolution of ejecta using the lag in the peak emission across wavelengths. 

\begin{table}[t!]
\centering
\caption{Number of polarized  sources, $N_{\rm src}$, expected above a 5$\sigma$ detection limit, $S_{5\sigma}$, for the LAT with $f_{sky}=0.4$.}
\label{tab:nsources}
\begin{tabular}{ccc} 
\hline
\hline
 $\nu$~[GHz] & $S_{5\sigma}$~[mJy] &$N_{\rm src}$ \\
\hline
{27} &20 & 80 \\
{39}& 10& 150\\
{93}&  7& 270 \\ 
{148}& 10 & 70\\  
\hline
\hline
\end{tabular}
\end{table}

In polarization we expect to detect as many as $\sim 270$ polarized radio sources at 93 GHz in a wide survey over 40\% of the sky.  Anticipated counts for other frequencies are shown in Table \ref{tab:nsources}, based on the baseline survey properties and the methodology of \citet{2017arXiv171209639P} that uses number counts and a log-normal model distribution of the polarization fraction. 

Such a large number of  blazars will enrich  polarized  source datasets  over a wide range of frequencies, where catalogs are still poor and incomplete, and will allow high-resolution follow-up with other instruments \citep{2017Galax...5...47P}. Polarization of sources detected in the mm/sub-mm wavelengths will shed light  on details of the magnetic field in unresolved   regions  closer to the active nucleus and to the jet. 

As discussed in Sec.~\ref{sec:highell}, the characterization of AGN sources' polarized spectral and spatial distribution properties is also critical for utilizing CMB polarization measurements at small angular scales, for example in the $E$-mode damping tail.

\subsection{Dusty star-forming galaxies}
\label{ssec:dsfg}
The DSFGs seen by SO will include both local galaxies ($z < 0.1$) and high redshift  galaxies (roughly $2 < z < 4$), with a population of strongly lensed sources extending to well beyond this range.
The strongly lensed galaxies are useful for studying properties of star formation near the nominal peak of the cosmic star formation history of the universe.  They even let us access the first billion years of cosmic history \citep[i.e.,~$z > 6, $][]{marrone2018}. The spatial distribution of the matter in the lenses themselves can also be studied in extraordinary detail, enabling the detection of structures of dark matter down to small ($10^9$~M$_{\odot}$) scales \citep[e.g.,][]{hezaveh2016}. These observations can provide crucial, direct information on the clumping of dark matter on scales at which cold dark matter cosmological simulations predict far more halos than are currently observed using faint galaxies as tracers. In addition to the lensed source population, some of the SO DSFG sample is expected to be groupings of multiple galaxies that could represent the precursors to present-day galaxy clusters. On the faint end of the source population, SO is likely to observe unlensed, intrinsically bright high-redshift dusty galaxies. Follow-up observations can differentiate these different sub-populations of DSFGs and will be crucial to realizing their scientific potential. 

To predict the number of DSFGs that SO will observe, we compute confusion and detection limits analytically from source count models and power spectra, considering two Poisson point source populations (AGN, DSFGs) and other extragalactic signals (CMB, tSZ, kSZ). We also include the baseline model for atmospheric noise and instrument sensitivity. The DSFG forecasts are based on models from \citet{Bethermin2012}, which include a strongly-lensed population component. Depending on modeling assumptions, in a single band SO is projected to find $\sim$8,500 DSFGs. Given the expected DSFG SED, the optimal SO band for finding DSFGs will be the 280~GHz band, at which the 5$\sigma$ sensitivity limit is expected to be $\sim$26~mJy.  Multi-frequency filtering will let us push to fainter source flux densities.

\subsection{Transient sources}
\label{ssec:transient}
SO provides a unique opportunity to continuously survey a very wide area of the sky (40\%), with a fast reobservation cadence of hours and long observing timescales of several years. The major expected source classes for SO are flares of AGN, discussed in Sec.~\ref{AGNs}, and the afterglows of gamma-ray bursts (GRBs), which are expected to peak at frequencies of several hundred GHz \citep{2002ApJ...568..820G} and to which SO will have particularly competitive sensitivity. Both of these source classes have expected fluctuation timescales of days to weeks, for which SO will survey an effective sky area of approximately 2 million square degrees.

The major science goal for transient sources with SO would be the identification of a GRB afterglow without the detection of accompanying gamma emission (a so-called orphan afterglow). These are a generic prediction of all GRB models, but none have yet been identified and their rate is so far unknown and subject to large uncertainties. An unexpectedly high, or low, rate of these type of objects would suggest that our ideas of the total energy budget of these objects is incorrect and impact both theories of gamma-ray bursts and models that rely on their aggregate contribution to the high-energy universe, in particular the origins of high-energy cosmic rays and the diffuse neutrino background. In standard models of GRB fireballs \citep{ghirlanda2014}, SO will have an expected number of detections larger than one, something not yet achieved in any band.

Even more exotic transient objects include Population III star GRBs \citep{2011ApJ...731..127T,2015MNRAS.453.2144M}, which have not yet been observed. These would be extremely high-redshift luminous objects (z $\sim$ 20) and are interesting in the contexts of structure formation, star formation, and reionization. Aside from these possibilities, the transient sky in the millimeter band is nearly unexplored, with only one survey  \citep{2016ApJ...830..143W} conducted to date covering less than 1\% of the sky for one year, and there is significant discovery space available to SO as a result.

\section{Forecast summary and conclusions}
\label{sec:summary}
\begin{table*}[ht!]
\centering
\caption[Simons Observatory Surveys]{Summary of SO key science goals\tablenotemark{a}} \small
\begin{tabular}{ l c |cc c  c |ll}
\hline
\hline
 & Parameter  & {SO-Baseline}\tablenotemark{b} & {\bf SO-Baseline}\tablenotemark{c} & SO-Goal\tablenotemark{d} & Current\tablenotemark{e} & Method & Sec.\\
&& (no syst) &&&&\\
\hline
&&&&&&\\
Primordial  &  $r$ & $0.0024$ & ${\bf 0.003}$ & $0.002$ & $0.03$ & $BB$ + ext delens & \ref{ssec:bb_results}\\
perturbations & $e^{-2\tau}\mathcal{P}(k=0.2 \rm{/Mpc})$ & $0.4$\% & ${\bf 0.5}$\% & 0.4\% & 3\% & $TT/TE/EE$ & \ref{sec:pk}\\
& $f^{\rm local}_{\rm{NL}}$& 1.8 & ${\bf 3}$ & 1 & 5 & $\kappa \kappa$ $\times$ LSST-LSS + 3-pt & \ref{sec:cross-corr}\\
&& 1 & ${\bf 2}$ & 1 &  & kSZ + LSST-LSS & \ref{ssec:nong_ksz}\\
&&&&&&\\
Relativistic species & \neff\ & 0.055 & ${\bf 0.07}$ & 0.05 & 0.2 & $TT/TE/EE$ + $\kappa \kappa$& \ref{ssec:neff}\\
&&&&&&\\
Neutrino mass & $\mnu$ & 0.033 & ${\bf 0.04}$ & 0.03 & 0.1 & $\kappa \kappa$ + DESI-BAO & \ref{sec:mnuForecast}\\
&& $0.035$ &${\bf 0.04}$ & 0.03 && tSZ-N $\times$ LSST-WL & \ref{sec:mnu-w-sz}\\
&& $0.036$ &${\bf 0.05}$ & 0.04 & & tSZ-Y + DESI-BAO & \ref{sec:tSZ_PS}\\
&&&&&&\\
Deviations from $\Lambda$ & $\sigma_8(z=1-2)$& 1.2\% & ${\bf 2}$\% & 1\% & 7\%& $\kappa \kappa$ + LSST-LSS & \ref{sec:cross-corr}\\
&& 1.2\% & ${\bf 2}$\% & 1\% & & tSZ-N $\times$ LSST-WL & \ref{sec:mnu-w-sz}\\
& $H_0$ ($\Lambda$CDM) & 0.3& ${\bf 0.4}$ & 0.3 & 0.5 & $TT/TE/EE$ + $\kappa \kappa$ & \ref{ssec:hubble}\\
&&&&&&\\
Galaxy evolution & $\eta_{\rm feedback}$ & 2\% & ${\bf 3}$\% & 2\%& 50-100\% & kSZ + tSZ + DESI & \ref{sec:sz_astro}\\
& $p_{\rm nt}$  & 6\% & ${\bf 8}$\% & 5\% & 50-100\% & kSZ + tSZ + DESI & \ref{sec:sz_astro}\\
&&&&&&\\
Reionization & $\Delta z$ & 0.4& ${\bf 0.6}$ & 0.3 & 1.4 & $TT$ (kSZ) & \ref{sec:kszreionization}\\
\hline
\hline
\end{tabular}
\begin{tablenotes}
\item \textsuperscript{a} 
All of our SO forecasts assume that SO is combined with {\it Planck} data.
\item \textsuperscript{b} 
This column reports forecasts from earlier sections (in some cases using 2 s.f.) and applies no additional systematic error.
\item \textsuperscript{c} 
This is the nominal forecast, increases the column (a) uncertainties by 25\% as a proxy for instrument systematics, and rounds up to 1 s.f. 
\item \textsuperscript{d} 
This is the goal forecast, has negligible additional systematic uncertainties, and rounds to 1 s.f. 
\item \textsuperscript{e} 
Primarily from \citealp{bkp} and \citealp{planck2018:parameters}.
\end{tablenotes}
\label{tab:goals}
\end{table*}

\begin{table*}[ht!]
\centering
\caption[]{Methods used for SO forecasts} \small
\begin{tabular}{c l c}
\hline
\hline
 Method & Description & Section  \\
\hline
& &\\
$TT$/$TE$/$EE$ & Temperature and $E$-mode polarization power spectra  & \ref{sec:highell} \\
$BB$ & $B$-mode polarization power spectrum & \ref{sec:bmodes},\ref{sec:lensing}\\
$\kappa \kappa$ & CMB lensing convergence power spectrum & \ref{sec:lensing} \\
3-pt & Bispectrum  & \ref{sec:bispec} \\
tSZ-N & Cluster catalog  & \ref{sec:sz} \\
tSZ-N$\times$$\kappa$ & Cluster masses calibrated using CMB lensing & \ref{sec:lensing},\ref{sec:sz} \\
tSZ-Y & tSZ Y-distortion map  & \ref{sec:sz} \\
$TT$ (kSZ) & kSZ effect measured via the temperature power spectrum & \ref{sec:highell} \\
&\\
\hline
&\\
ext delens & Delensing field estimated from large-scale structure surveys& \ref{sec:bmodes},\ref{sec:lensing} \\
tSZ-N$\times$LSST-WL & Cluster masses calibrated using LSST weak lensing data & \ref{sec:sz} \\
$\kappa \kappa$$\times$LSST-LSS & CMB lensing correlated with galaxy density in tomographic slices from 3 years of LSST & \ref{sec:methods},\ref{sec:lensing} \\
DESI-BAO & BAO measurements from the full DESI survey & \ref{sec:methods} \\
DESI & Galaxy positions from the full DESI survey & \ref{sec:sz} \\
&\\
\hline
\hline
\\
\end{tabular}
\label{tab:method}
\end{table*}

In this final section we draw together the forecasts from the set of SO probes, summarizing our key science goals, anticipated catalogs, and additional science. These forecasts include the impact of realistic atmospheric noise, foreground contamination, and other astrophysical uncertainties. Where more than one technique is available for measuring a parameter, we show these explicitly as independent measurements, rather than combining them. 

\subsection{Key science targets}

Our key science targets are summarized in Table~\ref{tab:goals}, where we specify the 1$\sigma$ forecast uncertainties on each parameter for the baseline and goal noise levels. For the baseline case, we show the forecast uncertainty as derived in the earlier sections, and also, for each case, we inflate the uncertainty by 25\% (rounding up to 1 significant figure) as a proxy for additional systematic errors, to be refined in future studies. This reflects the impact of effects including, for example, beam uncertainty as described in Sec. \ref{ssec:neff}, or bandpass uncertainty \citep{Ward2018}. As described in Sec. \ref{ssec:bblimits} for example, our imperfect knowledge of the instrument, and of additional external sources of noise, should be propagated through to our projected errors, and is the subject of our next study. For each science target we also give the current uncertainty, and specify which method is used. A key describing the methods is given in Table~\ref{tab:method}. In the following we describe each science target, focusing on the nominal baseline noise levels. \\

\begin{figure}[t!]
\includegraphics[width=\columnwidth]{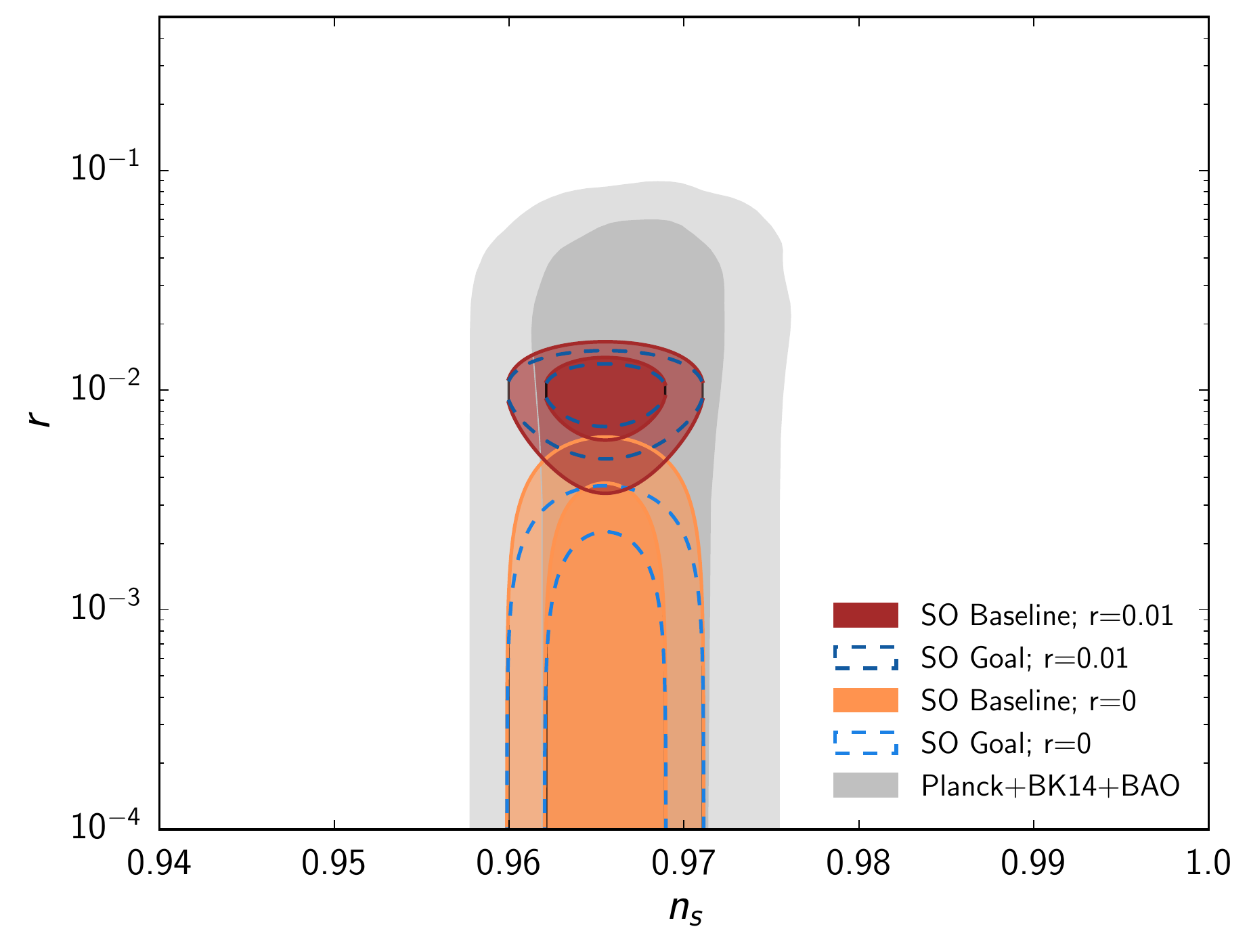}
\includegraphics[width=\columnwidth]{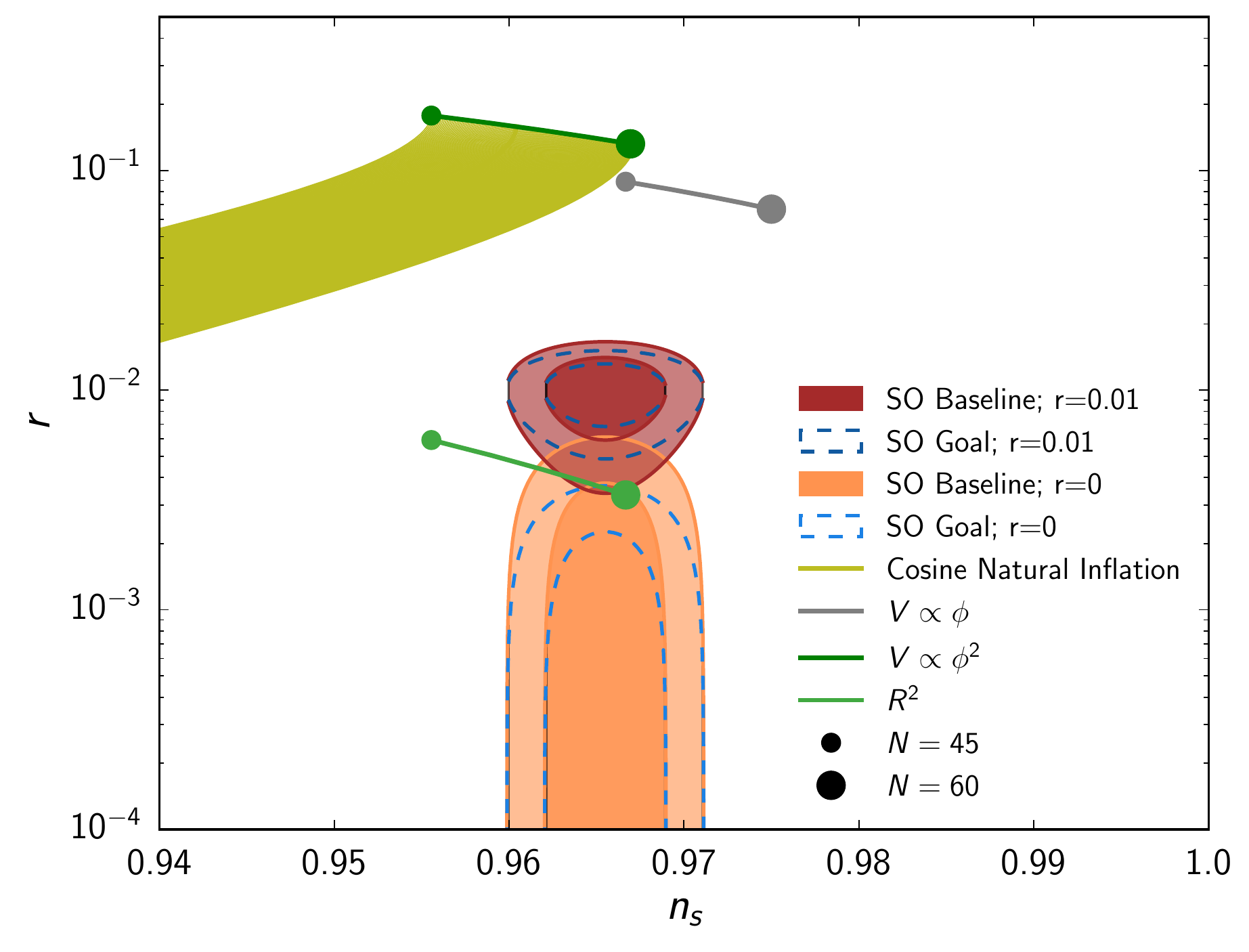}
\caption{
\emph{Top:} Summary of SO forecasts for the primordial power spectrum parameters $n_s$ and $r$ for two example cases: vanishingly small primordial tensor modes and primordial perturbations with $r=0.01$. The contours correspond to 68\% and 95\% confidence levels. For each model the SO baseline (goal) forecasts are shown as filled (dashed) contours. In gray we show the current most stringent constraint in this parameter space (\planck{} TT,EE,EE+lowE+lensing + BICEP2/Keck + BAO;~\citealp{planck2018:parameters})
\emph{Bottom:} \so{} forecasts (same as above) with predictions for $n_s$ and $r$ from some inflationary models for $N$ $e$-fold in the range [45--60] (Cosine Natural inflation~\citep{1990PhRvL..65.3233F}, Starobinsky ($R^2$) inflation~\citep{1980PhLB...91...99S,1992JETPL..55..489S}, and $\phi^n$ inflation~\citep{1983PhLB..129..177L,2010PhRvD..82d6003M}). For the two values of $r$ considered, SO could exclude, or detect, classes of models that are still in agreement with current data with $r<0.07$ \citep{bkp,planck2018:inflation}.}
\label{fig:nsrsum}
\end{figure}

\noindent
{\bf 1. Primordial perturbations}. \\
 {\bf (a) Tensor-to-scalar ratio:} SO aims to measure the tensor-to-scalar ratio, $r$, with $\sigma(r)=0.003$ for an $r=0$ model. This will enable at least a $3\sigma$ measurement of primordial gravitational waves if $r \ge 0.01$. If SO sees no signal, this would exclude models with $r \ge 0.01$ at more than 99\% confidence. Lowering the current limit by an order of magnitude will also better constrain non-inflation models. This constraint on $r$ will come from the large-scale $B$-modes measured by the SO SATs. The forecast constraints on $r$ and the spectral index of primordial perturbations, $n_s$, is shown in Fig.~\ref{fig:nsrsum}. 
\\
\noindent 
{\bf (b) Scalar perturbations:} Beyond improving the measurement of the spectral index of primordial perturbations, SO aims to estimate the primordial scalar amplitude at the half-percent level at scales ($k=0.2$/Mpc) smaller than those accessible to the \planck\ satellite. This will test the almost-scale-invariant prediction of inflation over a wider range of scales than yet probed to this precision, and identify possible deviations from a power law that are characteristic of some alternative models for the early universe. This constraint will come from the small-scale primary CMB temperature and $E$-mode polarization power spectra measured by the SO LAT.\\
 {\bf (c) Non-Gaussian perturbations:} SO targets a measurement of the non-Gaussianity of the primordial perturbations at the $\sigma(f^{\rm local}_{\rm NL})$=2 level, halving current constraints. 
 This constraint will be derived via two methods: (i) by correlating the CMB kinematic Sunyaev--Zel'dovich effect (derived from the temperature maps measured by the SO LAT) with the galaxy distribution in tomographic redshift bins from LSST; and (ii) by correlating the CMB lensing field (derived from the temperature and polarization maps measured by the SO LAT) with the LSST galaxy distribution. An additional cross-check will come from the bispectrum estimated from the SO LAT temperature and polarization maps.\\

\noindent
{\bf 2. Effective number of relativistic species.}\\
A universe with three neutrino species -- and no additional light species -- provides $3.046$ effective relativistic species. SO aims to measure $\sigma(N_{\rm eff})=0.07$, using the primary CMB temperature and polarization power spectrum measured from the SO LAT, more than halving the current limit from the \planck\ satellite. $\Delta N_{\rm eff} \ge 0.047$ is predicted for models containing additional light non-scalar particles that were in thermal equilibrium with the particles of the Standard Model at any point back to the time of reheating. SO's target would exclude at more than 95\% confidence any models with three or more such additional particles. The forecast limits are shown in Fig.~\ref{fig:Neff_targets}.\\

\begin{figure}[t!]
\includegraphics[width=\columnwidth]{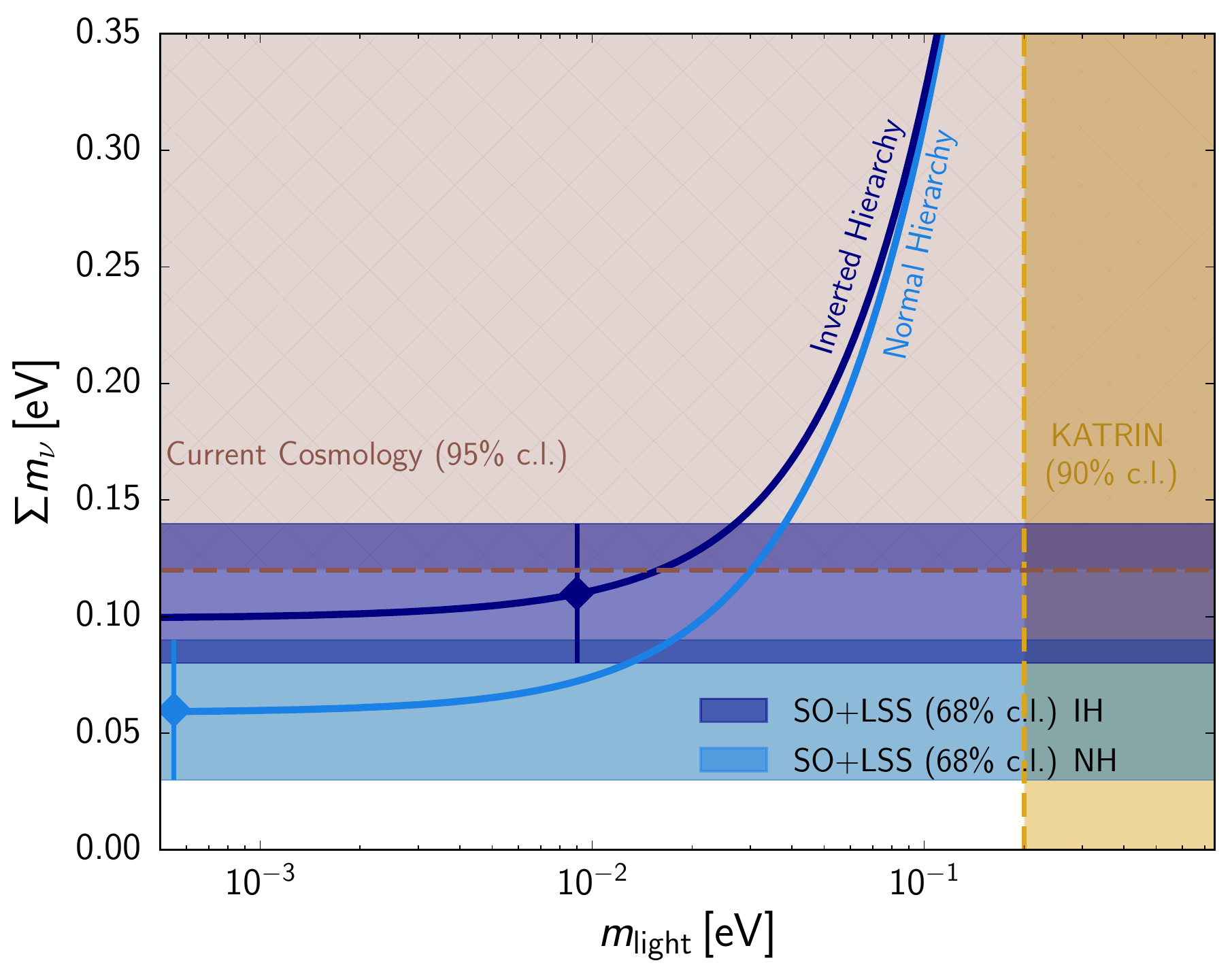}
\caption{
Summary of current limits on the neutrino mass scale, $\Sigma m_\nu$, and forecast sensitivity, from cosmological probes and laboratory searches. The mass sum is shown as a function of the mass of the lightest neutrino eigenstate, $m_\mathrm{light}$, for the normal and inverted hierarchy. Current cosmological bounds (TT,TE,EE+lowE+lensing+BAO,~\citealp{planck2018:parameters}) exclude at 95\% confidence the region above the horizontal brown dashed line. The $1\sigma$ sensitivity for \so{} (baseline with no systematic error or goal) combined with large scale structure measurements (LSS, as described in Table~\ref{tab:goals}) is shown for two example cases, enabling a ~$3\sigma$ measurement of $\Sigma m_\nu$ for the minimal mass scenario for the inverted ordering. The expected sensitivity from the $\beta$-decay experiment KATRIN~\citep{Angrik:2005ep} is indicated with a vertical yellow band on the right -- the projection here is done for NH, IH yields similar results with differences not visible on these scales.\\}
\label{fig:mnusum}
\end{figure}

\noindent
{\bf 3. Neutrino mass}. \\
The goal of SO is to measure the total mass in the three neutrino species with  $\sigma(\mnu)$=0.04~eV. If $\mnu \ge 0.1$~eV, such a measurement could give a clear indication of a non-zero mass sum. SO plans to achieve this measurement through three different methods: (i) CMB lensing from SO combined with new BAO measurements from DESI; (ii) SZ cluster counts from SO calibrated with weak lensing measurements from LSST; and (iii) thermal SZ distortion maps from SO combined with BAO measurements from DESI. The forecast constraints are shown in Fig.~\ref{fig:mnusum}, together with current limits and anticipated limits from laboratory measurements of neutrino $\beta$ decay~\citep{Angrik:2005ep}. The legacy SO dataset could be used in combination with a future cosmic variance-limited measurement of the optical depth to reionization (from $E$-mode polarization, measured for example by a CMB satellite or balloon experiment), which would enable a $6\sigma$ detection of the minimal allowed mass sum within the inverted hierarchy, and a 3$\sigma$ detection of the minimal mass sum within the normal hierarchy (which has $\mnu \ge 0.06$~eV).\\
  
 \noindent 
{\bf 4. Deviations from $\Lambda$}.\\
{\bf (a) Amplitude of matter perturbations at ${\bf z>1}$:} Upcoming optical data promise to constrain deviations from a cosmological constant at redshifts $z<1$. SO aims to provide complementary constraints by measuring the amplitude of matter perturbations, $\sigma_8$, out to $z=4$ with a 2\% constraint between $z=\textrm{1--2}$ (shown in Figs.~\ref{fig:s8} and \ref{fig:sz_s8}). The matter perturbation amplitude can be obtained in three ways: (i) SZ galaxy clusters in the LAT temperature maps, calibrated with LSST weak lensing measurements; (ii) SZ galaxy clusters calibrated with SO CMB lensing measurements estimated from the LAT maps; and (iii) CMB lensing maps from the SO LAT cross-correlated with the LSST galaxy number density in tomographic redshift bins. \\
{\bf (b) Derived Hubble constant:} SO aims to reduce the current uncertainty on the Hubble constant derived from the primary CMB within the \LCDM\ model, reaching a half-percent measurement on $H_0$ from the LAT temperature and polarization power spectra, shown in Fig.~\ref{fig:H0}. This will enhance the significance of any discrepancy in $H_0$ values inferred from the CMB  from Hubble diagram measurements at low redshift; a persistent discrepancy might indicate a departure from a cosmological constant, or other new physics.\\

\noindent
{\bf 5. Galaxy evolution: feedback efficiency and non-thermal pressure in massive halos.} \\
By measuring the thermal and kinematic Sunyaev--Zel'dovich effects in massive halos, SO aims to inform and refine models of galaxy evolution. It will do so by constraining the feedback efficiency in massive galaxies, groups, and clusters, $\eta_{\rm feedback}$ to 3\% uncertainty, and the degree of non-thermal pressure support, $p_{\rm nt}$, to 8\% uncertainty (shown in Fig.~\ref{fig:sz_energetics}). No strong limits on these important aspects of galaxy formation currently exist. These constraints will be derived from the LAT temperature maps, combined with galaxy positions measured by the DESI spectroscopic survey.\\

\noindent
{\bf 6. Reionization: measurement of duration}.\\ 
The reionization process is still poorly characterized. If the duration of reionization $\Delta z > 1$, SO aims to measure the average duration of reionization with a significance between $2\sigma$ and $3\sigma$, and thus constrain models for the ionizing process. Such a measurement would be among the first to probe the properties of the first galaxies, quasars, and the intergalactic medium in the reionization epoch. This measurement is derived from the power spectra of the temperature and polarization LAT maps, since patchy reionization adds excess variance to the temperature anisotropies through the kinematic SZ effect. The forecast constraints are shown in Fig.~\ref{fig:sz_reion}.\\

\begin{table*}[ht!]
\centering
\caption[]{Catalogs and additional science from SO} \small\begin{tabular}{ l c c ll}
\hline
\hline
 & Parameter  & {\bf SO-Baseline}  & Method & Section\\
\hline
&&&&\\
Legacy catalogs &  SZ clusters &  20,000 &tSZ & \ref{sec:sz} \\
& AGN  & 10,000 &    Sources & \ref{sec:source}\\
& Polarized AGN  & 300 &    Sources &\ref{sec:source}\\
& Dusty star-forming galaxies &  10,000 &   Sources &\ref{sec:source}\\
&&&&\\
Primordial perturbations & $f_{\rm NL}$ (equilateral) & 30 & $T$/$E$ & \ref{sec:bispec}\\
 & $f_{\rm NL}$ (orthogonal) &  10 &  &\ref{sec:bispec}\\
 & $n_s$ & 0.002 & $TT$/$TE$/$EE$ + $\kappa \kappa$ & \ref{sec:method}\\
 &&&\\
Big bang nucleosynthesis & $Y_P$ (varying \neff) & 0.007 & $TT$/$TE$/$EE$ + $\kappa \kappa$ & \ref{sec:highell}\\
& $\Omega_bh^2$ ($\Lambda$CDM) & 0.00005 & $TT$/$TE$/$EE$ + $\kappa \kappa$ &\ref{sec:method}\\
&&&&\\
Dark matter & DM--baryon interaction ($\sigma_p$, MeV) &  $5 \times 10^{-27}$ & $TT$/$TE$/$EE$+$\kappa \kappa$ & \ref{sec:highell}\\
 & UL axion fraction ($\Omega_a/\Omega_d$, $m_a = 10^{-26}~\mathrm{eV})$ &0.005 
 &$TT$/$TE$/$EE$+$\kappa \kappa$&\ref{sec:highell}\\
 &&&&\\
 Dark energy or & $w_0$ & 0.06& tSZ +LSST  &\ref{sec:sz}\\
\quad modified gravity & $w_a$ & 0.2 & tSZ + LSST & \ref{sec:sz}\\
 & Growth rate ($\Delta(\sigma_8 f_g)/\sigma_8f_g$) & 0.1 & kSZ + DESI &\ref{sec:sz}\\
 &&&&\\
Shear bias calibration & $m_{\rm z=1}$& $0.007$ &  $\kappa \kappa$+LSST&\ref{sec:lensing}\\
&&&\\
Reionization & $\log_{10}(\lambda_{\rm mfp})$ & 0.3&  $TT$/$TE$/$EE$ (kSZ)& \ref{sec:sz}\\
& Ionization efficiency ($\zeta$)  &40 & $TT$/$TE$/$EE$ (kSZ) & \ref{sec:sz}\\
\hline
\hline
\end{tabular}\label{tab:surveys}
\tablecomments{This table includes science forecast for SO thus far, described in earlier sections of this paper.\\}
\end{table*}

In addition to these key science goals, SO has a set of secondary science goals that have been summarized in the earlier sections of this paper. The forecasts are collected -- and rounded to one significant digit -- in Table \ref{tab:surveys} for reference. These include measuring additional non-Gaussian parameters describing the primordial perturbations, probing Big Bang Nucleosynthesis by measuring the primordial helium fraction, constraining interactions between dark matter particles and baryons, constraining the mass of ultra-light-axion dark matter, measuring the dark energy equation of state to cross-check constraints from optical surveys, calibrating the shear bias for LSST, and constraining the ionization efficiency in models of reionization. The SO data will also improve limits on other speculative extensions of the standard cosmology, including primordial isocurvature perturbations, modified gravity, cosmic strings, primordial magnetic fields, and cosmic birefringence.

\subsection{Legacy catalogs}

In addition to the science goals, SO's broader aim is to produce several high-level data products for use by the general astronomical community. This will include maps of the sky in temperature and polarization at 27, 39, 93, 145, 225, and 280 GHz, covering around $40\%$ of the sky at arcminute resolution, and around $10\%$ of the sky at degree-scale resolution and higher sensitivity. 
Combining these maps with data from the \planck\ satellite at large angular scales and higher frequencies, we will produce component-separated maps, including the blackbody CMB temperature and polarization, the Cosmic Infrared Background, the Compton-$y$ parameter from the SZ effect, and Galactic synchrotron and dust. We will also produce maps of the CMB lensing convergence and potential. 

In addition, we anticipate a legacy galaxy cluster catalog of 16,000 clusters detected via the SZ effect, and a point source catalog of 15,000 AGN and 8,500 dusty star-forming galaxies. These are summarized in Table \ref{tab:surveys}.

\subsection{Conclusions}

The Simons Observatory is due to start observations from the Atacama Desert in Chile in the early 2020s. Here we have summarized its broad science goals, and forecast its performance from a five-year survey. SO will improve measurements of the primary CMB polarization signal, and give unprecedented measurements of the secondary CMB signals including gravitational lensing and the thermal and kinematic Sunyaev--Zel'dovich effects. We will obtain new insight into a wide range of cosmological physics, from primordial tensor perturbations to feedback efficiency in galaxy formation. We have determined that the optimal configuration for SO to achieve its goals is the combination of a large 6-m telescope surveying $\approx 40\%$ of the sky that overlaps with optical surveys, and a set of small 0.5-m telescopes optimally designed to measure the largest angular scales attainable from Chile, surveying $\approx 10\%$ of the sky. 

SO will attain unprecedented levels of statistical uncertainty in its measurements. This statistical power can be fully exploited only if systematic errors are controlled sufficiently. Many known systematics in the current generation of experiments must be improved to attain the ambitious scientific goals presented here, and this requirement drives all aspects of the SO design. 
In the science projections presented here, systematic effects are modeled in a crude way by modestly inflating the baseline statistical error bars. The goal specifications assume that all systematics in total will be subdominant to statistical errors. The SO collaboration is undertaking an instrument modeling effort aimed at understanding the science impact of a range of possible systematic errors, and setting systematic error tolerances needed to achieve the ambitious science goals outlined here. Results, which we hope will be broadly useful for all future microwave background experiments, will be described in future papers.

\acknowledgments 
This work was supported in part by a grant from the Simons Foundation (Award \#457687, B.K.).
\ack{Alonso}{David}{DA aknowledges support from the Beecroft trust and from STFC through an Ernest Rutherford Fellowship, grant reference ST/P004474/1.}
\ack{Baccigalupi}{Carlo}{CB, NK, FP, DP  acknowledge support from the COSMOS Network (\href{cosmosnet.it}{cosmosnet.it}) from the Italian Space Agency (ASI), from the RADIOFOREGROUNDS project, funded by the European Commission’s H2020 Research Infrastructures under the Grant Agreement 687312, and from the INDARK initiative from the Italian Institute for Nuclear Physics (INFN).}
\ack{Boettger}{David}{DB thanks support from grant ALMA-CONICYT 31140004.}
\ack{Calabrese}{Erminia}{EC is supported by a STFC Ernest Rutherford Fellowship ST/M004856/2.}
\ack{Challinor}{Anthony}{AC acknowledges support from the UK Science and Technology Facilities Council (grant number ST/N000927/1).}
\ack{Chluba}{Jens}{JC is supported by the Royal Society as a Royal Society University Research Fellow at the University of Manchester, UK.}
\ack{Crichton}{Devin}{DC acknowledges the financial assistance of the South African SKA Project (SKA SA).}
\ack{Dobbs}{Matt}{Canadian co-authors acknowledge support from the Natural Sciences and Engineering Research Council of Canada.}
\ack{Errard}{Josquin}{JE and RSt acknowledge support from French National Research Agency (ANR) through project BxB no. ANR-17-CE31-0022.}
\ack{Fabbian}{Giulio}{GF acknowledges the support of the CNES postdoctoral fellowship.}
\ack{Freese}{Katherine}{AD, KF, MG, JEG, and SV acknowledge support by the Vetenskapsradet (Swedish Research Council) through contract No. 638-2013-8993 and the Oskar Klein Centre for Cosmoparticle Physics at Stockholm University. Further we acknowledge support from DoE grant DE- SC007859 and the LCTP at the University of Michigan.}
\ack{Gluscevic}{Vera}{VG gratefully
acknowledges the support of the Eric Schmidt Fellowship
at the Institute for Advanced Study.}
\ack{Hill}{Colin}{JCH is supported by the Friends of the Institute for Advanced Study.}
\ack{Hilton}{Matt}{MHi acknowledges support from the NRF and SKA-SA.}
\ack{Hlozek}{Renee}{The Dunlap Institute is funded through an endowment established by the David Dunlap family and the University of Toronto. The authors at the University of Toronto acknowledge that the land on which the University of Toronto is built is the traditional territory of the Haudenosaunee, and most recently, the territory of the Mississaugas of the New Credit First Nation. They are grateful to have the opportunity to work in the community, on this territory.} 
\ack{Huffenberger}{Kevin}{KH acknowledges support from NASA grant ATP-NNX17AF87G.}
\ack{Japan}{}{We acknowledge support from the JSPS KAKENHI Grant Number JP16K21744, JP17H06134, and JP17K14272.}
\ack{Kusaka}{Akito}{AK and KK acknowledges the support by JSPS Leading Initiative for Excellent Young Researchers (LEADER).}
\ack{L\'opez-Caraballo}{Carlos}{CL, RP and RD thanks CONICYT for grants Anillo ACT-1417 and QUIMAL-160009.}
\ack{Katayama}{Nobuhiko}{NK acknowledges that this work was supported by MEXT KAKENHI Grant Numbers JP17H01125, JSPS Core-to-Core Program and World Premier International Research Center Initiative (WPI), MEXT, Japan.}
\ack{LBNL}{Work at LBNL is supported in part by the U.S. Department of Energy, Office of Science, Office of High Energy Physics, under contract No. DE-AC02-05CH11231.}
\ack{Lewis}{Antony}{AL, JC, JP and MM acknowledge support from the European Research Council under the European Union's Seventh Framework Programme (FP/2007-2013) / ERC Grant Agreement No. [616170].}
\ack{Matsuda}{Fredrick}{FM acknowledges the support of the JSPS fellowship (Grant number JP17F17025).}
\ack{Maurin}{Loic}{LM thanks support from CONICYT, FONDECYT 3170846.}
\ack{Meerburg}{Pieter}{PDM acknowledges support from the Senior Kavli
Institute Fellowships at the University of Cambridge and from the Netherlands organization for scientific
research (NWO) VIDI grant (dossier 639.042.730).}
\ack{Niemack}{Michael}{MDN acknowledges support from NSF grants AST-1454881 and AST-1517049.}
\ack{Reichardt}{Christian}{FB and CR acknowledge support from the Australian Research Council through FT150100074 and the University of Melbourne.}
\ack{Remazeilles}{Mathieu}{MR was supported by the ERC Consolidator Grant CMBSPEC (No. 725456).}
\ack{Schmittfull}{Marcel}{MS acknowledges support from the Jeff Bezos Fellowship at the Institute for Advanced Study.}
\ack{Sehgal}{Neelima}{NS acknowledges support from NSF grant number 1513618.}
\ack{Sherwin}{Blake}{BDS acknowledges support from an Isaac Newton Trust Early Career Grant and an STFC Ernest Rutherford Fellowship.}
\ack{Takakura}{Satoru}{ST acknowledges the support of the JSPS fellowship (Grant number JP18J02133).}
\ack{Tajima}{Osamu}{ In Japan, this work was supported by JSPS KAKENHI Grant Number JP17H06134.}
\ack{Thorne}{Ben}{BT acknowledges the support of an STFC studentship.}
\ack{van Engelen}{Alexander}{AvE was supported by the Beatrice and Vincent Tremaine Fellowship at CITA.}
\ack{healpix}{}{Some of the results in this paper have been derived using the HEALPix~\citep{2005ApJ...622..759G} package.}
\ack{cmb}{s4}{We thank members of the CMB-S4 collaboration for useful discussions and interactions that have helped inform the SO design.}\\

\addcontentsline{toc}{section}{References}
\small
\setlength{\bibsep}{0.2cm}
\bibliographystyle{arxiv}
\bibliography{main_refs}  

\end{document}